\documentclass[type=doctor, zihao=-4, cjk-font=fandol, openany]{sjtuthesis}

\sjtusetup{
  info = {%
    zh / title           = {黑洞吸积的广义相对论磁流体与辐射转移数值模拟研究：耀发、进动及复杂时空背景},
    en / title           = {GRMHD and GRRT Simulations of Black Hole Accretion: Flares, Precession, and Complex Spacetimes},
    zh / keywords        = {超大质量黑洞、广义相对论磁流体模拟、吸积盘、辐射转移、双黑洞},
    en / keywords        = {Supermassive black hole, General relativistic magnetohydrodynamics, Accretion disk, Radiative transfer, Binary black hole},
    zh / author          = {蒋宏轩},
    en / author          = {Hongxuan Jiang},
    id              = {022426910020},
    zh / supervisor      = {Prof.\ Yosuke Mizuno},
    en / supervisor      = {Prof.\ Yosuke Mizuno},
    zh / department      = {李政道研究所},
    en / department      = {Tsung-Dao Lee Institute},
    zh / major           = {天文学},
    en / major           = {Astronomy},
    zh / degree          = {理学博士},
    en / degree          = {Doctor of Science},
  },
  style = {%
    page-number = {},
  },
  name = {
  },
}

\ctexset{
  contentsname   = {目\quad 录},
  listfigurename = {插\quad 图},
  listtablename  = {表\quad 格},
  figurename     = {图},
  tablename      = {表},
  bibname        = {参考文献},
  chapter = {
    name = {第,章}
  }
}

\usepackage[backend=biber,style=gb7714-2015, gbnamefmt=lowercase]{biblatex}
\DeclareSourcemap{
  \maps[datatype=bibtex]{
    \map{
      \step[fieldset=abstract, null]
      \step[fieldset=file, null]
      \step[fieldset=keywords, null]
    }
  }
}

\usepackage[perpage,bottom,hang]{footmisc}

\graphicspath{{figures/}}
\DeclareGraphicsExtensions{.pdf,.eps,.png,.jpg,.jpeg}

\usepackage{xcolor} 

\usepackage{booktabs}

\usepackage{multirow}

\usepackage{dcolumn}
\newcolumntype{d}[1]{D{.}{.}{#1}}

\usepackage{longtable}

\usepackage{threeparttable}

\usepackage{threeparttablex}

\usepackage[ruled,vlined,linesnumbered]{algorithm2e}

\usepackage{listings}
\lstdefinestyle{lstStyleCode}{%
  aboveskip         = \medskipamount,
  belowskip         = \medskipamount,
  basicstyle        = \ttfamily\zihao{-5}\setbaselineskip{12bp},
  commentstyle      = \slshape\color{black!60},
  stringstyle       = \color{green!40!black!100},
  keywordstyle      = \bfseries\color{blue!50!black},
  extendedchars     = false,
  upquote           = true,
  tabsize           = 2,
  showstringspaces  = false,
  xleftmargin       = 1em,
  xrightmargin      = 1em,
  breaklines        = false,
  framexleftmargin  = 1em,
  framexrightmargin = 1em,
  backgroundcolor   = \color{gray!10},
  columns           = flexible,
  keepspaces        = true,
}
\lstnewenvironment{codeblock}[1][]{%
  \lstset{style=lstStyleCode,#1}}{}

\usepackage{siunitx}

\usepackage{amsthm}

\usepackage{tikz}
\usetikzlibrary{arrows.meta, shapes.geometric}

\usepackage{pgfplots}
\pgfplotsset{compat=newest}

\usepackage{hologo}

\usepackage{hyperref}

\providecommand{\email}[1]{\href{mailto:#1}{\urlstyle{tt}\nolinkurl{#1}}}

\ctexset{
  contentsname   = {Contents},
  listfigurename  = {List of Figures},
  listtablename   = {List of Tables},
  bibname        = {References},
  chapter = {
    name = {Chapter,}
  }
}

\begin{document}

\frontmatter

\begingroup
\let\cleardoublepage\clearpage
\ExplSyntaxOn
\UseInstance { sjtu / page } { title / en } { en }
\ExplSyntaxOff
\endgroup

\makeatletter
\begingroup
\let\CTEX@chapter@break\clearpage
\makeatother
\begin{abstract}[en]
Horizon-scale imaging and multi-messenger astronomy now demand theoretical models that can interpret the extreme gravitational environments around supermassive black holes with substantially higher fidelity than before. In this dissertation, I combine state-of-the-art general relativistic magnetohydrodynamic simulations with covariant radiative-transfer calculations to study the rich electromagnetic phenomenology of these systems. By moving from isolated Kerr black holes to non-Kerr geometries and explicitly time-dependent binary spacetimes, this work develops a unified first-principles framework for black-hole accretion, jet production, and multi-band variability.

The first half of the thesis examines transient flares and global accretion dynamics in Kerr spacetimes. Introducing multiple magnetic loops into the initial field geometry shows that large-scale polarity inversions suppress the rapid accumulation of magnetic flux near the hole and keep the flow in a turbulent state in which violent magnetic reconnection occurs naturally. Macroscopic flux ropes form within those reconnecting regions, and I show that these structures can power the strongly variable near-infrared flares observed from Sagittarius A*. The same framework also explains the frequency-dependent delays seen in millimeter radio flares through the evolving optical depth of an expanding plasma. When the analysis is extended to spin-orbit misaligned systems, a new dynamical state appears: magnetically driven retrograde precession. In tilted magnetically arrested disks, the magnetic torque exerted by horizon-threading flux can exceed the relativistic Lense-Thirring torque, driving the disk and its jet to precess opposite to the black-hole spin.

The second half of the dissertation applies the same numerical strategy to spacetimes that are both more complex and more dynamical. To probe possible departures from general relativity, I simulate accretion in regular Loop Quantum Black Hole metrics. The loop-quantum corrections systematically enlarge the photon ring, modify the circular-polarization structure, and raise the effective horizon angular frequency, which in turn strengthens the Blandford-Znajek jet. Comparing these synthetic observables with Event Horizon Telescope measurements yields stringent new bounds on the corresponding quantum-gravity parameters. I then extend the framework to supermassive binary black holes by carrying out global three-dimensional simulations in a time-dependent superposed Kerr-Schild spacetime. These calculations show that localized shock activity can decouple from the observable light curves, while sharp multi-wavelength flares in coplanar binaries are driven primarily by gravitational self-lensing. In addition, spin-orbit coupling induces secular precession of the primary spin axis, offering a physical explanation for the twisted and wobbling jets inferred in candidate binary blazars. Overall, this dissertation translates key problems in relativistic plasma theory into observational tests that can be applied with current instruments and with future high-resolution facilities.
\end{abstract}
\endgroup

\makeatletter
\begingroup
\let\CTEX@chapter@break\clearpage
\makeatother
\tableofcontents
\endgroup
\listoffigures
\listoftables

\let\SJTUArxivOrigClearDoublePage\cleardoublepage
\let\cleardoublepage\clearpage
\mainmatter
\let\cleardoublepage\SJTUArxivOrigClearDoublePage

\chapter{Research Context and Scope}

\section{Scientific Drivers}

\subsection{Observational and Physical Background}

The study of supermassive black holes (SMBHs) has entered a new era of precision observation and multi-messenger astrophysics. In recent years, the Event Horizon Telescope Collaboration (EHTC) produced the first horizon-scale images of M87$^*$ and Sagittarius A$^*$ (Sgr~A$^*$), fundamentally changing how extreme gravitational environments are studied \cite{EventHorizonTelescope2019dse, 2022ApJ...930L..12E}. The observed asymmetric, crescent-like photon rings agree remarkably well with the predictions of general relativity (GR) for Kerr black holes, while their brightness asymmetries are naturally explained by relativistic Doppler beaming in the accreting plasma. Horizon-scale imaging has therefore shifted the field from indirect phenomenological inference toward direct tests of black-hole spacetimes and the turbulent magnetized plasma surrounding them.

At the same time, a horizon-scale interferometric image is only a time-averaged snapshot of a system that is intrinsically chaotic and highly dynamical. Together with the sub-millimeter view provided by the EHT, decades of multi-wavelength observations have shown that Sgr~A$^*$ is strongly variable \cite{2017A&A...602A..94G, Abuter2020}. In the near-infrared (NIR) and X-ray bands, rapid flares frequently raise the source flux by factors of 10 to 100 within only a few minutes \cite{2004A&A...427....1E}. During these flaring episodes, the GRAVITY instrument has detected continuous centroid motion and swings in the polarization position angle, pointing to compact emitting structures, often interpreted as magnetic flux ropes or hot spots, orbiting near the innermost stable circular orbit (ISCO) \cite{GRAVITY2018b}. These flares also show frequency-dependent time delays across the electromagnetic spectrum, with lower-frequency radio emission typically peaking after the NIR and X-ray activity \cite{Yusef-Zadeh2006, witzel_rapid_2021}.

Beyond isolated black holes, accretion flows are generically non-axisymmetric. The angular momentum carried by infalling gas is set by large-scale galactic processes, including chaotic cold accretion, galaxy mergers, and tidal disruption events (TDEs), and the orbital plane of the incoming material need not align with the black-hole spin axis \cite{2021ARA&A..59..117R, 2024Natur.630..325P}. This spin-orbit misalignment produces tilted accretion disks subject to relativistic frame dragging, which can drive Lense-Thirring precession and launch precessing or wobbling relativistic jets, as inferred in microquasars and active galactic nuclei (AGNs) \cite{Liska2018, Cui2023}.

The horizon-scale resolving power of the EHT also makes it possible to test the spacetime metric itself. Although the Kerr solution is the standard GR description of astrophysical black holes, the singularity at its center signals a breakdown of the classical theory and motivates quantum-gravity alternatives. Loop quantum gravity (LQG), for example, predicts regular black-hole solutions with modified near-horizon geometries \cite{Modesto2008im}. Since such modifications can change both the shadow size and the efficiency of the Blandford-Znajek (BZ) jet-launching mechanism, their observational signatures must be disentangled carefully from astrophysical plasma effects. Finally, recent evidence for a nanohertz stochastic gravitational-wave background from pulsar timing arrays (PTAs) \cite{2023ApJ...951L...8A} has made supermassive binary black holes (SMBBHs) prime multi-messenger targets. Identifying the electromagnetic counterparts of these binaries, such as the archetypal blazar OJ~287, requires a detailed understanding of how secondary black holes perturb the primary accretion flow and jet in the strong-field regime.

\subsection{Open Theoretical Problems}

Despite this rapid observational progress, major theoretical gaps still limit our ability to predict and interpret black-hole observables. First, existing models struggle to reconcile the relatively stable, time-averaged horizon-scale images of Sgr~A$^*$ with its violent, intermittent, multi-wavelength flaring behavior. State-of-the-art magnetically arrested disk (MAD) models initialized with a single magnetic loop often fail to reproduce both the observed variability constraints and the apparent absence of a prominent, persistent jet in Sgr~A$^*$. It remains unclear how sub-grid electron-heating mechanisms, such as turbulence and magnetic reconnection, map onto the observed light curves, spectra, and polarization signatures.

Second, the physical mechanisms that govern the precession of misaligned disks and jets remain incompletely understood. Lense-Thirring precession is usually assumed to enforce prograde rotation in tilted disks, but strong horizon-scale magnetic fields can alter or even reverse that behavior. Clarifying how magnetic torques compete with relativistic frame dragging is therefore essential for interpreting quasi-periodic jet wobbling.

Third, when EHT data are used to distinguish a Kerr black hole from a non-Kerr spacetime such as an LQG solution, the analysis is limited by severe degeneracies between spacetime geometry and astrophysical uncertainty in the accretion flow. Robust constraints require forward-modeling pipelines that evolve the plasma self-consistently within the candidate metric. Finally, predicting the electromagnetic signatures of SMBBHs requires relativistic, magnetized simulations in explicitly time-dependent spacetimes. Full numerical relativity is computationally prohibitive for the thousands of orbital periods needed to study secular disk evolution, leaving a major gap in our ability to generate realistic multi-wavelength light curves and synthetic images for binary systems.

\subsection{Dissertation Strategy}

The dissertation is organized around one central goal: connecting relativistic plasma physics near black holes to the observational constraints now available from multi-wavelength and multi-messenger campaigns. I pursue that goal with a first-principles program based on modern two- and three-dimensional GRMHD solvers, namely \texttt{BHAC}, \texttt{AthenaK}, and \texttt{KHARMA}, coupled to GRRT tools such as \texttt{BHOSS}, \texttt{ipole}, and \texttt{RAPTOR}.

Instead of treating the emission with overly simplified analytic prescriptions, I incorporate explicit electron thermodynamics into the modeling. The electron entropy is evolved together with the bulk flow so that turbulent and reconnection heating are followed self-consistently, and I also use hybrid thermal plus non-thermal electron distributions of $\kappa$ type constrained by PIC calculations. Replacing the usual single-loop magnetic initial state with multi-loop configurations allows the simulations to form plasmoids and large-scale polarity reversals naturally, which are both central to the flare problem. The same numerical framework is then generalized to more complicated backgrounds, including loop-quantum-inspired non-Kerr metrics and time-dependent superposed Kerr-Schild binary spacetimes. This end-to-end setup yields synthetic images, broadband light curves, and lag diagnostics that can be confronted directly with EHT, GRAVITY, and PTA observations.

\section{Main Research Questions}

Building on the methodological framework described above, the research presented in this thesis is organized around four main physical questions, progressing from isolated stationary black holes to misaligned, non-Kerr, and binary systems.

\paragraph{Magnetic Topology, Reconnection, and Broadband Flaring.}
Many GRMHD studies start from a single extended poloidal loop, a choice that rapidly pushes the flow into a jet-producing MAD state. Although useful, that setup does not capture the observed variability of Sgr~A$^*$ particularly well. In this thesis I instead initialize the torus with multiple radially ordered loops whose polarities alternate. That magnetic topology keeps the flow in a Standard and Normal Evolution (SANE) state for much longer and repeatedly drives strong reconnection in the current sheets separating neighboring loops. The resulting polarity inversions produce large flux ropes, the three-dimensional analogs of 2D plasmoids. With GRRT post-processing, I show that non-thermal electrons energized in those structures can account for the filamentary, rapidly varying morphology of the observed NIR flares. The same scenario also reproduces the lag between the NIR and millimeter bands, because the expanding flare plasma becomes optically thin at lower frequencies only after a delay set by self-absorption.

\paragraph{Misaligned Disks and Magnetically Induced Retrograde Precession.}
For a tilted accretion flow, the usual expectation is that Lense-Thirring torques make both the disk and the jet precess in the prograde sense. The simulations in this thesis show that this expectation breaks down once the system becomes strongly magnetized. Using high-resolution 3D GRMHD calculations, I identify a regime of retrograde precession driven by magnetic stresses. After the disk enters the MAD state, the large-scale poloidal field aligned with the black-hole spin axis applies a torque that can overtake the relativistic Lense-Thirring term and reverse the direction of precession. This mechanism offers a new way to interpret wobbling jets in sources such as M87 and Galactic microquasars.

\paragraph{Spacetime Tests from Horizon-Scale Imaging.}
To test the limits of general relativity, I extend the numerical framework to loop-quantum black holes (LQBHs). In these models, the polymeric function $P$ removes the classical singularity while modifying the near-horizon geometry. I perform both semi-analytical and full GRMHD calculations in this spacetime to extract observable signatures. The LQG corrections systematically enlarge the photon ring and alter the circular-polarization pattern, especially the $\beta_2$ mode. By comparing these synthetic observables with EHT data, I derive upper limits on the LQG parameter, namely $P \lesssim 0.2$ for Sgr~A$^*$ and $P \lesssim 0.07$ for M87$^*$. I also find that the LQG corrections effectively increase the horizon angular frequency, strengthen frame dragging, amplify the toroidal magnetic field in the funnel, and enhance the Blandford-Znajek jet power relative to a Kerr black hole with the same spin.

\paragraph{Electromagnetic Diagnostics of Supermassive Binary Black Holes.}
To explore electromagnetic counterparts of PTA gravitational-wave sources, this thesis develops global 3D GRMHD models of SMBBHs evolving in a time-dependent superposed Kerr-Schild spacetime constructed from high-order post-Newtonian orbits. I focus on binaries with mass ratio $q=0.1$ and survey several orbital setups. The calculations show that coplanar systems can generate pronounced self-lensing flares: during a primary transit across the secondary, the hotter and more compact NIR emission from the secondary is strongly magnified, and the reverse lensing geometry can also brighten the primary sub-millimeter disk. For high-spin, eccentric, and misaligned binaries, spin-orbit coupling produces strong Lense-Thirring precession of the primary spin axis and leads to twisted, wobbling jets reminiscent of the VLBI morphology reported for OJ~287.

\section{Organization of the Thesis}

Motivated by the observational and theoretical questions outlined above, this thesis is organized into two main parts. Part I focuses on time variability, flares, and global dynamics in the Kerr spacetime. Part II extends the same simulation-to-observable strategy to complex spacetimes, including both non-Kerr metrics and explicitly time-dependent binary environments.

\begin{itemize}
    \item \textbf{Chapter 2} introduces the numerical framework used throughout the thesis. It summarizes the GRMHD and GRRT methods, the coordinate and metric implementations (including MKS, FMKS, and Cartesian Kerr-Schild coordinates), the electron-thermodynamics prescriptions, and the time-dependent binary-metric infrastructure developed for SMBBH applications.

    \item \textbf{Part I: Flares and Precession of Accretion Flows in Kerr Metric}
    \begin{itemize}
        \item \textbf{Chapter 3} studies the role of the initial magnetic-field topology in setting the reconnection history, electron heating, and variability. Through two-dimensional two-temperature GRMHD calculations with multiple poloidal loops, it demonstrates that alternating polarities suppress the MAD state, repeatedly ignite reconnection, and generate plasmoids relevant to Sgr~A$^*$ variability.
        \item \textbf{Chapter 4} carries the same problem into three dimensions. It follows the development of large flux ropes inside current sheets and links their evolution, through GRRT post-processing, to intermittent flaring behavior, including NIR-dominant emission and filamentary 230 GHz structure.
        \item \textbf{Chapter 5} focuses on the time-lag behavior of Sgr~A$^*$ flares. Combining 3D two-temperature GRMHD with hybrid thermal and non-thermal electron populations, it shows that polarity reversals can ignite NIR flares, while changing optical depth in the expanding plasma delays the associated millimeter flare by as much as 50 minutes.
        \item \textbf{Chapter 6} turns to the global evolution of tilted accretion flows. It reports magnetically driven retrograde precession and explains how magnetic torques in the MAD regime can overwhelm the Lense-Thirring torque, offering a revised framework for interpreting precessing relativistic jets.
    \end{itemize}

    \item \textbf{Part II: Accretion and Shadows in Complex Spacetimes}
    \begin{itemize}
        \item \textbf{Chapter 7} examines observational tests of loop-quantum black holes. Using ray tracing in LQG-inspired metrics, it constructs synthetic shadow and polarimetric observables and combines them with EHT data for Sgr~A$^*$ and M87$^*$ to constrain the polymeric parameter $P$.
        \item \textbf{Chapter 8} presents full GRMHD evolutions in the LQBH spacetime. It analyzes how the modified geometry changes the dynamics, showing that the effective horizon angular frequency is enhanced, which strengthens the Blandford-Znajek mechanism and yields more powerful, more extended jets.
        \item \textbf{Chapter 9} addresses explicitly time-dependent spacetimes for supermassive binary black holes. Within a superposed Kerr-Schild metric, it studies how vertical impacts, coplanar embedded configurations, and spin-orbit precession shape the emission, and identifies signatures such as self-lensing flares and wobbling jets that can support future PTA and LISA searches.
    \end{itemize}

    \item \textbf{Chapter 10} summarizes the main conclusions of the dissertation and outlines future directions for connecting high-performance numerical simulations with the rapidly advancing landscape of horizon-scale imaging and multi-messenger astrophysics.
\end{itemize}

\chapter{Computational Framework for GRMHD and GRRT Modeling}

\section{General-Relativistic Magnetofluid Simulations}

Magnetized black-hole accretion flows in this thesis are described with the equations of general relativistic magnetohydrodynamics (GRMHD). In this description, the plasma is treated as a relativistic conducting fluid whose dynamics are evolved together with electromagnetic fields on a curved spacetime background. Unless noted otherwise, we adopt geometric units with $G=c=M=1$.

Numerically, these equations are evolved with conservative finite-volume shock-capturing methods derived from the formulation of \cite{2003ApJ...589..444G}. That framework underlies essentially all modern GRMHD solvers and serves as the basis for every simulation used in this dissertation.

\paragraph{Ideal MHD approximation.}
All simulations in this thesis adopt the ideal-MHD limit, in which the electrical conductivity is taken to be effectively infinite and the electric field vanishes in the comoving frame:
\begin{equation}
u_\mu F^{\mu\nu} = 0 .
\end{equation}
This condition enforces flux freezing, ensuring that magnetic field lines are advected with the plasma. While this approximation neglects kinetic-scale dissipation and particle acceleration, it robustly captures the large-scale dynamics of relativistic turbulence, jet launching, and magnetically driven variability relevant to horizon-scale observations.

\paragraph{Spacetime background.}
The fluid evolution is computed on a prescribed spacetime metric $g_{\mu\nu}$ with line element
\begin{equation}
ds^2 = g_{\mu\nu}\,dx^\mu dx^\nu .
\end{equation}
We neglect the self-gravity of the accretion flow, appropriate for systems in which the disk mass is tiny compared with the central black hole. Although many calculations are carried out in a stationary Kerr spacetime, later chapters generalize the same framework to time-dependent and non-Kerr metrics motivated by binary systems and quantum-gravity-inspired models. To keep the presentation general, the equations are written first in covariant form rather than in any particular coordinate system.

\subsection{Governing GRMHD Equations}

\subsubsection{Covariant Conservation Laws}

On macroscopic scales, a relativistic magnetized fluid obeys local conservation of rest mass and total energy--momentum together with Maxwell's equations. The rest-mass conservation law is
\begin{equation}
\label{eq:mass_cons_rewrite}
\nabla_\mu(\rho u^\mu)=0,
\end{equation}
where $\rho$ denotes the rest-mass density and $u^\mu$ is the fluid four-velocity, normalized so that $u^\mu u_\mu=-1$. In a coordinate basis, this equation can be written as
\begin{equation}
\label{eq:mass_cons_coord_rewrite}
\frac{1}{\sqrt{-g}}\partial_\mu\!\left(\sqrt{-g}\,\rho u^\mu\right)=0,
\end{equation}
with $g \equiv \det(g_{\mu\nu})$.

The joint evolution of matter and electromagnetic fields is further constrained by conservation of the total stress--energy tensor,
\begin{equation}
\label{eq:stress_cons_rewrite}
\nabla_\mu T^{\mu\nu}=0.
\end{equation}
In ideal GRMHD, the stress--energy tensor can be written as \cite{2003ApJ...589..444G}
\begin{equation}
\label{eq:Tmunu_rewrite}
T^{\mu\nu}=\left(w+b^2\right)u^\mu u^\nu
+\left(p+\frac{1}{2}b^2\right)g^{\mu\nu}
-b^\mu b^\nu,
\end{equation}
where $p$ denotes the gas pressure and
\begin{equation}
w \equiv \rho + u_{\rm int} + p
\end{equation}
defines the relativistic enthalpy density.

The magnetic field in the comoving frame is encoded in the four-vector $b^\mu$, which satisfies $b^\mu u_\mu = 0$. Its relation to the dual Faraday tensor ${}^\ast F^{\mu\nu}$ is
\begin{equation}
\label{eq:dualF_rewrite}
{}^\ast F^{\mu\nu} = b^\mu u^\nu - b^\nu u^\mu,
\end{equation}
with magnetic pressure given by $p_{\rm m} = b^2/2$, where $b^2 \equiv b^\mu b_\mu$.

\subsubsection{Magnetic-Field Evolution Equation}

Magnetic-field evolution follows from the homogeneous Maxwell equation,
\begin{equation}
\label{eq:homog_maxwell_rewrite}
\nabla_\mu\,{}^\ast F^{\mu\nu} = 0.
\end{equation}
In numerical implementations, a $3+1$ decomposition is adopted, and the spatial magnetic field components $B^i \equiv {}^\ast F^{it}$ are evolved. The induction equation then takes the form
\begin{equation}
\label{eq:induction_rewrite}
\partial_t\!\left(\sqrt{-g}\,B^i\right)
+ \partial_j\!\left[\sqrt{-g}\,\left(b^j u^i - b^i u^j\right)\right] = 0.
\end{equation}
The comoving magnetic field $b^\mu$ is reconstructed algebraically from $B^i$ and $u^\mu$, with explicit expressions depending on the chosen coordinate system.

\subsubsection{Divergence Control and Constrained Transport}

A fundamental constraint of electromagnetism is the absence of magnetic monopoles,
\begin{equation}
\label{eq:divB}
\nabla \cdot \mathbf{B}
= \frac{1}{\sqrt{-g}} \partial_i (\sqrt{-g} B^i) = 0.
\end{equation}
While this condition is preserved analytically by the induction equation, numerical discretization errors can introduce spurious magnetic divergence, leading to unphysical forces and instabilities.

To control this error, every simulation in this thesis uses a Constrained Transport (CT) scheme \cite{1988ApJ...332..659E}. In CT, magnetic components are stored as face-centered fluxes and updated through edge-centered electromotive forces. This staggered representation preserves the divergence-free condition to machine precision and therefore keeps the magnetic topology physically consistent over long integrations.

\subsection{Simulation Codes Employed in This Thesis}

The GRMHD calculations in this thesis rely on three complementary state-of-the-art codes, each chosen for a different numerical and physical regime.

\paragraph{BHAC.}
BHAC (Black Hole Accretion Code) \cite{2017ComAC...4....1P} is a finite-volume GRMHD solver for arbitrary stationary spacetimes. It includes horizon-penetrating coordinates, adaptive mesh refinement, and direct coupling to GRRT post-processing. In this dissertation, BHAC is used mainly for problems centered on horizon-scale emission and synthetic imaging, where an accurate treatment of the near-hole geometry is critical.

\paragraph{AthenaK.}
AthenaK \cite{2024arXiv240916053S} is the performance-portable successor to Athena++, built to use modern heterogeneous hardware efficiently. It combines high-order Godunov methods with constrained transport and flexible mesh refinement. Those features make it especially effective for large, time-dependent GRMHD calculations, including runs in evolving or externally prescribed spacetimes. Here it is employed primarily for binary-black-hole accretion and other dynamically varying gravitational backgrounds.

\paragraph{KHARMA.}
KHARMA \cite{2024arXiv240801361P} is a modular, portable GRMHD framework designed for efficient execution on both CPUs and GPUs. Based on the conservative HARM formulation, it handles complicated geometries while remaining suited to modern accelerator architectures. In this thesis, KHARMA is used for high-resolution studies of accretion-flow variability and magnetically driven flaring.

Despite differences in implementation and optimization, all three codes solve the same underlying GRMHD equations using conservative finite-volume methods and constrained transport. This consistency enables robust cross-validation of physical results across independent numerical frameworks.

\section{General-Relativistic Radiative Transfer}

To compare the GRMHD models with data, we generate synthetic images and spectra by solving the radiative-transfer problem in curved spacetime. In this dissertation, GRRT is applied as a post-processing step: snapshots from the GRMHD evolution supply the plasma state, including density, internal energy, velocity, and magnetic-field structure. The observable radiation is then obtained by integrating photon geodesics together with the covariant transfer equation. This procedure enables direct quantitative comparisons between theoretical accretion models and multi-wavelength interferometric observables, including both total intensity and polarization.

Established GRRT codes are used in complementary roles: \texttt{RAPTOR} \cite{Bronzwaer2018} is employed for selected cross-checks, \texttt{ipole} \cite{2018MNRAS.475...43M} is used for semi-analytic modeling, and the main polarized-transfer calculations for binary systems are carried out with \texttt{BHOSS} \cite{2020IAUS..342....9Y}. One technical contribution of this thesis is the construction of a dedicated interface between GPU-accelerated \texttt{AthenaK} outputs and \texttt{BHOSS}, which makes accurate radiative-transfer calculations possible in adaptive-mesh binary black-hole spacetimes.

\subsection{Ray Tracing and Geodesic Integration}

Synthetic observables are generated by tracing rays backward from a distant camera. Each detector pixel is associated with a photon geodesic specified by a position $x^\mu$ and a wavevector $k^\mu$. Their evolution is governed by the geodesic equation:
\begin{align}
\frac{d x^{\alpha}}{d \lambda} &= k^{\alpha}, \\
\frac{d k^{\alpha}}{d \lambda} &= -\Gamma^{\alpha}_{\ \mu\nu}\,k^{\mu}k^{\nu},
\end{align}
Here $\lambda$ is an affine parameter and $\Gamma^{\alpha}_{\ \mu\nu}$ are the Christoffel symbols of the spacetime metric $g_{\mu\nu}$. The integration runs from the observer screen inward until the ray reaches either the event horizon or the edge of the simulation domain.

In practice, the camera is defined by a viewing geometry, namely inclination $\theta_{\rm obs}$, azimuth $\phi_{\rm obs}$, and distance $r_{\rm obs}$, together with an orthonormal tetrad that maps pixel coordinates to initial photon directions. High-order integrators are used to preserve numerical accuracy, especially for rays that pass close to the photon shell where lensing is strongest.

\subsection{Covariant Polarized Radiative Transport}

The radiation field evolves along geodesics according to the covariant radiative transfer equation. For polarized transport, the state of the radiation is described by the Lorentz-invariant Stokes vector $\boldsymbol{\mathcal{S}} = (I, Q, U, V)^{\rm T}/\nu^3$. The evolution of $\boldsymbol{\mathcal{S}}$ includes emission, absorption, and Faraday rotation/conversion:
\begin{equation}
\frac{d\boldsymbol{\mathcal{S}}}{d\lambda}
= -k_\mu u^\mu\left[
\boldsymbol{\overline{\varepsilon}}
- \mathbf{K}\,\boldsymbol{\mathcal{S}}
\right],
\label{eq:pgrrt}
\end{equation}
where $u^\mu$ is the four-velocity of the fluid, $\boldsymbol{\overline{\varepsilon}}$ denotes the invariant emissivity vector, and $\mathbf{K}$ is the transfer matrix that encodes the absorption and Faraday coefficients.

To ensure numerical robustness, the emission coefficients are evaluated in the local fluid frame, where the magnetic field is aligned with a principal axis. The result is then transformed to the observer's frame. This transformation requires parallel transport of a polarization basis along the geodesic so that the computed electric vector position angle (EVPA) consistently reflects the relativistic rotation of the polarization plane induced by the spacetime geometry.

\subsection{The \texttt{AthenaK–BHOSS} Coupling}

An important technical development of this thesis is a custom interface connecting \texttt{AthenaK} to the \texttt{BHOSS} radiative-transfer engine. Its design addresses two main requirements: efficient handling of block-structured adaptive-mesh-refinement (AMR) data without intermediate-format conversion, and support for the time-dependent metrics needed in binary-black-hole simulations.

The interface reads simulation data directly from the standard \texttt{AthenaK} HDF5 output, which provides a portable and widely supported I/O route. From those files it extracts eight primitive MHD variables: the rest-mass density ($\rho$), the normal-frame three-velocity components ($\tilde{v}^i$), the gas pressure ($P$), and the cell-centered magnetic-field components ($B^i$).

Efficient ray tracing through the \texttt{AthenaK} AMR hierarchy requires rapid localization of photon positions. To achieve this, we implemented an FNV-1a hash-table search that gives $O(1)$ identification of the memory block associated with a given spatial location. Performance under OpenMP is improved further through thread-local caching, which exploits the spatial coherence of neighboring rays.

To support high-order interpolation for geodesics that cross meshblock boundaries, the interface automatically fills ghost zones for all 26 topological neighbors, including faces, edges, and corners. The reconstruction procedure depends on the refinement level of the neighbor:
\begin{itemize}
    \item \textbf{Same-level neighbors:} Direct data copy.
    \item \textbf{Coarser neighbors:} Linear prolongation with gradient reconstruction.
    \item \textbf{Finer neighbors:} Volume-weighted restriction (8-cell averaging).
\end{itemize}

\section{Coordinate Choices and Metric Implementation}

Accurate simulations of accretion flows that pass through the event horizon require a suitable coordinate system. Boyer-Lindquist coordinates are not convenient for this purpose because they contain a coordinate singularity at the horizon radius $r_+$, where $g_{rr} \to \infty$. Modern GRMHD calculations therefore adopt horizon-penetrating coordinates built from the Kerr-Schild (KS) form. In this thesis, I use members of the modified Kerr-Schild (MKS) family, including the Funky MKS (FMKS) and wide-pole Kerr-Schild (WKS) variants, to balance spatial resolution against timestep stability.

\subsection{Horizon-Penetrating Kerr-Schild Coordinates}

The foundation for all coordinate systems used in this thesis is the Kerr-Schild form, which expresses the spacetime metric $g_{\mu\nu}$ as a perturbation of the Minkowski metric $\eta_{\mu\nu}$ via a null vector $l_\mu$:
\begin{equation}
g_{\mu\nu} = \eta_{\mu\nu} + 2H l_\mu l_\nu.
\end{equation}
The key property of this decomposition is that the metric determinant matches that of the Minkowski background, $\sqrt{-g} = \sqrt{-\eta}$. As a consequence, the metric components remain regular across the event horizon, allowing fluid fluxes to be integrated explicitly through $r_+$ without imposing artificial inner boundary conditions.

The simulations are evolved on a logically Cartesian grid $x^\mu = (t, x^1, x^2, x^3)$ that maps onto the physical Kerr-Schild coordinates $(t, r, \theta, \phi)$. The radial coordinate is usually chosen as $r = \exp(x^1)$ so that the grid concentrates resolution near the black hole, where the relevant dynamical timescales are shortest.

\subsection{Spherical-Grid Layouts}

While the radial mapping is generally consistent across codes (eKS), the poloidal mapping $\theta(x^2)$ varies significantly to address specific physical or numerical constraints.

\subsubsection{Standard Modified Kerr-Schild (MKS)}
The standard MKS transformation, widely used in \texttt{BHAC} and early \texttt{HARM} versions, is designed to resolve the thin accretion disk. The mapping concentrates grid cells toward the equatorial plane:
\begin{equation}
\theta = \pi x^2 + \frac{1-h}{2} \sin(2\pi x^2),
\end{equation}
where $x^2 \in [0, 1]$ is the uniform logical coordinate and $h \in [0, 1]$ controls the concentration of cells toward the equator (typically $h=0.3$). This choice provides high resolution near the midplane ($\theta \sim \pi/2$) but leaves the polar regions relatively under-resolved.

\subsubsection{Funky Modified Kerr-Schild (FMKS)}
A major limitation of spherical grids is the Courant-Friedrichs-Lewy (CFL) timestep constraint. In standard MKS, the physical cell width $\Delta x_{phy} \approx r \Delta \theta$ becomes extremely small near the poles ($\theta \to 0$) at small radii ($r \to r_+$), which forces the global timestep to become prohibitively small.

To alleviate this problem, \texttt{KHARMA} uses the ``Funky'' MKS (FMKS) coordinates. FMKS adds a cylindrifying term that enlarges the physical size of cells near the polar axis close to the horizon. The mapping blends the standard midplane-focused grid ($\theta_g$) with a cylindrified grid ($\theta_j$):
\begin{equation}
\theta = \theta_g + e^{-s \Delta x^1} (\theta_j - \theta_g).
\end{equation}
Here, $\theta_g$ is the standard MKS mapping defined above. The cylindrified component $\theta_j$ takes the form:
\begin{equation}
\theta_j = N (2x^2 - 1) \left[ 1 + \left( \frac{2x^2 - 1}{B(1+\alpha)^{1/\alpha}} \right)^\alpha \right] + \frac{\pi}{2},
\end{equation}
with normalization factor $N = \frac{\pi}{2} [1 + B^{-\alpha}(1+\alpha)^{-1}]^{-1}$. This transformation essentially "warps" the grid lines at small radii (controlled by the blending factor $e^{-s \Delta x^1}$) to behave more like a cylinder, ensuring that polar cells maintain a reasonable width to relax the CFL condition.

\subsubsection{Wide-pole Kerr-Schild (WKS)}

For simulations that follow the jet from the black hole out to galactic scales, high resolution is needed along the polar axis even at large radii. Standard MKS and FMKS grids often relax the pole too strongly at those distances. To address that limitation, \cite{2025ApJ...995..122C} introduced the WKS system.

The WKS mapping divides the grid into a linear region (uniform resolution) and a non-linear region (stretched resolution) using a smooth transition function:
\begin{align}
\theta = \frac{\pi}{2} \Bigg[ & 1 + f_{\rm lin}(2x^2 - 1) \nonumber \\
& + (1 - f_{\rm lin}) \left\{ \tanh\left(\frac{x^2 - 1}{\lambda}\right) + 1 \right\} \nonumber \\
& - (1 - f_{\rm lin}) \left\{ \tanh\left(-\frac{x^2}{\lambda}\right) + 1 \right\} \Bigg].
\end{align}
Here, $f_{\rm lin}$ (typically $0.6$) defines the fraction of the grid covering the central region with uniform spacing, while the remaining fraction $(1-f_{\rm lin})$ uses the $\tanh$ function to smoothly stretch cells. The parameter $\lambda$ controls the smoothness of the transition.

Unlike FMKS, which focuses on timestep optimization at the horizon, WKS ensures that the ghost cells and boundary regions retain reasonable aspect ratios and that high angular resolution is preserved along the jet axis far from the black hole. This makes WKS particularly advantageous for multizone simulations where the jet propagates into a coarse ambient medium.

\subsubsection{Internal Static Mesh Refinement}
A fundamental limitation of standard analytical mappings (such as MKS or FMKS) is the rigid coupling between the logical grid and the physical resolution. Although logarithmic radial spacing efficiently covers vast spatial scales, from the event horizon ($r_g$) to the Bondi radius ($R_B$), maintaining sufficient angular resolution in the jet region at large radii typically requires an excessive number of grid cells at the horizon and therefore leads to severe timestep penalties.

To mitigate this, \texttt{KHARMA} supports Internal Static Mesh Refinement (ISMR). Unlike dynamic AMR, which responds to evolving flow features, ISMR allows for the pre-definition of hierarchical refinement zones based on geometric criteria. As demonstrated in recent work by \cite{2025ApJ...995..122C}, this technique enables selective enhancement of the grid resolution in critical regions, such as the polar jet axis or the equatorial disk plane, while maintaining a coarser mesh in the ambient medium. This approach effectively decouples the local grid density from the global coordinate transformation, permitting rigorous convergence studies of relativistic jets extending to galactic scales without the prohibitive computational cost associated with a uniformly high-resolution spherical grid.

\subsubsection{Symmetry Assumptions and Their Limits}
A critical optimization in many spherical GRMHD codes (including standard configurations of \texttt{HARM}, \texttt{BHAC}, and \texttt{KHARMA}) is the assumption of metric axisymmetry. To accelerate calculations, these codes often neglect the computation of connection coefficients (Christoffel symbols) associated with the azimuthal direction ($\partial_\phi g_{\mu\nu} = 0$). While efficient for stationary Kerr spacetimes, this approximation strictly limits their applicability to symmetric systems. They are less suitable for binary black holes (BBH) or strongly tilted disks where the metric or the flow structure breaks axisymmetry significantly near the polar coordinate singularity.

\subsection{Cartesian Implementation in {\tt AthenaK}}

\texttt{AthenaK} \cite{2024arXiv240916053S} abandons spherical topology in favor of Cartesian Kerr-Schild (CKS) coordinates $(t, x, y, z)$. The metric is constructed without reference to a grid pole, using the standard Kerr-Schild decomposition:
\begin{equation}
g_{\mu\nu} = \eta_{\mu\nu} + 2H l_\mu l_\nu,
\end{equation}
where $\eta_{\mu\nu}$ is the Minkowski metric and $l_\mu$ denotes a null vector. The transformation from the Kerr-Schild radius $r$ and polar angle $\theta$ to Cartesian coordinates is
\begin{equation}
x = (r \cos\phi + a \sin\phi) \sin\theta, \quad y = (r \sin\phi - a \cos\phi) \sin\theta, \quad z = r \cos\theta.
\end{equation}

\subsubsection{Geometric Flexibility and Computational Cost}
The primary advantage of the CKS framework in \texttt{AthenaK} is its geometric flexibility. Since the grid is Cartesian and the code computes full 3D metric derivatives (including all $\partial_\phi$ terms), it naturally handles complex, non-axisymmetric spacetimes such as Binary Black Holes (BBH) and tilted accretion disks without polar singularities or axis-stability issues.

Nevertheless, this flexibility comes at a steep computational cost. Unlike MKS grids, which logarithmically resolve the horizon with a modest cell count, a uniform Cartesian grid is extremely inefficient for spherical geometries. \texttt{AthenaK} overcomes this through aggressive Block-Structured Adaptive Mesh Refinement (AMR), using a "box-in-box" refinement strategy to resolve the horizon. This demands significantly larger GPU memory footprints than spherical codes, because ensuring $\Delta x \ll r_g$ at the center while extending to large radii generates far more meshblocks than the efficient logarithmic stretching used in MKS.

\section{Electron Thermodynamics and Nonthermal Populations}
\label{sec:2T-c1}

In low-luminosity AGNs such as Sgr A$^*$, the Coulomb-coupling timescale between ions and electrons is much longer than the accretion timescale. The plasma therefore remains in a two-temperature state, with ions typically much hotter than electrons ($T_{\rm i} \gg T_{\rm e}$). To model this behavior self-consistently, we evolve the electron entropy equation alongside the GRMHD flow, following \cite{Ressler2015}:
\begin{equation}
    \rho T_{\rm e} u^\mu \partial_\mu s_{\rm e} = f_{\rm e} Q,
\end{equation}
where $\rho$ is the fluid density, $s_{\rm e}$ is the electron entropy per unit mass, $u^\mu$ is the four-velocity, $Q$ is the total heating rate derived from dissipation of grid-scale energy (via shocks and reconnection), and $f_{\rm e}$ is the fraction of that heat assigned to electrons.

We set $f_{\rm e}$ with sub-grid heating prescriptions motivated by turbulence and magnetic reconnection \citep[e.g.,][]{Rowan2017, 2019PNAS..116..771K}. For orientation, it is useful to compare these physically motivated kernels with the widely used parameterized R-$\beta$ prescription \citep{Moscibrodzka2016}. In the formulation adopted by \citet{Mizuno2021}, the ion-to-electron temperature ratio is written as a function of the pressure ratio $\beta \equiv p_{\rm gas}/p_{\rm mag}$:
\begin{equation}
    \frac{T_{\rm i}}{T_{\rm e}} = R_{\rm low} \frac{1}{1+\beta^2} + R_{\rm high} \frac{\beta^2}{1+\beta^2},
\end{equation}
Here $R_{\rm low}$ governs magnetically dominated regions (low $\beta$, such as the jet), whereas $R_{\rm high}$ sets the temperature ratio in gas-pressure-dominated zones (high $\beta$, such as the disk). In many applications, one fixes $R_{\rm low}=1$ and adjusts $R_{\rm high}$ to reproduce the observed spectrum \citep{Mizuno2021}. In the calculations presented here, by contrast, $T_{\rm e}$ comes directly from the simulation thermodynamics through the turbulent and reconnection heating prescriptions, so the spatial pattern of $T_{\rm i}/T_{\rm e}$ is determined self-consistently rather than imposed through a single fitted value of $R_{\rm high}$.

For the NIR flares of Sgr A$^*$, a purely thermal emissivity is unlikely to be sufficient \citep[e.g.,][]{Scepi2022}. We therefore adopt a kappa ($\kappa$) electron distribution function in the radiative-transfer calculation. This eDF joins a thermal Maxwell-J\"uttner core smoothly to a non-thermal power-law tail:
\begin{equation}
    \frac{dn_{\rm e}}{d\gamma_{\rm e}}=\frac{N}{4\pi}\gamma_{\rm e}\sqrt{\gamma_{\rm e}^2-1}\left(1+\frac{\gamma_{\rm e}-1}{\kappa w}\right)^{-(\kappa+1)},
\end{equation}
where $\gamma_{\rm e}$ is the electron Lorentz factor, $w$ sets the energy width and is related to the temperature $\Theta_{\rm e}$, and $\kappa$ controls the slope of the non-thermal tail. Neither $\kappa$ nor the acceleration efficiency is treated as a fixed constant. Instead, we use sub-grid models calibrated with Particle-in-Cell (PIC) simulations of turbulence \citep{Meringolo2023} and magnetic reconnection \citep{Ball2018}. These models provide local values of $\kappa(\beta, \sigma)$ and of the non-thermal energy fraction $\tilde{\epsilon}(\beta, \sigma)$ as functions of the local magnetization $\sigma$ and plasma beta $\beta$.

Finally, the total emissivity $j_{\rm tot}$ and absorptivity $\alpha_{\rm tot}$ used in our GRRT calculations are a weighted sum of the thermal and non-thermal components:
\begin{equation}
\begin{aligned}
    j_{\rm tot}&=(1-\tilde{\epsilon})j_{\rm Th}+\tilde{\epsilon}j_{\rm kappa},\\
    \alpha_{\rm tot}&=(1-\tilde{\epsilon})\alpha_{\rm Th}+\tilde{\epsilon}\alpha_{\rm kappa}.
\end{aligned}
\label{Eq: emi&alpha-c1}
\end{equation}
This hybrid approach ensures that regions undergoing strong magnetic dissipation contribute significantly to the non-thermal emission, while the bulk accretion flow remains dominated by thermal emission.

\section{Binary-Black-Hole Simulation Framework}

Modeling the long-term accretion dynamics of binary black-hole (BBH) systems requires a time-dependent background metric that evolves consistently with the binary orbit. Full numerical-relativity (NR) solutions are computationally too expensive for the thousands of orbital periods needed to study secular disk evolution, so we adopt a semi-analytic approach. Our framework couples a high-order post-Newtonian (PN) trajectory integrator to a superposed Kerr-Schild metric solver optimized for GPU architectures.

\subsection{Orbital Dynamics and Trajectory Construction}

To evolve the metric, we require accurate histories of the black-hole positions $\mathbf{x}_A(t)$, velocities $\mathbf{v}_A(t)$, and spin vectors $\mathbf{S}_A(t)$, where $A=1,2$. We therefore developed a dedicated trajectory integrator that evolves the coupled ordinary differential equations (ODEs) for relative acceleration and spin precession from large separations down to the late-inspiral regime.

The equations of motion are based on the harmonic-coordinate expansion of the Einstein field equations. To model both the inspiral rate and the precessional dynamics of tilted disks realistically, we include terms up to high post-Newtonian order, following Blanchet (2014) \cite{2014LRR....17....2B} and Levi \& Steinhoff (2015) \cite{2015JHEP...09..219L}. The total relative acceleration $\mathbf{a}$ is decomposed into conservative and dissipative (radiation-reaction) terms:
\begin{equation}
\mathbf{a} = \mathbf{a}_{\rm Newt} + \mathbf{a}_{\rm cons} + \mathbf{a}_{\rm diss}.
\end{equation}

Our implementation includes the following specific PN corrections:
\begin{itemize}
    \item \textbf{Conservative Dynamics (Orbital Phase):} We include non-spinning corrections to the Newtonian force up to 2PN order. This captures the relativistic periastron advance and the shaping of the effective potential.
    \item \textbf{Spin-Orbit (SO) Coupling:} To model the Lense-Thirring precession of the orbital plane accurately, which is critical for studying warped accretion disks, we include both the Leading Order (1.5PN) and Next-to-Leading Order (2.5PN) conservative corrections.
    \item \textbf{Spin-Spin (SS) Coupling:} We account for interactions between the two black hole spins, including the Leading Order (2PN) and the Next-to-Leading Order (3PN) terms.
    \item \textbf{Radiation Reaction (Dissipative):} The orbital decay is driven by energy and angular momentum loss to gravitational waves. We implement dissipative terms up to 4PN for non-spinning contributions and 3.5PN for spin-orbit tail effects.
\end{itemize}

At the same time, the orientation of each black hole's spin is evolved through the precession equation $\dot{\mathbf{S}}_A = \mathbf{\Omega}_A \times \mathbf{S}_A$, where the precession frequency $\mathbf{\Omega}_A$ includes geodetic, Lense-Thirring, and quadrupole-monopole self-interaction terms. The system of ODEs is integrated using an explicit order-8 Runge-Kutta method (\texttt{DOP853}) with adaptive time-stepping to maintain a relative tolerance of $10^{-11}$.

\subsection{Time-Dependent Superposed Kerr-Schild Spacetime}

To model the dynamical spacetime of a binary system, we use an analytic superposition of two moving black holes. The metric $g_{\mu\nu}$ is constructed by treating the curvature sourced by each hole as a perturbation of a common Minkowski background $\eta_{\mu\nu}$. In the laboratory frame, which is also the simulation frame, the metric is written as the sum of the background and the two black-hole perturbations:

\begin{equation}
g_{\mu\nu}(t, \mathbf{x}) = \eta_{\mu\nu} + h_{\mu\nu}^{(1)}(t, \mathbf{x}) + h_{\mu\nu}^{(2)}(t, \mathbf{x}).
\label{eq:superposition_def}
\end{equation}

Here, $h_{\mu\nu}^{(A)}$ denotes the gravitational perturbation generated by black hole $A$ ($A \in \{1, 2\}$) in the laboratory frame. Because the holes are moving with velocities $\mathbf{v}_A$, the stationary solution for each object must be Lorentz transformed. The explicit metric construction is

\begin{equation}
g_{\mu\nu} = \eta_{\mu\nu} 
+ \left[ \left( \Lambda_1 \, g^{\rm KS}_{(1)} \, \Lambda_1^{\top} \right)_{\mu\nu} - \eta_{\mu\nu} \right] 
+ \left[ \left( \Lambda_2 \, g^{\rm KS}_{(2)} \, \Lambda_2^{\top} \right)_{\mu\nu} - \eta_{\mu\nu} \right],
\label{eq:explicit_boost}
\end{equation}
where $g^{\rm KS}_{(A)}$ is the stationary Kerr-Schild metric tensor of black hole $A$ evaluated in its own rest frame, and $\Lambda_A \equiv \Lambda(\mathbf{v}_A)$ is the Lorentz boost matrix that transforms covariant tensor components from the comoving frame to the laboratory frame. The term $\left( \Lambda_A \, g^{\rm KS}_{(A)} \, \Lambda_A^{\top} \right)$ represents the full metric of the single black hole after boosting it to the laboratory frame; subtracting $\eta_{\mu\nu}$ isolates the perturbation $h_{\mu\nu}^{(A)}$ required for the superposition in Eq.~\eqref{eq:superposition_def}.

In index notation, the boosted metric components for object $A$ are calculated as:
\begin{equation}
\left(g_{\rm boosted}^{(A)}\right)_{\mu\nu} = \Lambda^{\alpha}_{\ \mu}(\mathbf{v}_A) \, \Lambda^{\beta}_{\ \nu}(\mathbf{v}_A) \, g^{\rm KS}_{\alpha\beta}\big(M_A, a_A; \bar{x}^\gamma \big),
\end{equation}
where $\Lambda^{\alpha}_{\ \mu}$ are the components of the standard Lorentz boost. It is crucial to note that the rest-frame metric $g^{\rm KS}_{\alpha\beta}$ depends on the rest-frame coordinates $\bar{x}^\gamma$. These are obtained by performing the inverse Lorentz transformation on the laboratory coordinates relative to the black hole's instantaneous position $\mathbf{x}_A(t)$:
\begin{equation}
\bar{x}^\gamma = (\Lambda^{-1})^\gamma_{\ \delta}(\mathbf{v}_A) \left[ x^\delta_{\rm lab} - x^\delta_A(t) \right].
\end{equation}

The intrinsic Kerr-Schild metric in the rest frame takes the familiar form:
\begin{equation}
g^{\rm KS}_{\alpha\beta} = \eta_{\alpha\beta} + 2 H(r) l_\alpha l_\beta,
\end{equation}
where $H(r)$ is the scalar potential and $l_\alpha$ is the ingoing null co-vector. Substituting this expression back into Eq.~\eqref{eq:explicit_boost} ensures that the event horizons and ergospheres of both black holes are Lorentz-contracted and distorted appropriately according to their instantaneous orbital velocities.

\subsection{Optimized ``Vector Boost'' Implementation}

A naive implementation of the boosted metric requires transforming the full rank-2 tensor $h_{\mu\nu}$ at every grid point and time step:
\begin{equation}
h_{\alpha\beta}^{\rm lab} = \Lambda^\mu_{\ \alpha} \Lambda^\nu_{\ \beta} h_{\mu\nu}^{\rm rest}.
\end{equation}
This tensor operation scales as $O(16^2)$ floating-point operations and represents a significant bottleneck for long-duration simulations. To address this, we implemented a "Vector Boost" optimization that exploits the algebraic structure of the Kerr-Schild metric ($h_{\mu\nu} \propto l_\mu l_\nu$).

Instead of boosting the tensor, we boost the null vector $l_\mu$ directly:
\begin{equation}
l^{\mu}_{\rm lab} = \Lambda^\mu_{\ \nu}(\mathbf{v}_A) \, l^{\nu}_{\rm rest}.
\end{equation}
The metric is then reconstructed in the laboratory frame via:
\begin{equation}
g_{\mu\nu}^{\rm lab} = \eta_{\mu\nu} + \sum_{A=1}^{2} 2 H_A \, (l^{\rm lab}_{\mu})_A \, (l^{\rm lab}_{\nu})_A.
\end{equation}
This optimization reduces the computational complexity by approximately an order of magnitude (from tensor operations to vector operations), resulting in an 80\% faster GRMHD simulation.

It is crucial to emphasize that this optimization is exact. Since the Lorentz transformation is linear and the Kerr-Schild form is algebraic, the "Vector Boost" method introduces no approximations. It yields results identical to the full tensor transformation to within machine precision, while significantly reducing the computational cost of the GRMHD and GRRT calculations.

\part{Flaring Activity and Precession in Kerr Accretion Flows}
\chapter[Multi-loop two-temperature GRMHD]{Magnetic-Topology Control of Accretion, Plasmoids, and Electron Heating in Multi-loop Two-Temperature GRMHD Flows}

An earlier form of this work was published by Jiang H.-X., Mizuno Y., Fromm C. M., and Nathanail A. in MNRAS, volume 522, page 2307 (2023) \cite{2023MNRAS.522.2307J}.

\section{Physical Motivation}

Why Sgr A$^*$ and M 87 produce such different large-scale outflows, despite both being low-luminosity systems accreting far below Eddington \citep[e.g.,][]{Yuan2014, 2008ARA&A..46..475H, Ho2009, Marrone2007}, remains a central unsolved problem. M 87 powers a relativistic jet that stays collimated on kiloparsec scales, whereas Sgr A$^*$ still lacks an unambiguous jet detection. Even models that do allow outflow from Sgr A$^*$ \cite{Yusef-Zadeh2020} generally predict much slower motion than is inferred for M 87. The contrast suggests that the flow geometry and magnetic structure near the hole must matter strongly.

Numerical studies often interpret M 87 as a MAD source \citep[e.g.,][]{2019ApJ...875L...5E, Cruz-Osorio2021, Yuan2022}. Several analyses of the EHT data for Sgr A$^*$ also prefer MAD-like models over standard SANE solutions \cite{2022ApJ...930L..16E}. That conclusion is still provisional, however, because no available model family reproduces the imaging, spectral, and timing constraints simultaneously. The mismatch is most obvious in the variability diagnostics \cite{2022ApJ...930L..16E}.

Targeted simulations of Sgr A$^*$ make the tension even clearer. In the wind-fed calculations of \cite{Ressler2020}, gas supplied by roughly thirty Wolf-Rayet stars pushes the inner flow toward a MAD state, and \cite{Murchikova2022} showed that the resulting 230 GHz structure function can agree reasonably well with sub-millimeter observations. Those models inherit their feeding conditions from larger-scale wind calculations \cite{10.1093/mnras/stz3605}, and the adopted wind magnetization remains consistent with the limits summarized by \cite{Michail2021}. MAD flows are also attractive flare candidates because they naturally create energetic plasmoids \cite{2020MNRAS.497.4999D, 2021MNRAS.502.2023P, Ripperda2021, Scepi2022}. The difficulty is that they also tend to launch strong Blandford-Znajek jets \cite{Blandford1977}, a feature that resembles M 87 more closely than Sgr A$^*$.

This makes the initial magnetic topology a natural control parameter. \cite{Narayan2012} contrasted single large poloidal loops, which accumulate horizon-threading flux efficiently, with alternating-polarity loop patterns, which keep the near-hole field weaker. In the BZ picture \cite{Blandford1977}, an efficient jet requires both rapid spin and substantial magnetic flux. If neighboring loops reconnect before that flux reaches the horizon, the funnel remains less magnetized and the flow stays farther from a fully arrested state. Axisymmetric calculations \cite{Nathanail2020} and later 3D extensions \cite{Chashkina2021, Nathanail2021} showed that mixed-polarity multi-loop tori can sustain repeated reconnection, abundant plasmoids, and even striped jets \cite{Chashkina2021}, while remaining highly time dependent rather than relaxing into a steady jet solution.

That possibility is especially relevant for Sgr A$^*$ because several observations point to compact activity near the hole. GRAVITY measured an NIR flare component whose centroid follows a smooth orbital path \cite{GRAVITY2018b, 2017A&A...602A..94G}, placing the emitting region only a few gravitational radii from the horizon \cite{GRAVITY2018b}. Variability statistics favor a persistent background plus a flare-dominated bright tail \citep[e.g.,][]{TheGRAVITYCollaboration2020, Abuter2020}. At radio frequencies, \cite{Michail2021} reported simultaneous 8 GHz and 10 GHz observations separated by about twenty minutes, which they interpreted as adiabatic expansion. \cite{Nathanail2022} further argued that the absence of a clear jet signal at 43 GHz and 86 GHz can help distinguish among MAD, SANE, and alternating multi-loop scenarios. EHT-ALMA polarimetry \cite{2022A&A...665L...6W} is likewise consistent with a hot spot inside a vertically magnetized arrested region. Even so, some MAD models that otherwise fit Sgr A$^*$ still fail to reproduce the observed 230 GHz variability \cite{2022ApJ...930L..12E}.

Semi-analytic models can match several global observables of Sgr A$^*$, including the broadband spectrum and multi-band variability \cite{Yuan2003}, but they do not determine how flares are triggered or how the gas behaves immediately around the horizon. \cite{Yuan2009} proposed an episodic-jet interpretation analytically, and \cite{2020MNRAS.499.1561Z} later explored related behavior numerically. More recently, \cite{2022ApJ...933...55C} found in 3D GRMHD calculations that plasmoids produced very near the hole usually remain trapped in the disk, while structures created farther out can be expelled magnetically.

Several plasma processes may contribute to plasmoid formation, including tearing-mediated plasmoid instability \cite{Ripperda2021}, pressure and pinch modes \cite{McKinney2006, 2018ApJ...868..146N, Chatterjee2019}, Parker instability \cite{2022ApJ...933...55C}, and mixed Kelvin-Helmholtz plus tearing behavior \cite{2022ApJ...929...62B}. Because the plasmoid instability is fundamentally a tearing process \cite{2017ApJ...850..142C}, I use the shorter label tearing instability throughout this chapter. Current sheets are still essential, but turbulence can seed KH-like and pinch-like disturbances that help organize plasmoid-rich regions.

Motivated by that observational and theoretical picture, I analyze a set of two-dimensional two-temperature GRMHD simulations of thick magnetized tori threaded by multiple poloidal loops. By changing both the loop polarity and the loop wavelength, I test how magnetic topology modifies the accretion state, the plasmoid population, and the electron-heating outcome.

Section 3.2 describes the numerical setup and the imposed magnetic geometry. Section 3.3 follows the accretion evolution. Section 3.4 examines plasmoid formation. Section 3.5 compares the self-consistent two-temperature results with the parameterized $R-\beta$ model. Section 3.6 summarizes the main findings.

\section{Simulation Configuration} \label{Sec: setup}

\begin{table}
\begin{tabular*}{\columnwidth}{@{\extracolsep{\fill}}llll}
\hline
\multicolumn{1}{l}{Case} & a      & Polarity    & $\lambda_{\rm r}$ \\ 
\hline
\tt M0                        & 0.9375 & Same           & 0                    \\
\tt M6a                       & 0.9375 & Alternating   & 6                \\
\tt M6b                       & 0.9375 & Same          & 6                       \\
\tt M6c                       & 0      & Alternating    & 6                 \\
\tt M6d                       & 0      & Same         & 6                        \\
\tt M20a                      & 0.9375 & Alternating    & 20                \\
\tt M20b                      & 0.9375 & Same          & 20                       \\
\tt M20c                      & 0      & Alternating    & 20                \\
\tt M20d                      & 0      & Same          & 20                       \\
\tt M50a                      & 0.9375 & Alternating    & 50                \\
\tt M80a                      & 0.9375 & Alternating    & 80                \\
\tt M80b                      & 0.9375 & Same           & 80                       \\
\tt M80c                      & 0      & Alternating    & 80                \\
\tt M80d                      & 0      & Same           & 80                       \\
\hline                       
\end{tabular*}
\caption{Simulation set used in this chapter. Equation~\ref{Eq: MAD_A} defines the seed magnetic field, and Figure~\ref{fig: initial condition-c2} shows representative examples of the initial loop arrangement.}
\label{Table: MAD}
\end{table}
\begin{figure}
    \centering
         \includegraphics[width=.4\linewidth]{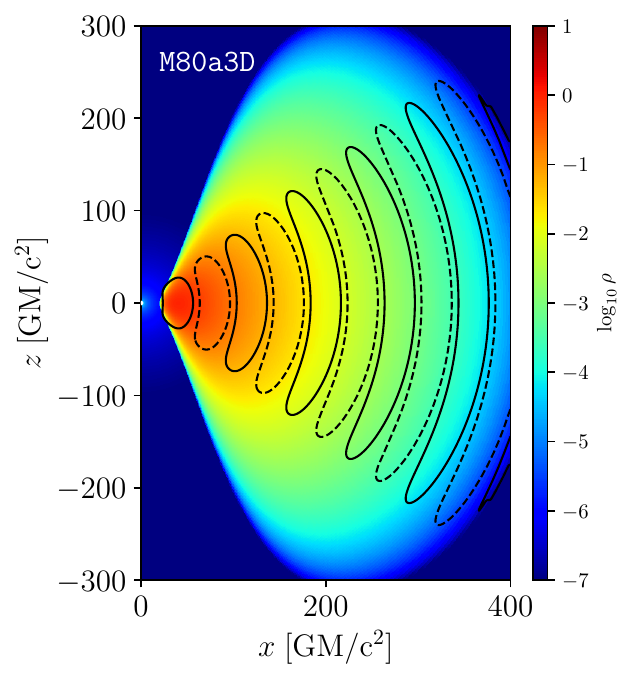}
    \caption{Representative starting configurations for the alternating- and same-polarity families at $\lambda_{\rm r}=80$. The color scale shows $\log \rho$, and the black contours follow the imposed magnetic loops; solid and dashed curves mark opposite polarities.}
    \label{fig: initial condition-c2}
\end{figure}
All calculations in this chapter are performed with BHAC \cite{Porth2017, Olivares2019}, which evolves axisymmetric two-temperature GRMHD tori on a fixed black-hole background. Unless noted otherwise, I work in units with $GM=c=1$ and absorb the conventional factor $1/\sqrt{4\pi}$ into the magnetic field. In that convention, the ideal-GRMHD system solved by BHAC is
\begin{equation}
	\nabla_{\mu}\left(\rho u^{\mu}\right)=0,\,\,\,
	\nabla_{\mu}T^{\mu\nu}=0,\,\,\,
	\mathrm{and}\,\,\,
	\nabla_{\mu}^*F^{\mu\nu}=0,
\end{equation} 
In these equations, $\rho$ is the rest-mass density, $u^{\mu}$ is the fluid four-velocity, $T^{\mu\nu}$ is the stress-energy tensor, and ${}^*F^{\mu\nu}$ is the dual Faraday tensor. In the ideal-MHD limit, the stress-energy tensor becomes
\begin{equation}
    T^{\mu\nu} = \left( \rho h_{\rm tot} \right) u^{\mu}u^{\nu}+ \left( p+{1 \over 2} b^2 \right) g^{\mu\nu}-b^{\mu}b^{\nu},
\end{equation}
where $h_{\rm tot} = h + b^2/\rho$ is the total specific enthalpy, $p$ is the gas pressure, and $g^{\mu\nu}$ is the metric tensor.

The background metric is written in spherical Modified Kerr-Schild coordinates; the full construction is given by \cite{Porth2017}. In this coordinate system, the magnetic four-vector can be written as
\begin{equation}
    b^t = \frac{\gamma\left(B^i v_{i}\right)}{\alpha}, \,\,\,b^{i}=\frac{B^i +\alpha b^t u^i}{\gamma},
\end{equation}
where $\gamma$ is the Lorentz factor, $B^i$ is the Eulerian magnetic field, $v^i$ is the fluid three-velocity, and $\alpha$ is the lapse. Throughout this chapter I denote the plasma beta and magnetization by $\beta=p/b^2$ and $\sigma=b^2/\rho$, respectively, with $b^2$ given by \cite{Porth2017}
\begin{equation}
    b^2 = \frac{B^2+\alpha^2(b^t)^2}{\gamma^2} = \frac{B^2}{\gamma^2}+\left(B^i v_i\right)^2,
\end{equation}
where $B^2 = B^i B_i$. The system is closed with an ideal-gas equation of state that relates the specific enthalpy $h$ to the pressure $p$ and density $\rho$:
\begin{equation}
    h = 1 + \frac{\Gamma_{\rm g}}{(\Gamma_{\rm g}-1)}\frac{p}{\rho},
\end{equation}
where $\Gamma_{\rm g}$ is the fluid adiabatic index.

The initial gas distribution is a Fishbone-Moncrief equilibrium torus \cite{1976ApJ...207..962F}. I place its inner edge at $r_{\rm in}=20\,r_{\rm g}$ and its pressure maximum at $r_{\rm max}=40\,r_{\rm g}$, where $r_{\rm g}\equiv GM/c^2$ is the gravitational radius of a black hole with mass $M$. A constant adiabatic index $\Gamma_{\rm g}=4/3$ is adopted \cite{Rezzolla2013}.

Each run is initialized with a purely poloidal magnetic field defined through the vector potential
\begin{equation}
    A_{\rm \phi} \propto \left(\rho-0.01\right)\left(r/r_{\rm in}\right)^3\sin^3\theta\exp\left(-r/400\right).
    \label{Eq: MAD_A}
\end{equation}
To create multiple loops, I modulate $A_{\phi}$ by $\cos\left((N-1)\theta\right)$ in the meridional direction and by $\sin\left(2\pi(r-r_{\rm in})/\lambda_{\rm r}\right)$ radially. The parameter $\lambda_{\rm r}$ sets the spacing between adjacent radial loops, while $N$\footnote{The torus opening angle spans approximately $-\pi/4$ to $\pi/4$. In the $\theta$ direction, $A_{\phi}\propto \cos((N-1)\theta)$. Choosing $N=3$ leaves the polarity unchanged across the torus thickness and therefore produces only one loop in that direction.} determines the meridional pattern. I keep $N=3$ for all models. The total field strength is fixed through the minimum plasma beta, $\beta \equiv p_{\rm g}/p_{\rm mag}$ with $p_{\rm mag}=b^2/2$ and $b^2=b^\mu b_\mu$. Throughout this chapter I adopt $\beta_{\rm min}=100$, so the initial field is weaker than in several earlier studies \cite{Chashkina2021, Ripperda2021}.

MRI growth is initiated by applying a random 2\% perturbation to the gas pressure.

Table~\ref{Table: MAD} summarizes the model suite. The suffix ``a'' labels rapidly rotating black holes threaded by alternating-polarity loops, while ``b'' labels spinning runs with same-polarity loops. The suffixes ``c'' and ``d'' denote the matching alternating- and same-polarity cases for non-spinning black holes. Model M0 is the single-loop reference run and evolves toward a standard MAD state. Most simulations use a $1024\times512$ grid in radius and polar angle; model M80a is also repeated at $4096\times2048$ for a convergence test. The Appendix compares the accretion history and density structure of those two resolutions (Figure~\ref{fig: 2D_high_low_res}). Because the large-scale agreement is good, I use the lower-resolution grid for the main discussion.

At the radial boundaries, the primitive variables are copied in the usual inflow/outflow manner. At the pole, reflecting conditions are imposed so that $B^\theta \to -B^\theta$ and $v^\theta \to -v^\theta$, while the remaining quantities are mirrored evenly. The azimuthal direction is periodic.

Following \cite{Mizuno2021}, I evolve the electron thermodynamics on top of the single-fluid GRMHD background. In this two-temperature extension, the ion and electron temperatures, $T_{\rm i}$ and $T_{\rm e}$, are evolved separately, while the total gas temperature satisfies $T_{\rm gas}=T_{\rm i}+T_{\rm e}$. As in earlier studies \citep[e.g.,][]{Ressler2015, Chael2018, Dihingia2023}, Coulomb coupling, anisotropic heat transport, and radiative cooling are omitted. Electron heating is supplied by grid-scale dissipation, and I compare two sub-grid channels: turbulent heating and magnetic-reconnection heating. For the turbulent contribution, I adopt the fit to damped MHD turbulence from \cite{Kawazura2019}, which gives the electron-heating fraction
 \begin{equation}
     f_{\rm e} = \frac{1}{1+Q_{\rm i}/Q_{\rm e}},\label{eq2}
 \end{equation}
 where
 \begin{equation}
     \frac{Q_{\rm i}}{Q_{\rm e}} = \frac{35}{1 + \left(\beta/15\right)^{-1.4}\exp\left(-0.1 T_{\rm e}/T_{\rm i}\right)}.
 \end{equation}
For the reconnection contribution, I use the fitting formula calibrated against the particle-in-cell calculations of \cite{Rowan2017}:
 \begin{equation}
     f_{\rm e} = \frac{1}{2} \exp\left[\frac{-(1-\beta/\beta_{\rm max})}{0.8+\sigma_{\rm h}^{0.5}}\right],
     \label{Eq: reconnection heating-c2a}
 \end{equation}
 where $\beta_{\rm max} = 1/4\sigma_{\rm h}$, and $\sigma_{\rm h} = b^2/\rho h$ is the magnetization defined with respect to the fluid specific enthalpy $h = 1 + \Gamma_{\rm g} p_{\rm g}/(\Gamma_{\rm g}-1)$. 

 \begin{figure*}
     \centering
          \includegraphics[height=0.4\linewidth]{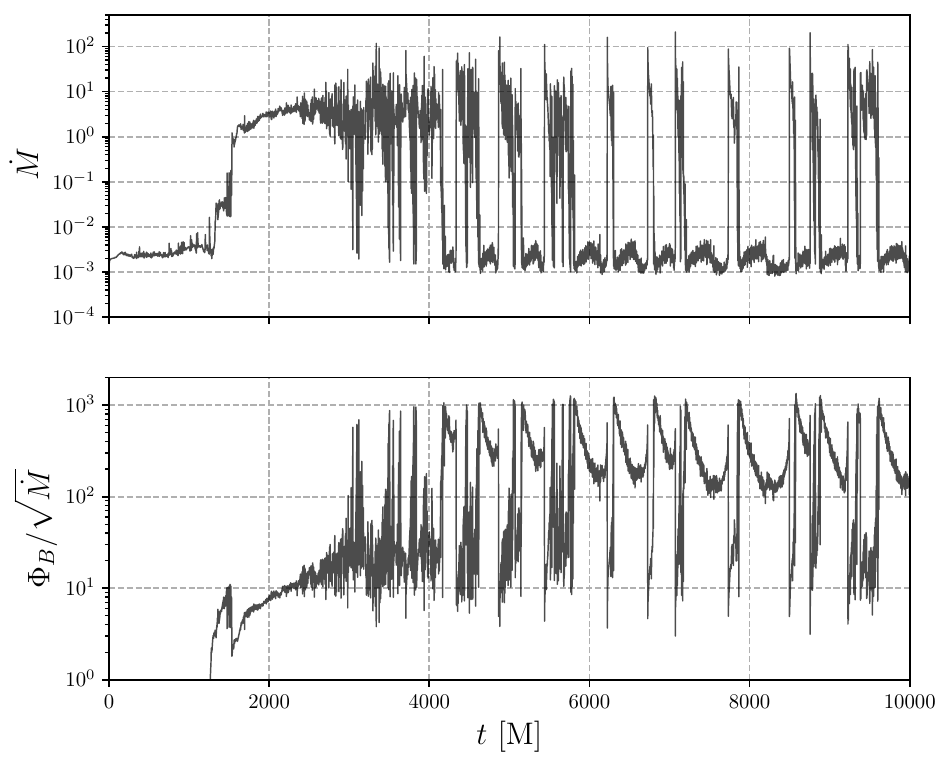}
          \includegraphics[height=0.4\linewidth]{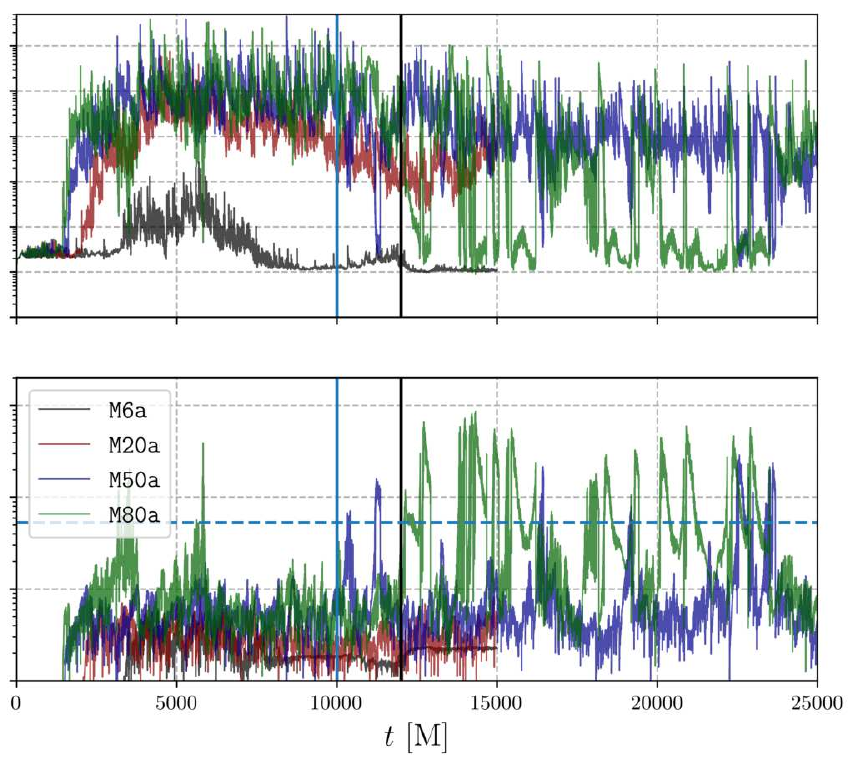}
     \caption{Mass-accretion histories and horizon magnetic fluxes for the $a=0.94$ runs. The left panel shows the single-loop reference model M0. The right panel compares alternating-polarity multi-loop models with $\lambda_{\rm r}=6$ (M6a, black), 20 (M20a, red), 50 (M50a, blue), and 80 (M80a, green). Vertical markers indicate the identified SANE-to-MAD transition times for M50a and M80a, and the blue dashed horizontal line marks the nominal MAD threshold. Longer loops evolve more slowly and were therefore followed for a longer duration.}
     \label{fig: high_spin_alternating}
 \end{figure*}
%
The electron temperature is computed with the sub-grid framework introduced by \cite{Ressler2015}. Because the electron sector does not feed back dynamically on the bulk GRMHD evolution in the present setup, the same GRMHD solution can be post-processed with either the turbulent-heating \citep[e.g.,][]{Kawazura2019} or reconnection-heating \citep[e.g.,][]{Rowan2017} prescription.

For completeness, I summarize the thermodynamic closure inherited from \cite{Mizuno2021}. The GRMHD system itself remains single fluid, but the electron entropy is evolved as an auxiliary variable. I assume equal number densities and a common bulk four-velocity for ions and electrons, namely $n_{\rm e}=n_{\rm i}=n$ and $u_{\rm e}^{\mu}=u_{\rm i}^{\mu}=u^\mu$. The electron temperature is then reconstructed from the entropy equation
\begin{equation}
    \rho T_{\rm e} \partial_{\rm \mu} s_{\rm e} = f_{\rm e} Q, \label{Eq: entropy_equ}
\end{equation}
where $\rho$ is the rest-mass density, $Q$ is the volumetric dissipation rate, $s_{\rm e}=(\Gamma_{\rm e}-1)^{-1}\log(p_{\rm e}/\rho^{\Gamma_{\rm e}})$ is the electron entropy, $\Gamma_{\rm e}$ is the electron adiabatic index, $p_{\rm e}$ is the electron pressure, and $f_{\rm e}$ is the fraction of dissipated energy assigned to electrons. As in \cite{Mizuno2021}, Coulomb coupling, anisotropic conduction, and radiative cooling are omitted. The dissipation rate $Q$ is evaluated following \cite{Ressler2015}.

Within BHAC, Eq.~\ref{Eq: entropy_equ} is evolved in the equivalent conservative form
\begin{equation}
\partial_{\rm \mu}\left(\sqrt{-g}\rho u^{\rm \mu} \kappa_{\rm e}\right) = \frac{\sqrt{-g}\left(\Gamma_{\rm e}-1\right)}{\rho^{\Gamma_e-1}}f_{\rm e}Q,
\end{equation}
where $\kappa_{\rm e}\equiv \exp\left[\left(\Gamma_{\rm e}-1\right)s_{\rm e}\right]$. Further implementation details are given in \cite{Mizuno2021, Ressler2015}.

The turbulent closure described above is applied directly to one branch of the calculation. For the reconnection branch, I repeat the adopted fitting formula here for convenience,
\begin{equation}
    f_{\rm e} = \frac{1}{2} \exp\left[\frac{-(1-\beta/\beta_{\rm max})}{0.8+\sigma_{\rm h}^{0.5}}\right],
    \label{Eq: reconnection heating-c2b}
\end{equation}
where $\beta_{\rm max}=1/4\sigma_{\rm h}$ and $\sigma_{\rm h}=b^2/\rho h$ is the magnetization defined with respect to the fluid specific enthalpy $h=1+\Gamma_{\rm g}p_{\rm g}/(\Gamma_{\rm g}-1)$. Because the prefactor limits the heating fraction to $f_{\rm e}=1/2$, the reconnection prescription keeps funnel electrons cooler than the turbulent prescription in strongly magnetized regions, as discussed later in Section 3.4.

Throughout this chapter, I set $\Gamma_{\rm e}=4/3$ and initialize the electron internal-energy density as $u_{\rm e}=0.1u_{\rm g}$ relative to the fluid internal-energy density.

\section{Plasma Evolution for Different Initial Magnetic Topologies}


Following \cite{Porth2019}, we evaluate the mass-accretion rate at the event horizon as
\begin{equation}
    \dot M = \int_0^{2\pi}\int_0^{\pi} \rho u^r \sqrt{-g}d\theta d\phi, \label{Eq: a_rate-c2}
\end{equation}
The dimensionless magnetic flux threading the event horizon is defined as
\begin{equation}
    \Phi_{\rm B} = \frac{1}{2}\int_0^{2\pi}\int_{0}^{\pi}\left|B^r\right|\sqrt{-g}d\theta d\phi. \label{Eq: B-flux-c2}
\end{equation}

In the present unit convention, this horizon magnetic-flux measure differs from the definition used by \citet{tchekhovskoy_efficient_2011} by a factor of $\sqrt{4\pi}$.
The left panel of Figure~\ref{fig: high_spin_alternating} shows the reference single-loop model M0. Because there are no additional small loops that can reconnect away the field, the torus evolves into the standard MAD state. The repeated dips in both $\dot M$ and $\Phi_{\rm B}$ are the characteristic arrested-flow signature. I use this solution as the baseline for the multi-loop sequence.

\subsection{Multiple Magnetic Loops with Alternating Polarity}
\begin{figure}
    \centering
         \includegraphics[width=\linewidth]{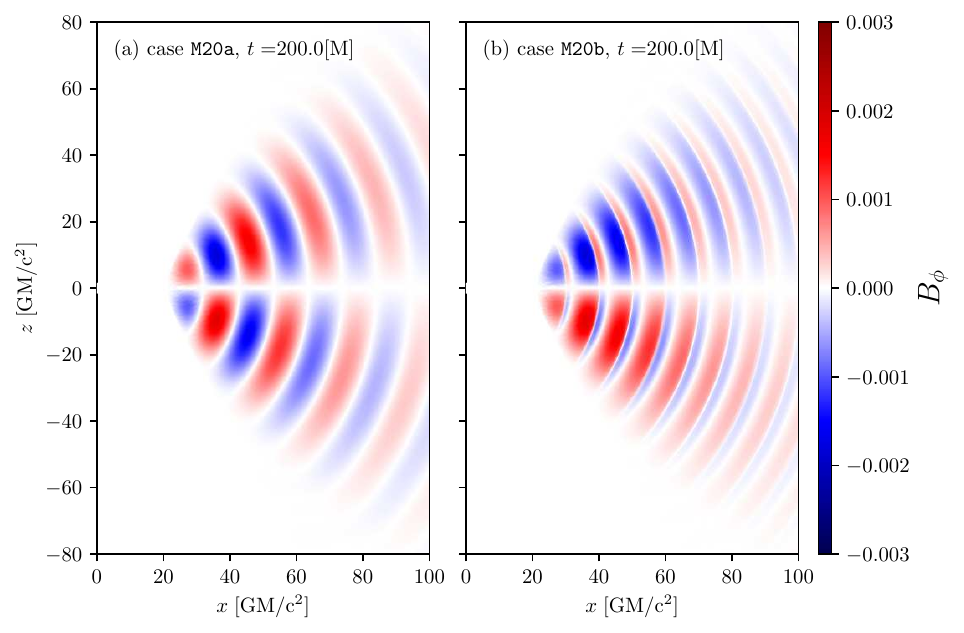}
    \caption{Early-time toroidal-field maps produced by differential rotation. The left panel shows the alternating-polarity model M20a, and the right panel shows the same-polarity model M20b.}
    \label{fig: M80_B}
\end{figure}
\begin{figure}
    \centering
	\includegraphics[height=.6\columnwidth]{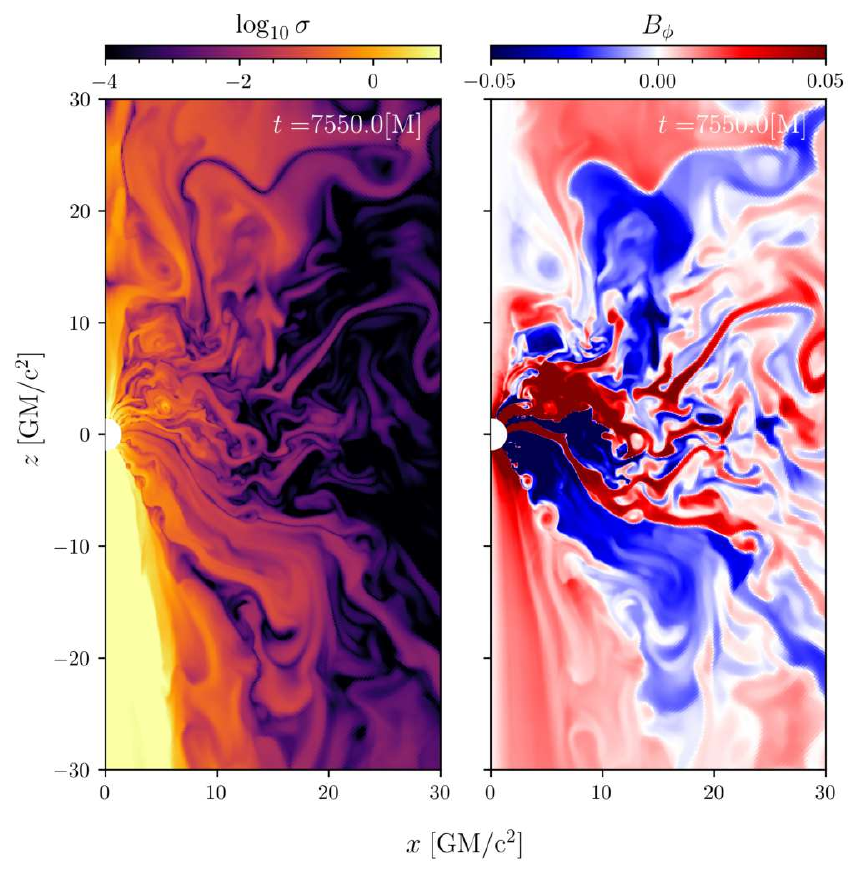}
    \caption{Snapshot of model M20a at $t=7550\,M$. The left panel shows the magnetization $\sigma$, and the right panel shows the toroidal field $B_{\rm \phi}$. By this time the upper jet branch has vanished because the flow has become strongly chaotic.}
    \label{fig: M20a_2D_magnetization}
\end{figure}

\begin{figure*}
    \centering
         \includegraphics[width=\linewidth]{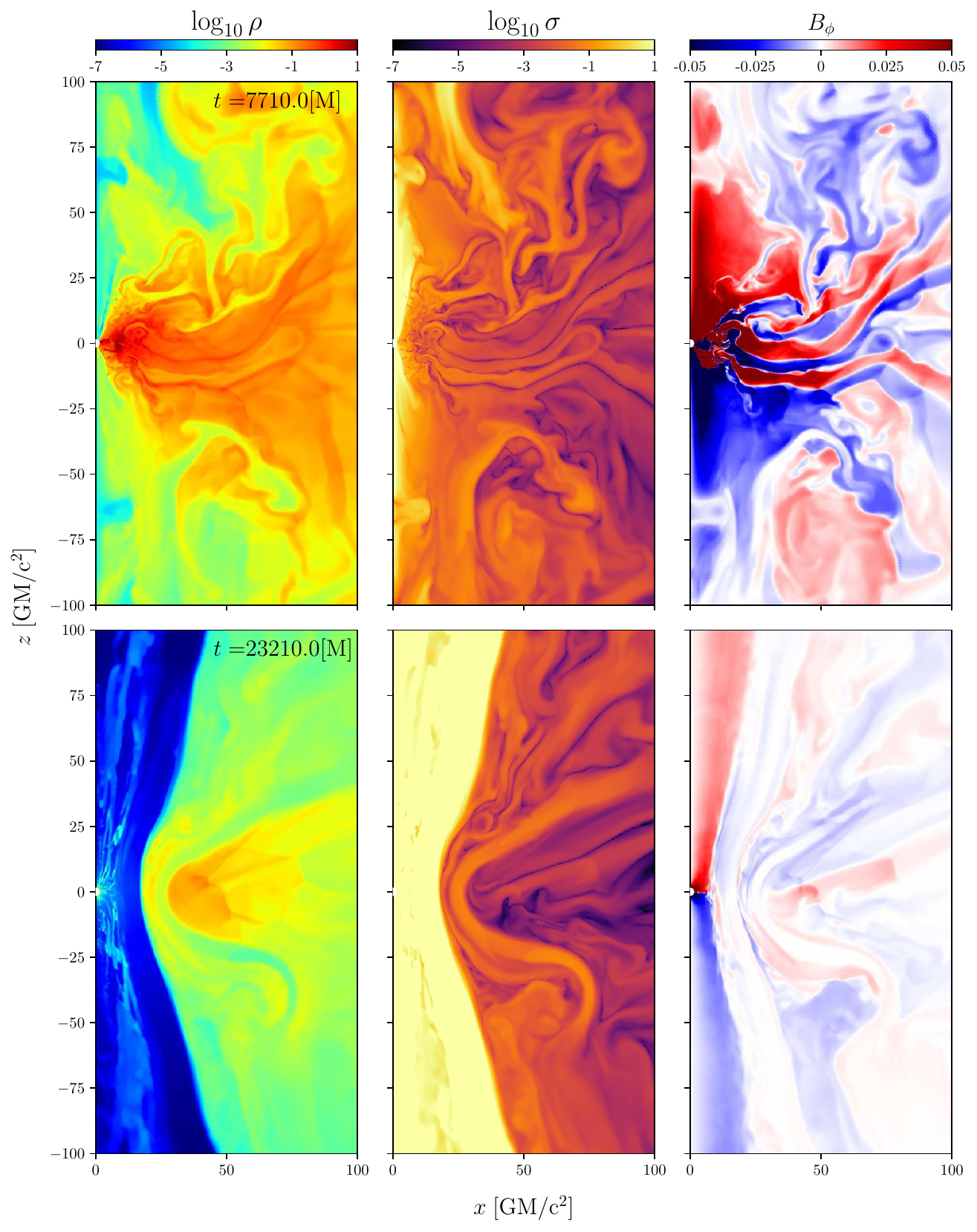}
         \caption{Representative states from the long-wavelength alternating-polarity run M80a with $\lambda_{\rm r}=80$. From left to right, the columns show $\log\rho$, $\sigma$, and $B_{\rm \phi}$. The upper row corresponds to the earlier SANE-like stage at $t=7710\,M$, whereas the lower row shows the later MAD-like stage at $t=23210\,M$.}
    \label{fig: 2D_M80a}
\end{figure*}

%
The right panel of Figure~\ref{fig: high_spin_alternating} shows the same diagnostics for rapidly rotating black holes ($a=0.9375$) initialized with alternating-polarity loops. The four curves correspond to M6a, M20a, M50a, and M80a, whose radial wavelengths are $\lambda_{\rm r}=6$, 20, 50, and 80.

The shortest-loop model, M6a, dissipates magnetic energy too quickly to build a strong inner field. Reconnection starts across the torus before MRI can grow fully, so both the accretion rate and the horizon flux remain low. Even though the nominal MRI resolution is adequate ($Q_{\rm r}\gg10$), the instability is effectively weakened by losses at the loop interfaces.
Models M20a and M50a initially behave more like the reference run. Their larger loops allow MRI to remain active for longer, and the early accretion histories broadly resemble M0. The later evolution separates the two cases. Model M20a never develops the robust oscillatory drops in $\dot M$ and $\Phi_{\rm B}$ that characterize a fully arrested state. When the first two loops accrete between 2000 and 4000 M, the horizon flux does increase temporarily and the inflow weakens in response, but the effect does not persist. I therefore do not classify M20a as a genuine MAD solution.
Figure~\ref{fig: M20a_2D_magnetization} illustrates the magnetic structure of model M20a at $t=7550\,M$. A strongly magnetized lower funnel produces a one-sided jet. Near the horizon, a polarity reversal creates a current sheet in the lower hemisphere, and that sheet seeds plasmoids, visible most clearly in the right panel. Similar structures were noted by \cite{Nathanail2020}. The corresponding formation mechanism is discussed in the next section.

Increasing the loop wavelength to $\lambda_{\rm r}=50$ changes the balance. In model M50a, dissipation inside the torus is weaker, MRI survives longer, and the accretion rate rises to roughly $\dot M\sim10$. The horizon flux also grows more efficiently. For roughly the first half of the evolution the flow is still SANE-like, but near $t\approx11500\,M$ the flux history develops a saw-tooth pattern, showing that the near-horizon magnetic field has become strong enough to intermittently impede the inflow.

Model M80a continues this trend even more clearly. Its accretion rate approaches that of the single-loop MAD reference shown in the left panel of Figure~\ref{fig: high_spin_alternating}. The upper row of Figure~\ref{fig: 2D_M80a} shows the earlier SANE-like stage at $t=7710\,M$. Because alternating loops continue to reach the hole, the funnel remains denser than in the single-loop case. At the same time, the longer wavelength reduces internal dissipation enough for magnetic flux to accumulate efficiently on the horizon. After about $t=12000\,M$, the near-horizon field repeatedly suppresses accretion, indicating a transition to a MAD-like state. The detailed time series still differ from M0, however, because the funnel polarity keeps changing as new loops enter. At late times, the lower row of Figure~\ref{fig: 2D_M80a} shows that the MAD-like state has become fully established.
%
\subsection{Multiple Magnetic Loops with the Same Polarity}\label{sec: GRMHD_sam}
\begin{figure}
	\centering
	\includegraphics[height=.8\columnwidth]{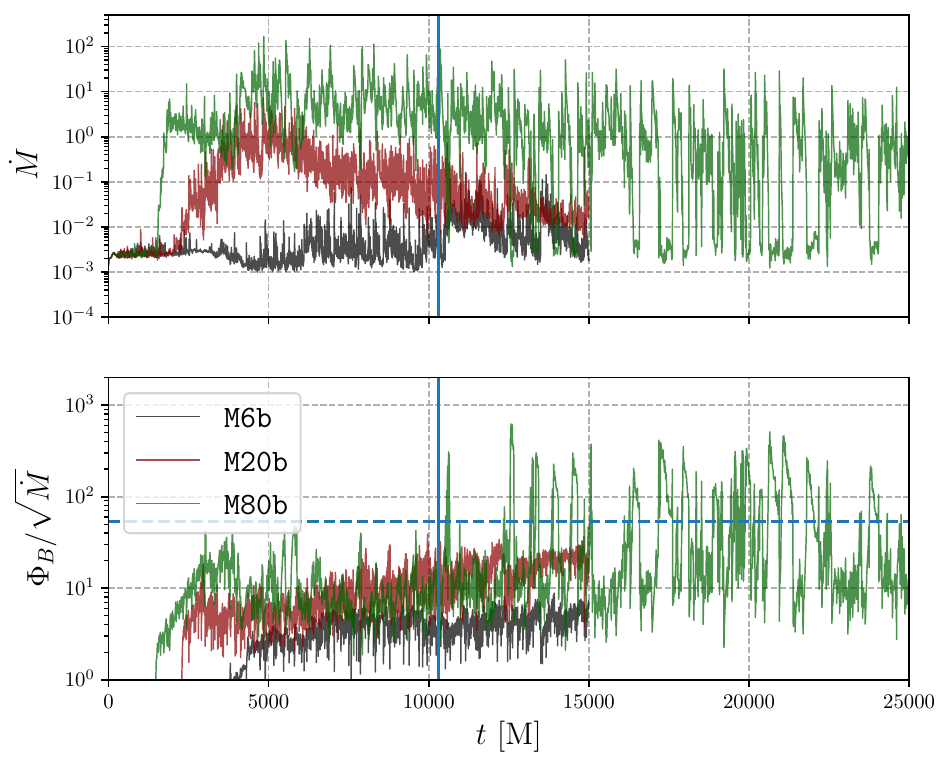}
    \caption{Accretion-rate and horizon-flux histories for the rapidly rotating same-polarity runs. The black, red, and green curves correspond to M6b, M20b, and M80b. The blue dashed horizontal line marks the MAD threshold, while the blue vertical line shows the transition time identified for M80b. Because M80b evolves most slowly, it is extended to 25,000 M.}
    \label{fig:Mdot_high_sam}
\end{figure}


Figure~\ref{fig:Mdot_high_sam} compares the horizon accretion rate and magnetic-flux growth for the spinning models initialized with same-polarity loops. The black, red, and green curves correspond to M6b, M20b, and M80b. Relative to the alternating-polarity sequence in Figure~\ref{fig: high_spin_alternating}, the histories of both $\dot M$ and $\Phi_{\rm B}$ differ substantially because the loops now interact without immediate cancellation between neighboring opposite-polarity regions.

Model M6b contains the shortest same-polarity loops, so the torus loses magnetic energy rapidly at early times. As a result, the accretion rate stays strongly suppressed between about 2,000 M and 6,000 M, when the field has been depleted too strongly for vigorous MRI growth. A similar temporary suppression appears in the short-loop alternating-polarity models. Once that internal dissipation subsides, the flow is effectively left with a weaker residual loop, MRI strengthens, accretion resumes, and noticeable horizon-flux growth begins only after $t\sim5,000\,M$.

Model M20b follows the same overall sequence, but the broader loops change the details. Differential rotation amplifies the field more efficiently, and the dissipation inside the torus is both weaker and shorter-lived than in M6b. MRI therefore develops earlier and drives a more extended turbulent structure. Even though $\Phi_{\rm B}$ rises steadily as the field strengthens, the flow still resembles a SANE disk rather than a fully arrested one.

As in the alternating-polarity sequence, increasing the loop size reduces magnetic dissipation inside the torus. The early evolution of M80b therefore stays much closer to the single-loop reference run. Before $t\approx10,000\,M$, the disk remains SANE-like while the horizon flux grows steadily. Once enough same-polarity flux has accumulated, the system crosses into a MAD state, and the later portion of Figure~\ref{fig:Mdot_high_sam} shows the corresponding arrested behavior clearly.

\subsection{Transition from SANE to the MAD State}
\begin{figure}
    \centering
         \includegraphics[width=.6\linewidth]{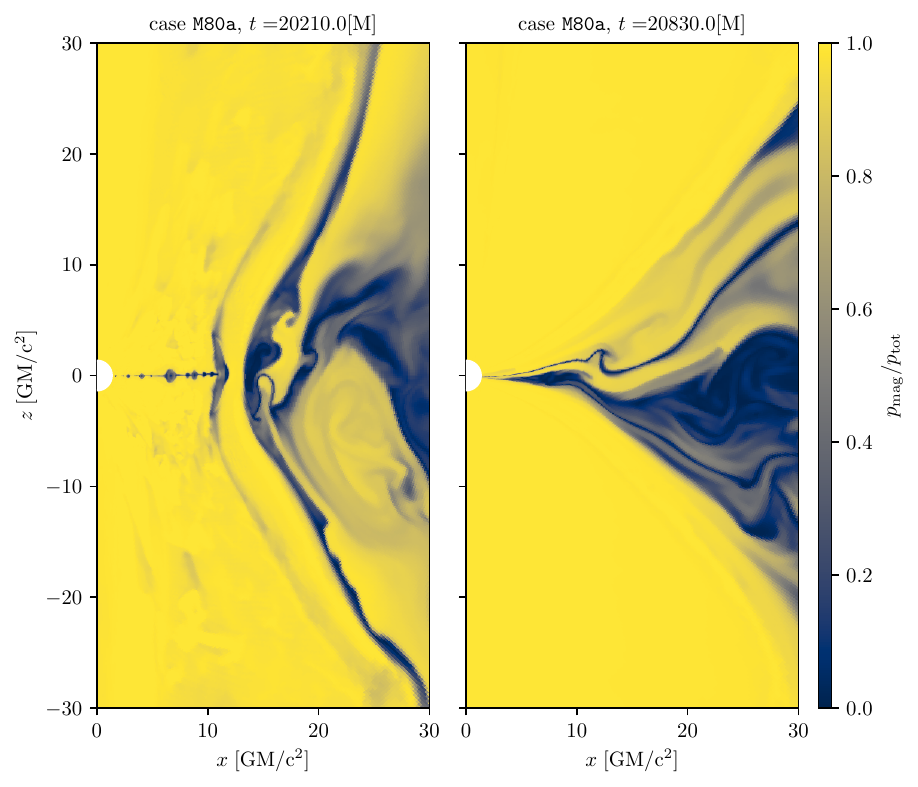}
    \caption{Magnetic-pressure fraction $p_{\rm mag}/p_{\rm tot}$ for two representative states: a fully arrested flow in the left panel and a later post-MAD stage in the right panel.}
    \label{fig: B_phi_ratio}
\end{figure}
\begin{figure}
    \centering
         \includegraphics[width=.8\linewidth]{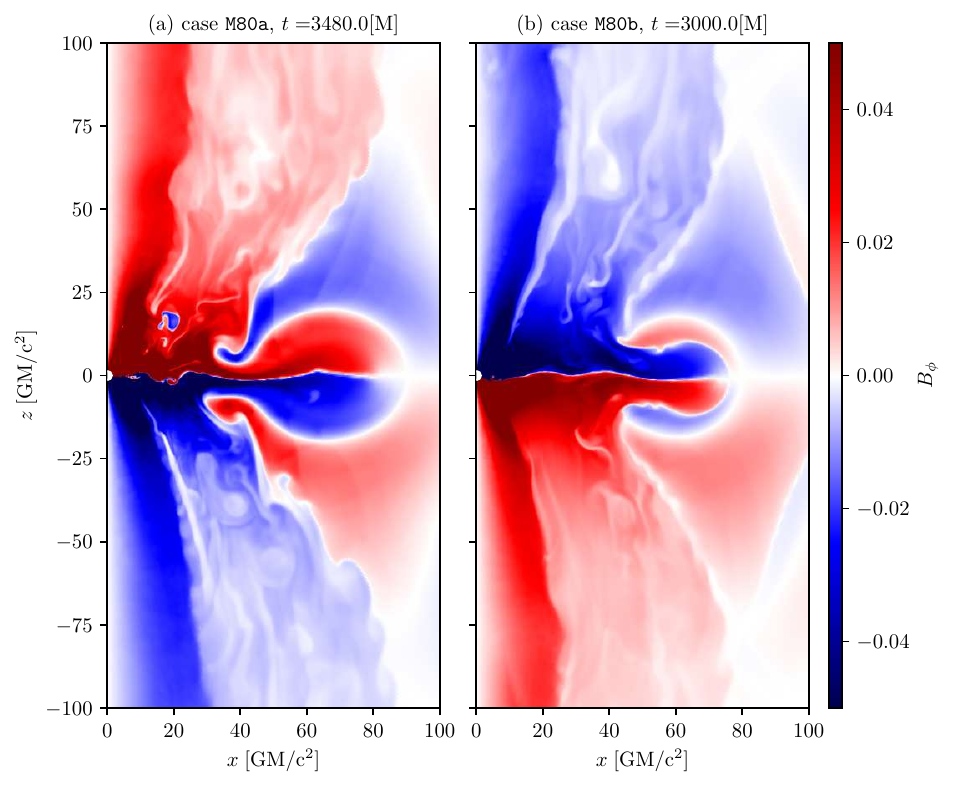}
    \caption{Toroidal-field maps for the long-loop models. The left panel shows the alternating-polarity run M80a at $t=3480\,M$, and the right panel shows the same-polarity run M80b at $t=3000\,M$.
    }
    \label{fig: M80_RT_instability}
\end{figure}

%

Earlier 2D GRMHD studies \cite[e.g.,][]{tchekhovskoy_efficient_2011} typically classify a flow as MAD once the normalized horizon flux exceeds about 15. In our unit system, the equivalent value is roughly 53 because of the extra factor of $\sqrt{4\pi}$. I also use the onset of accretion quenching as a practical indicator, since the MAD state is accompanied by saw-tooth drops in $\dot M$ and strong suppression of the inflow. Figure~\ref{fig: 2D_M80a} contrasts the density, magnetization, and toroidal field of model M80a during its SANE phase (upper) and its MAD phase (lower). When the inflowing field keeps the same polarity, flux can continue to load the horizon, while frame dragging around the rotating black hole strengthens the toroidal component. In the MAD stage, magnetic pressure dominates the near-horizon region. Figure~\ref{fig: B_phi_ratio} shows that $p_{\rm mag}/p_{\rm tot}$ there approaches unity \cite{2022ApJ...939...31M}, with $p_{\rm tot}=p_{\rm mag}+p_{\rm gas}$. Once magnetic support reaches that level, the inflow is intermittently choked and the system becomes arrested.

Although sufficiently long-loop models in both families evolve from SANE-like behavior toward MAD-like behavior after $t\approx10,000\,M$ (see Figures~\ref{fig: high_spin_alternating} and \ref{fig:Mdot_high_sam}), the route to that transition is not the same in the two sequences, even when their normalized horizon flux exceeds the nominal threshold.

As noted above, the same-polarity models first lose part of their magnetic energy through an early dissipation phase inside the torus. That initial loss delays both the gas inflow and the accumulation of ordered horizon-threading flux. If enough field survives, however, the remaining flux can still grow until the flow reaches the MAD regime.

The alternating-polarity models follow a different path because no single field sign dominates the torus. Instead, adjacent radial loops carry opposite polarity. Reconnection between neighboring loops therefore tangles the field much more strongly inside the torus. During this stage, the flow develops abundant small-scale mixed-polarity structure (see the right panel of Figure~\ref{fig: M20a_2D_magnetization}).

In the left panel of Figure~\ref{fig: M80_RT_instability}, the second loop reaches the hole through a narrow channel with a toroidal polarity opposite to that of the first loop (blue in the upper hemisphere and red in the lower hemisphere). As those opposite-sign loops arrive, the jet-sheath polarity flips repeatedly. Each reversal resets the ordered magnetic buildup at the horizon. The flow therefore remains SANE-like until the surviving large-scale flux finally becomes strong enough to cross the MAD threshold, which happens only in the longest-loop cases such as M50a and M80a. In the alternating-polarity sequence, the loop size is therefore the key control parameter for whether a MAD state can ultimately emerge.

\cite{Chashkina2021} reported a MAD-to-SANE transition in their single-loop model 2D1. Their initial field was stronger than ours because they adopted a smaller value of $\beta_{\rm min}(=1)$. Even so, their torus was smaller, so the long-term supply of large-scale magnetic flux was weaker. Once that flux reservoir was exhausted, the flow could no longer remain magnetically arrested and relaxed back to a SANE state. The late-time behavior of our M80a/b models is driven by a different detailed mechanism, but it is likewise controlled by the initial magnetic-field configuration.

%

\subsection{Nonrotating Black-Hole Cases}

 \begin{figure}
     \centering
	\includegraphics[height=.6\columnwidth]{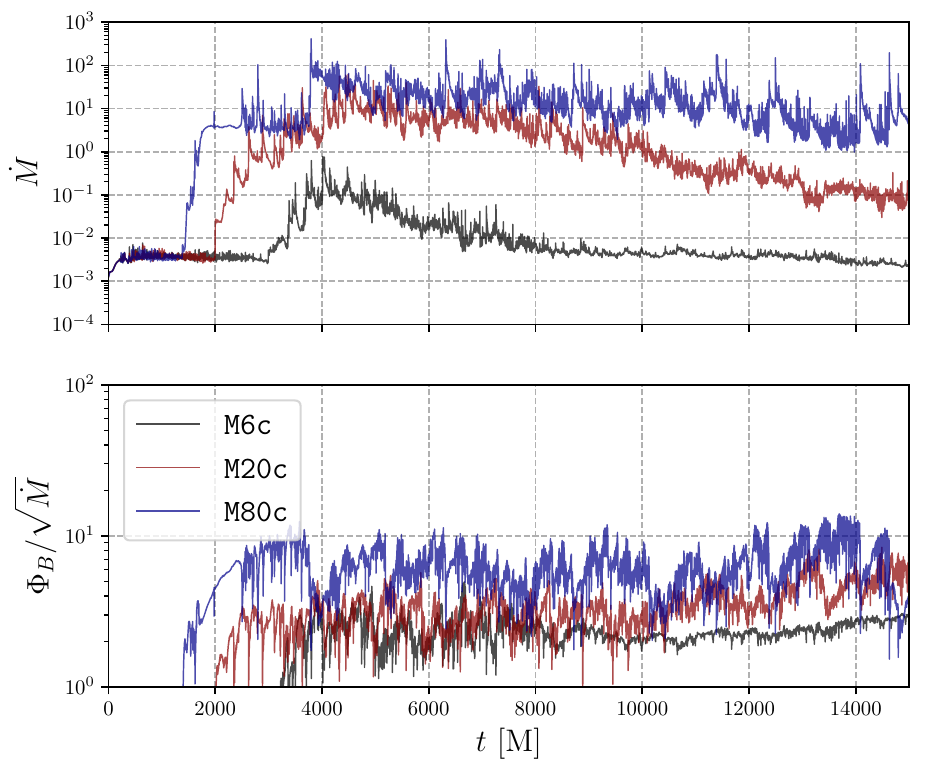}
	\includegraphics[height=.6\columnwidth]{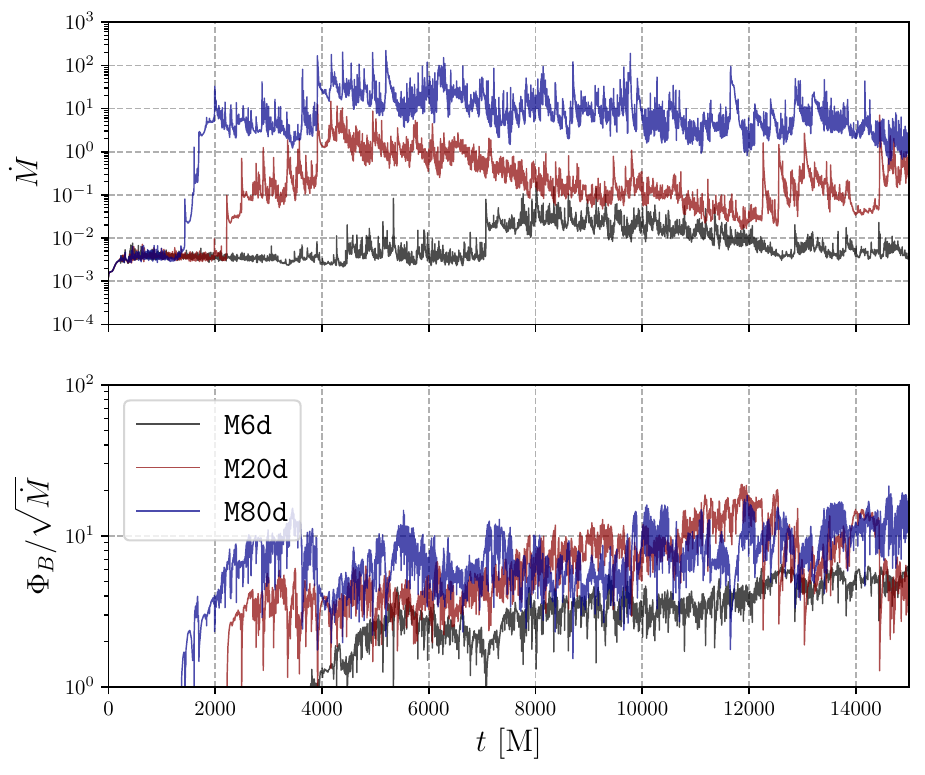}
    \caption{The same diagnostics as Figure~\ref{fig: high_spin_alternating}, now shown for the non-spinning multi-loop runs. The upper panels correspond to alternating polarity, and the lower panels correspond to same polarity.}
     \label{fig:Mdot_zero_alt}
 \end{figure}
Figure~\ref{fig:Mdot_zero_alt} indicates that, except for the large-loop models M80c and M80d, the histories of both mass accretion rate and magnetic flux in the non-spinning runs remain close to those in the spinning cases. Overall, black-hole spin does not strongly modify the accretion process inside the torus itself. Its influence is instead much more pronounced in the jet power and in the jet morphology \cite{2019ApJ...875L...5E, Nathanail2021}. For the short-wavelength configurations, the torus structure stays very similar to that seen in the rotating-black-hole simulations. As in the rotating alternating-polarity models, the horizon-threading field in M6c and M20c flips sign rapidly, so maintaining a persistent jet is difficult; the outcome is instead a weak, transient, and frequently one-sided outflow. By contrast, the same-polarity models M6d and M20d produce weaker but more sustained outflows. In these runs, MRI-driven turbulence dissipates magnetic flux inside the torus. At the horizon, however, flux with the same sign continues to accumulate, raising the funnel magnetization to a modest value. The resulting magnetic pressure is then sufficient to launch jets in those cases. For the alternating-polarity non-spinning models, the funnel never reaches the same magnetization level because loops of opposite sign keep accreting through that region.

For the longer-wavelength loop models, M80c and M80d, the behavior departs somewhat from that of the spinning runs. Figure~\ref{fig:Mdot_zero_alt} shows that their mass-accretion and magnetic-flux histories still broadly resemble those of the rotating large-loop models (see Figures~\ref{fig: high_spin_alternating} and \ref{fig:Mdot_high_sam}). Even so, neither non-spinning model reaches the MAD state, independent of the polarity pattern. This points to black-hole spin as a crucial ingredient for efficient field accumulation on the horizon. Through frame dragging, a rotating black hole effectively strengthens the field near the hole and raises the horizon magnetization above the non-spinning level. That extra amplification appears to be required for the long-wavelength multi-loop models to enter the MAD state.

\section{Flux-Rope and Plasmoid Formation}


In our simulations, short-wavelength multi-loop configurations produce accretion flows that remain SANE-like even though the initial torus is comparatively large. MRI-driven turbulence, together with reconnection inside the torus, creates a highly tangled magnetic structure. That disordered field geometry causes abrupt polarity reversals and favors one-sided jet activity (see Figure~\ref{fig: M20a_2D_magnetization}). The same overall picture was also reported by \cite{Nathanail2021}. When loops with opposite polarity come into contact, reconnection dissipates magnetic energy and triggers plasmoid formation. In addition to direct reconnection, turbulent plasma instabilities can also produce plasmoids, especially Kelvin-Helmholtz (KH) and tearing modes \cite{2022ApJ...929...62B}. Earlier work has already examined the connection between plasmoid formation and these instabilities \cite[e.g.,][]{Loureiro2012, Ni2017}. KH modes create plasmoids at the shear interface between the funnel and the sheath, whereas tearing modes arise in thin extended current sheets and generate plasmoid chains. \cite{Ripperda2021} discussed this chain formation in the equatorial plane of MAD flows. We find an analogous plasmoid-chain behavior during the MAD phase of our large-loop models. I now turn to a more detailed analysis of the plasmoid properties.

\begin{figure}
    \centering
	\includegraphics[width=0.7\columnwidth]{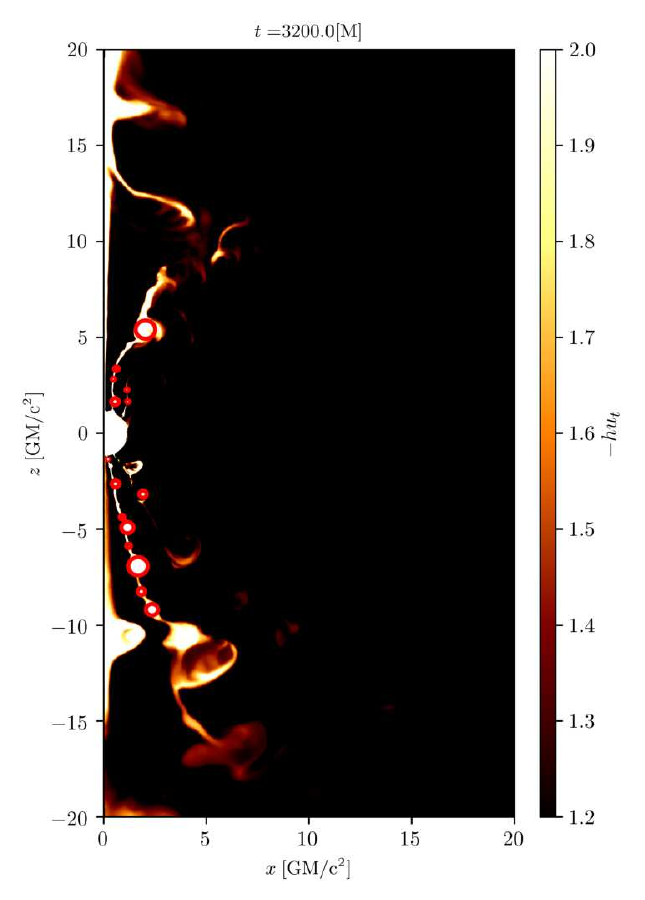}
    \caption{Distribution of the geometric Bernoulli parameter $-hu_{\rm t}$ in the rotating alternating-polarity model M50a with $\lambda_{\rm r}=50$ at $t=3200\,M$. Red circles indicate the detected plasmoids.}
    \label{fig: case_M50a_hut}
\end{figure}

\begin{figure*}
	\includegraphics[width=0.8\textwidth]{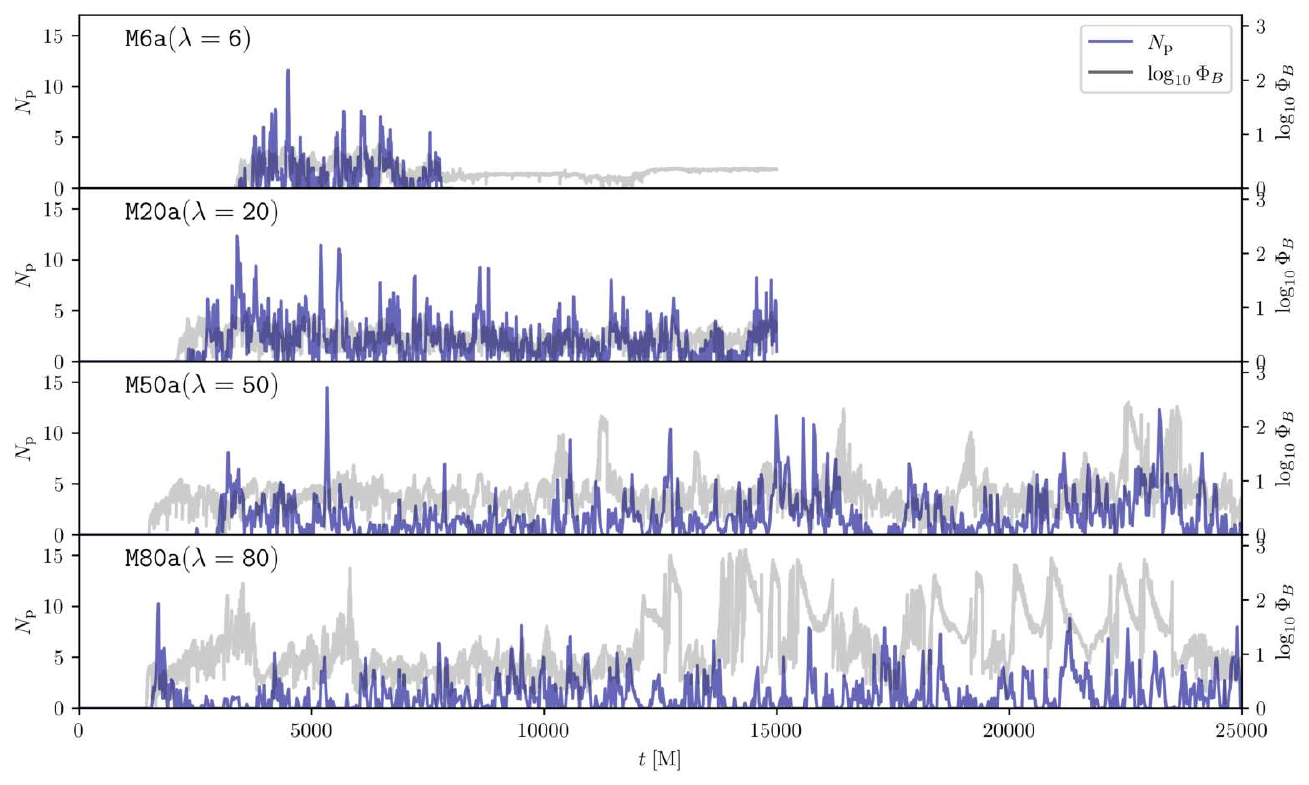}
    \caption{Time evolution of the plasmoid count for rotating black-hole models with alternating polarity and different loop wavelengths. The blue curves show the plasmoid number, while the gray curves show the magnetic-flux rate $\Phi_{\rm B}$.}
    \label{fig: plasmoid_amount_alt}
\end{figure*}
\begin{figure*}
	\includegraphics[width=0.8\textwidth]{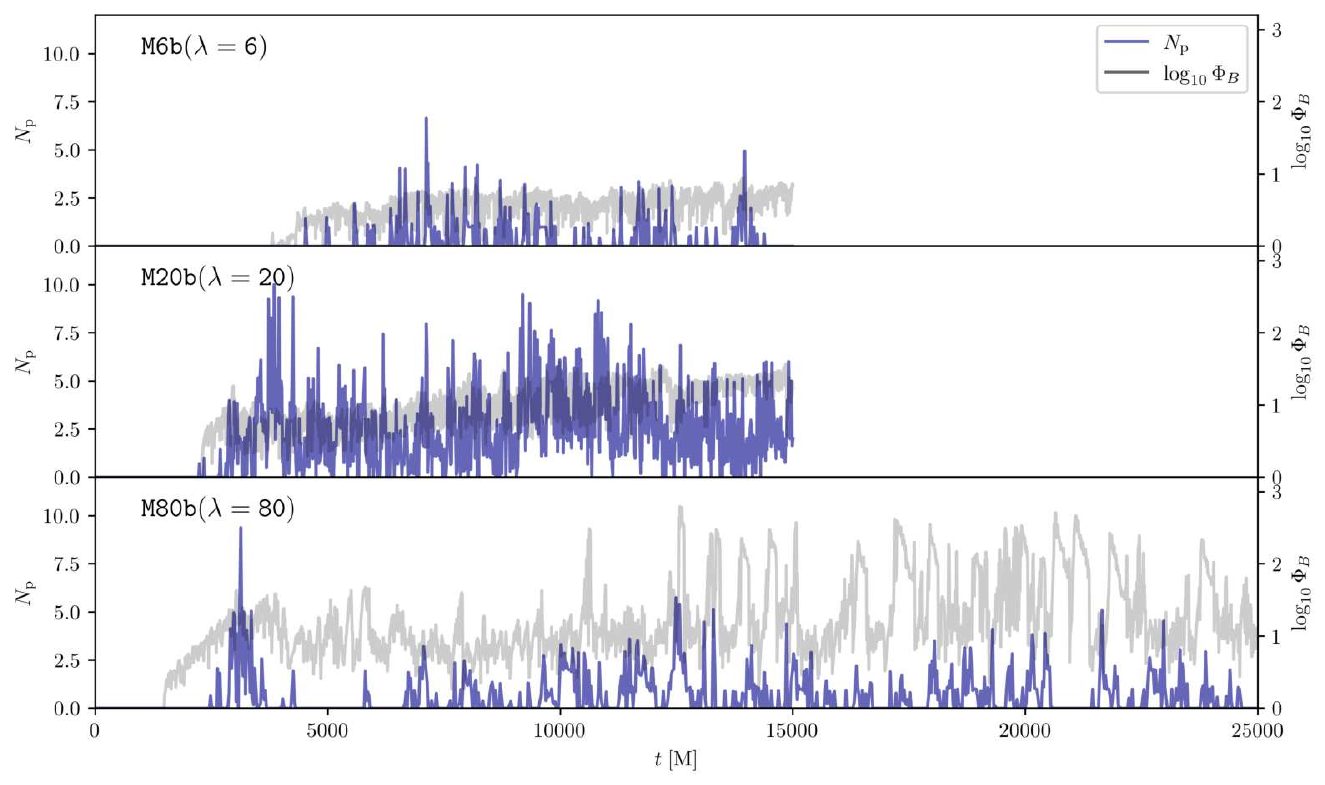}
    \caption{The same diagnostic as Figure~\ref{fig: plasmoid_amount_alt}, but now shown for rotating models with same-polarity magnetic loops.}
    \label{fig: plasmoid_amount_sam}
\end{figure*}

\begin{figure*}
    \centering
	\includegraphics[height=.6\textwidth]{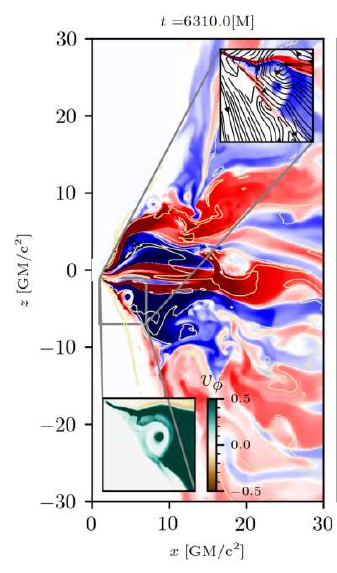}
	\includegraphics[height=.6\textwidth]{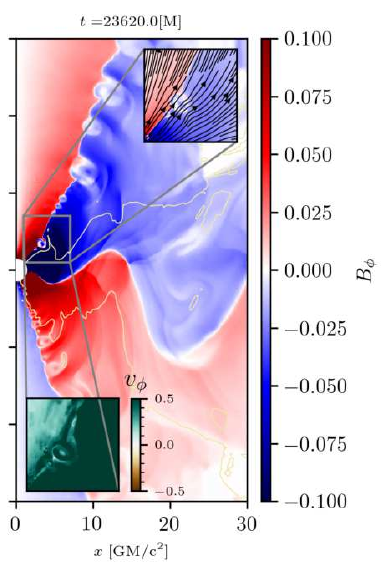}
    \caption{Toroidal-field maps for two representative cases: M80d at $t=6310\,M$ in the left panel and M80a at $t=23620\,M$ in the right panel. The zoomed regions highlight plasmoids produced by different mechanisms. Streamlines in the zoomed panels trace the poloidal fluid velocity, and the green-brown color scale gives the toroidal velocity. The left panel illustrates plasmoid formation involving both KH instability and reconnection. The right panel shows plasmoids generated by tearing instability in the current sheet. Yellow contours in each panel mark the transition between gravitationally bound and unbound regions through the geometric Bernoulli parameter $u_{\rm t}=-1$.
    }
    \label{fig: KHI and reconnection}
\end{figure*}

To analyze individual plasmoids, an objective identification procedure is required. I follow the strategy introduced by \cite{Nathanail2020}. At each stored snapshot, the scikit-image tools \cite{VanDerWalt2014} are used to locate coherent structures and measure their positions and sizes within a search domain spanning 0--20 M in radius and $-20$--20 M vertically. Snapshots are sampled every 10 M in time. Because the focus here is on jet plasmoids, I apply a Bernoulli cut based on $-hu_{\rm t}$ and retain only cells with $-hu_{\rm t}>1.02$, corresponding to unbound material. This filter removes the torus and most of the slowly moving disk wind. Figure~\ref{fig: case_M50a_hut} shows one example for model M50a at $t=3200\,M$, where the jet plasmoids stand out clearly. By construction, this procedure rejects most plasmoids produced by equatorial tearing because those structures usually remain bound. For each detected structure, I then estimate its centroid and characteristic radius. To avoid counting marginal features set by the grid scale, only plasmoids with radii larger than $1/30\,M$ are kept. The retained sample therefore emphasizes the larger and more energetic plasmoids, which are also the ones most likely to leave observable flare-like signatures.

The rate at which plasmoids appear depends on how often reconnection and KH instability are triggered along the jet-sheath interface. Figures~\ref{fig: plasmoid_amount_alt} and \ref{fig: plasmoid_amount_sam} show the time history of the plasmoid count $N_{\rm p}$ for the rotating-black-hole runs with alternating and same loop polarity, while the horizon magnetic flux $\Phi_{\rm B}$ is plotted as a light-gray reference curve. All rotating-black-hole models launch appreciable jets despite their different initial magnetic topologies. That makes plasmoid production efficient: a fast funnel outflow meets sheath inflow across a sharp shear layer, which encourages KH growth, while the multi-loop field geometry also builds strong current sheets in the sheath. Reconnection and tearing in that region therefore provide the main channel for plasmoid production.



The time evolution of the alternating-polarity sequence shows that the long-loop cases can pass from SANE into MAD, whereas the short-loop cases remain in the SANE state throughout the run. In the short-loop models M6a and M20a, the initially ordered field pattern is disrupted soon after the start of the simulation. In model M20a, for example, the second panel of Figure~\ref{fig: plasmoid_amount_alt} shows that ongoing reconnection maintains roughly 5--10 plasmoids for most of the run. The same qualitative behavior is also seen in M6a. The third and fourth panels of Figure~\ref{fig: plasmoid_amount_alt} show that, during the early stage ($t\le4000\,M$), the horizon magnetic flux rises and then falls again. This increase-decrease cycle is clearly visible at least twice. Whenever the flux curve turns from rising to declining, the polarity of the near-horizon field begins to reverse, which triggers strong magnetic reconnection and a burst of plasmoid production. Owing to the intermediate size and stronger magnetic field of the loops in model M50a, about 15 plasmoids are formed near $t\sim3000\,M$, which is the largest number of plasmoids among the alternating-polarity cases. Later, as the system approaches the MAD phase, reconnection becomes less frequent and only a small number of plasmoids is produced. The reason is that MRI continues to sustain loop structure inside the torus even at late times, so some reconnection between small-scale loops still persists. Model M80a retains more large-scale loops and therefore sustains enhanced plasmoid formation for a longer interval. Between $t=7000$ and $12000\,M$, the torus field becomes strongly tangled and reconnects frequently, which leads to abundant plasmoid production. After $t=12000\,M$, the MAD stage begins. In that phase, reconnection occurs less often, just as in M50a. The larger-scale field suppresses the kind of continuous plasmoid generation driven by rapid repeated reconnection. Even so, occasional polarity reversals still occur in the funnel, and those events continue to produce some plasmoids; in the time-history curve, they appear as the remaining spikes in the plasmoid count.
%

\begin{table*}
\centering
\begin{tabular*}{0.85\textwidth}{@{\extracolsep{\fill}}llllll}
\hline
\multicolumn{1}{l}{Case}             &Average Range & Tur, $R_{\rm low}$ & Tur, $R_{\rm high}$ & Rec, $R_{\rm low}$ & \multicolumn{1}{l}{Rec, $R_{\rm high}$} \\ 
\hline
$\tt M0$          & 6000-10000\,M & 1.59  & 6.66  & 2.62  & 6.69  \\
$\tt M6a$         & 5000-9000\,M & 2.11  & 3.63  & 1.94  & 3.87   \\
$\tt M20a$        & 8000-12000\,M & 1.03  & 4.24  & 1.45  & 3.76  \\
$\tt M80a$ (SANE) & 6000-10000\,M & 1.05  & 2.61  & 2.54  & 4.24  \\
$\tt M80a$ (MAD)  & 20000-24000\,M & 2.00  & 5.00  & 2.42  & 3.46  \\
$\tt M6b$         & 5000-9000\,M & 2.54  & 4.78  & 1.11  & 5.93  \\
$\tt M20b$        & 6000-10000\,M & 0.87  & 4.70  & 2.16  & 5.28  \\
$\tt M80b$ (SANE) & 6000-10000\,M & 0.96  & 2.45  & 2.51  & 4.09  \\
$\tt M80b$ (MAD)  & 18000-22000\,M & 1.25  & 5.35  & 2.36  & 4.07  \\
\hline
\end{tabular*}
\caption{Best-fit values of $R_{\rm low}$ and $R_{\rm high}$ for the turbulent-heating and reconnection-heating models.}
\label{Table: R_beta_fitting}
\end{table*}
\begin{figure*}
    \centering
	\includegraphics[height=.45\textwidth]{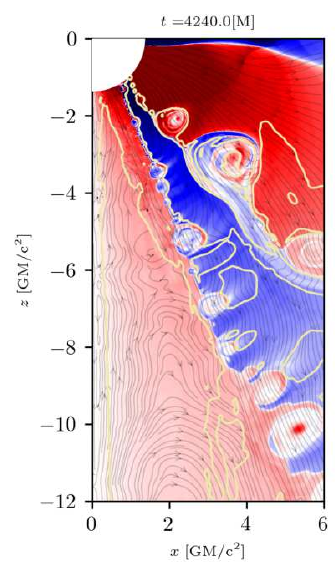}
	\includegraphics[height=.45\textwidth]{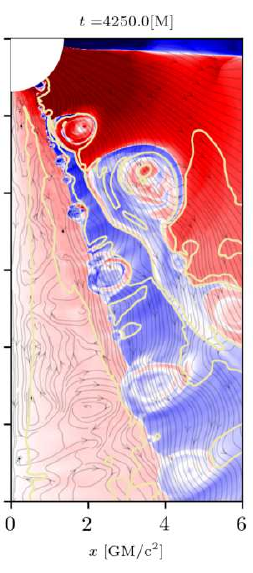}
	\includegraphics[height=.45\textwidth]{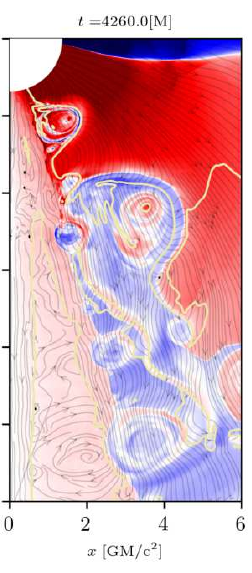}
	\includegraphics[height=.45\textwidth]{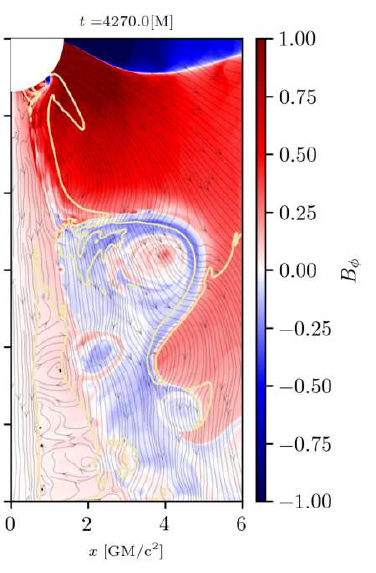}
    \caption{Time sequence of the near-horizon toroidal-field structure in the high-resolution run of model M80a. Streamlines trace the poloidal velocity, and yellow contours mark the boundary between bound and unbound plasma through the Bernoulli condition $u_{\rm t}=-1$. The sequence highlights the onset of tearing in the current sheet and the subsequent emergence of plasmoids.
    }
    \label{fig: tearingI}
\end{figure*}

Figure~\ref{fig: plasmoid_amount_sam} shows how the plasmoid number evolves in the same-polarity sequence, which differs substantially from the alternating-polarity sequence. In model M6b, the short loops lead to early dissipation between neighboring structures inside the torus. Even so, that reconnection does not generate many plasmoids. Only after the magnetic field has been amplified sufficiently does noticeable plasmoid formation appear, around $t=8000\,M$. Once the early dissipation stage is over, the plasmoid-production rate declines again. Model M20b follows the same overall trend, but because its magnetic field is stronger, it is harder to dissipate completely. Continuous reconnection within and between loops therefore creates plasmoids more steadily. As discussed above, dissipation is weaker in M80b; a few plasmoids still form before the flow turns MAD, but once the MAD phase is established, plasmoid formation is nearly shut off.

The non-rotating models produce far fewer detectable plasmoids than the rotating ones, mainly because most of their plasmoids remain gravitationally bound and are therefore missed by our selection procedure. As discussed above, black-hole spin does not greatly modify the accretion flow itself. It does, however, matter strongly for jet production, since a non-spinning hole has no ergosphere and therefore supports much weaker outflows \cite{Blandford1977}. Reconnection inside the torus is also reduced, which in turn lowers the number of tearing-driven plasmoids.

Among the non-rotating long-loop models, M80c and M80d form plasmoids through different channels. In the alternating-polarity model M80c, the weaker velocity contrast suppresses KH growth, while relatively strong current sheets and reconnection zones arise at the funnel-sheath interface and dominate plasmoid production. Model M80d shows the opposite balance: reconnection in the jet sheath is weaker, but the gradual accumulation of magnetic flux on the horizon strengthens the funnel outflow and therefore enhances KH instability. The plasmoid-production rates in these two models end up being similar, even though the underlying mechanism is different. The smaller-loop non-rotating models follow the same general pattern. Because non-rotating black holes do not amplify the magnetic field through frame dragging, KH instability is significantly weaker in these runs. The lower funnel magnetization also limits the strength of the current sheets. Together, these effects reduce the number of plasmoids relative to the rotating-black-hole models.

Previous work \cite{Ripperda2021, 2009PhPl...16k2102B} has shown that tearing instability requires a sufficiently large Lundquist number in order to grow and generate plasmoids.
\begin{equation}
    S = v_{\rm A} L/\eta_{\rm num} \gtrsim 10^4,
\end{equation}
where $v_{\rm A}\sim c$ is the Alfv\'en speed, $L$ is the current-sheet length, and $\eta_{\rm num}$ is the effective numerical resistivity. In our ideal-GRMHD calculations, the resistivity is purely numerical and scales as $\eta_{\rm num}\propto \Delta x^p$ with $p\approx2$ for the second-order scheme used in BHAC \cite{Porth2017, Ripperda2021}. The current sheets in our simulations are typically longer than $\sim5\,r_{\rm g}$ (see Figure~\ref{fig: tearingI}). Most tearing events occur close to the horizon, where the standard-run spatial resolution is about $0.025\,r_{\rm g}$. These values imply that tearing can develop once the current sheet exceeds a length of roughly $5\,r_{\rm g}$. In the high-resolution run, $\Delta x$ is four times smaller, so the Lundquist number increases by about an order of magnitude and tearing becomes correspondingly stronger. Although the exact numerical resistivity slightly changes the plasmoid-formation rate, it does not alter the main conclusion: tearing instability generates many plasmoids a few gravitational radii from the black hole in the sheath region.

Figure~\ref{fig: tearingI} illustrates how tearing instability develops and generates plasmoids in the high-resolution M80a run. Many small plasmoids first appear in the current sheet close to the black hole ($\lesssim5\,r_{\rm g}$), as seen in the first panel of Figure~\ref{fig: tearingI}. The later panels then show their growth and mergers over time. Most of these plasmoids are unbound and propagate outward. Two current sheets are visible in the sheath region; the smaller plasmoids born in the left sheet expand and merge as the evolution proceeds. The fourth panel captures a merger between two large plasmoids.

\begin{figure}
   \centering
        \includegraphics[width=\linewidth]{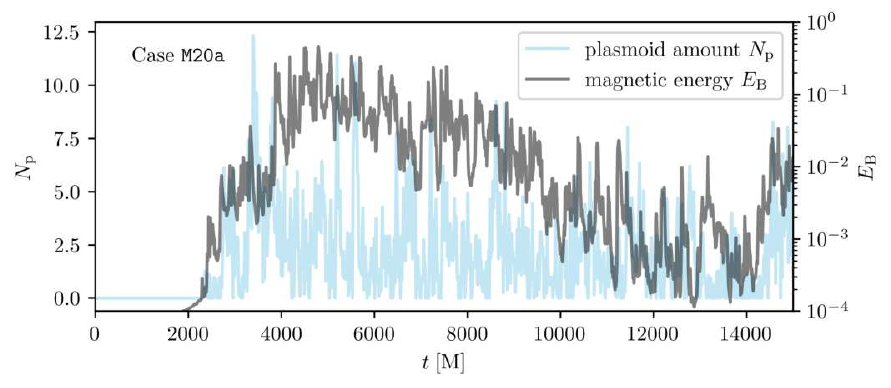}
   \caption{Comparison between the density-weighted mean magnetic energy (black) and the plasmoid count (cyan) for model M20a.
   }
   \label{fig: B_energy}
\end{figure}

Figure~\ref{fig: B_energy} compares the plasmoid count from Figure~\ref{fig: plasmoid_amount_alt} with the density-weighted magnetic-energy measure $E_{\rm B}$ for model M20a, defined as
\begin{equation}
    E_{\rm B} = \frac{B^2\rho}{\bar \rho},
\end{equation}
where $\bar{\rho}$ is the mean density. Since most plasmoids form close to the hole, I evaluate the magnetic energy over the region $x<30\,r_{\rm g}$ and $-30\,r_{\rm g}<y<30\,r_{\rm g}$. MRI first raises the magnetic-energy content between roughly 2,000 and 6,000 M. Near $t=3,500\,M$, an opposite-polarity loop reaches the black hole, violent reconnection begins, and plasmoid production surges. At that moment Figure~\ref{fig: B_energy} shows a sharp drop in magnetic energy together with a peak in the plasmoid number. The same anti-correlation appears again later and further supports the connection between reconnection and plasmoid formation.

Both Kelvin-Helmholtz and tearing instabilities are capable of producing plasmoids \cite{2022ApJ...929...62B}, and our simulations show examples of each. In the left panel of Figure~\ref{fig: KHI and reconnection}, the streamlines wrap around the structure and reveal clear vorticity, identifying a KH-driven plasmoid. Time-dependent inspection also shows that some plasmoids of this type rotate rapidly, and the same region exhibits strong poloidal-velocity shear. The tearing-mediated plasmoids in the right panel of Figure~\ref{fig: KHI and reconnection}, by contrast, arise inside current sheets and do not show the same obvious vortex motion or rapid internal spin. The KH-dominated plasmoids are also cooler than the tearing-dominated ones, so their weaker electron heating should make them less conspicuous observationally.

For spinning black holes, most plasmoids first emerge where the jet core meets the sheath, and the outflow then advects them away from the hole. During that outward transport, they expand and steadily cool. The non-rotating models behave differently: their weaker outflows fail to remove the structures efficiently, so the plasmoids linger inside the sheath instead. The later evolution therefore diverges markedly between rotating and non-rotating cases.

\section{Electron-Temperature Structure}

\begin{figure*}
    \centering
	\includegraphics[height=0.4\textwidth]{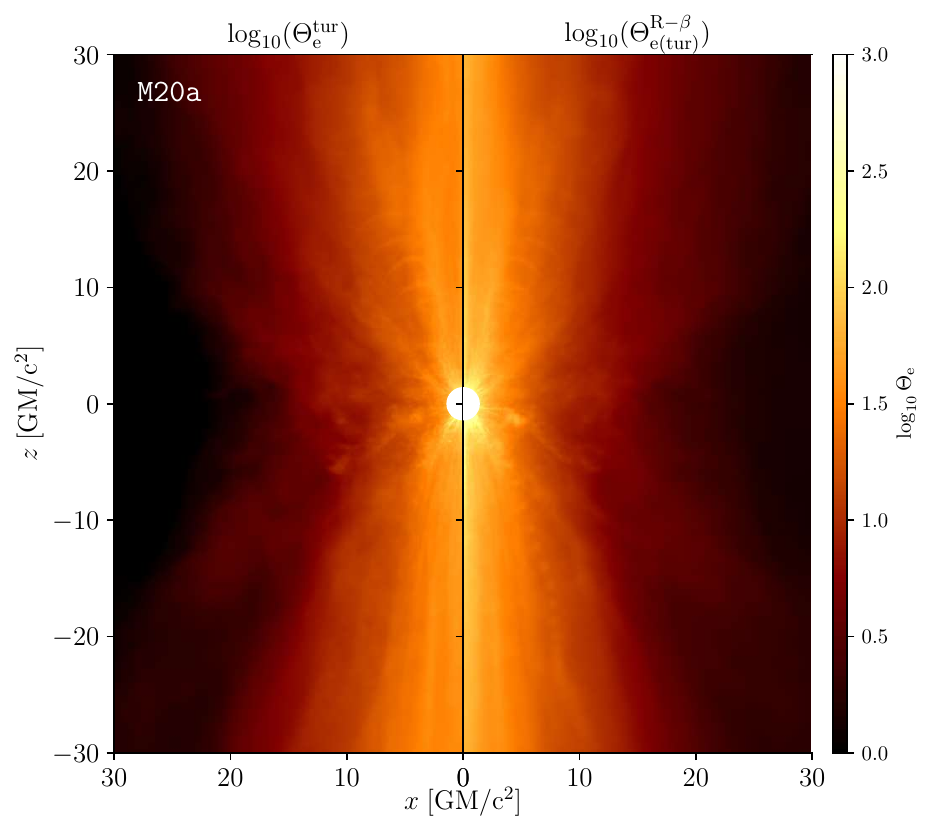}
	\includegraphics[height=0.4\textwidth]{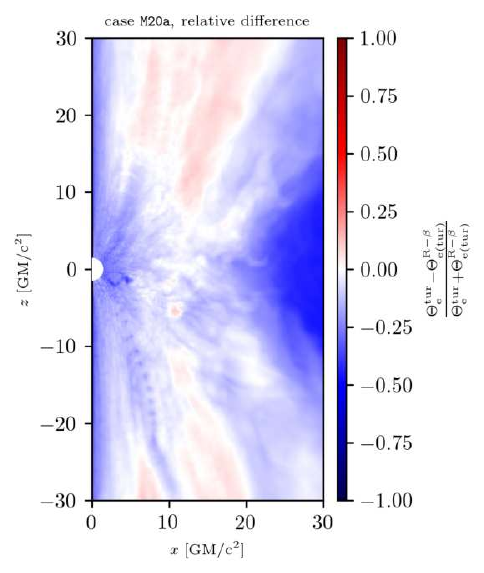}
    \caption{Three views of the mean electron-temperature structure in the rotating alternating-polarity run M20a, each scaled by its own maximum over $t=8000$--$12000\,M$. The left panel displays the two-temperature turbulent-heating solution, the middle panel shows the matched $R-\beta$ reconstruction with best-fit parameters $R_{\rm low}=1.03$ and $R_{\rm high}=4.24$, and the right panel maps their fractional difference.}
    \label{fig: averaged_electron_temperature}
\end{figure*}

\begin{figure*}
    \centering
         \includegraphics[width=\linewidth]{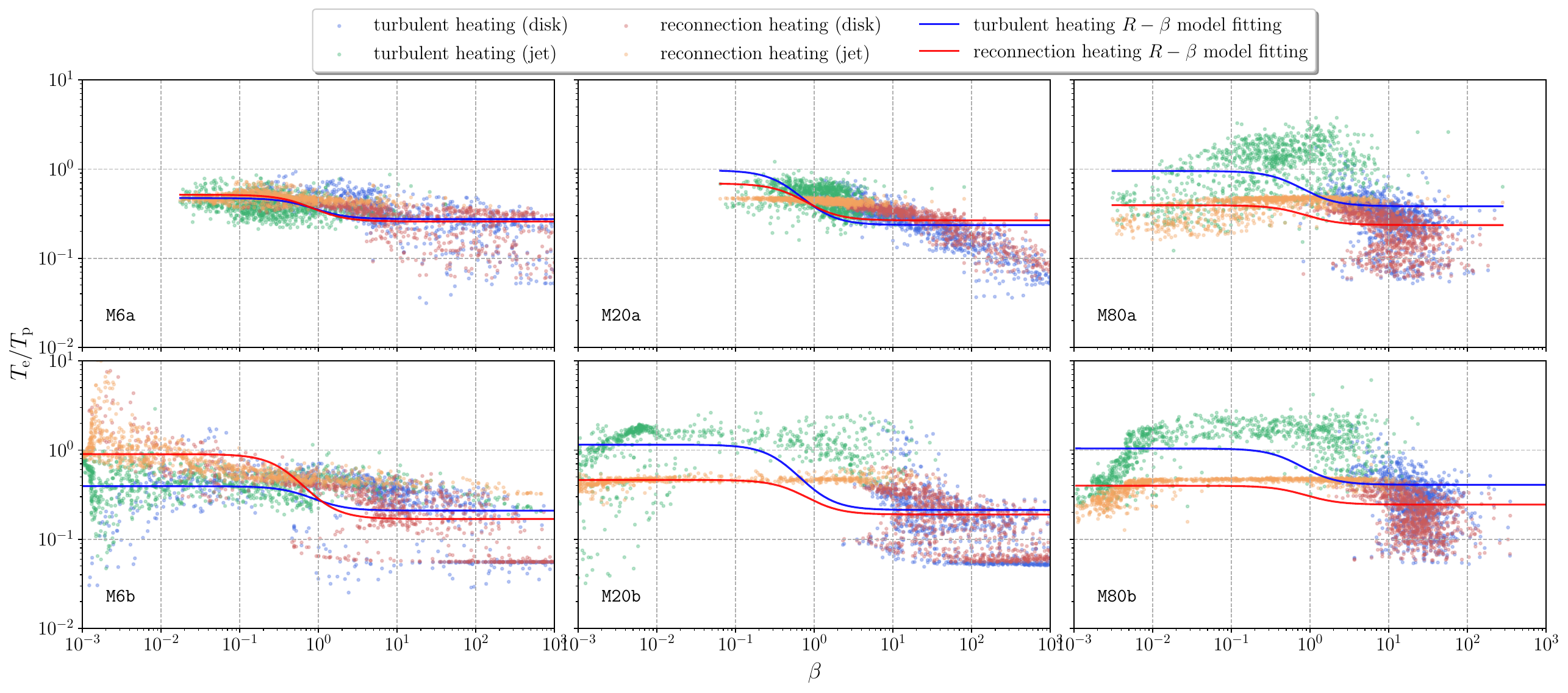}
    \caption{$T_{\rm e}/T_{\rm p}$ versus $\beta$ for rotating multi-loop models with different loop wavelengths. The upper row corresponds to alternating-polarity runs, and the lower row to same-polarity runs. For turbulent heating, disk and jet cells are plotted in red and blue; for reconnection heating, the corresponding regions appear in green and orange. Each sample is averaged over a 4000 M window within $r\le100\,M$, with the exact intervals listed in Table~\ref{Table: R_beta_fitting}. For M80a and M80b, which traverse both SANE and MAD stages, the SANE interval is used here. Solid blue and red curves mark the best-fit $R-\beta$ relation.}
    \label{fig: TeTp-beta-c2}
\end{figure*}
\begin{figure*}
    \centering
         \includegraphics[width=\linewidth]{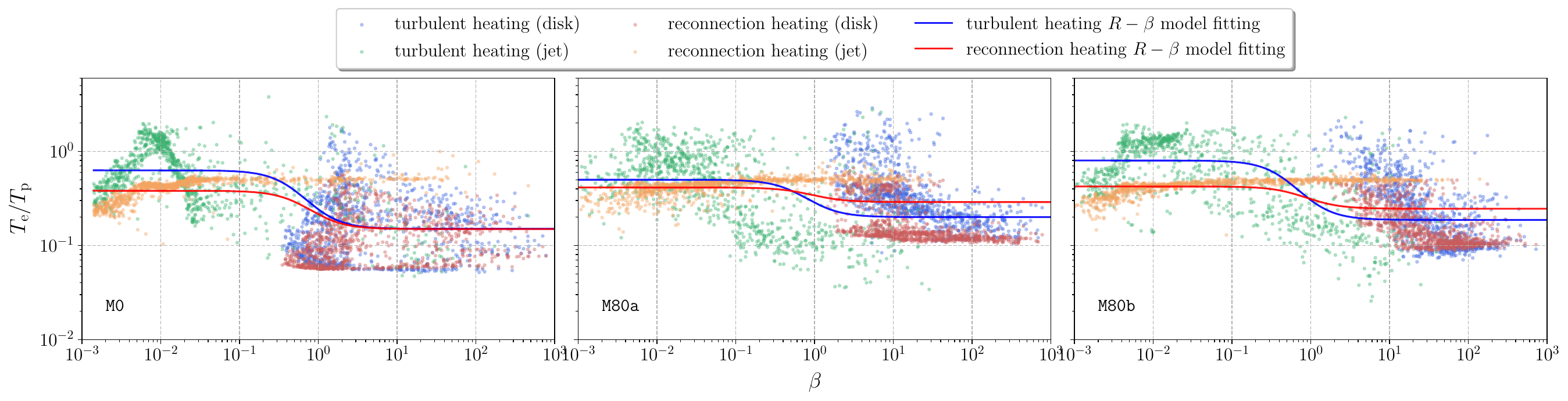}
    \caption{MAD-stage version of the diagnostic shown in Figure~\ref{fig: TeTp-beta-c2}. The left, middle, and right panels correspond to the single-loop reference model M0, the alternating-polarity multi-loop model M80a, and the same-polarity multi-loop model M80b.}
    \label{fig: TeTp-beta_MAD}
\end{figure*}

The quantity that most directly connects the plasma dynamics to the observable emission is the electron temperature. In the two-temperature GRMHD calculations, that field is evolved self-consistently instead of being introduced only during post-processing. Figure~\ref{fig: averaged_electron_temperature} contrasts the time-averaged turbulent-heating temperature map with the corresponding $R-\beta$ reconstruction derived from the same simulation and includes the residual between the two descriptions. The fitted $R-\beta$ prescription has the form
\begin{equation}
    \frac{T_{\rm p}}{T_{\rm e}} = \frac{1}{1+\beta^2}R_{\rm low}+\frac{\beta^2}{1+\beta^2}R_{\rm high},
    \label{Eq: R-beta-c2}
\end{equation}
where $R_{\rm low}$ and $R_{\rm high}$ are free parameters. For this comparison, I use the best-fit values obtained directly from the $T_{\rm e}/T_{\rm i}$ distributions measured in the two-temperature simulations themselves (Figure~\ref{fig: TeTp-beta-c2}). In a broad sense, the two methods agree: both place the hottest material in the funnel and therefore imply brighter jet emission than disk emission. The difference appears in the small-scale structure. The turbulent-heating map contains compact, patchy hot features, whereas the $R-\beta$ reconstruction is smoother and more extended. This indicates that the self-consistent two-temperature treatment preserves localized hot current sheets and related high-temperature substructure more faithfully. On a cell-by-cell basis, turbulent heating gives larger normalized electron temperatures in the funnel and the shear layer, while the $R-\beta$ prescription gives higher values across much of the disk. Because the time-averaged turbulent- and reconnection-heating maps are qualitatively similar, Figure~\ref{fig: averaged_electron_temperature} shows only the turbulent-heating example.

\begin{figure*}
    \centering
	\includegraphics[height=0.4\textwidth]{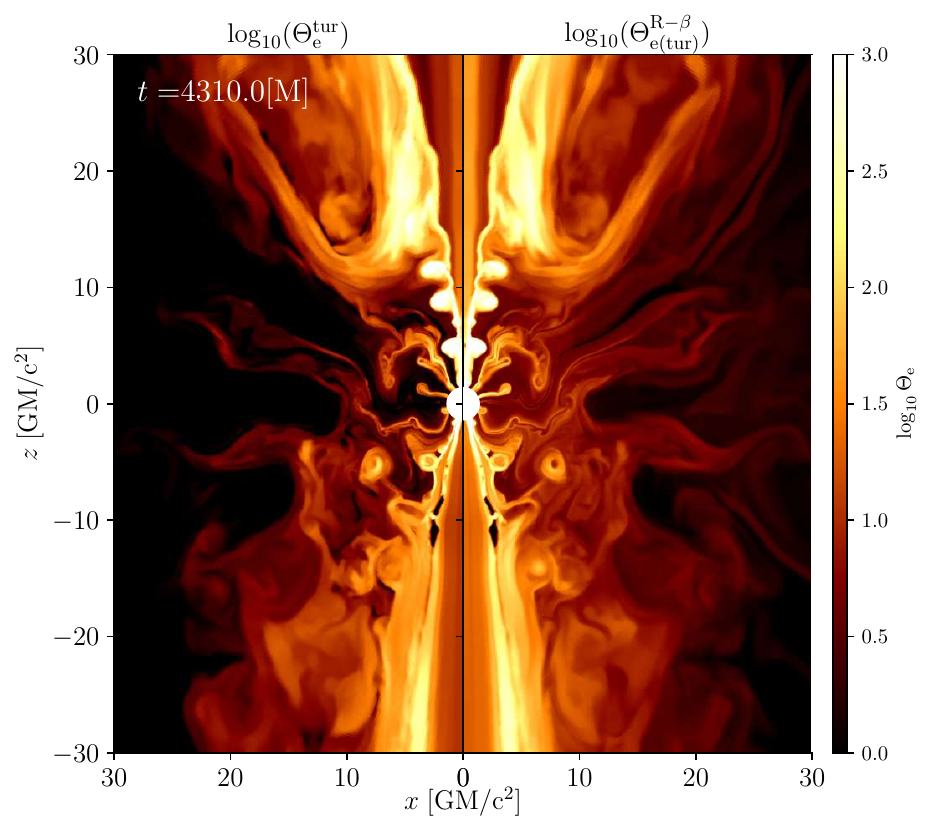}
	\includegraphics[height=0.4\textwidth]{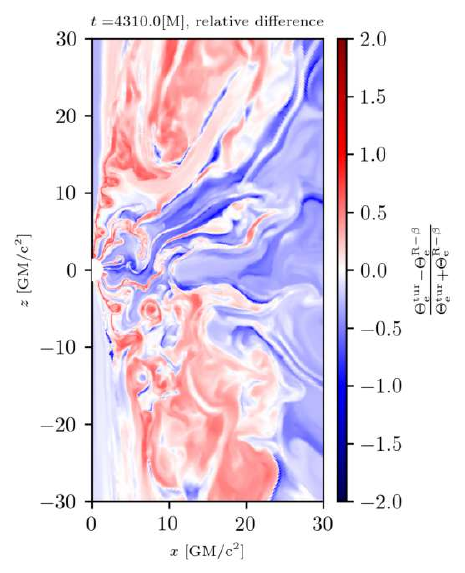}
    \caption{The same layout as Figure~\ref{fig: averaged_electron_temperature}, but shown for a single snapshot of the rotating alternating-polarity model M20a at $t=4310\,M$.}
    \label{fig: plasmoid electron temperature}
\end{figure*}


Figure~\ref{fig: TeTp-beta-c2} plots the relation between the electron-to-ion temperature ratio, $T_{\rm e}/T_{\rm i}$, and the plasma beta during the SANE stage of the multi-loop models. Each panel uses data averaged over a 4000 M window, with the corresponding intervals listed in Table~\ref{Table: R_beta_fitting}. For both turbulent and reconnection heating, the fitted $R-\beta$ relation reproduces the point clouds reasonably well. Even so, the detailed trend of $T_{\rm e}/T_{\rm i}$ with beta still depends on both the magnetic topology and the black-hole spin. In broad terms, lower-$\beta$ regions host larger electron and ion temperatures, and in some runs the electron temperature even exceeds the ion temperature at the smallest beta values.

In single-fluid GRMHD post-processing, the $R-\beta$ prescription is commonly used under the simplifying assumption $T_{\rm e}\approx T_{\rm i}$ when electron temperatures are inferred. Within that framework, $T_{\rm e}/T_{\rm i}=1$ marks the maximum value allowed by the prescription. These largest ratios occur in the strongly magnetized funnel, but that region is also the part most affected by numerical floor treatment, so the inferred electron temperature there is likely biased high \cite[e.g.,][]{Mizuno2021}.

Earlier studies \cite[e.g.,][]{Moscibrodzka2016, 2019ApJ...875L...5E, Mizuno2021} typically fixed $R_{\rm low}=1$ and varied only $R_{\rm high}$. Here both parameters are fitted simultaneously through least squares for every model, and the resulting values are listed in Table~\ref{Table: R_beta_fitting}. The preferred ranges, approximately $1<R_{\rm low}<3$ and $5<R_{\rm high}<10$, remain close to those obtained previously for single-loop configurations \cite{Mizuno2021}. This suggests that changes in magnetic topology do not drastically alter the overall electron-heating behavior.

Overall, the turbulent-heating fits favor smaller $R_{\rm low}$ values than the reconnection-heating fits, and this separation becomes even clearer in the MAD state. Figure~\ref{fig: TeTp-beta_MAD} shows the corresponding MAD-phase scatter plots for the single-loop model M0, the alternating-polarity case M80a, and the same-polarity case M80b. In low-$\beta$ gas, which usually traces the funnel, turbulent heating produces larger electron temperatures than reconnection heating. At high beta, the ordering reverses and reconnection heating becomes hotter. This simply reflects the limiting efficiency of the two prescriptions: turbulent heating can approach $f_{\rm e}\sim1$, whereas reconnection heating is capped at 0.5 (Eqs.~\ref{eq2} and \ref{Eq: reconnection heating-c2b}). The contrast can reach roughly a factor of two, especially in the lowest-$\beta$ regions. Observationally, this suggests that turbulent heating should favor brighter jet emission, while reconnection heating should weight the accretion flow more strongly. In reality, both channels likely operate together, so detailed GRRT comparisons may eventually constrain their relative importance. Because radiative cooling is omitted here, the absolute temperatures should still be regarded as somewhat overestimated \cite{Ryan2018, Chael2018, Yoon2020, Dihingia2023}.
 


\subsection{Comparing Two-Temperature and $R-\beta$ Prescriptions in Plasmoids}

As noted above, the two-temperature solution yields hotter funnel electrons than the fitted $R-\beta$ prescription. This mismatch highlights where the self-consistent thermodynamic treatment becomes especially valuable. Figure~\ref{fig: plasmoid electron temperature} further shows that the turbulent-heating map has much stronger contrast than the $R-\beta$ reconstruction. Within the torus, the two-temperature result is cooler and therefore more variable, whereas in the current sheet and inside the plasmoids it produces substantially higher electron temperatures.

\begin{figure}
    \centering
	\includegraphics[height=1\columnwidth]{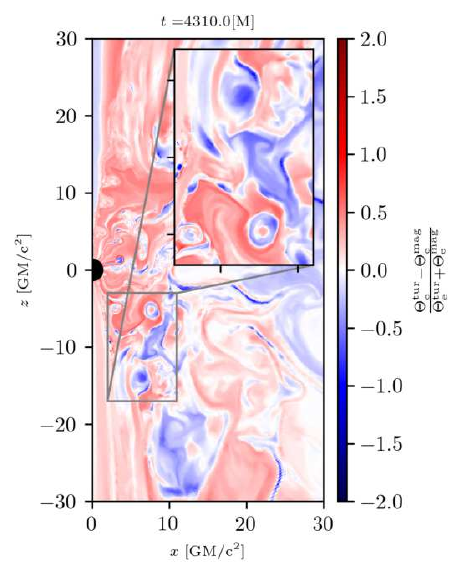}
    \caption{Relative-difference map comparing the electron temperatures produced by the turbulent-heating and reconnection-heating prescriptions in the plasmoid-rich region of the rotating alternating-polarity model M20a at $t=4310\,M$.}
    \label{fig: difference_plasmoid}
\end{figure}

The way plasmoids are heated also depends on which dissipation prescription is adopted. Their internal structure often separates into three zones: a dense magnetized core, a narrow current sheet, and an outer layer with weaker field. This stratification is especially clear for KH-related plasmoids (left panel of Figure~\ref{fig: KHI and reconnection}), whereas tearing plasmoids usually lack such a pronounced dense core. In general, turbulent heating raises the electron temperature more strongly in highly magnetized regions (Figures~\ref{fig: TeTp-beta-c2} and \ref{fig: TeTp-beta_MAD}), so it preferentially heats the funnel and many of the hottest plasmoids, especially those produced by tearing. Figure~\ref{fig: difference_plasmoid}, however, shows that reconnection heating can exceed turbulent heating in KH-related plasmoids. Together with the fitted $R-\beta$ trends, this indicates that turbulent heating dominates in magnetically strong zones such as jets and many plasmoids, whereas reconnection heating is more effective in weakly magnetized regions such as the torus.

\begin{figure}
    \centering
    \includegraphics[width=.6\columnwidth]{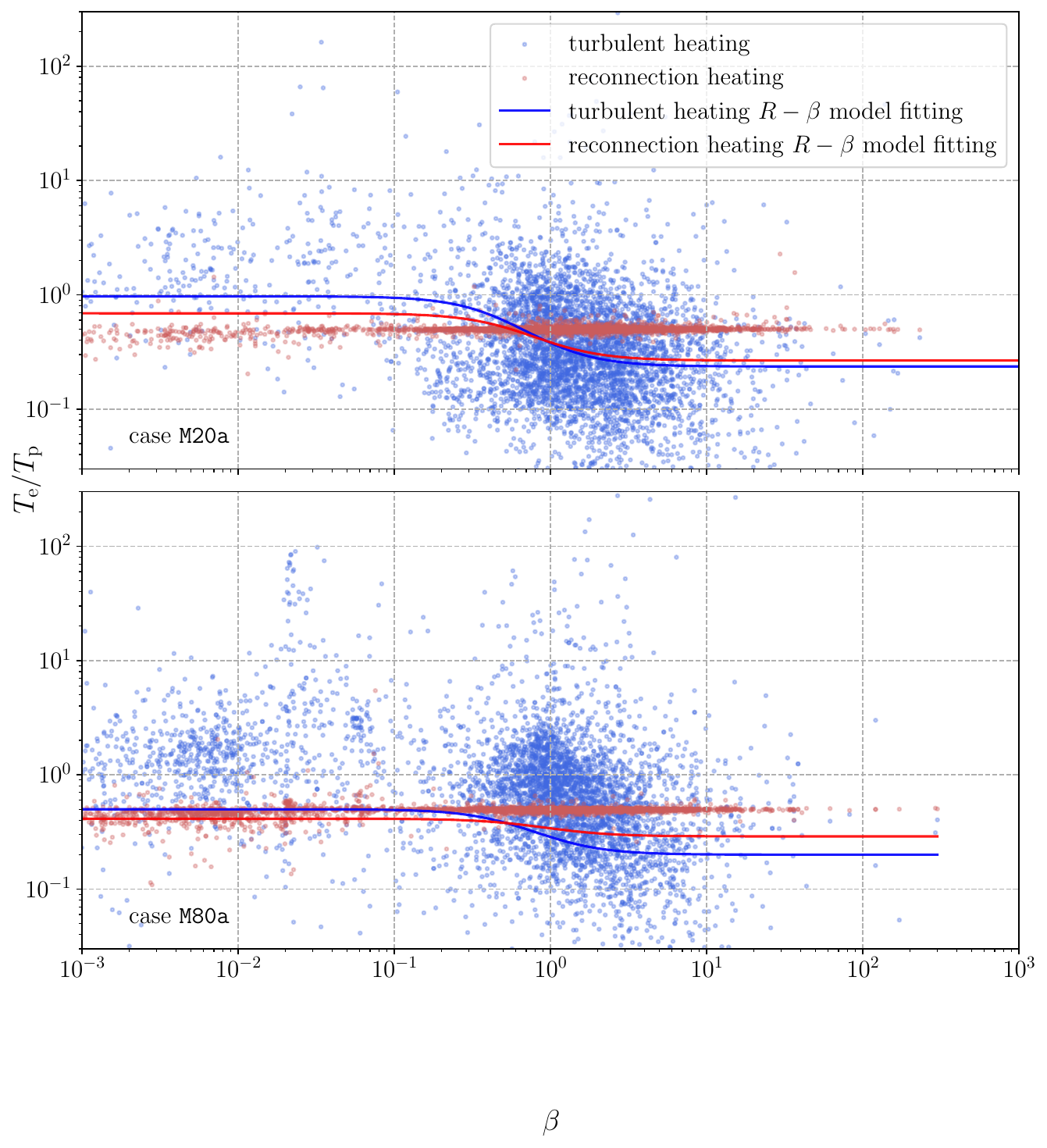}
    \caption{Plasmoid-only version of the $T_{\rm e}/T_{\rm p}$-$\beta$ diagnostic. All plotted cells belong to regions classified as plasmoid material throughout the time series, and the overlaid $R-\beta$ curves use the parameters listed in Table~\ref{Table: R_beta_fitting}.}
    \label{fig: plasmoid_scatter}
\end{figure}
Figure~\ref{fig: plasmoid_scatter} extracts the $T_{\rm e}/T_{\rm p}$-$\beta$ relation only within plasmoid material for the representative models M80a and M20a. In the reconnection-heating case, the points gather close to the upper limit implied by Eq.~\ref{Eq: reconnection heating-c2b}. Turbulent heating instead produces a much broader cloud, which indicates that the electron-heating efficiency varies substantially from one plasmoid to another. The fitted $R-\beta$ curves shown here use the parameters in Table~\ref{Table: R_beta_fitting}, with the MAD-phase fit adopted for M80a. Those curves follow the reconnection-heating branch reasonably well at $\beta<1$, but they do not capture the broader turbulent-heating distribution and also miss the high-$\beta$ portion of the reconnection-heating sample.
\begin{figure}
    \centering
         \includegraphics[width=\linewidth]{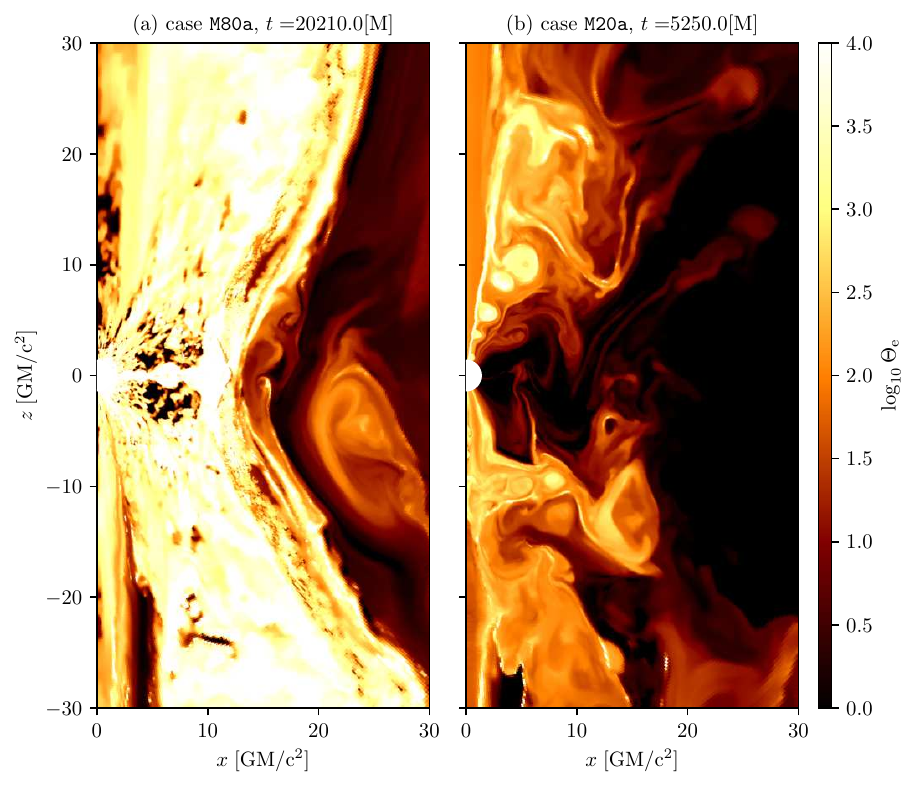}
    \caption{Electron-temperature maps, normalized to the maximum value in each panel, for two contrasting states: the MAD-stage snapshot of M80a at $t=20210\,M$ on the left and the SANE-stage snapshot of M20a at $t=5250\,M$ on the right.}
    \label{fig: jet_plasmoid}
\end{figure}

Figure~\ref{fig: jet_plasmoid} shows that once the flow becomes MAD, the jet reaches electron temperatures $\Theta_{\rm e}$ high enough to outshine individual plasmoids. The plasmoids are still hot, but their radiative imprint is likely to be buried beneath the jet emission. Model M20a behaves differently because its funnel does not contain a comparably powerful jet. There the plasmoids become the hottest structures in the domain and should dominate the emission variability. Future work will compute radiation light curves directly from the multi-loop GRMHD outputs to test that connection more quantitatively.

\section{Discussion and Summary}

Taken together, the 2D two-temperature GRMHD models analyzed here show how thick tori threaded by multiple poloidal loops evolve when the loop scale is varied. For the short-wavelength cases, $\lambda_{\rm r}=6$ (M6a,b,c,d) and $\lambda_{\rm r}=20$ (M20a,b,c,d), none of the simulations settles into a persistent MAD state. In the alternating-polarity sequence, MRI-driven turbulence repeatedly pushes adjacent loops together and triggers strong reconnection within the torus. A substantial fraction of the magnetic flux is then dissipated before it can accumulate on the horizon, and the original topology is rapidly mixed away. In the same-polarity sequence, by contrast, most of the early reconnection remains concentrated near the loop boundaries, after which one sign gradually dominates the surviving field.

The long-wavelength cases begin in a similar way but later move toward MAD-like behavior. In the alternating-polarity runs, the larger loop spacing weakens the MRI-driven dissipation inside the torus, allowing more magnetic flux to survive long enough to reach the horizon. Whenever a loop of opposite sign arrives, reconnection becomes vigorous again and fills the region between neighboring loops with plasmoids. The same process also disrupts the jet, leaving either a one-sided outflow or a weaker striped configuration. Once enough flux reaches the hole, the flow becomes MAD. Even in that phase, however, the multi-loop solutions remain less dense and less magnetized than the single-loop MAD reference model. Plasmoid production decreases after the system becomes strongly arrested, but it rises again when another oppositely directed loop reaches the horizon and reactivates reconnection. The flow can then drift back toward SANE before sufficient flux builds up once more. Over long durations, the system therefore cycles between SANE-like and MAD-like states, with the recurrence time governed mainly by the imposed loop wavelength.

The plasmoid statistics and internal properties also depend strongly on the loop size. In these models, plasmoids are produced mainly by magnetic reconnection together with plasma instabilities, especially tearing and Kelvin-Helmholtz modes. In the rotating-black-hole runs, frame dragging strengthens the funnel field and intensifies reconnection along the jet-sheath interface. For that reason, the short-wavelength cases M6a,b and M20a,b generate such events more often than the long-loop models. The non-rotating runs are less efficient plasmoid producers because reconnection is weaker without frame-dragging amplification, and the lack of a strong jet also reduces the shear that would otherwise encourage KH growth. The calculations directly demonstrate that KH instability and magnetic reconnection can operate together to produce plasmoids \cite{2022ApJ...929...62B}. KH-generated objects are usually bound and have comparatively low electron temperatures. Their rapid rotation reflects the strong shear between neighboring layers, and the surrounding flow develops the expected vortex structure. Reconnection-generated plasmoids, by contrast, emerge from extended current sheets that fragment into chains through tearing.

Electron temperature is equally important because it controls the radiation directly. I therefore compared two self-consistent heating prescriptions, turbulent heating and reconnection heating, and examined the resulting $T_{\rm e}/T_{\rm i}$ distributions. Across the different magnetic topologies, the usual $R-\beta$ parameterization still reproduces both prescriptions fairly well, in line with \cite{Mizuno2021}. When the fits are carried out separately, however, turbulent heating generally favors larger values of both $R_{\rm low}$ and $R_{\rm high}$. The models predict hot electrons in both the funnel and the plasmoids, although the plasmoid temperatures are likely somewhat overestimated because radiative cooling is neglected \cite{Ryan2018, Chael2018, Yoon2020, Dihingia2023}. In cases without a persistent strong jet, the plasmoids should dominate the emission and behave like hot spots close to the black hole, naturally producing flare-like variability similar to that observed in Sgr A$^*$.

Inside the plasmoids, the self-consistent temperatures differ markedly from those produced by the parameterized $R-\beta$ model. In most cases, the $R-\beta$ prescription underestimates the electron temperature inside the plasmoid. The two heating prescriptions also separate clearly from one another: reconnection heating raises the plasmoid temperature more efficiently and produces an almost single-valued $T_{\rm e}/T_{\rm i}$ with only weak beta dependence, whereas turbulent heating leaves a much broader distribution. These differences show that two-temperature simulations retain physical information that cannot always be compressed into a simple parameterized prescription.

Recent work by \cite{Nathanail2021} emphasized that reconnection sites in multi-loop systems are intrinsically three dimensional. In that regime, the funnel field becomes twisted and sheared at the same time, a tendency already hinted at by the 2D result in the right panel of Figure~\ref{fig: jet_plasmoid} but impossible to capture fully without the $\phi$ direction. In three dimensions, the analogous structure should be a flux rope wrapped around the torus axis rather than a strictly two-dimensional plasmoid. A complete characterization of these objects will therefore require future 3D two-temperature GRMHD calculations with multiple poloidal loops.


\newpage
{
\section*{Appendix Materials}  
\setcounter{section}{0}
\renewcommand{\thesection}{\thechapter.\Alph{section}}
\renewcommand{\theHsection}{appendix.\thechapter.\Alph{section}}
\renewcommand{\theHsubsection}{appendix.\thechapter.\Alph{section}.\arabic{subsection}}

\section{Resolution Comparison}

To evaluate how strongly the results depend on numerical resolution, I reran the 2D GRMHD calculation for a rotating black hole threaded by long alternating-polarity loops ($\lambda_{\rm r}=80$) at two grid sizes while keeping all initial conditions fixed. The baseline model uses $(N_r,N_\theta)=(1024,512)$, whereas the refined calculation adopts $(N_r,N_\theta)=(4096,2048)$. Figure~\ref{fig: Mdot_high_low_res} compares the horizon accretion-rate and magnetic-flux histories for the two grids and shows that the overall time evolution is very similar. Figure~\ref{fig: 2D_high_low_res} provides the corresponding comparison of the time-averaged density, velocity, and magnetic-field structure. The finer grid resolves somewhat sharper small-scale features and slightly larger contrast, but the global morphology is nearly unchanged. I therefore conclude that the main accretion-flow properties discussed in this chapter depend only weakly on numerical resolution and are already captured by the lower-resolution setup.

\begin{figure}
    \centering
    \includegraphics[width=0.6\linewidth]{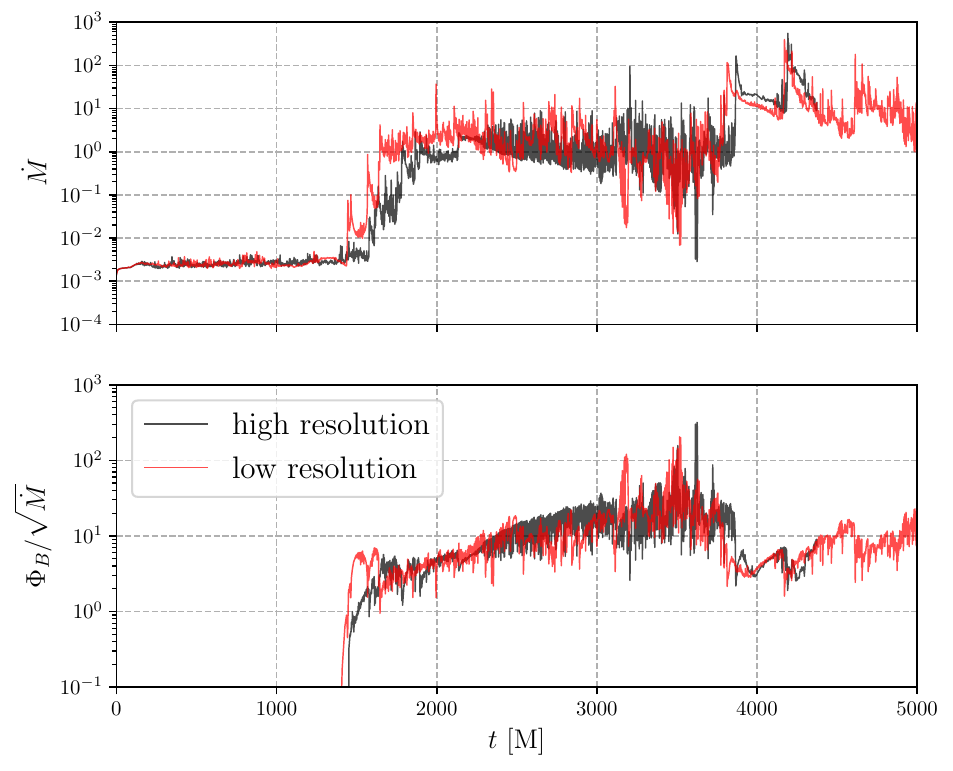}
    \caption{Evolution of the mass-accretion rate (upper panel) and the horizon-threading magnetic flux (lower panel) for the rotating long-loop alternating-polarity model M80a. The refined calculation is shown in black, and the standard-resolution run is shown in red.
    }
    \label{fig: Mdot_high_low_res}
\end{figure}
\begin{figure}
    \centering
    \includegraphics[width=0.6\linewidth]{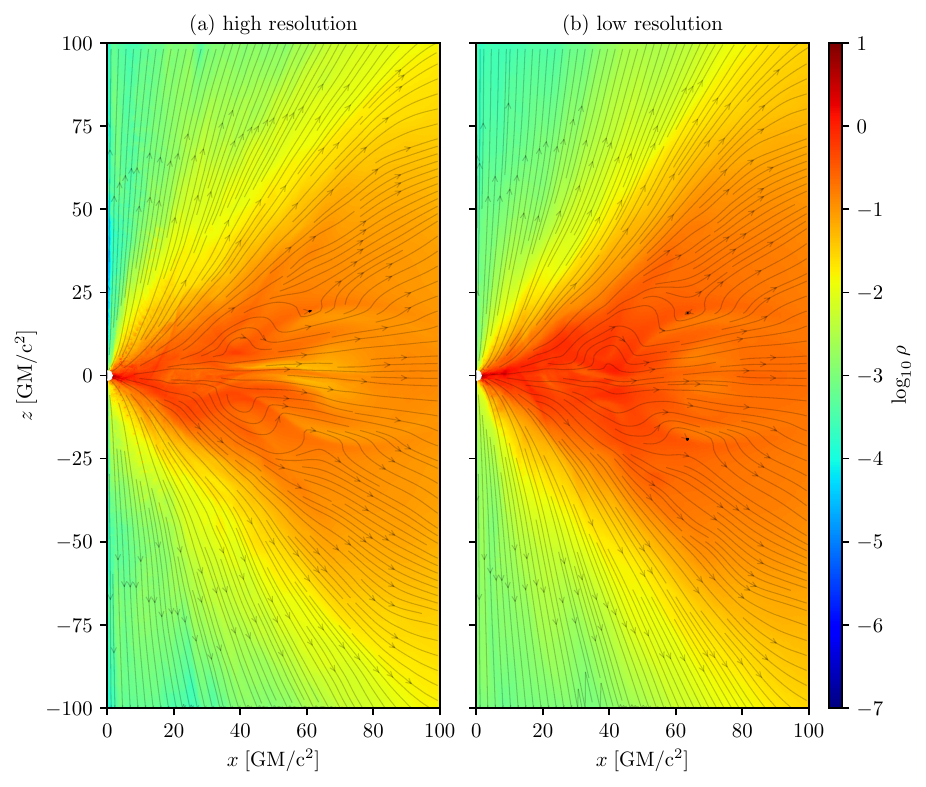}
    \includegraphics[width=0.6\linewidth]{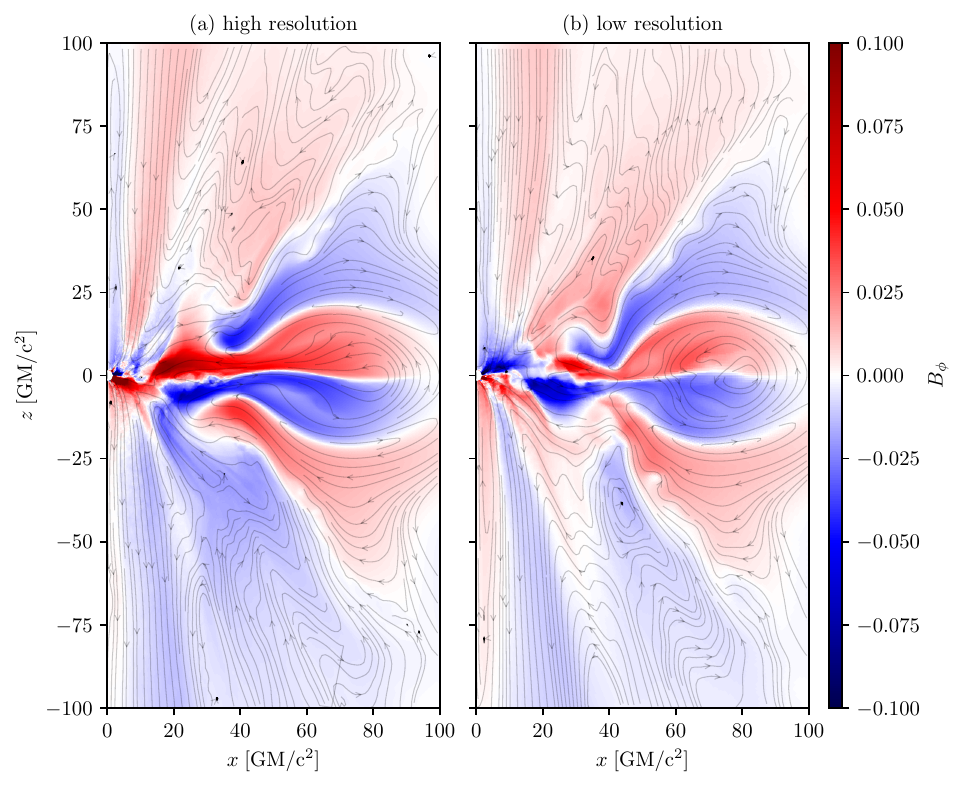}
    \caption{Time-averaged logarithmic density (upper row) and toroidal magnetic field (lower row) for the rotating long-loop alternating-polarity model M80a. The refined run appears on the left, and the standard-resolution run appears on the right. Black streamlines trace the fluid velocity in the upper maps and the poloidal magnetic field in the lower maps.}
    \label{fig: 2D_high_low_res}
\end{figure}
}

\chapter[Flux ropes in multi-loop GRMHD]{Flux-Rope Dynamics and Emission in Three-Dimensional Multi-loop Two-Temperature GRMHD Flows}

An earlier version of this study appeared in A\&A (2024, 688, A82), authored by Jiang H.-X., Mizuno Y., I. K. Dihingia, A. Nathanail, Z. Younsi, and C. M. Fromm \cite{2024A&A...688A..82J}.

\section{Observational Motivation and Physical Context}
Sgr~A$^*$ is the best observed nearby supermassive black hole, yet the physical state of its accretion flow remains uncertain \cite[e.g.,][]{2002apa..book.....F, 2010RvMP...82.3121G, 2012RAA....12..995M}. The EHT campaign removed much of the previously allowed parameter space \cite{2022ApJ...930L..12E, 2022ApJ...930L..13E, 2022ApJ...930L..14E, 2022ApJ...930L..15E, 2022ApJ...930L..16E}, but no currently tested model family reproduces the full observational set simultaneously \cite{2022ApJ...930L..16E}. In those comparisons, MAD solutions often perform better than standard SANE disks, tilted flows \cite{Liska2018}, or wind-fed scenarios \cite{Ressler2020}. Because both Sgr~A$^*$ and M~87$^*$ are LLAGNs \cite[e.g.,][]{Yuan2014, 2008ARA&A..46..475H} accreting at highly sub-Eddington rates \cite{Ho2009, Marrone2007}, while M~87$^*$ is also commonly interpreted as MAD \cite{2019ApJ...875L...5E, Cruz-Osorio2021, Yuan2022}, it is still unclear whether Sgr~A$^*$ belongs to the same accretion class.

The nature of the Sgr~A$^*$ flare hot spot is equally unsettled \cite{2023MNRAS.522.2307J, TheGRAVITYCollaboration2020, Lin2023}. GRAVITY measured NIR motion from a compact region orbiting near the ISCO \cite{GRAVITY2018b}, placing the flare source only a few gravitational radii outside the horizon. Comparable variability also appears in the submillimeter and NIR bands \cite[e.g.][]{Do2019}. During bright episodes, the NIR luminosity can exceed the quiescent level by several factors, and the flux statistics are consistent with a persistent background plus intermittent flares \cite{Abuter2020}. Joint EHT-ALMA polarimetry \cite{2022A&A...665L...6W} is compatible with a vertical magnetic flux rope expelled from an arrested inner flow \cite[e.g.,][]{2020MNRAS.497.4999D, 2021MNRAS.502.2023P, Scepi2022}. One example is the clockwise hot-spot motion reported by \cite{2020MNRAS.497.4999D} in MAD simulations of Sgr~A$^*$. Both EHT image modeling and independent flare fits tend to prefer low inclinations \cite{2019ApJ...875L...5E, VonFellenberg2023}, while polarimetric tomography places the flare motion in a low-inclination plane \cite{Levis}. Flux ropes and plasmoid chains in the jet sheath naturally provide that kind of trajectory \cite{2023MNRAS.522.2307J, Nathanail2020, Nathanail2021, Aimar2023, Mellah2023}.

Earlier GRMHD studies of multi-loop tori already showed that plasmoids form naturally in such flows \cite{Nathanail2020, 2023MNRAS.522.2307J}. Neighboring loops merge, drive turbulence, and repeatedly trigger reconnection. Fully kinetic 3D GRPIC calculations likewise reveal plasmoid chains in the jet sheath \cite{Mellah2023}. Because loop mergers continually regenerate reconnecting current sheets, these systems can combine weak and time-dependent jets with efficient plasmoid production \cite[see][]{Nathanail2020, Chashkina2021, Nathanail2021, 2023MNRAS.522.2307J}. In our previous axisymmetric study \cite{2023MNRAS.522.2307J}, most chains formed where opposite magnetic polarities met, but the imposed symmetry confined them to poloidal structures. \cite{Nathanail2021} later emphasized that realistic chains are intrinsically three dimensional and require the toroidal field to evolve self-consistently. That is the main reason for the present 3D extension. Although the current resolution still misses the smallest structures, the two-temperature treatment yields a more reliable electron temperature and therefore a firmer basis for emission modeling.

Mean-field dynamos can also modify the magnetic topology and may naturally create polarity reversals \cite{DelZanna2022, Mattia2020, Mattia2022}. Earlier simulations \cite{Mizuno2021} nevertheless found that reversals generated inside a MAD flow do not, by themselves, strongly suppress a powerful jet. Limited resolution may contribute to that outcome by weakening the dynamo, which is one reason why explicit dynamo terms are often added \cite[e.g.,][]{Mattia2020, Mattia2022, 2024MNRAS.527.3018Z}. Even very high-resolution MAD simulations still launch strong persistent jets \cite{Ripperda2021}. If the goal is instead to obtain a weaker jet together with frequent sheath reconnection, a deliberately multi-loop magnetic configuration remains the more natural option because it continually injects mixed-polarity plasma into the torus.

Previous work treated these systems in a single-fluid approximation and therefore could not predict the radiation self-consistently \cite{2023MNRAS.522.2307J}. Radiatively inefficient accretion flows require a two-temperature treatment for realistic synchrotron modeling \cite{Ressler2015}. In our earlier study, we used two-temperature GRMHD simulations of magnetized multi-loop tori together with the electron-heating prescriptions of \cite{Rowan2017, Kawazura2019} to examine the electron-temperature structure. That analysis showed that plasmoids and current sheets near the hole can reach temperatures very different from those implied by simple parameterized electron-to-ion temperature-ratio models once the thermodynamics are evolved self-consistently \cite{2023MNRAS.522.2307J}. It also suggested that the flare phenomenology of Sgr~A$^*$ depends on the distinct plasmoid populations produced by different magnetic topologies. Here I extend that program to fully 3D two-temperature GRMHD simulations of multi-loop accretion flows. The models remain mostly in the SANE regime. Although current observations of Sgr~A$^*$ often favor MAD solutions, many SANE models still satisfy the EHT constraints \cite{2022ApJ...930L..16E}. At the same time, none of the single-loop MAD or SANE models tested there reproduces the observed variability \cite{2022ApJ...930L..16E}. Multi-loop flows are therefore attractive because they combine repeated reconnection with more accumulated magnetic flux than a standard SANE torus \cite{Fromm2022}. The goal of this chapter is to determine how loop size changes the accretion dynamics and whether such flows provide a plausible explanation for Sgr~A$^*$. I then compute thermal-synchrotron GRRT observables and compare them with the GRMHD emissivity structure to identify which regions dominate the emission at different frequencies, especially during hot-spot formation and variability episodes.

Section 4.2 describes the numerical setup and initial magnetic configuration. Section 4.3 presents the GRMHD and GRRT results. Section 4.4 summarizes the main conclusions.

\section{Simulation Configuration}

The setup follows our earlier studies \cite{Mizuno2021, 2023MNRAS.522.2307J}, so I summarize only the ingredients needed here. The 3D GRMHD calculations are carried out with {\tt BHAC}, which solves the $3+1$ ideal-GRMHD equations in geometric units ($G=M=c=1$, with the factor $1/\sqrt{4\pi}$ absorbed into the magnetic field) \cite{Porth2017, Olivares2019}. The spacetime is fixed to Kerr, and the self-gravity of the gas is neglected.

Each model begins from a rotationally supported Fishbone-Moncrief hydrodynamic torus \cite{1976ApJ...207..962F}. The adopted parameters are as follows: the inner edge is placed at $r_{\rm in}=20\,r_{\rm g}$ and the pressure maximum at $r_{\rm max}=40\,r_{\rm g}$, where $r_{\rm g}=GM/c^2$ is the gravitational radius. The adiabatic index is fixed at $\Gamma=4/3$ \cite{Rezzolla2013}, and all runs use a black-hole spin of $a=0.9375$. Because the torus mass is negligible relative to the hole, the spacetime remains stationary Kerr. MRI is seeded by applying a 2\% random pressure perturbation inside the torus.

\begin{figure}
    \centering
         \includegraphics[width=.8\linewidth]{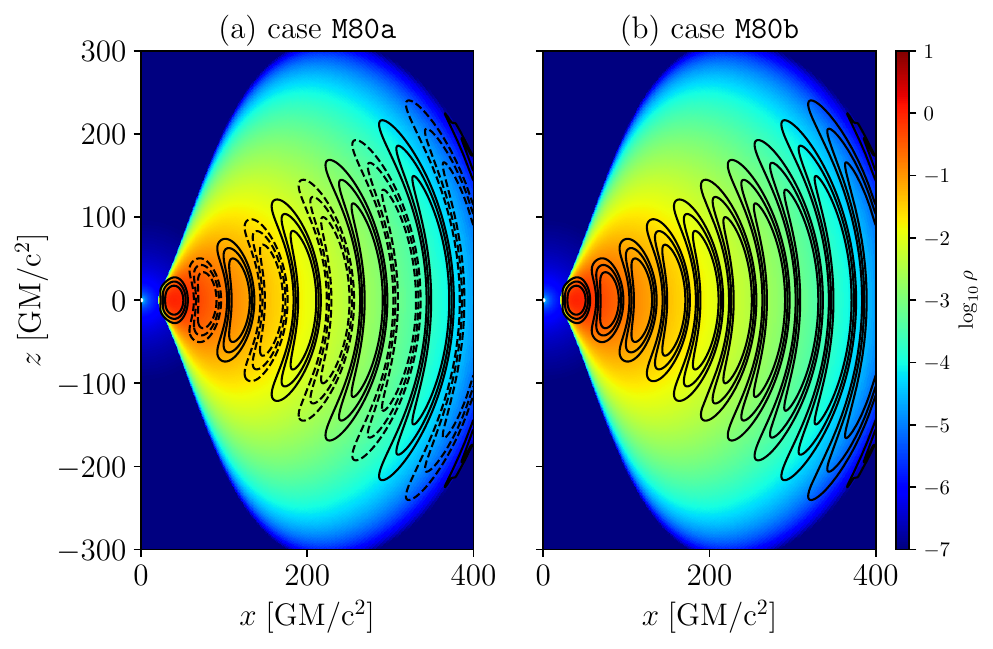}
    \caption{Initial state of the 3D simulations. The color scale shows logarithmic density, while the black contours trace the imposed magnetic loops; solid and dashed curves distinguish opposite polarities.}
    \label{fig: initial condition-c3}
\end{figure}

At the initial time, the magnetic field is purely poloidal and is introduced through the vector potential
\begin{equation}
\begin{aligned}
    A_{\rm \phi}\propto& \exp{(-r/400)}(\rho - 0.01)(r/r_{\rm in})^3\sin^3\theta\\
    &\cos((N-1)\theta)\sin(2\pi(r-r_{\rm in})/\lambda),
\label{eq:A_phi}
\end{aligned}
\end{equation}
where $N=3$ in every model and $\lambda$ controls the radial wavelength of the loop pattern. The overall normalization of $A_{\rm \phi}$ is chosen so that the minimum plasma beta is $\beta_{\rm min}=100$, with $\beta\equiv p_{\rm g}/p_{\rm mag}$. Here $p_{\rm g}$ is the gas pressure and $p_{\rm mag}=b^2/2$ is the magnetic pressure, with $b^2=b_{\mu}b^{\mu}$ and $b^\mu$ the four-magnetic field. Figure~\ref{fig: initial condition-c3} displays the initial density together with the seeded poloidal loops traced by black contours. Solid and dashed curves distinguish the two magnetic polarities. The example corresponds to model M80a3D; model M20a3D uses the same torus but shorter individual loops.

Here I extend two representative models from our earlier 2D study \cite{2023MNRAS.522.2307J}, namely M20a and M80a, to fully 3D simulations. Their three-dimensional counterparts are labeled M20a3D and M80a3D, with $\lambda=20$ and $\lambda=80$, respectively.

In this work, I employ two prescriptions for electron heating: turbulent heating and reconnection heating. The turbulent-heating model, derived from the damping of MHD turbulence, is given by \cite{Kawazura2019}
\begin{equation}
     f_{\rm e} = \frac{1}{1+Q_{\rm i}/Q_{\rm e}},
\end{equation}
where
\begin{equation}
     \frac{Q_{\rm i}}{Q_{\rm e}} = \frac{35}{1 + \left(\beta/15\right)^{-1.4}\exp\left(-0.1 T_{\rm e}/T_{\rm i}\right)}.
\end{equation}
For reconnection heating, we adopt the fitting relation calibrated with PIC simulations \cite{Rowan2017},
\begin{equation}
     f_{\rm e} = \frac{1}{2} \exp\left[\frac{-(1-\beta/\beta_{\rm max})}{0.8+\sigma_{\rm h}^{0.5}}\right],
     \label{Eq: reconnection heating-c3}
\end{equation}
where $\beta_{\rm max}=1/4\sigma_{\rm h}$ and $\sigma_{\rm h}=b^2/\rho h$ is the magnetization defined with respect to the fluid specific enthalpy $h=1+\Gamma_{\rm g}p_{\rm g}/(\Gamma_{\rm g}-1)$. The turbulent and reconnection heating prescriptions are applied simultaneously by evolving the electron entropy for the two cases separately and reconstructing the corresponding electron temperatures afterward. This procedure avoids nonlinear feedback on the underlying GRMHD evolution and allows a direct comparison between the two heating models on the same simulation data.

The computational domain extends to $r=2,500\,r_{\rm g}$. The models are evolved in spherical Modified Kerr-Schild coordinates (see \cite{Porth2017}) with an effective resolution of $384\times192\times192$ over the full $2\pi$ azimuthal range and three levels of static mesh refinement. Model M20a3D is evolved to $t=15,000\,M$, whereas model M80a3D is followed to $20,000\,M$. Both integrations are long enough for the flow to reach a quasi-steady state.

\begin{figure}
    \centering
         \includegraphics[width=.8\linewidth]{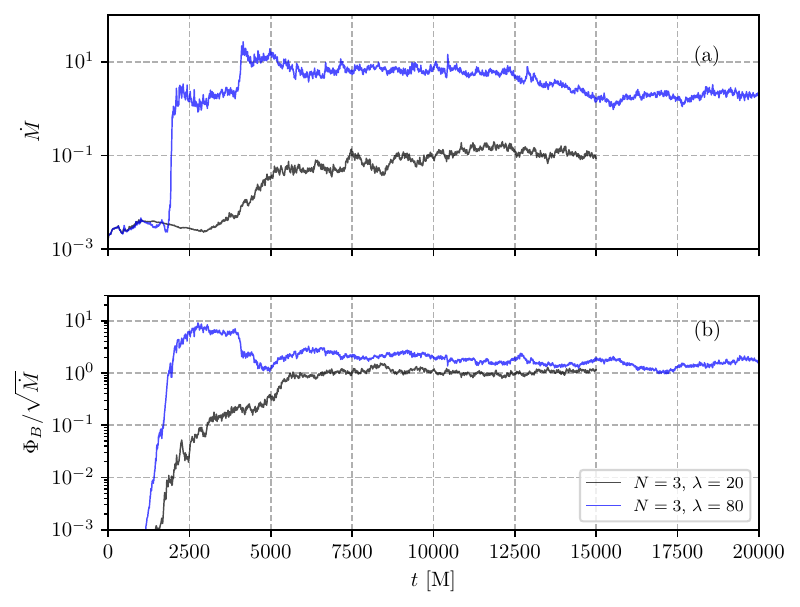}
    \caption{Horizon diagnostics for the two 3D multi-loop models. Panel (a) gives the mass-accretion rate $\dot M$, and panel (b) gives the normalized magnetic flux $\Phi_{\rm B}/\sqrt{\dot M}$. Black curves correspond to M20a3D, and blue curves correspond to M80a3D.}
    \label{fig: Mdot-c3} 
\end{figure}

To obtain synthetic observables, I post-process selected GRMHD snapshots with {\tt BHOSS} \cite{2012A&A...545A..13Y, 2020IAUS..342....9Y}. The GRRT solver follows photon geodesics through the 3D simulation output and returns images, spectra, and light curves for a camera specified by the observing angle $i$ and frequency $\nu$. In all calculations presented here, the line of sight is fixed to $i=30^\circ$, in line with \cite{2022ApJ...930L..16E}. Emission from zones strongly influenced by numerical floor treatment is suppressed by imposing a magnetization cutoff, $\sigma=b^2/\rho<1$. I also adopt the fast-light approximation, neglecting the light-travel time across the source. Sgr~A$^*$ serves as the fiducial target, with $M_\bullet=4.5\times10^6\,M_\odot$ and $D_{\rm SgrA}=8.5\times10^3\,{\rm pc}$, and the camera spans $160\,\mu{\rm as}$ on a $1,000\times1,000$ pixel grid. For the electron temperature, I compare the evolved two-temperature solution with the parameterized $R-\beta$ prescription. The mass unit is chosen so that the time-averaged 230 GHz flux equals 2.4 Jy. Because the discussion below concentrates on the submillimeter regime, where thermal emission dominates, only thermal synchrotron radiation is included \cite[e.g.,][]{2023arXiv230816740N}.

\begin{figure*}
	\includegraphics[height=.4\linewidth]{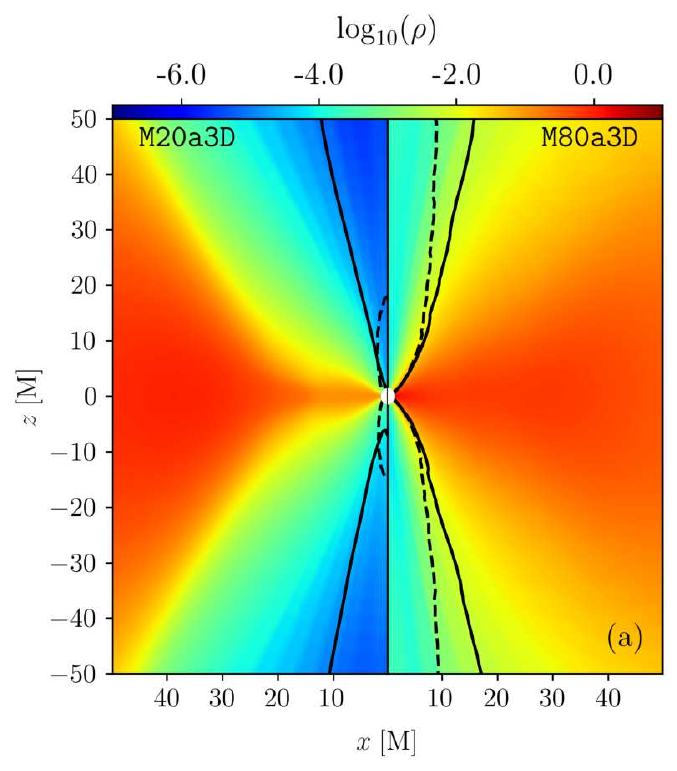}
	\includegraphics[height=.4\linewidth]{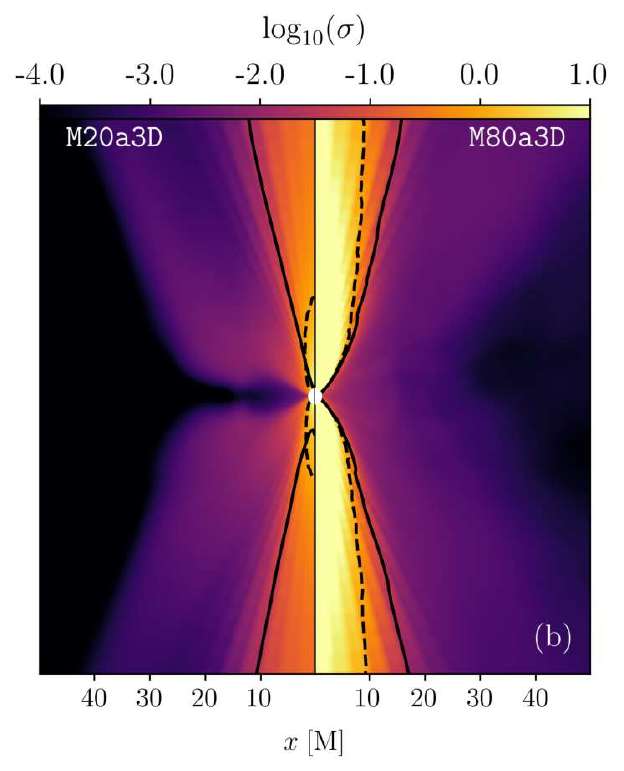}
    \includegraphics[height=.4\linewidth]{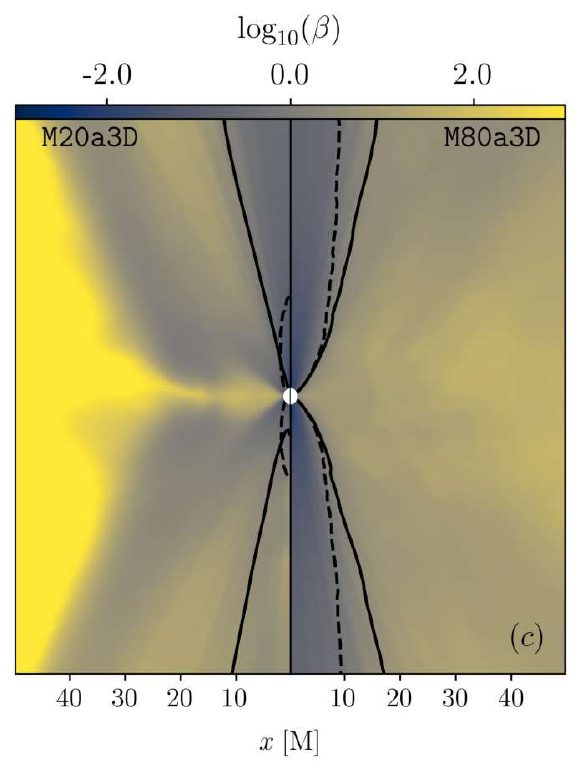}
    \includegraphics[width=.9\linewidth]{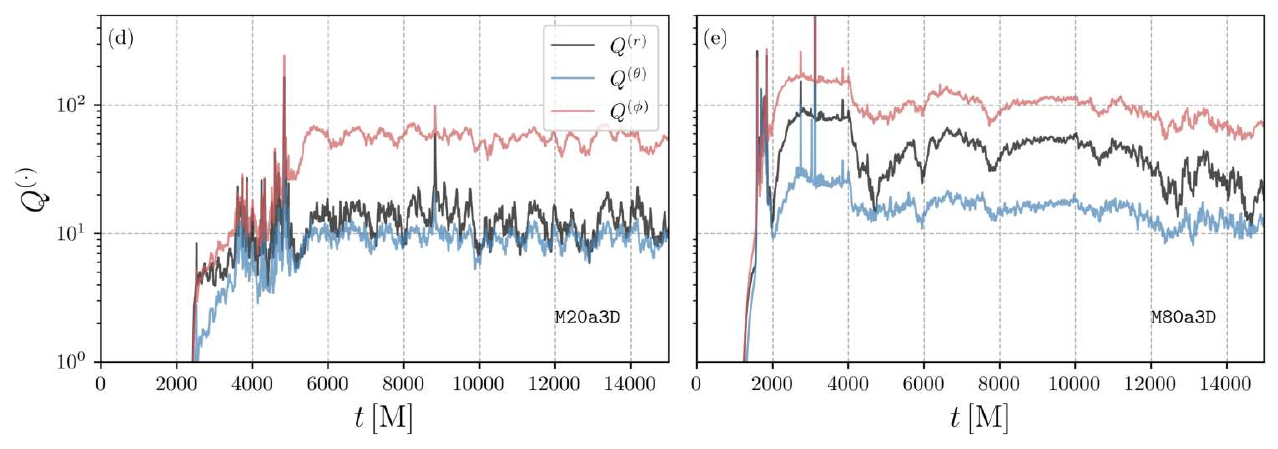}
    \caption{Time- and azimuth-averaged flow structure over $t=10,000\,M$ to $15,000\,M$. Panels (a)--(c) show $\rho$, $\sigma$, and the plasma beta. The left column corresponds to the $\lambda=20$ model M20a3D, and the right column corresponds to the $\lambda=80$ model M80a3D. In each map, the solid contour marks $-hu_t=1$ and the dashed contour marks $\sigma=1$. Panels (d) and (e) show the corresponding time evolution of the averaged MRI quality factors for the two runs.}
    \label{fig: avg_GRMHD}
\end{figure*}
\begin{figure}
    \centering
         \includegraphics[width=.6\linewidth]{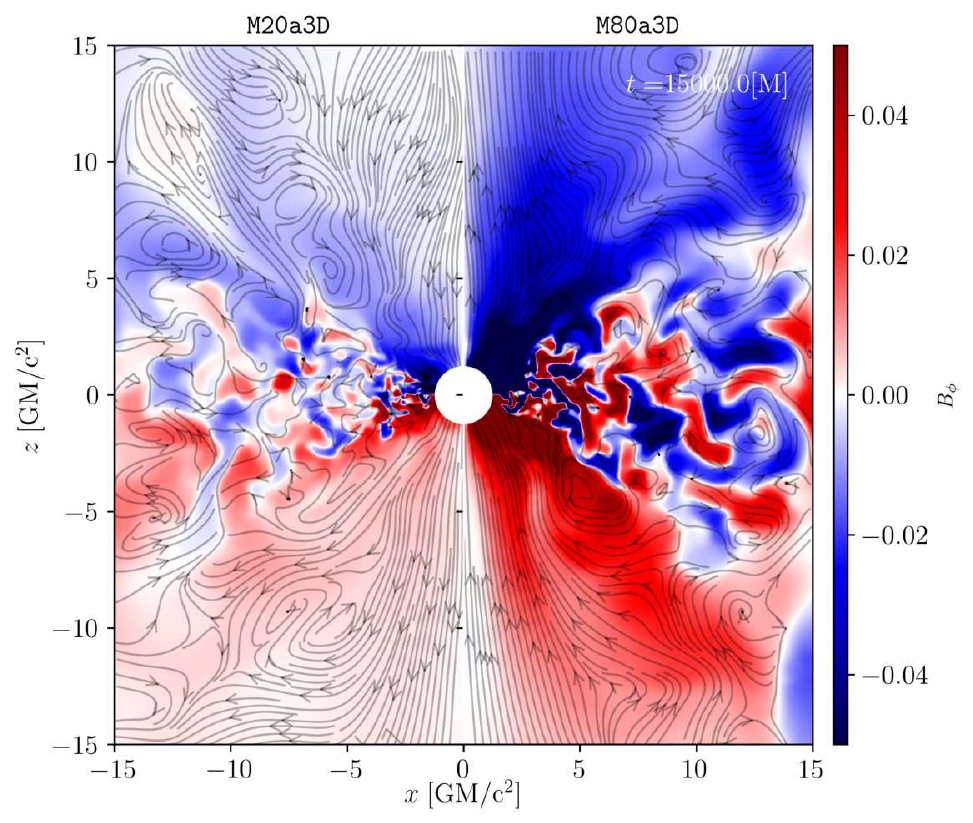}
    \caption{Toroidal magnetic-field structure in the two 3D models. The left panel corresponds to M20a3D, and the right panel corresponds to M80a3D. Streamlines trace the accompanying poloidal field geometry.}
    \label{fig: Bphi}
\end{figure}
\section{Simulation Results}
\subsection{Accretion-Flow Dynamics}

To characterize the accretion behavior in the 3D GRMHD runs with multiple poloidal loops, I follow \cite{Porth2019} and evaluate the mass accretion rate at the event horizon as
\begin{equation}
    \dot M = \int_0^{2\pi}\int_0^{\pi} \rho u^r \sqrt{-g}d\theta d\phi, \label{Eq: a_rate-c3}
\end{equation}
and the magnetic flux rate measured at the event horizon as,
\begin{equation}
    \Phi_{\rm B} = \frac{1}{2}\int_0^{2\pi}\int_{0}^{\pi}\left|B^r\right|\sqrt{-g}d\theta d\phi. \label{Eq: B-flux-c3}
\end{equation}
Figure~\ref{fig: Mdot-c3} follows the mass-accretion rate $\dot M$ and normalized magnetic flux $\Phi_{\rm B}/\sqrt{\dot M}$ for models M20a3D (black) and M80a3D (blue). Because the larger loops in M80a3D evolve more slowly, that run is continued to $20,000\,M$ so that its late-time behavior can be tracked. In M20a3D, the initial magnetic pattern is largely erased and the flow becomes strongly nonlinear after about $5,000\,M$, so that run is stopped at $15,000\,M$.

The 3D model M20a3D broadly retains the accretion behavior already seen in the earlier 2D model M20a of \cite{2023MNRAS.522.2307J}. During the quasi-steady interval ($10,000$--$15,000\,M$), it never develops a persistent jet, similar to the smaller-torus models of \cite{Nathanail2022}. The main new ingredient in 3D is interchange-driven turbulence in the $\phi$ direction, which is absent in axisymmetry \cite{Begelman2022}. That extra mixing disrupts the initially ordered loops before they reach the horizon. The resulting dissipation weakens MRI growth, erodes the poloidal field, and lowers the horizon accretion rate, as shown by the black curve in Figure~\ref{fig: Mdot-c3}(a). In M20a3D, the mean accretion rate stays near $\dot M\sim10^{-1}$, about two orders of magnitude below the single-loop models of \cite{Mizuno2021}. Larger initial loops alter the balance: M80a3D accretes roughly ten times faster than M20a3D, even though both multi-loop 3D runs retain substantially less magnetic flux than the single-loop reference solutions. Since the seed field follows the torus density profile (Eq.~\eqref{eq:A_phi}), the outer loops begin weaker than the inner ones. Even so, at least four loops in M80a3D and more than ten in M20a3D still carry more than 10\% of the peak torus magnetic energy and therefore remain dynamically relevant. Relative to the single-loop simulations of \cite{Mizuno2021}, M80a3D develops richer small-scale magnetic structure, preserves only about half as much magnetic flux, and still does not become fully MAD. Because dissipation is weaker than in M20a3D, MRI remains more effective and the accretion rate is correspondingly higher.

Figure~\ref{fig: avg_GRMHD} clarifies this contrast by showing time- and azimuth-averaged distributions between $t=10,000\,M$ and $15,000\,M$ of (a) density $\log_{10}(\rho)$, (b) magnetization $\log_{10}(\sigma)$, and (c) plasma beta $\log_{10}(\beta)$ for both models. In panels (a)--(c), the solid and dashed contours correspond to $-hu_t=1$ and $\sigma=1$. Panels (d) and (e) give the matching histories of the averaged MRI quality factors $Q^{(r,\theta,\phi)}$ for M20a3D and M80a3D, measured over the dense torus region $r<50\,r_{\rm g}$ and $60^\circ<\theta<120^\circ$. The density maps in Figure~\ref{fig: avg_GRMHD}(a) show that M20a3D retains a more extended bound region, traced by the solid contour, and a smaller disk than M80a3D. Larger seed loops leave behind a less disordered magnetic configuration and reduce reconnection-driven dissipation inside the torus \cite{2023MNRAS.522.2307J}. As a result, M80a3D develops both a stronger jet and a lower plasma beta, hence a stronger magnetic field, than M20a3D; see Figure~\ref{fig: avg_GRMHD}(b) and (c). Because the remnant field after loop mergers differs between the two runs, the effective MRI strength also differs. In Figure~\ref{fig: avg_GRMHD}(d) and (e), all MRI quality factors remain above 8, which indicates adequate resolution \cite{Porth2019, 2011ApJ...738...84H}. Even so, the values in M20a3D are slightly smaller than in M80a3D, consistent with the weaker magnetic field in its torus. That difference feeds directly into the contrasting accretion rates seen in Figure~\ref{fig: Mdot-c3} and also explains why M20a3D develops a thinner dense equatorial layer.

Figure~\ref{fig: Bphi} compares the toroidal magnetic field $B_\phi$ and the associated poloidal streamlines in the two 3D models. In each snapshot, patches of one sign are embedded inside broader regions of the opposite sign, showing that MRI-driven field reversals remain active. Near the event horizon, M80a3D sustains a stronger toroidal component than M20a3D, consistent with the more developed jet structure already seen in Figure~\ref{fig: avg_GRMHD}(b).

\subsection{Radiative Signatures}
I now turn to the observable emission of the multi-loop models. Single-loop accretion flows have already been explored extensively with both self-consistent two-temperature treatments and parameterized $R-\beta$ prescriptions \cite[e.g.,][]{Moscibrodzka2016, Mizuno2021, Fromm2022}. By contrast, the multi-loop configuration undergoes much more repeated reconnection, generates a more irregular sheath, and drives jets whose strength varies substantially with time. The purpose of this section is to determine how those dynamical differences appear in the radiation.

In order to isolate the thermal contribution, the radiative-transfer calculation in this chapter includes only thermal synchrotron emission. The local emissivity is evaluated with the approximate expression of \cite{Leung2011}:
\begin{equation}
    j_{\rm \nu}= \frac{2\pi e^2 \nu_{\rm s} n_{\rm e}}{2c K_{\rm 2}\left(1/\Theta_{\rm e}\right)}\left(X^{1/2}+2^{11/12}X^{1/6}\right)^2\exp{\left(-X^{1/3}\right)}, \label{Eq: MBS_emissivity}
\end{equation}
where 
\begin{equation}
    X\equiv \frac{\nu}{\nu_{\rm s}},\quad \nu_{\rm s} \equiv \frac{2}{9}\nu_{\rm c}\Theta_{\rm e}^2\sin{\theta} \label{Eq: X}.
\end{equation}
Here $\Theta_{\rm e}=k_{\rm B} T_{\rm e}/m_{\rm e} c^2$ denotes the dimensionless electron temperature, $k_{\rm B}$ is Boltzmann's constant, $T_{\rm e}$ is expressed in cgs units, $K_2$ is the modified Bessel function of the second kind, and $\nu_{\rm c}$ is the cyclotron frequency,
\begin{equation}
    \nu_{\rm c} \equiv \frac{eB}{2\pi m_{\rm e} c}=2.8\times 10^6B\,\rm{Hz}, \label{Eq: nu_c}
\end{equation}
where $e$ is the electron charge, $m_{\rm e}$ is the electron mass, $B=\sqrt{B^iB_i}$ is the magnetic-field strength, and $B^i$ is the Eulerian three-field.

For the electron thermodynamics, I compare the evolved two-temperature solutions with the usual $R-\beta$ prescription. In that parameterized model, the electron temperature is reconstructed from the proton temperature through \cite{Moscibrodzka2016}
\begin{equation}
    \frac{T_{\rm p}}{T_{\rm e}} = \frac{1}{1+\beta^2}R_{\rm l}+\frac{\beta^2}{1+\beta^2}R_{\rm h},
    \label{Eq: R-beta-c3}
\end{equation}
where $R_{\rm l}$ and $R_{\rm h}$ are dimensionless constants. Throughout this chapter I fix $R_{\rm l}=1$ and consider $R_{\rm h}=1$ and 160 in order to bracket the two-temperature heating results.

\begin{figure}
    \centering
         \includegraphics[width=\linewidth]{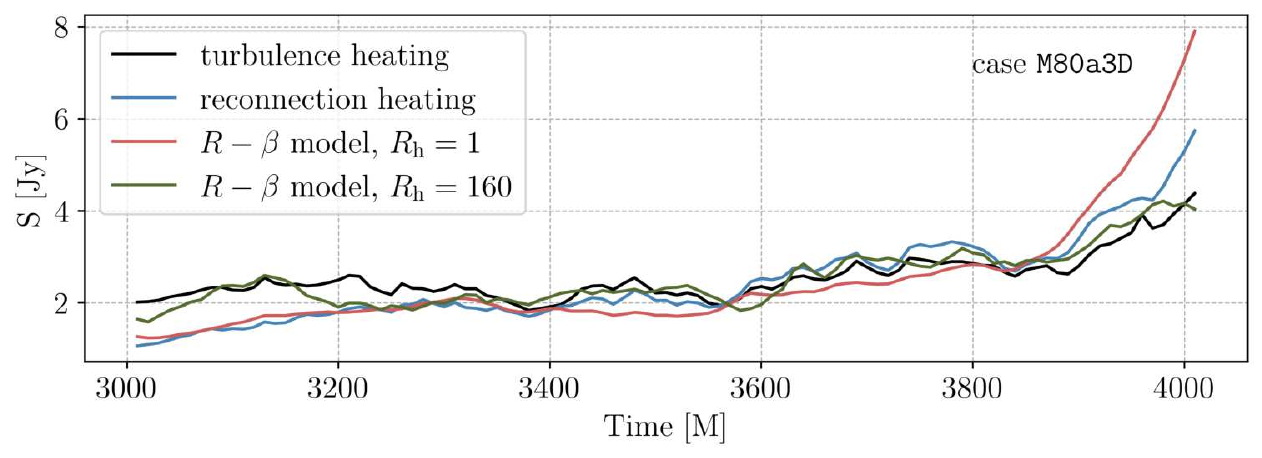}
    \caption{Early-time 230 GHz light curves for model M80a3D. The black and blue solid curves show the turbulence-heating and reconnection-heating solutions, while the red and dark-green curves give the two $R-\beta$ reconstructions with $R_{\rm h}=1$ and $R_{\rm h}=160$.}
    \label{fig: case6_lc_300-400}
\end{figure}
\begin{figure}
	\includegraphics[width=\linewidth]{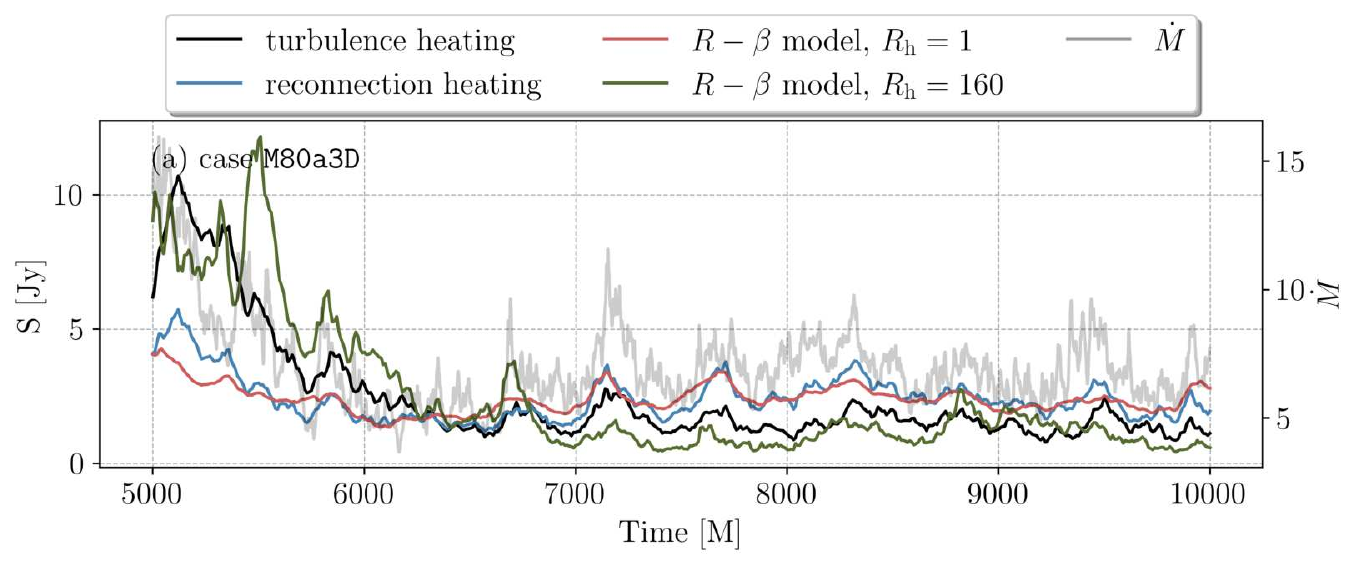}
	\includegraphics[width=.99\linewidth]{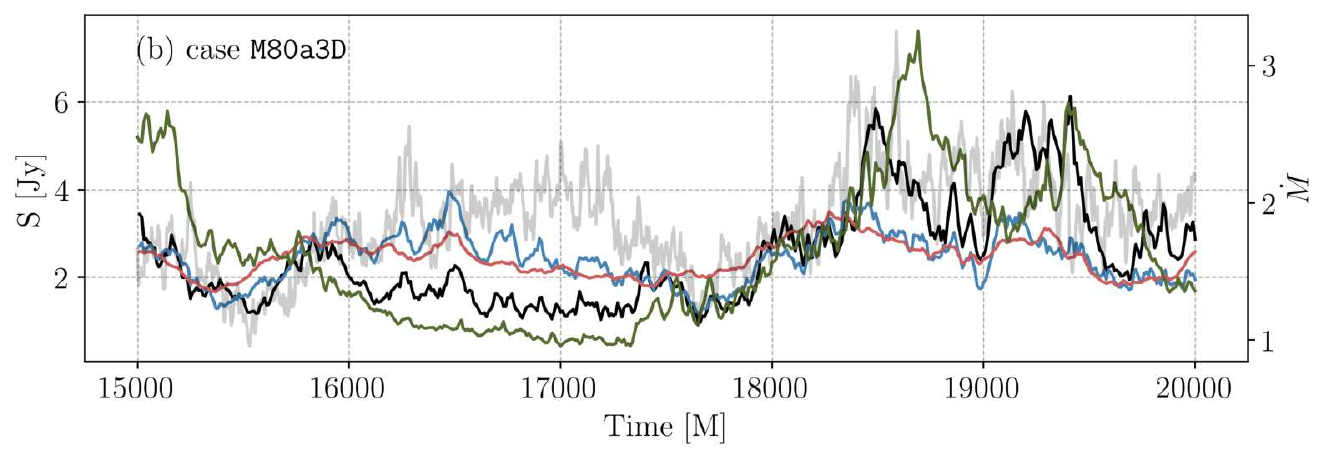}
	\includegraphics[width=\linewidth]{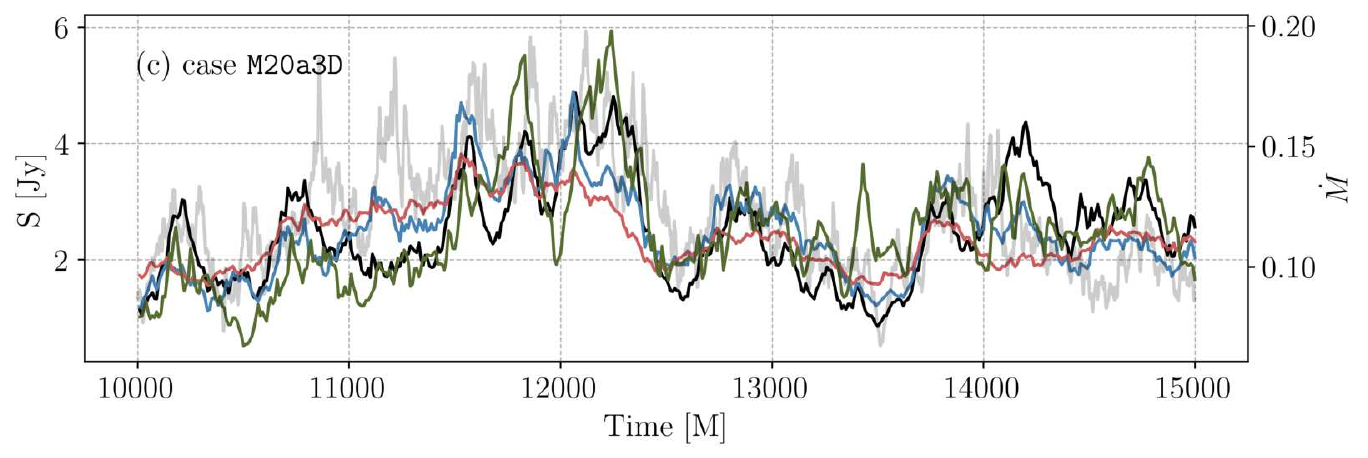}
    \caption{Later-stage 230 GHz light curves for the same four temperature prescriptions. Panels (a) and (b) show M80a3D over two separate intervals using the same color coding as Figure~\ref{fig: case6_lc_300-400}, and panel (c) shows the corresponding result for M20a3D. The gray solid curve in each panel gives the contemporaneous mass-accretion rate.}
    \label{fig: flare_lc}
\end{figure}
\begin{figure*}
\centering
	\includegraphics[width=.9\linewidth]{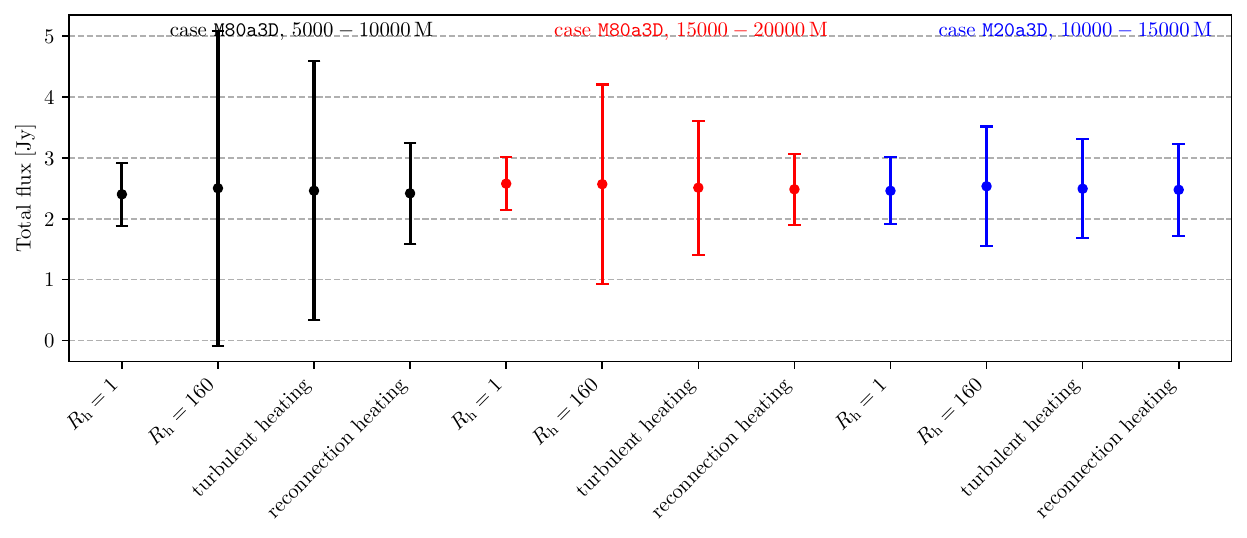}
    \caption{Standard deviations of the 230 GHz light curves for the different heating prescriptions. The error bars show the standard deviation itself. Black symbols correspond to M80a3D over $5,000$--$10,000\,M$, red symbols to the same model over $15,000$--$20,000\,M$, and blue symbols to M20a3D over $10,000$--$15,000\,M$.}
    \label{fig: std}
\end{figure*}
%

\subsubsection{Light-Curve Behavior}
Figure~\ref{fig: case6_lc_300-400} shows the early 230 GHz light curves of model M80a3D. The black, blue, red, and green curves correspond to turbulence heating, reconnection heating, and the two $R-\beta$ choices with $R_{\rm h}=1$ and $R_{\rm h}=160$. During $t=3,000$--$4,000\,M$, only the first loop has reached the hole, so the flow still behaves much like a standard single-loop torus. All four prescriptions therefore produce nearly identical light curves. Before $t\gtrsim4,000\,M$, oppositely directed loops have not yet begun to reconnect strongly, consistent with the single-loop-like stage discussed by \cite{Mizuno2021}. Once the next loop of opposite sign arrives, the curves separate because the first major polarity inversion releases magnetic energy efficiently and brightens the source.

Figure~\ref{fig: flare_lc} collects the later 230 GHz light curves for the same four temperature prescriptions in M80a3D and M20a3D. The spread between models is now much larger than in Figure~\ref{fig: case6_lc_300-400}. The mean luminosity still follows the broad evolution of the accretion flow, but the detailed flare pattern depends sensitively on how the electrons are heated. The turbulence-heating solution remains closest to the $R_{\rm h}=160$ reconstruction, while reconnection heating lies nearer to the $R_{\rm h}=1$ case. Those two $R_{\rm h}$ limits therefore bracket the two-temperature results by representing, respectively, weak and strong disk-electron heating in high-$\beta$ gas. When $R_{\rm h}$ is large, the disk electrons are cooler and the disk contributes less to the total flux. Before the flow enters the strongly turbulent stage, the self-consistent two-temperature models and the $R-\beta$ reconstructions should still give nearly the same 230 GHz signal. That expectation is exactly what Figure~\ref{fig: case6_lc_300-400} shows, and it is also consistent with earlier single-loop calculations \cite{Mizuno2021}.

Figure~\ref{fig: std} measures the variability amplitude directly through the standard deviation of the 230 GHz light curves. The two left groups correspond to M80a3D over $t=5,000$--$10,000\,M$ and $t=15,000$--$20,000\,M$, and the right group shows the same statistic for M20a3D over $t=10,000$--$15,000\,M$. Among the four prescriptions, the $R-\beta$ model with $R_{\rm h}=160$ is the most variable, while the $R_{\rm h}=1$ case is the least. Figures~\ref{fig: flare_lc}(a) and (b) also show that the variability weakens as the simulation progresses and repeated loop mergers gradually erase the original magnetic pattern. The same decline appears in Figure~\ref{fig: std}, where every prescription becomes less variable at later times. Comparing Figures~\ref{fig: flare_lc}(a) and (c) further shows that M80a3D, whose initial loops are larger and stronger, remains more variable overall than M20a3D. Figure~\ref{fig: std} shows the same ordering in the averaged standard deviations.

\begin{figure}
        \centering
	\includegraphics[width=.7\linewidth]{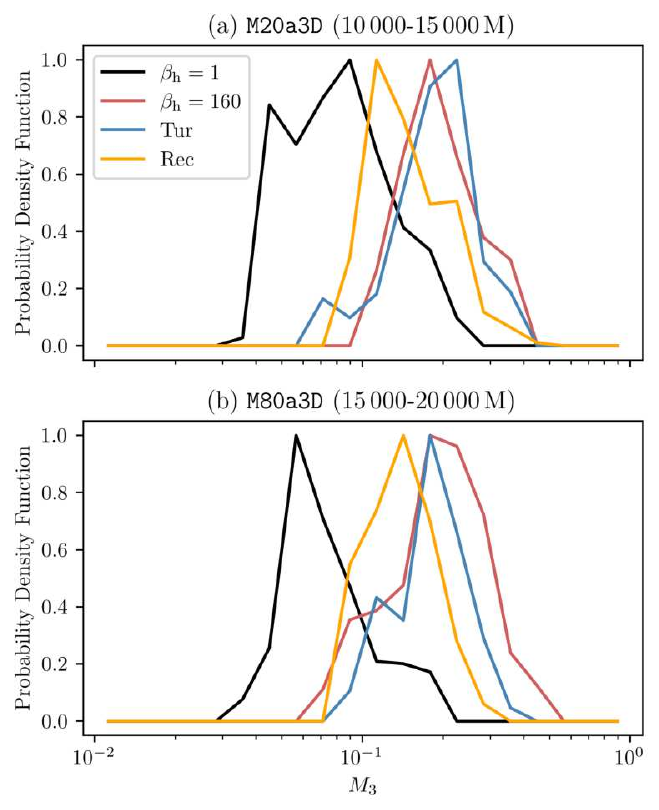}
    \caption{Probability distribution functions of $M_3$ for M20a3D in panel (a) and M80a3D in panel (b). Each curve is normalized to unit peak. Black and red correspond to the two $R-\beta$ choices, while blue and yellow denote the turbulence-heating and reconnection-heating solutions.}
    \label{fig: PDF_M3}
\end{figure}

Besides the standard deviation, \citet{2022ApJ...930L..16E} introduced a 3 hr modulation index, $M_3$, as an additional variability diagnostic. It is defined as the ratio of the standard deviation $\sigma_{\rm 3\,hr}$ to the mean $\mu_{\rm 3\,hr}$ within an approximately three-hour interval:
\begin{equation}
    M_3\equiv \frac{\sigma_{3\,\rm hr}}{\mu_{3\,\rm hr}}
\end{equation}
This statistic provides a second way to summarize the light-curve variability. Historical estimates suggest $M_3\lesssim0.1$, although many of the models examined in \citet{2022ApJ...930L..16E} lie above that level. Figure~\ref{fig: PDF_M3} plots the resulting $M_3$ distributions for the four heating prescriptions applied to M80a3D and M20a3D. The hierarchy mirrors Figure~\ref{fig: std}: the turbulence-heating solution tracks the $R-\beta$ case with $R_{\rm h}=160$, while reconnection heating and the $R_{\rm h}=1$ prescription are less variable. The $R-\beta$ model with $R_{\rm h}=1$ yields $M_3\approx0.06$, close to the historical value quoted for Sgr~A$^*$.

These light-curve differences already imply that the dominant emitting region changes with the electron-heating prescription through its effect on the temperature structure. I now examine that connection more directly.

\subsection{GRRT Images During Flares}
\subsubsection{Emission Region at 230 GHz}

\begin{figure*}
\centering
        \includegraphics[height=.275\linewidth]
        {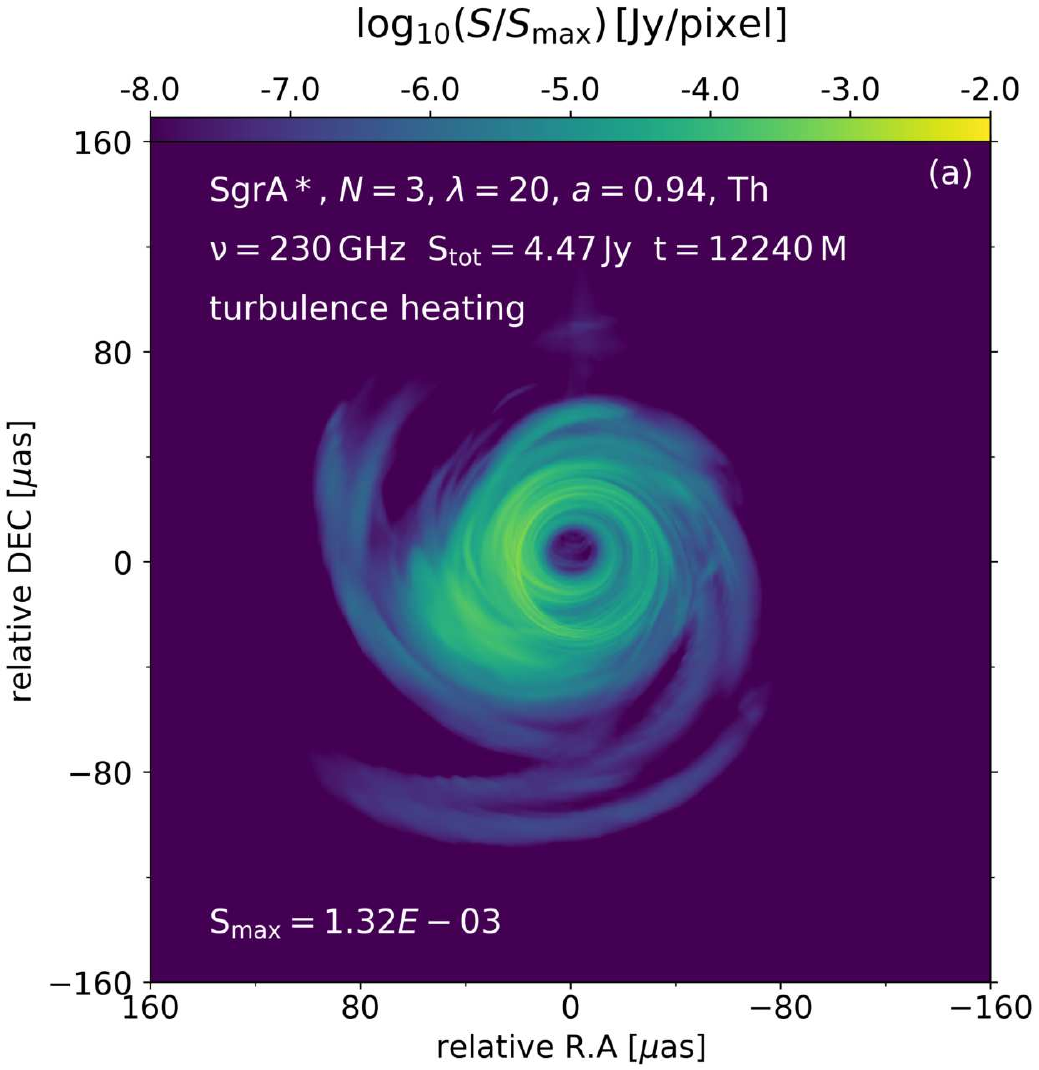}
	\includegraphics[height=.275\linewidth]{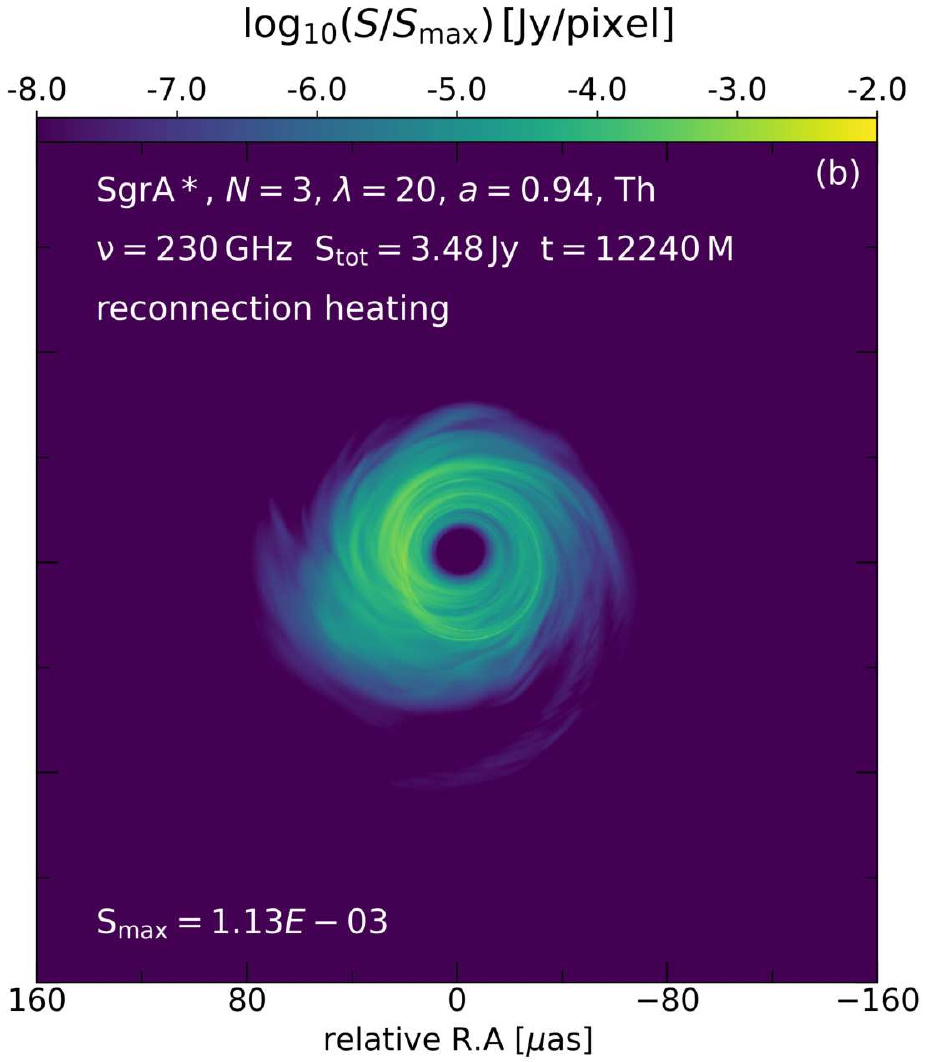}
	\includegraphics[height=.275\linewidth]{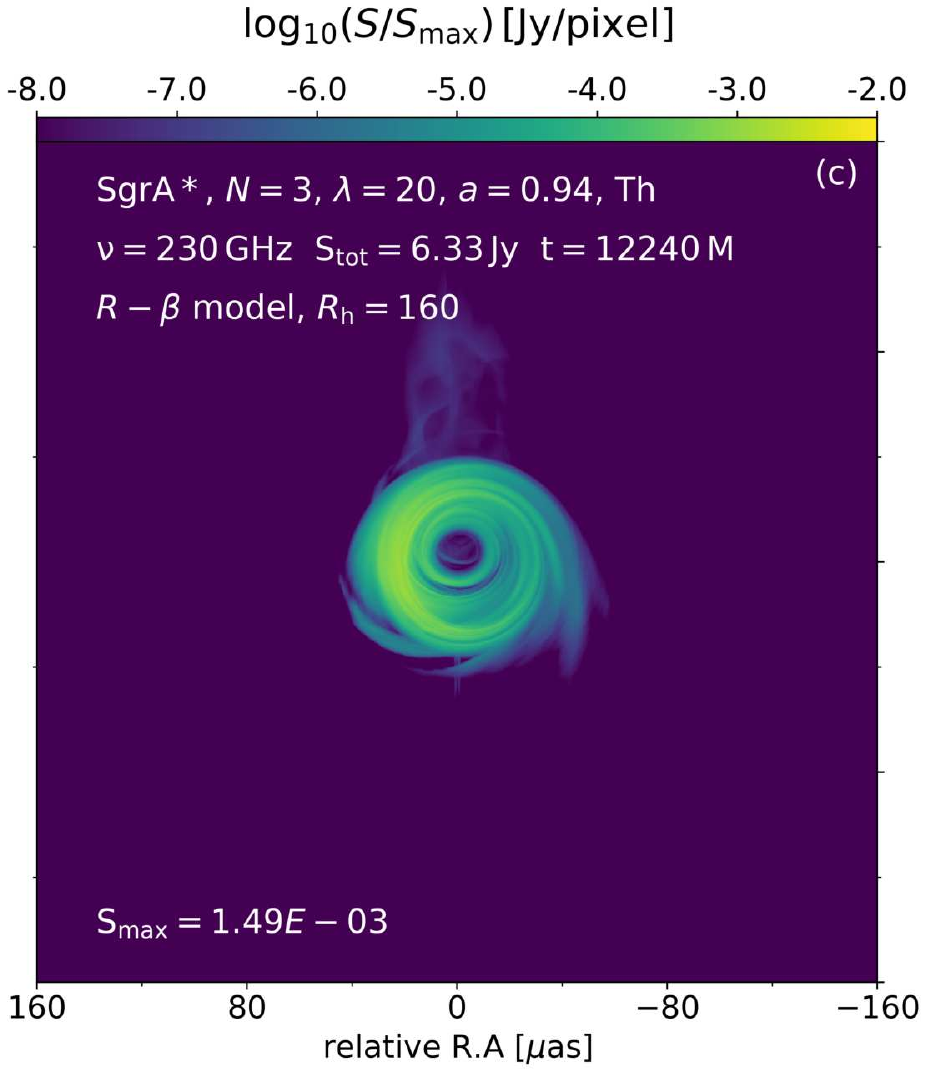}
	\includegraphics[height=.275\linewidth]{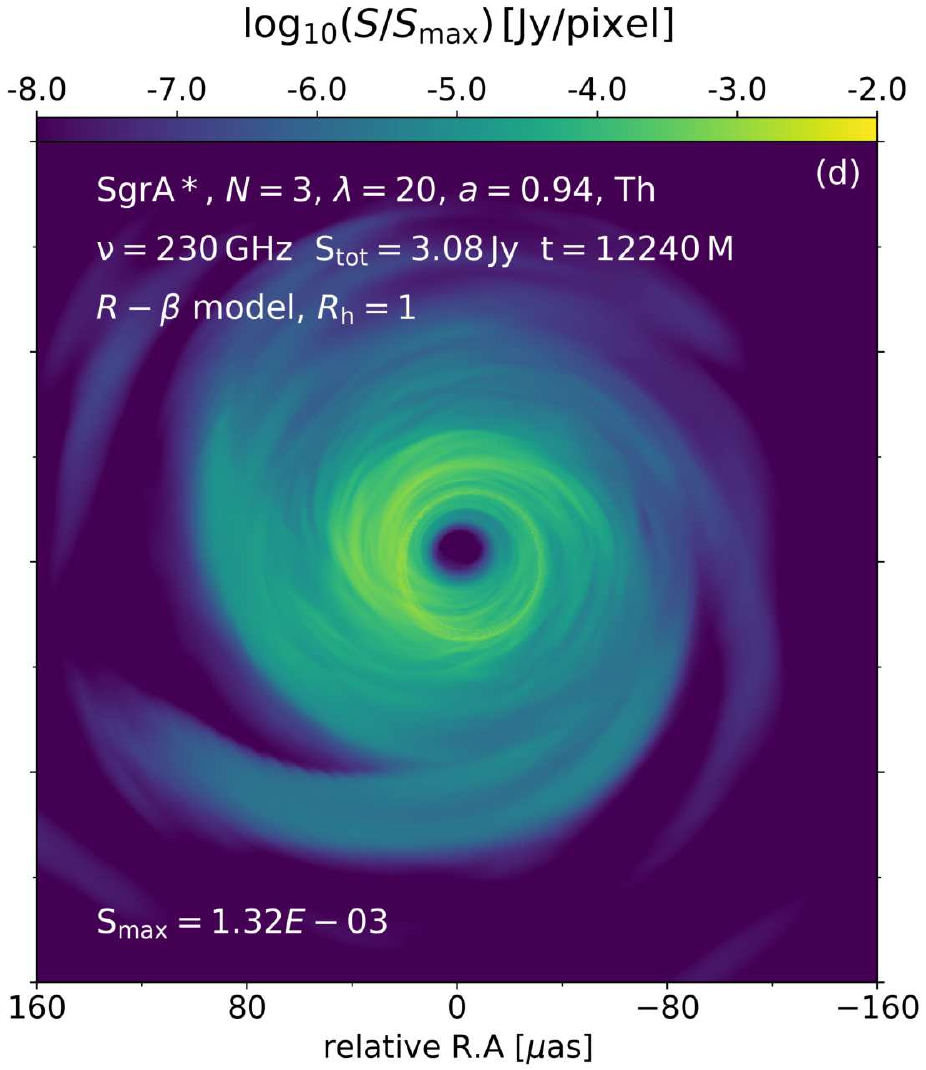} 
 \centering
	\includegraphics[height=.3\linewidth]{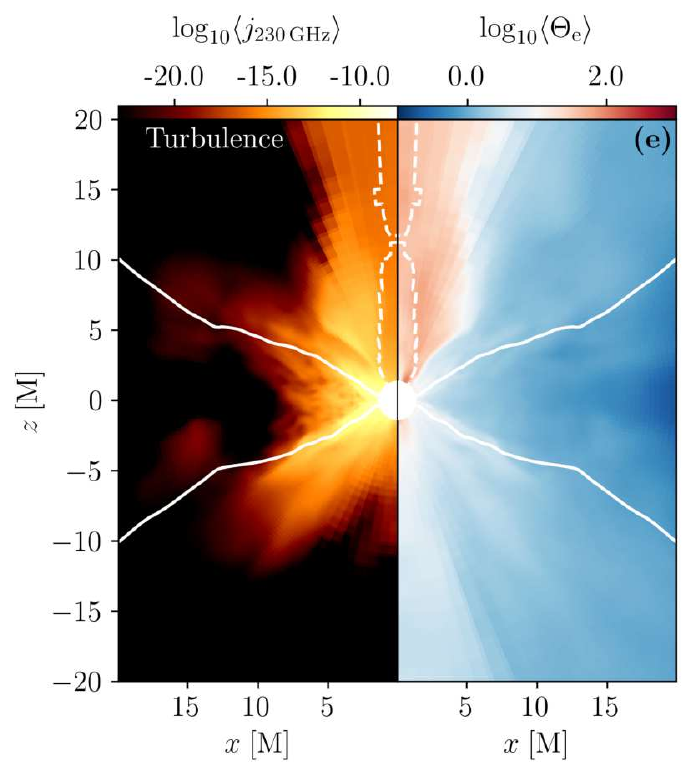}
	\includegraphics[height=.3\linewidth]{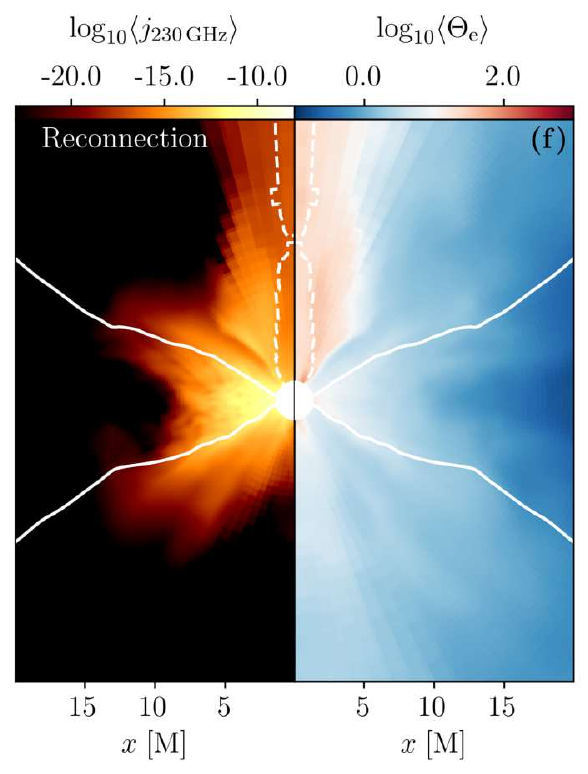}
	\includegraphics[height=.3\linewidth]{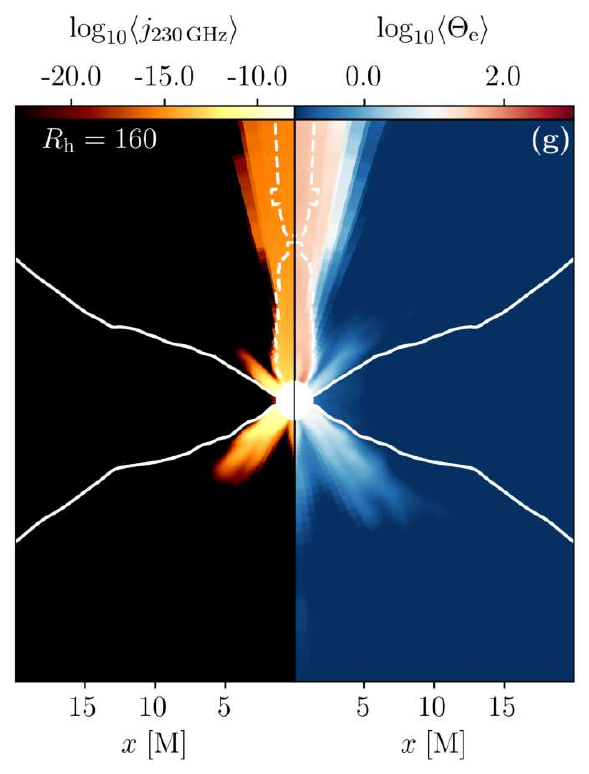}
	\includegraphics[height=.3\linewidth]{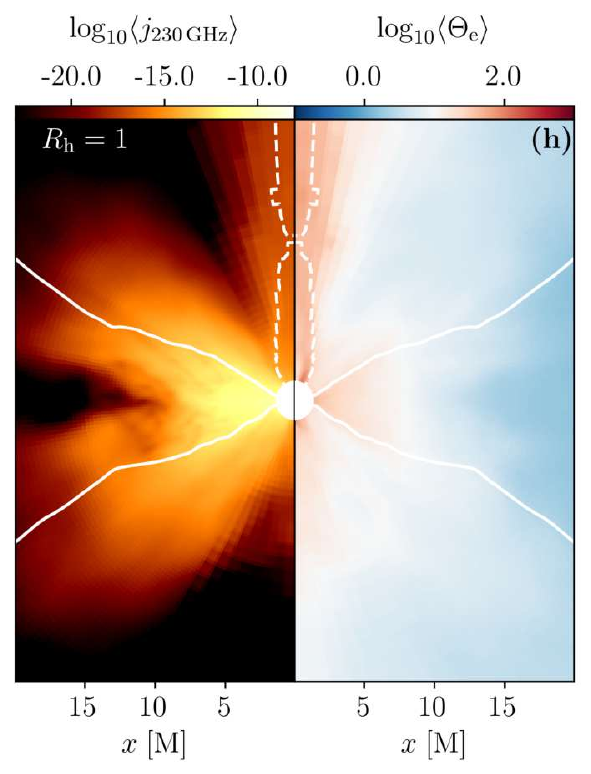}
	
    \caption{Top row: GRRT snapshots of Sgr A$^*$ at $t=12,240\,M$ and $i=30^\circ$ for four heating prescriptions: turbulence heating in panel (a), reconnection heating in panel (b), and the two $R-\beta$ reconstructions with $R_{\rm h}=160$ and $R_{\rm h}=1$ in panels (c) and (d). Bottom row: the matching 230 GHz emissivity maps and electron-temperature structures. In panels (e)--(h), the solid white contour marks the torus boundary at $\rho=0.01$, while the dashed contour encloses the $\sigma>1$ region excluded from the GRRT calculation.}
    \label{fig: Thetae}
\end{figure*}

The previous subsection showed that the heating prescription changes the light-curve morphology in a way that matters directly for the observed variability of Sgr~A$^*$ \cite{2022ApJ...930L..16E}. That difference cannot be understood from the bulk dynamics alone; it also depends on which part of the flow contributes most of the radiation. I therefore next link the GRRT images directly to the underlying GRMHD emissivity structure.

Figure~\ref{fig: Bphi} already indicated that M20a3D is more turbulent and reconnects more often than M80a3D, and therefore produces more erupting flux ropes \cite{2022ApJ...933...55C}. Since those structures are the most plausible counterparts of the observed Sgr~A$^*$ flares, the rest of this discussion concentrates on M20a3D.

The top row of Figure~\ref{fig: Thetae} shows 230 GHz snapshots at $t=12,240\,M$ for the four heating prescriptions applied to M20a3D. Panels (a) and (b) correspond to the two-temperature turbulence and reconnection models. The turbulence-heating case produces a broader image with more obvious filamentary structure than the reconnection-heating case. Panels (c) and (d) show the two $R-\beta$ reconstructions. Of the four images, the $R_{\rm h}=160$ case is the most compact, whereas $R_{\rm h}=1$ gives the broadest emitting region.

Those morphological differences are controlled by the electron-temperature distribution. To show this explicitly, the second row of Figure~\ref{fig: Thetae} presents the corresponding azimuthally averaged emissivity maps and temperature structure for each prescription. The two two-temperature solutions remain similar in their large-scale geometry, but turbulence heating extends the emitting sheath much farther, naturally producing the broader and more filamentary image. In the reconnection-heating case, the dominant contribution is concentrated more tightly in the equatorial disk.

Within the $R-\beta$ family, the image depends strongly on the choice of $R_{\rm h}$. For $R_{\rm h}=160$, the disk is too cool to contribute much to the radiation (Figure~\ref{fig: Thetae}(g)), so the sheath dominates the image in much the same way as in the turbulence-heating model. For $R_{\rm h}=1$, by contrast, the hotter disk broadens the emitting zone substantially (Figure~\ref{fig: Thetae}(h)), producing the diffuse image seen in Figure~\ref{fig: Thetae}(d). Appendix A compares these temperature structures in more detail.

\begin{figure*}
    \centering
	\includegraphics[width=.8\linewidth]{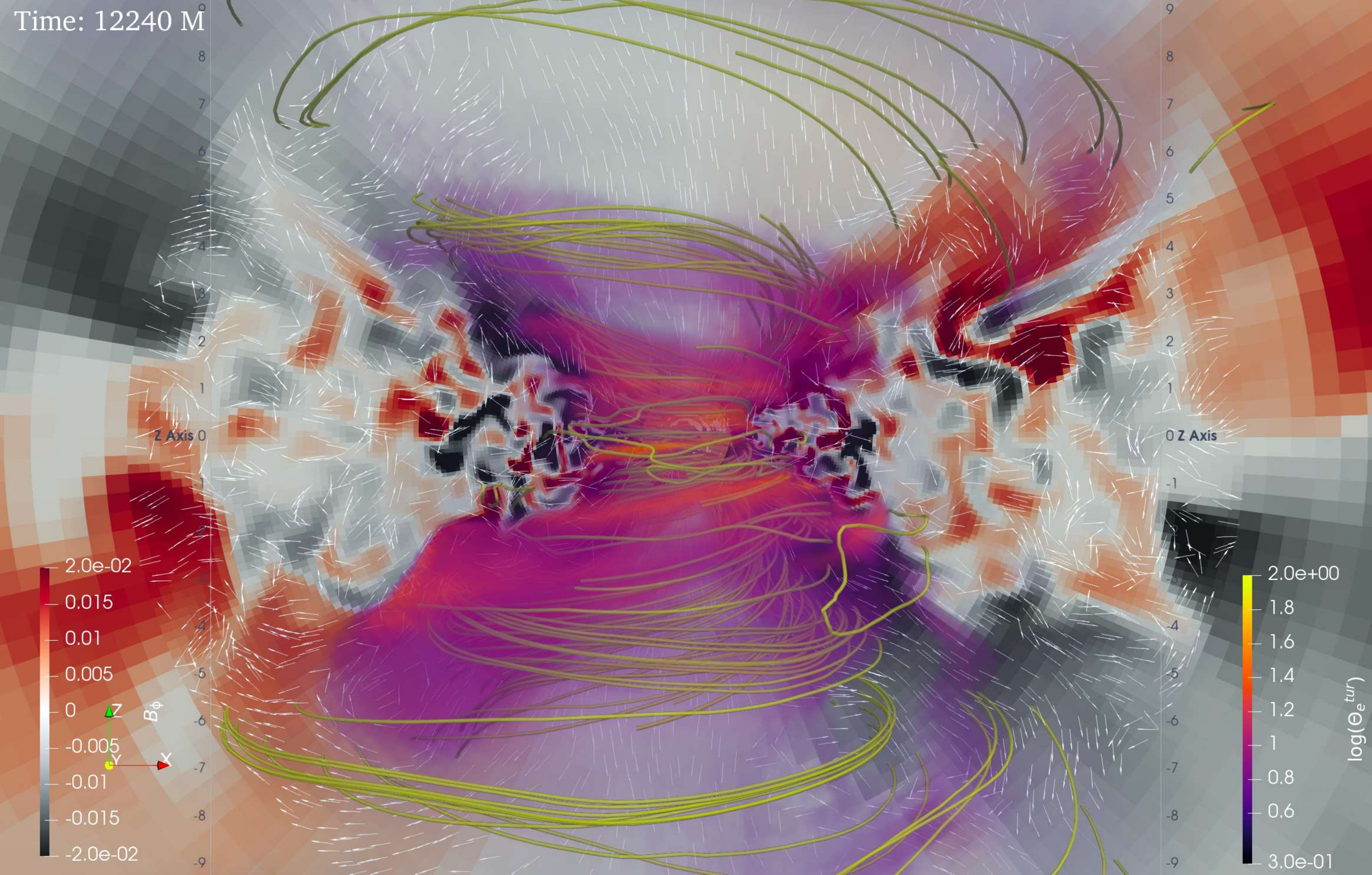}
    \caption{Three-dimensional rendering of the electron temperature (yellow to black) and toroidal magnetic field (red to black) in the turbulence-heating version of M20a3D at 12,240 M. White arrows indicate the local poloidal-field direction, and yellow tubes trace representative magnetic-field lines. The rendered cube spans $10\,r_{\rm g}$.}
    \label{fig: flare_Thetae}
\end{figure*}

The jet sheath is especially important because it hosts most of the plasmoid chains \cite{2023MNRAS.522.2307J, Nathanail2020} and flux ropes \cite{2022ApJ...933...55C}; similar structures also appear in the GRPIC calculations of \cite{Mellah2023}. These ropes combine comparatively ordered magnetic fields with elevated density and electron temperature. Figure~\ref{fig: flare_Thetae} shows a 3D rendering of the electron temperature in the turbulence-heating version of M20a3D at $t=12,240\,M$. The filamentary sheath stands out clearly and coincides with a flux rope threaded by ordered magnetic bundles, drawn as yellow field lines, while the red-to-black background indicates the toroidal field. These ropes are generated near the horizon by reconnection in the sheath and later evolve into helical structures. I interpret them as the 3D counterparts of the plasmoid-chain structures seen in higher-resolution 2D calculations \cite{2023MNRAS.522.2307J}. The same rendering also emphasizes the strong fields at the interfaces between torus regions of opposite polarity, where reconnection is strongest. Because the present 3D runs remain coarser than the 2D ones, the tearing cascade does not separate into individually resolved plasmoids. At sufficiently high 3D resolution, explicit plasmoid chains should emerge \cite{Ripperda2021}; on the current grid, the flux ropes are best regarded as unresolved plasmoid complexes.

\begin{figure}
    \centering
	\includegraphics[width=.6\linewidth]{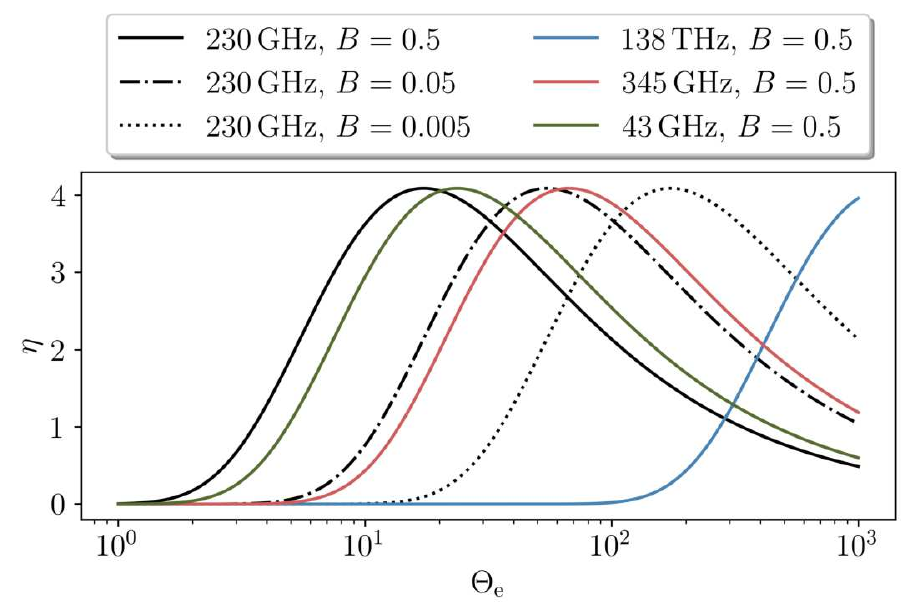}
    \caption{Emissivity coefficient $\eta$ as a function of electron temperature. The solid curves show the 230 GHz dependence for several magnetic-field strengths. The purple dash-dotted, red dashed, and dotted curves show the corresponding behavior at 46 GHz, 86 GHz, and NIR frequencies for a representative flux-rope field strength of $B=0.05$.}
    \label{fig: emissivity}
\end{figure}

\subsubsection{Filamentary Structure and Flux Ropes}

The turbulence-heating images in the previous subsection displayed a pronounced filamentary morphology, and those filaments are tied directly to the flux ropes. To make that connection explicit, I now examine how the emissivity depends on electron temperature. Equation~\ref{Eq: MBS_emissivity} shows that the thermal synchrotron emissivity $j_{\rm \nu}$ is proportional to the electron number density $n_{\rm e}$, which follows the gas density $\rho$ in the simulations. Figure~\ref{fig: Thetae} also shows that the polar funnel reaches high electron temperature, but its density is so low that its total emissivity remains modest.

In the ultra-relativistic limit, $\Theta_{\rm e}\gg1$, one has \cite{Leung2011}
\begin{equation}
    K_{\rm 2}\left(1/\Theta_{\rm e}\right)\sim \Theta_{\rm e}^2.
\end{equation}
Since each heating model produces a different temperature distribution, it is helpful to isolate the $\Theta_{\rm e}$ dependence explicitly. Holding the other local quantities, such as $n_{\rm e}$ and $B$, fixed, Eq.~\ref{Eq: MBS_emissivity} may be rewritten as
\begin{equation}
    j_{\rm \nu}\propto \eta(\Theta_{\rm e}, B)=\left(X^{1/2}+2^{11/12}X^{1/6}\right)^2\exp{\left(-X^{1/3}\right)}, \label{Eq: gamma}
\end{equation}
where $X\equiv \nu/\nu_{\rm s}$ and therefore scales as $\Theta_{\rm e}^{-2}$. Here $\eta$ simply denotes the factor proportional to the emissivity. Equations~\ref{Eq: X} and \ref{Eq: nu_c} show that the dependence of $j_{\rm \nu}$ on $\Theta_{\rm e}$ is set entirely by the observing frequency $\nu$ and the local field strength $B$ (see also \cite{2022ApJ...930L..16E}). Figure~\ref{fig: emissivity} plots this coefficient for several frequencies and field strengths, which makes the temperature dependence of the radiation easy to read directly.

%
\begin{figure}
    \centering
	\includegraphics[width=.6\linewidth]{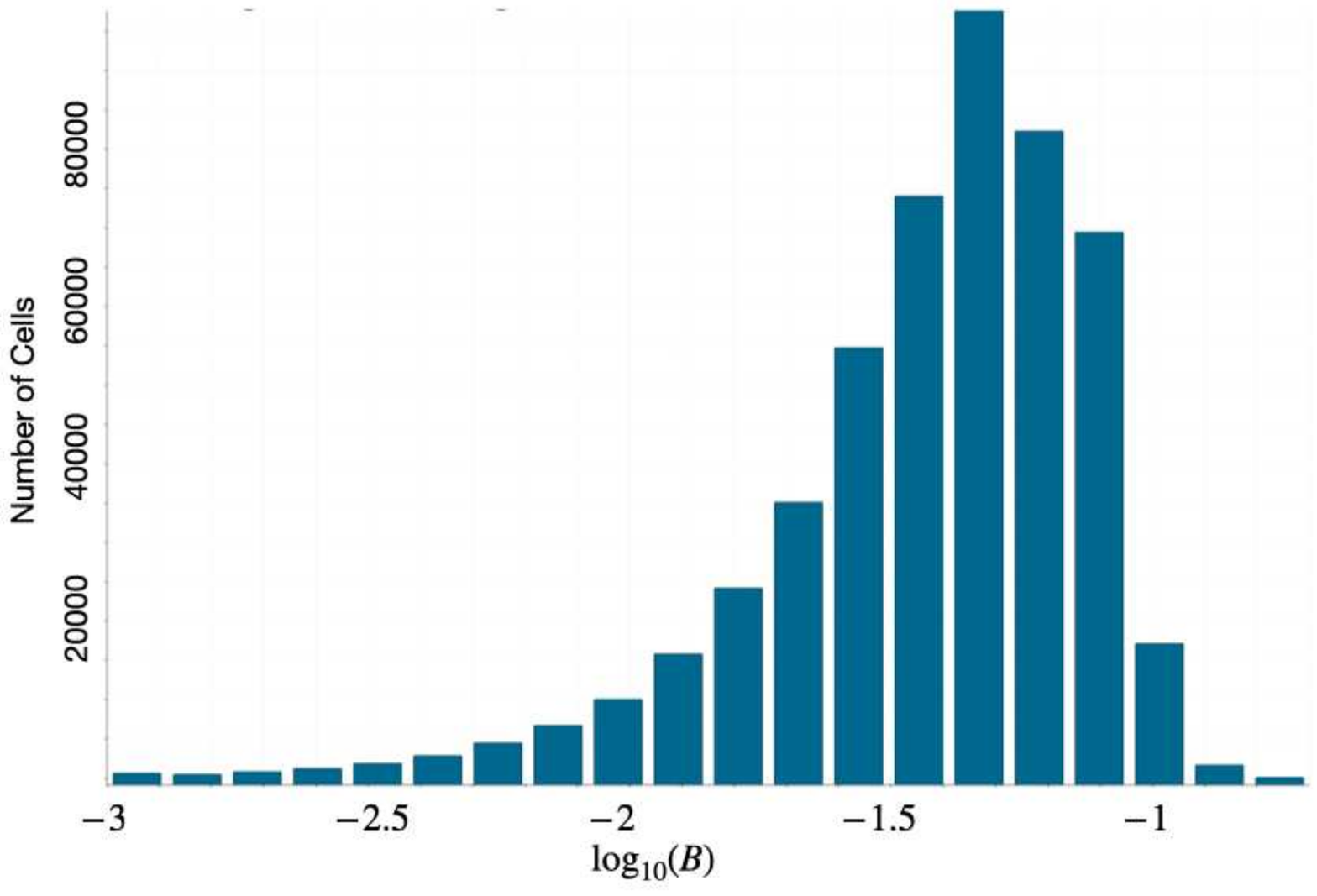}
    \caption{Histogram of the magnetic-field strength in the flaring region, in code units. The sample is drawn from the hot ($\Theta_{\rm e}>10$), dense ($\rho>0.001$) part of the turbulence-heating version of M20a3D at $t=12,240\,M$.}
    \label{fig: B_distribution}
\end{figure}

\begin{figure}
    \centering
    \includegraphics[width=.4\linewidth]
{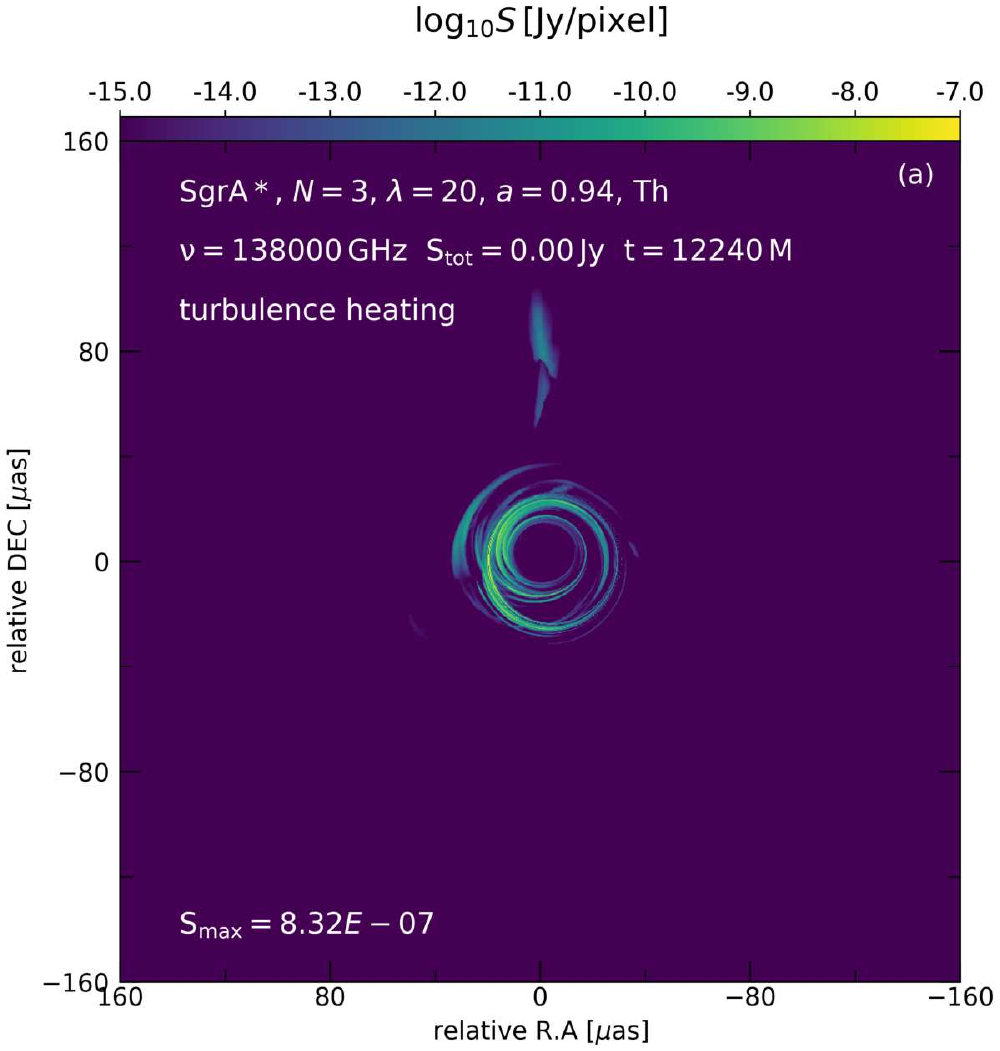}
	\includegraphics[width=.4\linewidth]{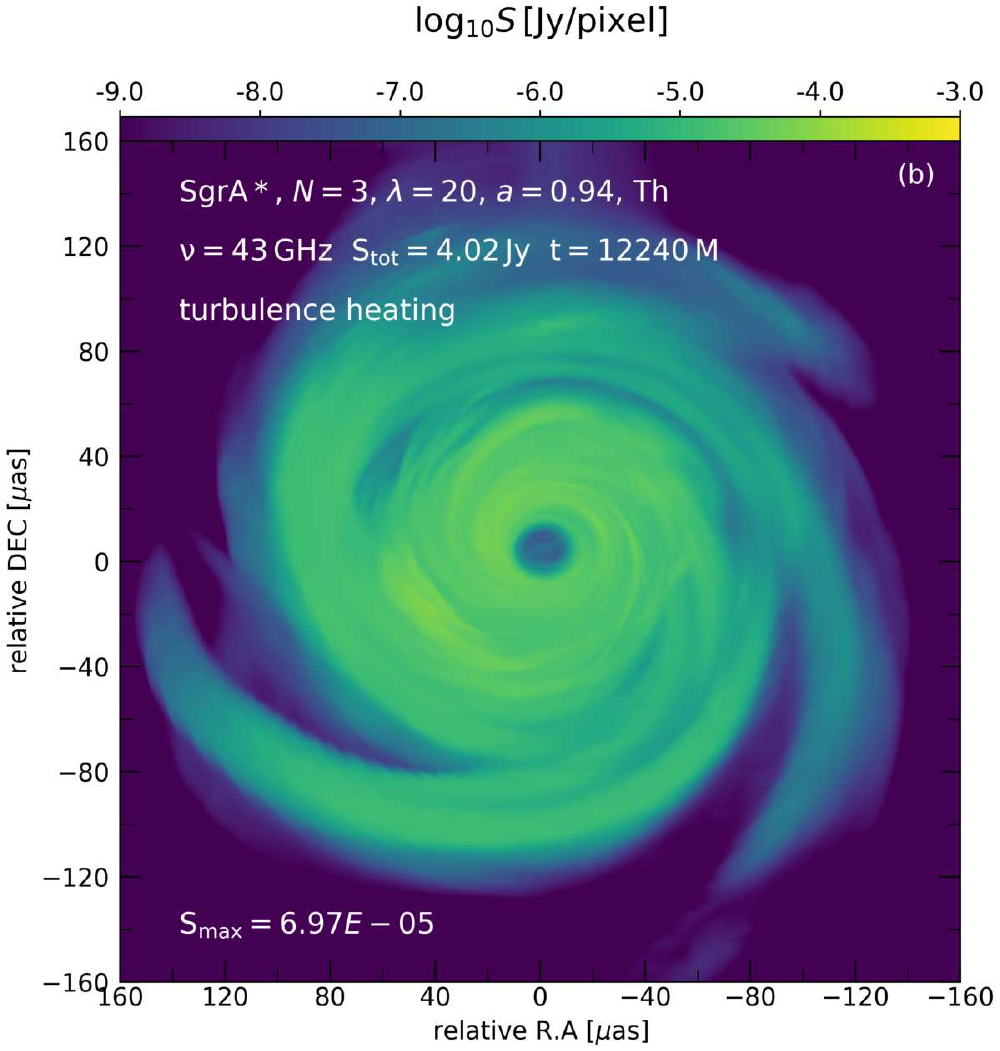}
    \caption{GRRT images from the turbulence-heating model at $t=12,240\,M$. Panel (a) shows the NIR image, and panel (b) shows the 43 GHz image.}
    \label{fig: 43GHz}
\end{figure}

To estimate the field strength inside the flux ropes, I sample the hot ($\Theta_{\rm e}>10$) and dense ($\rho>0.001$) part of M20a3D at $t=12,240\,M$ in the turbulence-heating solution. The resulting distribution is shown in Figure~\ref{fig: B_distribution}. It indicates that the ropes are more strongly magnetized than most of the surrounding flow. A large fraction of the selected cells cluster near $B=10^{-1.3}\approx0.05$ in code units. Using that value as a representative field strength, Figure~\ref{fig: emissivity} shows how the emissivity varies with electron temperature at different observing frequencies. For fixed $\nu$ and $B$, the emissivity rises with $\Theta_{\rm e}$, reaches a maximum, and then declines again. The location of the peak depends on both the frequency and the field strength.

At 230 GHz, a stronger magnetic field shifts the emissivity peak toward lower electron temperature. This explains why the funnel, although both hot and strongly magnetized, still contributes only weakly to the observed flux. The filamentary structures in Figure~\ref{fig: flare_Thetae} typically reach $\Theta_{\rm e}\sim10$--100, which coincides with the high-emissivity range at 230 GHz for the representative value $B=0.05$. That is why the turbulence-heating image in Figure~\ref{fig: Thetae}(a) looks so filamentary.

Figure~\ref{fig: 43GHz} shows images from the same snapshot at two additional frequencies, 43 GHz and 134 THz. At 43 GHz, the emissivity peaks at lower temperature, so in the optically thin limit the hottest ropes contribute less and the filamentary pattern becomes weaker. In the NIR, the peak shifts instead to $\Theta_{\rm e}>1,000$. Since the maximum temperatures in the simulation are of that order, the NIR emissivity stays on the rising part of the curve shown in Figure~\ref{fig: emissivity}. The hottest zones therefore dominate the 138 THz radiation. As the observing frequency increases from 43 GHz to the NIR, the emitting region also becomes more compact because the cooler outer flow contributes less, allowing the ropes to stand out much more clearly.

\begin{figure}
    \centering
    \includegraphics[width=.6\linewidth]{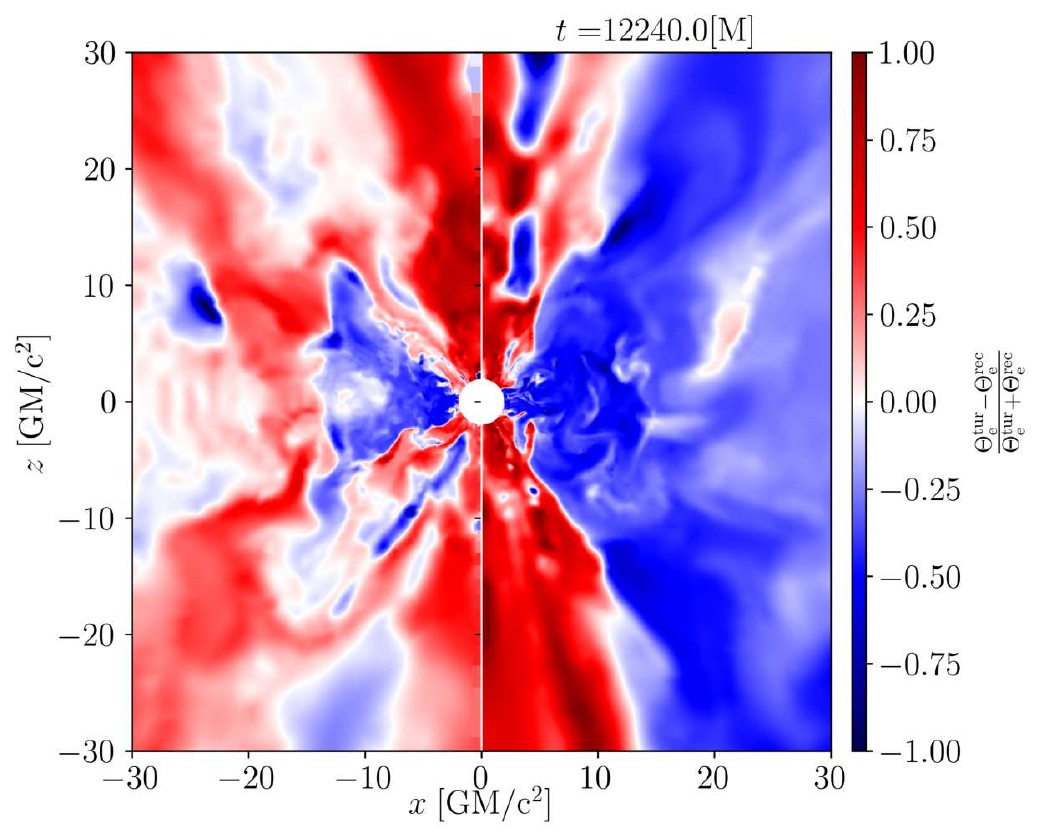}
    \caption{Relative-difference map between the electron-temperature distributions obtained with the turbulence-heating and reconnection-heating prescriptions for M20a3D at $t=12,240\,M$.}
    \label{fig: eta}
\end{figure}

\begin{figure*}
\centering
	\includegraphics[width=.9\linewidth]{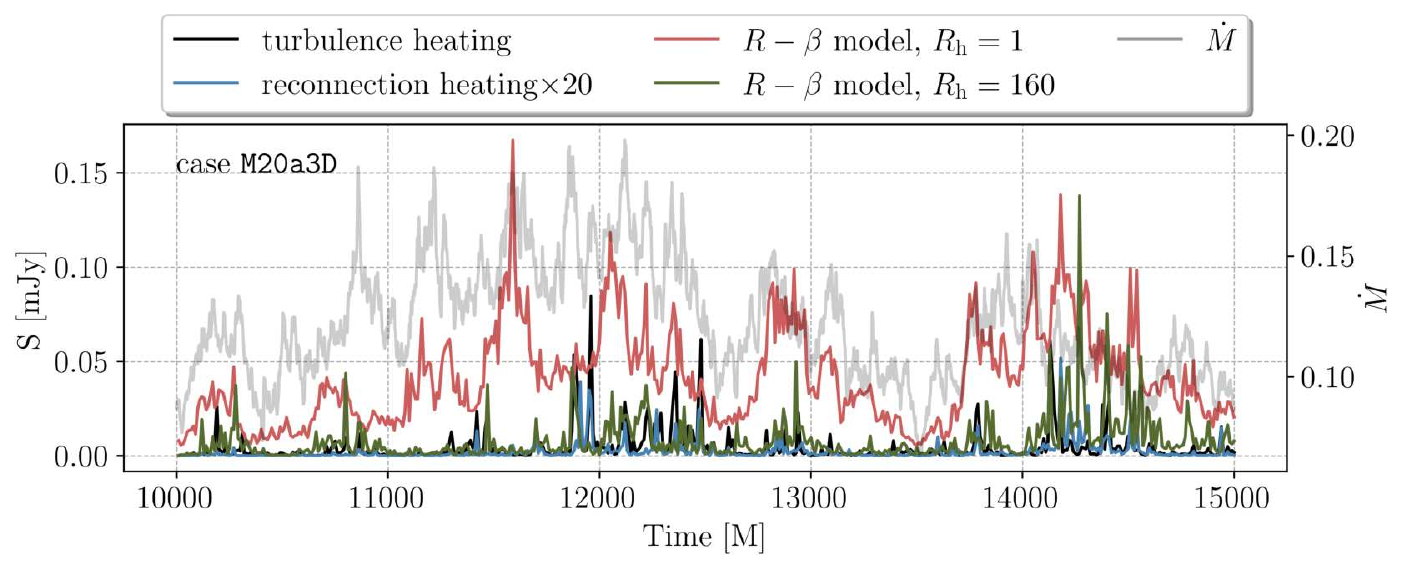}
    \caption{Same format as Figure~\ref{fig: flare_lc}, now shown for the NIR band.}
    \label{fig: NIR}
\end{figure*}

To compare turbulence and reconnection heating more directly, Figure~\ref{fig: eta} contrasts their electron temperatures cell by cell for M20a3D at $t=12,240\,M$. The left-right asymmetry simply reflects the azimuthal asymmetry of the underlying flux-rope-bearing accretion flow. The comparison shows that turbulence heating raises the temperature more strongly in the jet and sheath, whereas reconnection heating favors the dense disk. Since the two models have similar mass-accretion rates ($\dot{M}_{\rm tur}=2.19\times10^{-8}\,M_{\odot}\,{\rm yr}^{-1}$ and $\dot{M}_{\rm rec}=1.19\times10^{-8}\,M_{\odot}\,{\rm yr}^{-1}$), this temperature difference is the main reason why the turbulence-heating case is much brighter at 138 THz than the reconnection-heating case (Figure~\ref{fig: NIR}). In the reconnection-heating model, even the hottest ropes remain cooler than their counterparts in the turbulence-heating run, despite the hotter torus.

\begin{figure*}
    \centering \includegraphics[height=0.37\linewidth]{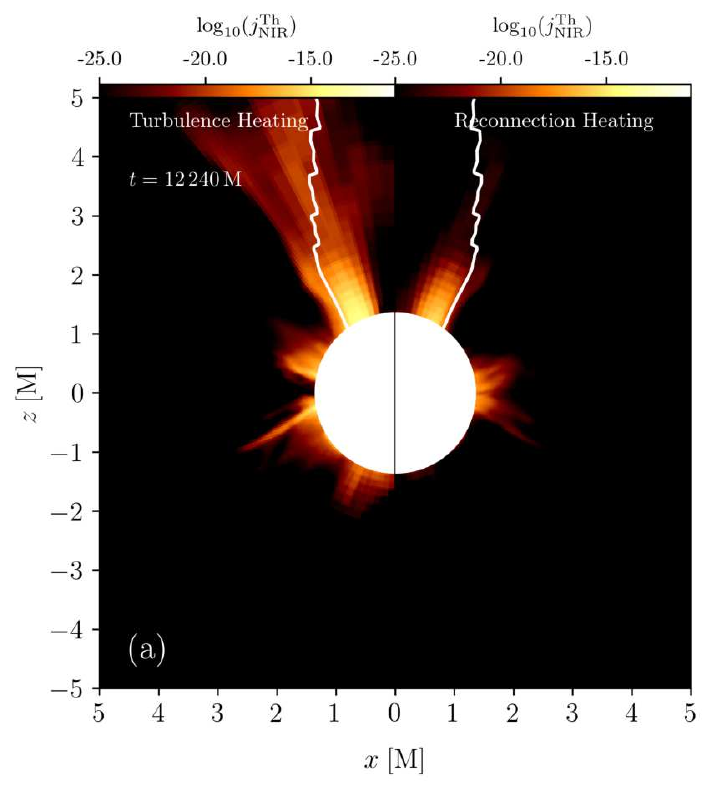}
\includegraphics[height=0.37\linewidth]{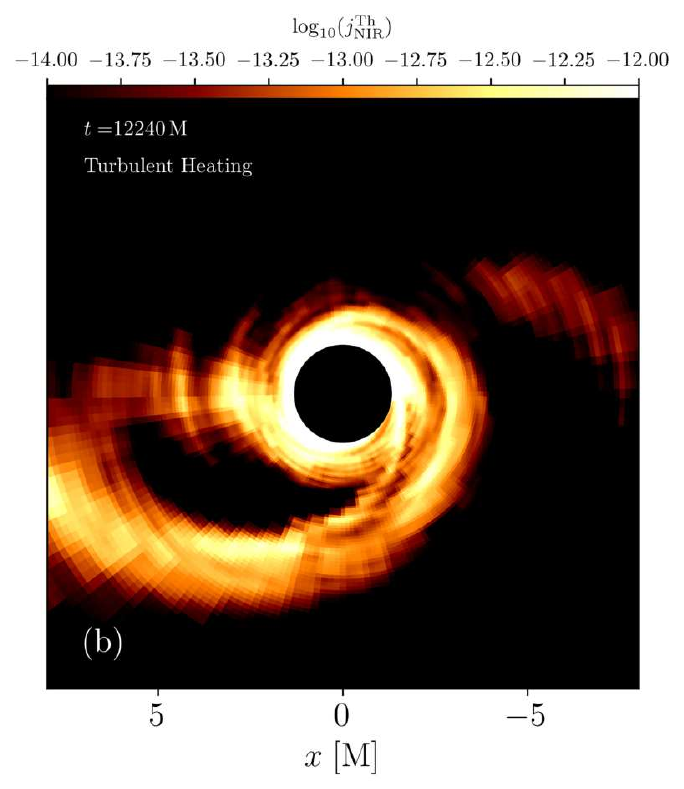}
\includegraphics[height=0.37\linewidth]{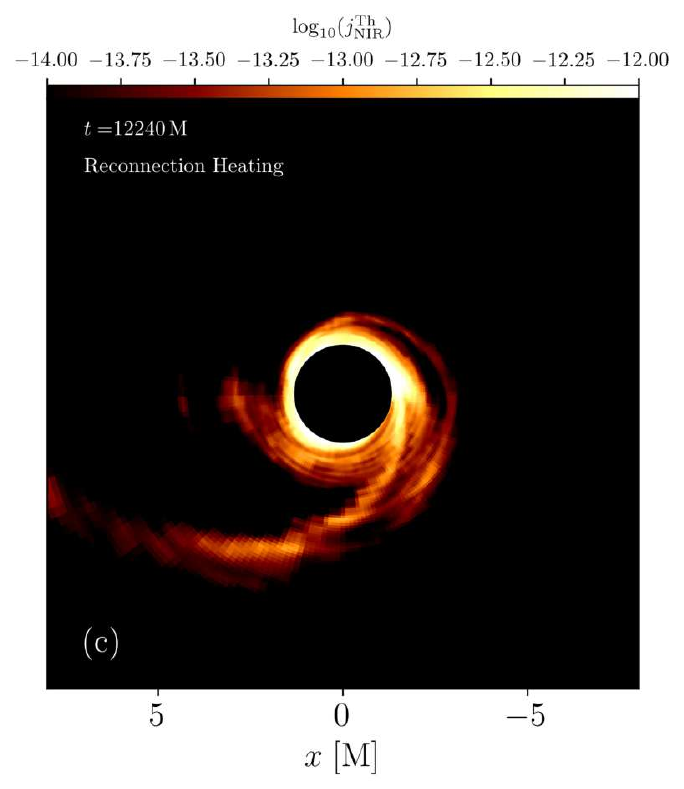}
    \caption{Panel (a) shows azimuthally averaged NIR emissivity maps for the turbulence-heating model (left) and the reconnection-heating model (right). Panels (b) and (c) show the corresponding zenith-averaged emissivity distributions for the turbulence and reconnection cases.}
    \label{fig:phi_avg}
\end{figure*}

\subsubsection{Emission Region at NIR Frequencies}

All models are normalized at 230 GHz through the accretion-rate calibration, so their mean NIR levels are free to differ. Figure~\ref{fig: NIR} shows the resulting NIR light curves for M20a3D under the different heating prescriptions\footnote{Because the reconnection-heating model is much fainter in the NIR, its light curve is multiplied by a factor of 20 in the figure.}. In contrast to the 230 GHz case, the two-temperature models are now much more clearly separated from the parameterized $R-\beta$ prescriptions. The reconnection-heating curve is roughly an order of magnitude fainter than the other three, whereas the $R-\beta$ model with $R_{\rm h}=1$ is the brightest. The turbulence-heating result remains closest to the $R-\beta$ model with $R_{\rm h}=160$. During the most flare-active intervals, $t=12,000$--$13,000\,M$ and $t=14,000$--$15,000\,M$, the turbulence-heating light curve also develops sharper spikes than either $R-\beta$ model. The $R_{\rm h}=1$ curve is brighter, but it is also much smoother. These differences show that the dominant NIR-emitting region depends very strongly on the heating prescription. Even the brightest modeled peaks remain one to two orders of magnitude below the observed Sgr A$^*$ NIR range of 5 to 25 mJy \cite{Abuter2020}, which is expected because real NIR flares are thought to be dominated by non-thermal radiation. The comparison therefore reinforces the conclusion that a non-thermal component is needed to reproduce the observed flare amplitudes \cite{2006AAS...20911207Y, dodds-eden_time-dependent_2010, witzel_rapid_2021, zhao_impact_2023}, although that extension is left to future work.

Even so, the thermal NIR emission remains diagnostically useful. In every model, the NIR variability is governed mainly by the hottest plasma, with typical temperatures $\Theta_{\rm e}\gtrsim100$ and peak values around $\Theta_{\rm e}\sim500$. Figure~\ref{fig: emissivity} shows that, for fixed magnetic field, the thermal synchrotron emissivity continues to increase with electron temperature up to roughly $\Theta_{\rm e}\lesssim1,000$, so the brightest NIR zones naturally coincide with the hottest gas. As already emphasized by \cite{2023MNRAS.522.2307J}, the $R-\beta$ prescription cannot reproduce the fine substructure of hot plasmoids and current sheets. The present 3D calculations lead to the same conclusion: the self-consistent two-temperature solutions retain much more small-scale structure than the parameterized model (Figure~\ref{fig: Thetae}). Figure~\ref{fig:phi_avg} identifies the dominant NIR-emitting zones by showing the emissivity distributions of the turbulence- and reconnection-heating models after averaging over $\phi$ and $\theta$ at $t=12,240\,M$. In the turbulence-heating run, the NIR emission is concentrated in the sheath. In the reconnection-heating case, the much weaker NIR component is centered instead on the torus. The $\theta$-averaged maps in Figure~\ref{fig:phi_avg}(b) and (c) also show a bright spiral-like structure associated with the flux rope. Relative to reconnection heating, the turbulence-heating prescription produces a brighter and more extended high-emissivity region, which naturally yields a larger total NIR flux.

Because the flux ropes contribute so strongly in the NIR, they are plausible counterparts of the observed NIR flares \cite[e.g.][]{Abuter2023}. Reconstructed 3D flare orbits also favor a low-inclination plane \cite{Levis}. This differs from MAD-based flare models, where the emitting structure remains close to the equatorial plane and therefore corresponds to an inclination nearer $90^\circ$ \cite[e.g.][]{Scepi2022, 2020MNRAS.497.4999D, 2021MNRAS.502.2023P}. In the present simulations, most flux ropes occupy the sheath, where the characteristic inclination is lower, around $\sim30^\circ$, and where the structures evolve rapidly with time. Their orbital timescale is about $\sim50\,M$, and most survive for only about one orbit before dissipating. The implied flare duration is therefore also of order 10--100 M, as suggested by Figure~\ref{fig:phi_avg}. In the turbulence-heating model, repeated energy release from these zones drives the strongest light-curve variability because the ropes themselves dominate the radiation (Figure~\ref{fig:phi_avg}(b)). The variability pattern therefore maps directly onto the evolution of the hottest plasma.

 

\section{Summary and Conclusions}

I performed 3D two-temperature GRMHD simulations of magnetized accretion flows around a rotating black hole initialized with multiple magnetic loops. Two loop scales, $\lambda=20$ and $\lambda=80$, were studied. For the electron thermodynamics, I compared turbulent heating, reconnection heating, and a parameterized $R-\beta$ prescription. The main results are summarized below.

\begin{enumerate}
    \item The initial loop scale controls both magnetic dissipation and MRI activity in the torus. Shorter loops dissipate more efficiently and weaken the MRI, whereas longer loops permit a partial transition from SANE toward MAD. This matches the earlier 2D trend \cite{2023MNRAS.522.2307J}, although the 3D transition remains incomplete because the field never becomes strong enough to choke the inflow entirely.
    \item While only the first loop is accreted, the flow behaves much like the single-loop solutions of \cite{Mizuno2021}. In that stage, the 230 GHz light curves are nearly identical across all heating prescriptions, whether they come from the evolved two-temperature model or from the $R-\beta$ reconstruction. Once loops of opposite sign begin to arrive, reconnection strengthens and the flow becomes strongly turbulent.
    \item After the turbulence is fully developed, the light curves depend strongly on the heating prescription. The turbulence-heating model and the $R_{\rm h}=160$ $R-\beta$ case are the most variable because they receive a larger fraction of their flux from the strongly varying sheath. The reconnection-heating model and the $R_{\rm h}=1$ case are less variable because a larger share of their radiation comes from the calmer equatorial flow.
    \item The electron temperature regulates the thermal emissivity in a highly frequency-dependent fashion. At higher frequencies, the radiation is dominated by hot structures such as erupting flux ropes. Since those ropes combine enhanced density with elevated electron temperature, they are more variable than the equatorial flow and produce both the filamentary image morphology and the strong high-frequency variability, especially in the NIR. At lower frequencies such as 43 GHz, the filamentary pattern is much less prominent.
    \item Compared with reconnection heating, the turbulence-heating model produces a broader electron-temperature distribution. It heats the jet and flux ropes more strongly while leaving the disk cooler. The same qualitative behavior already appeared in the 2D calculations \cite{2023MNRAS.522.2307J}. Although the large-scale temperature structure still resembles the $R-\beta$ result, the most important differences arise on smaller spatial scales.
\end{enumerate}

Overall, the 3D multi-loop calculations show that such accretion flows can produce thermal synchrotron behavior that is relevant to the observed variability of Sgr A$^*$. The models do not settle into the cleanest MAD states, even though many present EHT interpretations favor MAD-like solutions for Sgr A$^*$ \cite{2022ApJ...930L..16E}. More generally, however, no currently available model family, whether SANE or MAD, matches the full set of EHT constraints while also reproducing the required variability. That tension motivates exploring multi-loop accretion as an alternative flare scenario. In these simulations, reconnection in the jet sheath emerges as a plausible driver of the flaring activity. Thermal radiation alone still fails to supply enough NIR flux, so the next step will be to add non-thermal emission from the flux ropes and measure its contribution directly. A further limitation is the present 3D resolution: tearing and plasmoid formation are not yet fully resolved, which likely biases the electron temperature low inside those structures. Higher-resolution calculations, together with additional diagnostics such as visibility-amplitude morphology and M-ring fitting, will be needed to test the multi-loop picture more rigorously for Sgr A$^*$.

I also find that flux ropes with comparatively ordered magnetic fields contribute an increasing fraction of the total flux at higher observing frequencies. Their polarized emission should therefore become progressively more important. Testing that expectation will require polarized GRRT calculations in future work.


\newpage
{
\section*{Supplementary Appendix}  
\setcounter{section}{0}
\renewcommand{\thesection}{\thechapter.\Alph{section}}
\renewcommand{\theHsection}{appendix.\thechapter.\Alph{section}}
\renewcommand{\theHsubsection}{appendix.\thechapter.\Alph{section}.\arabic{subsection}}

\section{Electron-Temperature Profile}

\begin{figure}
    \centering
    \includegraphics[width=.6\linewidth]{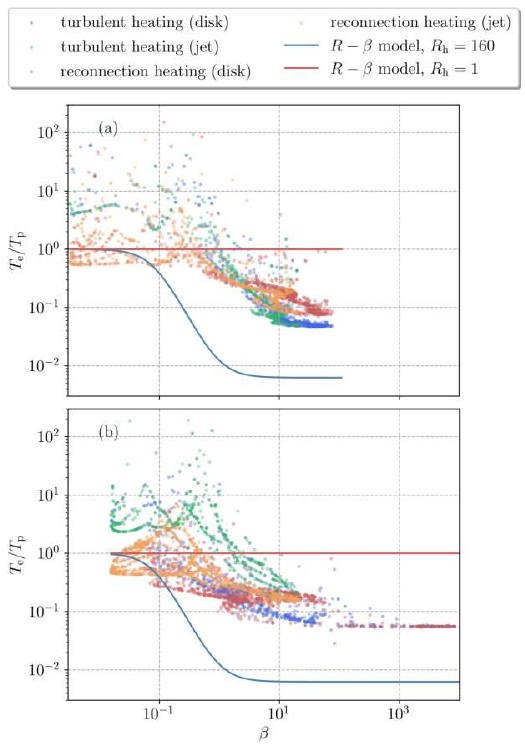}
    \caption{Time-averaged $T_{\rm e}/T_{\rm p}$-$\beta$ distributions for the 3D models. Panel (a) shows M80a3D over 15,000--20,000 M, and panel (b) shows M20a3D over 10,000--15,000 M. Colors distinguish jet and disk cells and also separate turbulent from reconnection heating; the jet is defined by $\sigma>1$. The red and blue reference curves correspond to the $R-\beta$ prescriptions with $R_{\rm h}=1$ and 160.}
    \label{fig: TeTp-beta-c3}
\end{figure}
The two-temperature models imply a more complicated relation between $T_{\rm e}/T_{\rm p}$ and plasma beta than a simple $R-\beta$ fit can describe. Figure~\ref{fig: TeTp-beta-c3} therefore plots the time- and azimuthally averaged $T_{\rm e}/T_{\rm p}$-$\beta$ relation for M80a3D and M20a3D. In the jet regions of both models, some points reach $T_{\rm e}/T_{\rm p}>1$, showing that the electron temperature can dominate locally. At low beta the scatter is substantial, again indicating that the simple $R-\beta$ prescription misses part of the underlying physics there. In the weakly magnetized, high-$\beta$ regime, the two-temperature models move toward similar values in M20a3D (lower panel of Figure~\ref{fig: TeTp-beta-c3}). Even so, near $\beta\sim10$, the reconnection-heating model gives a larger $T_{\rm e}/T_{\rm p}$ ratio, corresponding to the bluish disk region in Figure~\ref{fig: eta}. The same qualitative trend already appeared in the earlier 2D study \cite{2023MNRAS.522.2307J}.
}

\chapter[Multi-band flares from Sgr A$^*$]{Magnetic-Reconnection Flares in Sgr A$^*$ Across Multiple Wavebands}

A prior version of this article appeared in Jiang H.-X., Mizuno Y., Dihingia I. K., Yuan F., Lin X., Fromm C. M., Nathanail A., and Younsi Z., 2025, ApJ, 990, 81 \cite{2025ApJ...990...81J}.


\section{Astrophysical Motivation for Sgr A$^*$ Flares} \label{sec:intro-c4}
Sagittarius~A$^*$, hereafter Sgr~A$^*$, is the supermassive black hole at the Galactic center and the nearest object of its kind that can be studied in comparable detail. EHT measurements place it at a distance of about $D\sim 8\,\rm kpc$ and imply a mass of roughly $M_{\rm SgrA}\sim 4\times 10^6\,M_\odot$ \cite{2022ApJ...930L..12E}. Its surrounding accretion flow forms a bright ring with an observed diameter of $51.8\pm2.3\,\rm \mu as$ \cite{2022ApJ...930L..12E}. Because the source is so close, Sgr~A$^*$ provides an unusually clean laboratory for strong-gravity accretion physics. It is usually interpreted within the radiatively inefficient accretion-flow framework introduced by \citet{Yuan2003}.

The emission from Sgr~A$^*$ is strongly time variable, which makes the source observationally challenging \cite{2022ApJ...930L..15E}. Most of the time the system remains in a quiescent state, but it intermittently produces flares from the X-ray band down to the radio, with flux increases that can approach an order of magnitude above the baseline level \cite[e.g.,][]{Abuter2020, 2001Natur.413...45B, Genzel2003, Schodel2011, Mossoux2020, Murchikova2021}. With the angular resolution of GRAVITY \cite{Abuter2017}, \cite{GRAVITY2018b} detected orbital hot-spot motion during NIR flares near the last stable orbit of Sgr~A$^*$. ALMA polarimetry further shows that the sub-millimeter flare component is strongly polarized \cite{2022A&A...665L...6W}. Comparisons of NIR and radio light curves further suggest that the variability amplitude grows toward higher observing frequency \cite{Abuter2020, Subroweit2017}.

Multi-band observations offer some of the strongest clues to the flare origin. In most reported events, the X-ray and NIR flares are simultaneous \cite{2004A&A...427....1E, 2006A&A...450..535E}. The accompanying sub-millimeter variability usually appears later \cite{2006A&A...450..535E, 2003ApJ...586L..29Z, 2006ApJ...644..198Y}. The statistical study of \cite{witzel_rapid_2021} found a characteristic delay of roughly $\sim20$ minutes between 230 GHz emission and the 138 THz NIR band, although some individual events show lags as long as $\sim4.5$ hours. This is consistent with the earlier measurements of \citet{Yusef-Zadeh2006}, who showed that the 43 GHz peak precedes the 22 GHz peak. Taken together, these results are commonly interpreted as signatures of expanding plasmoids \cite{Yusef-Zadeh2006, dodds-eden_time-dependent_2010, Rauch2016}.

Many different scenarios have been invoked to explain the flares of Sgr~A$^*$, including shocks in jets \cite[e.g.,][]{2001A&A...379L..13M, 2008A&A...492..337E} and magnetic reconnection operating directly inside the accretion flow \cite[e.g.,][]{2004ApJ...606..894Y, Yuan2009, 2014MNRAS.440.2185D, Nathanail2020, Nathanail2022, dodds-eden_time-dependent_2010, 2017MNRAS.468.2447P, 2020MNRAS.497.4999D, 2020MNRAS.494.5923P, 2021MNRAS.507.5281C, 2022ApJ...926..136W, Ripperda2017, Ripperda2021}. Reconnection is often considered the leading explanation for the observed hot-spot behavior, but neither the field geometry nor the exact flare site is known with confidence.

A common class of models places the dissipation inside a MAD flow during episodes when magnetic flux erupts from the inner disk \cite{Scepi2022, 2020MNRAS.497.4999D, Ripperda2021, 2021MNRAS.502.2023P, 2024arXiv240410982A, 2024A&A...689A.112V, 2025arXiv250107521A}. After the accumulated flux saturates, plasma is expelled from the near-horizon region and forms a rotating spiral. Reconnection then proceeds at the interface between that strongly magnetized ejecta and the surrounding gas. Polarization calculations for such vertically expelled flux ropes can reproduce loops in the Stokes $\mathcal{Q-U}$ plane \cite{2023arXiv230816740N}, one of the signatures observed during Sgr~A$^*$ flares \cite{2022A&A...665L...6W}. In that picture, however, the hot spot remains embedded in the broader accretion flow, so pronounced outward propagation is not automatically obtained.

An alternative picture moves the dissipation site into the coronal flow, where flux ropes can form and rise away from the hole. \citet{Yuan2009} argued that turbulence together with differential rotation can generate both reconnection and flux ropes even without an initially multi-loop field. Once such structures appear, magnetic-pressure forces can push them outward and connect the flare more directly to a plasmoid ejection. Later 3D simulations \cite{2022ApJ...933...55C} supported that scenario by producing flux ropes even from single-loop initial conditions. \citet{Xi2024} then identified a flux rope whose projected motion and super-Keplerian orbit are consistent with the GRAVITY measurements. The radiative behavior of those structures has since been investigated in several later studies \cite[e.g.,][]{Aimar2023, Lin2023, Mellah2023, Xi2024, 2024arXiv240714312D}.

High-resolution calculations also suggest that toroidal magnetic flux can be transformed into poloidal loops through the joint action of the MRI and a mean-field dynamo \cite{2020MNRAS.494.3656L, 2023ApJ...954...40K, 2023ApJ...954L..21G, 2024ApJ...960...97R}. That process can generate polarity reversals inside the torus and produce the familiar butterfly pattern \cite{DelZanna2022, Mattia2020, Mattia2022, 2018ApJ...861...24H, 2024MNRAS.527.3018Z}. In wind-fed models such as \cite{2023MNRAS.521.4277R, Ressler2020}, the toroidal flux is usually assumed to come from the stellar wind, but the wind itself may already carry mixed polarities. If so, the torus could inherit a multi-loop magnetic geometry even in the wind-fed case, which remains only weakly explored. Motivated by that possibility, several studies have examined multi-loop field configurations with radial polarity reversals. When flux of alternating sign is swallowed by the hole, the system cannot maintain a large net magnetic flux efficiently, so the flow remains in the Standard And Normal Evolution (SANE) regime. Jet production is then weakened, while magnetic energy is dissipated preferentially through reconnection \cite{2023MNRAS.522.2307J, Nathanail2020, Nathanail2022, 2015MNRAS.446L..61P, 2019MNRAS.487.4114Y, 2019MNRAS.484.4920Y, Mahlmann2020}. Large polarity reversals can also strongly affect jet stability and may arise from toroidal flux generated out of the original poloidal field \cite{2024ApJ...975...57P, 2024MNRAS.532.1522J}. In addition, such inversions can repeatedly trigger reconnection and thereby accelerate non-thermal electrons \cite{Ball2018, 2017MNRAS.468.2447P, 2004ApJ...606.1083H, 2016ApJ...826...77B}. The multi-loop SANE picture therefore remains a plausible alternative to the better-known MAD flux-eruption scenario for Sgr~A$^*$.

In our previous work \cite{2023MNRAS.522.2307J, Jiang2024}, we showed that multi-loop magnetic fields inside the torus generate many plasmoid chains through polarity inversions. In that study, the NIR emission was computed by GRRT post-processing with a purely thermal electron distribution inferred from the GRMHD data. That setup produced NIR emission that was too weak to match the observed flares of Sgr~A$^*$. Here I extend the analysis by including both thermal and non-thermal electrons when modeling the radiative output of polarity-inversion events. With that addition, the same magnetic configuration can produce strong NIR flares compatible with the observations and can also generate natural time delays between different bands.

The rest of this chapter is organized as follows. Section~5.2 describes the GRMHD model together with the GRRT post-processing setup, Section~5.3 presents the results, and Section~5.4 summarizes the conclusions.

\section{Simulation Model and Radiation Setup} \label{sec: method}

\subsection{GRMHD and GRRT Setup}
To connect directly with the earlier study, I adopt a setup closely related to the 3D GRMHD model of \cite{Jiang2024}. Very short loops dissipate magnetic energy too quickly, whereas very long loops launch jets that are too strong, so the present calculation uses an alternating-polarity multi-loop field with an intermediate wavelength following \cite{Nathanail2020}:
%
\begin{equation}
\begin{aligned}
    A_{\rm \phi}\propto& \exp{(-r/400)}(\rho - 0.01)(r/r_{\rm in})^3\sin^3\theta\\
    &\cos((N-1)\theta)\sin(2\pi(r-r_{\rm in})/\lambda_{\rm r}),
\end{aligned}
\end{equation}
The parameter $\lambda_{\rm r}$ sets the radial scale of the magnetic loops. In the present model I adopt $\lambda_{\rm r}=30\,r_{\rm g}$, where $r_{\rm g}\equiv GM/c^2$ is the gravitational radius. The normalization of $A_{\rm \phi}$ is chosen so that the minimum plasma beta is $\beta_{\rm min}=100$, with $\beta \equiv p_{\rm g}/p_{\rm mag}$. The initial gas distribution is a Fishbone-Moncrief hydrostatic torus \cite{1976ApJ...207..962F}. I set $r_{\rm in}=20\,r_{\rm g}$, $r_{\rm max}=40\,r_{\rm g}$, and fix the black-hole spin to $a=0.9375$. Further details of the GRMHD setup and numerical resolution are collected in Appendix~\ref{Sec: kharma} and \ref{sec:res_check}. The electron temperature is evolved with the two-temperature method of \cite{Ressler2015}; the specific prescription used here is summarized in Sec.~\ref{sec:2T-c4}.

To compute the observables, the GRMHD snapshots are post-processed with the GRRT code \texttt{BHOSS} \cite{2012A&A...545A..13Y, 2020IAUS..342....9Y}. Transfer coefficients are evaluated following \cite{Marszewski2021}. Non-thermal radiation is modeled with a $\kappa$ eDF restricted to $3.5\leq\kappa\leq7.5$. The local value of $\kappa$ is assigned through a sub-grid prescription calibrated against turbulent plasma simulations (see Sec.~\ref{sec:2T-c4}). Whenever the fit falls outside the allowed interval, which can happen for $\sigma(=b^2/\rho)>5$ or for very weak magnetization $\sigma\ll1$, the non-thermal synchrotron emissivity $j_{\rm kappa}$ is set to zero (Appendix~\ref{sec: sigma_cut}). The thermal emissivity $j_{\rm Th}$ is removed only in strongly magnetized regions with $\sigma>5$. Accordingly, the total emissivity is written as $j_{\rm tot}=(1-\tilde{\epsilon})j_{\rm Th}+\tilde{\epsilon}j_{\rm kappa}$, where $\tilde{\epsilon}$ measures the fraction of the local energy budget associated with magnetic energy; its full definition is given in Sec.~\ref{sec:2T-c4} and in \cite{Fromm2022}. At high frequencies, the cooling correction proposed by \cite{Scepi2022} is also applied; the implementation is summarized in Appendix~\ref{sec: cooling}. Throughout this chapter, Sgr~A$^*$ is modeled with mass $M_\bullet=4.14\times 10^6\,M_{\odot}$ and distance $D_{\rm SgrA}=8.127\,\rm kpc$ \cite{Abuter2019_BHMASS}. The ray-traced images cover a field of view of $30\,r_{\rm g}\times 30\,r_{\rm g}$, equivalent to $150\,\rm \mu as\times150\,\rm \mu as$, on a $1024\times1024$ grid. The inclination relative to the spin axis is fixed at $25^\circ$. The GRMHD mass unit is normalized by requiring a flux of $3.5\,\rm Jy$ at 230 GHz. With that choice, the model approximately reproduces the observed SED from millimeter to NIR frequencies (Appendix~\ref{sec: sed}). The GRRT analysis uses snapshots between 8,000 and $11,000\,GM/c^3$, when polarity reversals are most frequent and the NIR activity is strongest.

\subsection{Thermal and Nonthermal Turbulence/Reconnection Models}
\label{sec:2T-c4}

We compute the electron temperature by evolving the electron entropy while including sub-grid heating from both turbulence and magnetic reconnection \cite{2019PNAS..116..771K, Rowan2017}; a broader review of these ideas is provided by \cite{Dihingia2023}.

The NIR flares of Sgr~A$^*$ are unlikely to be reproduced with a purely thermal model alone \cite{Scepi2022, Yuan2003}. We therefore supplement the thermal component with sub-grid non-thermal prescriptions calibrated by Particle-In-Cell simulations of turbulence and reconnection. The synchrotron emissivity is evaluated with a kappa eDF, which connects a Maxwell-J\"uttner-like core smoothly to a power-law tail \cite{Fromm2022, 2006PPCF...48..203X, Davelaar2019}. In the high-energy tail, one may roughly identify $p=\kappa-1$, so $\kappa$ acts as an effective spectral slope. The local $\kappa$ value therefore stands in for unresolved plasma microphysics. In this chapter, we use a sub-grid prescription based on PIC turbulence studies \cite{Meringolo2023}.

The $\kappa$ eDF is written explicitly as
\begin{equation}
    \frac{dn_{\rm e}}{d\gamma_{\rm e}}=\frac{N}{4\pi}\gamma_{\rm e}\sqrt{\gamma_{\rm e}^2-1}\left(1+\frac{\gamma_{\rm e}-1}{\kappa w}\right)^{-(\kappa+1)},
\end{equation}
where $n_{\rm e}$ is the electron number density, $\gamma_{\rm e}$ is the electron Lorentz factor, $w$ measures the width of the $\kappa$ distribution, and $N$ is a normalization constant (see \cite{Pandya2016} for details). The role of $\kappa$ is analogous to the power-law index $p$ in a standard power-law eDF. When $\kappa\gg10$, the $\kappa$ eDF approaches the Maxwell-J\"uttner form (see Figure 4 of \cite{Fromm2022}). The width parameter $w$ controls the energy spread \cite{Fromm2022}; in the limit $\kappa\rightarrow\infty$, it reduces to the dimensionless electron temperature $\Theta_{\rm e}$. We then evaluate the non-thermal synchrotron emissivity using the fitting formulae of \cite{Marszewski2021}.

For $\kappa$, we adopt two fitting expressions motivated by PIC calculations, one for special-relativistic turbulence \cite[$\kappa_{\rm tur}$]{Meringolo2023} and one for magnetic reconnection \cite[$\kappa_{\rm rec}$]{Ball2018}. They are written as
\begin{equation}
    \kappa_{\rm tur}(\beta,\sigma)=2.8+\frac{0.2}{\sqrt{\sigma}} + 1.6\sigma^{-6/10}\tanh{\left(2.25\beta\sigma^{1/3}\right)}, \label{Eq: kappa_tur-c4}
\end{equation}
\begin{equation}
    \kappa_{\rm rec}(\beta,\sigma)=2.8+\frac{0.7}{\sqrt{\sigma}} + 3.7\sigma^{-0.19}\tanh{\left(23.4\beta\sigma^{0.26}\right)},
    \label{Eq: kappa_rec}
\end{equation}

Here, $\sigma$ denotes the magnetization and $\beta$ is the plasma-beta parameter.

Following \cite{Davelaar2019, 2018A&A...612A..34D}, we use the non-thermal acceleration efficiency to determine the width $w$ of the kappa distribution. In \cite{Davelaar2019}, the corresponding expression is
\begin{equation}
    w=\frac{\kappa-3}{\kappa}\Theta_{\rm e}+\tilde{\epsilon}\frac{\kappa-3}{6\kappa}\frac{m_{\rm p}}{m_{\rm e}}\sigma, \label{Eq: w}
\end{equation}
The first term describes the thermal contribution, and the second encodes the non-thermal one; $m_{\rm p}$ and $m_{\rm e}$ are the proton and electron masses. We assume that the non-thermal population draws its energy from the magnetic reservoir. Because the fitting formula for the emissivity of a kappa eDF is valid only for $3.5\leq\kappa\leq7.5$ \cite{Marszewski2021}, we enforce that interval in both prescriptions. The parameter $\tilde{\epsilon}$ quantifies how much of the magnetic energy is used to broaden the kappa distribution \cite{Fromm2022}, namely the magnetic-energy fraction within the total energy budget. Earlier studies often fixed $\tilde{\epsilon}$ to constant values such as 0, 0.5, or 1.0 \cite{Cruz-Osorio2021, Fromm2022}. In a realistic flow, however, the non-thermal acceleration efficiency should depend on the local plasma conditions. The turbulence-based and reconnection-based fits used here are taken from \cite{Meringolo2023} and \cite{Ball2018}:
\begin{equation}
    \tilde{\epsilon}_{\rm tur} = 1.0 - \frac{0.23}{\sqrt{\sigma}} + 0.5 \sigma^{1/10}\tanh{\left[-10.18\beta \sigma^{1/10}\right]}, \label{Eq: epsilon_tur}
\end{equation}
\begin{equation}
    \tilde{\epsilon}_{\rm rec} = 1.0 - \frac{1}{4.2\sigma^{0.55}+1} + 0.64 \sigma^{0.07}\tanh{\left[-68\beta \sigma^{0.13}\right]}, \label{Eq: epsilon_rec}
\end{equation}
Accordingly, we use $\tilde{\epsilon}$ in Eq.~\ref{Eq: w} to set the local weight of the non-thermal electron population.

Following \cite{Fromm2022}, we express the total emissivity $j_{\rm tot}$ and absorptivity $\alpha_{\rm tot}$ as the sum of thermal and kappa-based terms:
\begin{equation}
\begin{aligned}
    j_{\rm tot}&=j_{\rm Th}(1-\tilde{\epsilon})+\tilde{\epsilon}j_{\rm kappa},\\
    \alpha_{\rm tot}&=\alpha_{\rm Th}(1-\tilde{\epsilon})+\tilde{\epsilon}\alpha_{\rm kappa},
\end{aligned}
\label{Eq: emi&alpha-c4}
\end{equation}
The same efficiency factor $\tilde{\epsilon}$ from Eqs.~\ref{Eq: epsilon_tur} and \ref{Eq: epsilon_rec} is used here. In practice, we first form the combined emissivity and absorptivity and then provide those total coefficients to the GRRT solver.

In weakly magnetized gas with $\sigma<0.01$, we retain only the thermal emissivity and absorptivity. The hybrid prescription of Eq.~\ref{Eq: emi&alpha-c4} is used only above that threshold.

The sub-grid prescriptions are activated through thresholds tied to the large-scale plasma state. Non-thermal particles are added only where the resolved fluid variables indicate conditions favorable for reconnection, because the global simulation cannot resolve the current sheets and other microphysical structures in which the acceleration physically takes place. In that sense, the method transfers the results of earlier small-scale studies into the global calculation, while a fully self-consistent cross-scale treatment remains beyond the resolution of the present model.

\section{Results and Physical Interpretation}\label{sec: results}
\subsection{Multi-Frequency Light Curves}

\begin{figure*}
    \centering
    \includegraphics[width=.9\linewidth]{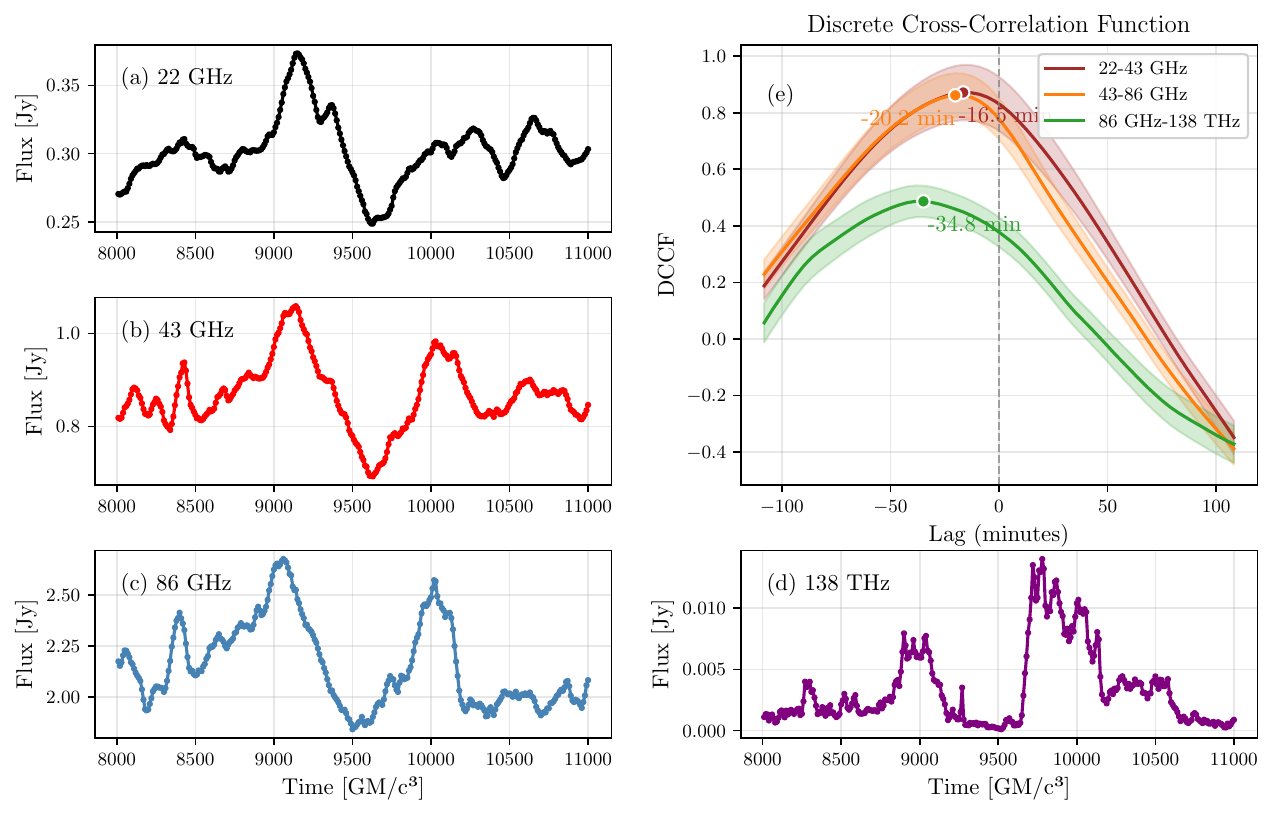}
    \caption{Multi-band light curves and the corresponding discrete cross-correlation functions (DCCFs). Panels (a)--(d) show the GRRT light curves at 22 GHz (black), 43 GHz (red), 86 GHz (blue), and 138 THz (NIR, purple). Panel (e) shows the DCCFs for the pairs 22--43 GHz (blue), 43--86 GHz (orange), and 86--138 THz (green), with shaded regions marking the $1\sigma$ uncertainty.}
    \label{fig: f1}
\end{figure*}

Earlier work has shown that multi-loop accretion flows can undergo repeated reconnection as successive polarity inversions pass through the system \cite[e.g.,][]{Nathanail2020, 2017MNRAS.468.2447P}. Those inversions create many current sheets and plasmoid-rich structures and therefore offer a natural route to the kind of flare activity observed in Sgr~A$^*$ \cite[e.g.,][]{Abuter2020}. In such regions, electrons may be accelerated into a non-thermal tail, allowing magnetic energy to be transferred into the eDF and to power bright NIR events.

After adding non-thermal electrons in the GRRT post-processing, I compute light curves at multiple frequencies over the interval 8,000--$11,000\,GM/c^3$, when polarity inversions occur most often; the result is shown in Figure~\ref{fig: f1}. Panels (a)--(d) display the emission at 22 GHz, 43 GHz, 86 GHz, and 138 THz (NIR) in black, red, blue, and purple. The 138 THz curve shows three major flares near $t\sim8320$, 9000, and $9780\,GM/c^3$. Their peaks reach about 14 mJy, whereas the minima fall below 0.1 mJy. In that sense, the modeled 138 THz variability is broadly consistent with the observed NIR behavior of Sgr~A$^*$ \cite{Abuter2020}.

Panels (a)--(d) of Figure~\ref{fig: f1} also reveal a clear frequency-dependent lag between the NIR and the millimeter-radio bands. As in earlier studies \cite[e.g.,][]{witzel_rapid_2021, 2024arXiv240410982A}, I quantify that lag with the cross-correlation function (CCF); the implementation is summarized in Appendix~\ref{Sec: CCF} and Sec.~\ref{sec: statistics}. Panel (e) shows the corresponding DCCFs for 22 GHz, 43 GHz, 86 GHz, and 138 THz, with shaded bands indicating the uncertainty in each lag bin. The inferred delays are about 16.5 minutes between 22 and 43 GHz, 20.2 minutes between 43 and 86 GHz, and 34.8 minutes between 86 GHz and 138 THz. The lag therefore becomes larger toward lower observing frequency. The flare near $9,000\,GM/c^3$ illustrates this particularly clearly, with the radio peak following the higher-frequency NIR event. That ordering matches the observational picture in which lower-frequency flares lag behind higher-frequency ones \cite[e.g.,][]{Yusef-Zadeh2006, 2008ApJ...682..361Y}.

\begin{figure*}
\centering
 	\includegraphics[width=.49\linewidth]{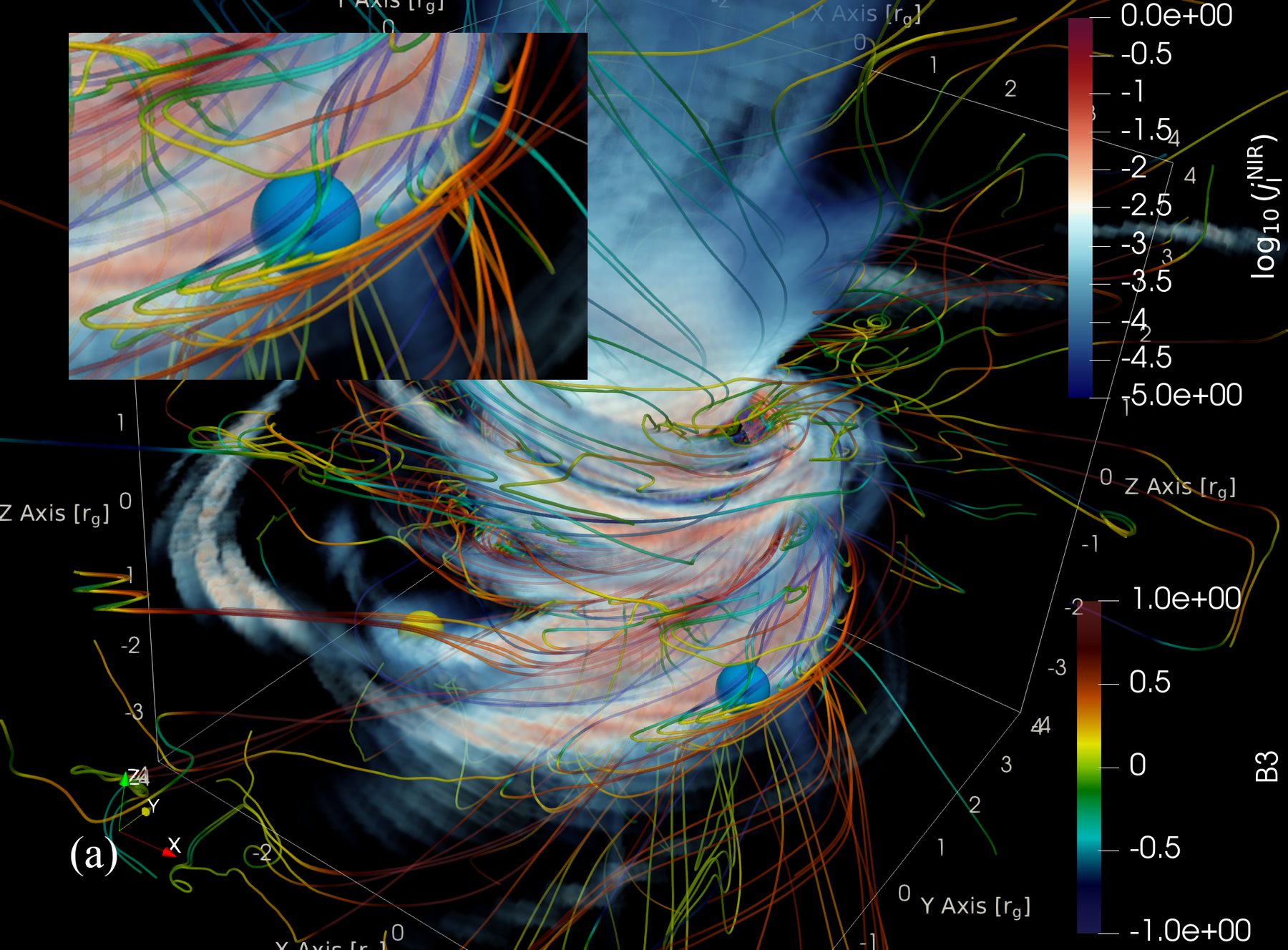}
 	\includegraphics[width=.49\linewidth]{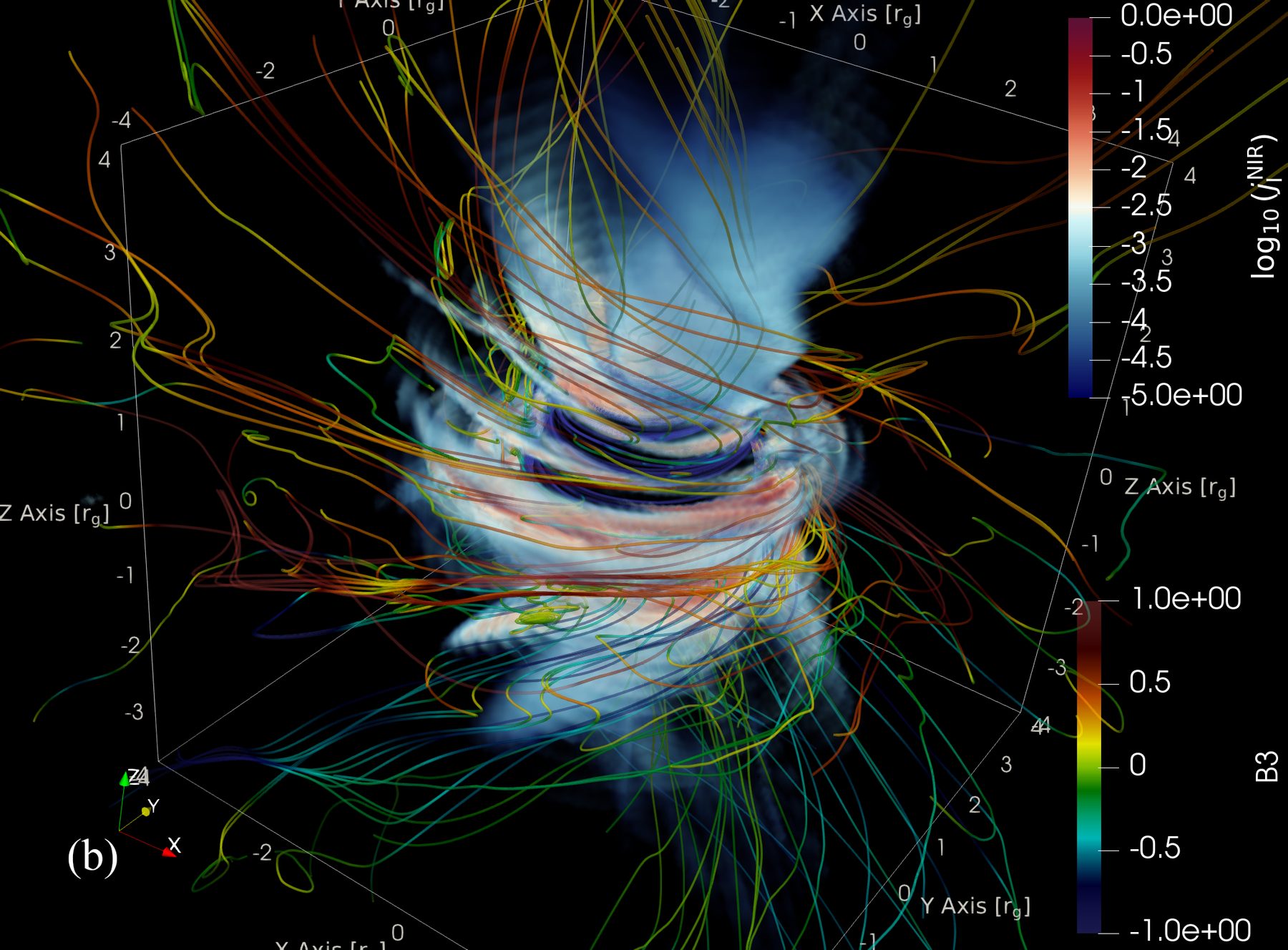}
 	\includegraphics[height=.4\linewidth]{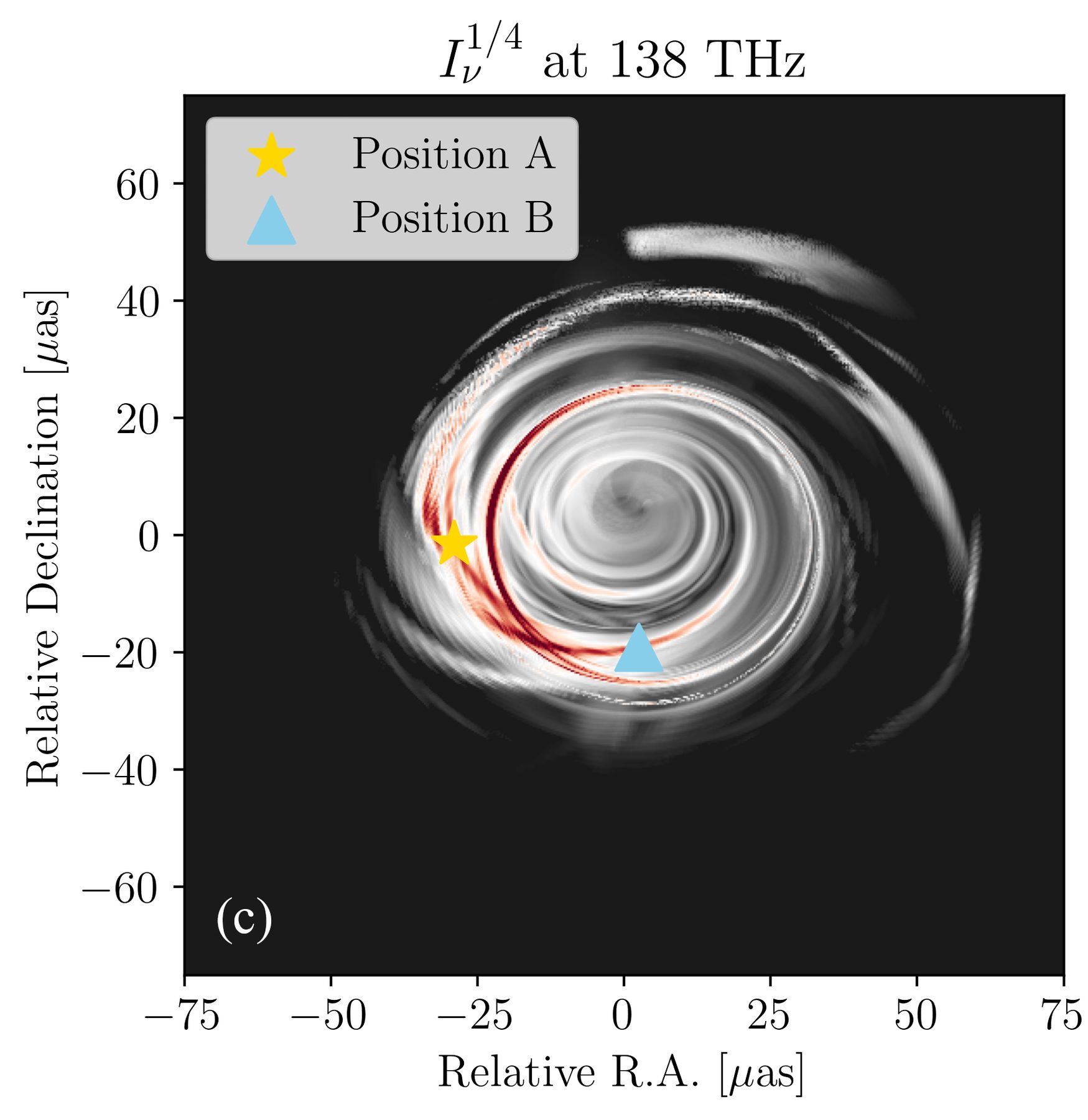}
 	\includegraphics[height=.4\linewidth]{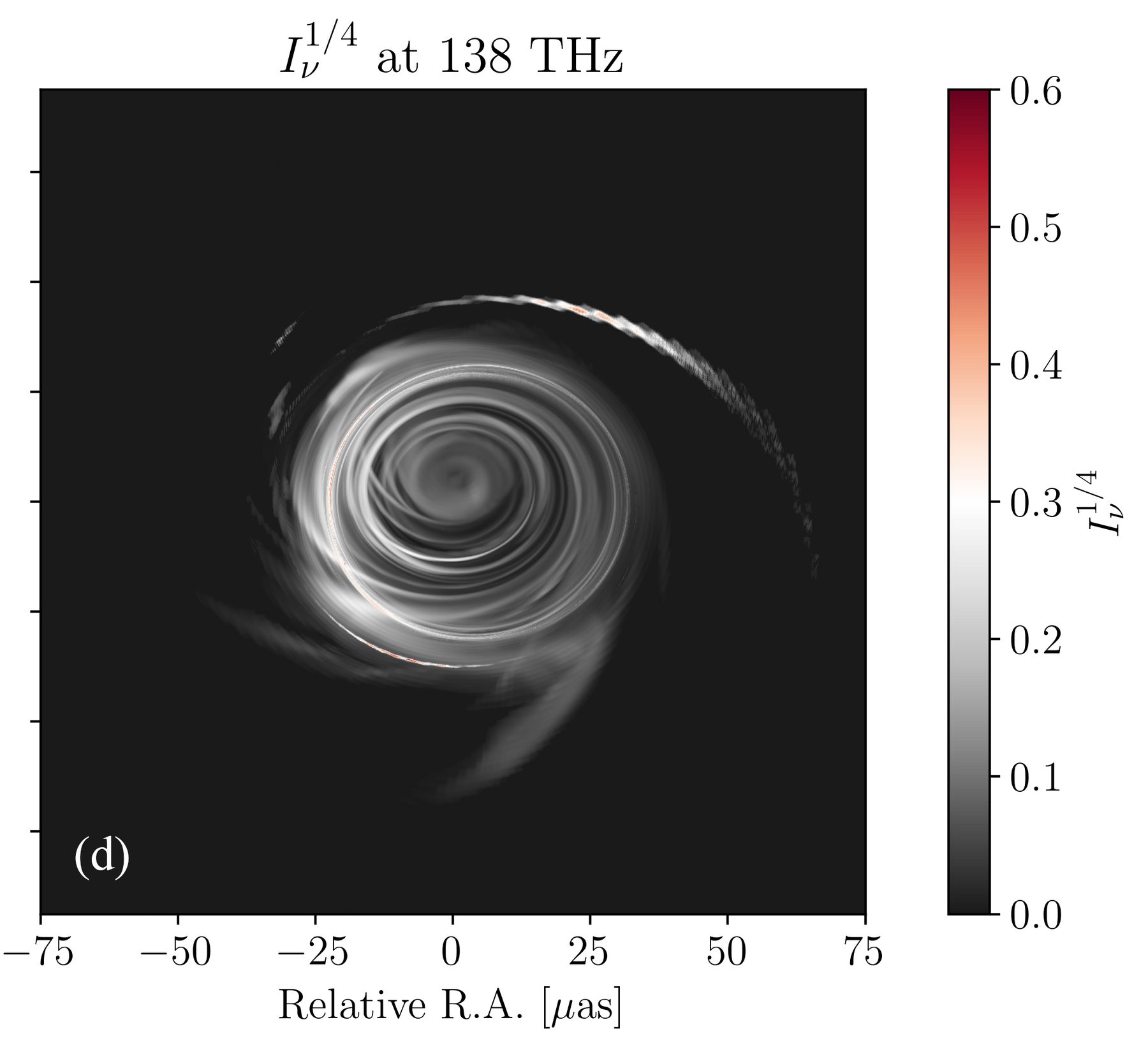}
 	\caption{Large-scale polarity inversion associated with an NIR flare. Panels (a) and (b) show the normalized NIR emissivity ($j_{\rm I}^{\rm NIR}$, scaled to its maximum) together with magnetic-field lines for a flaring state at $t=8,320\,M$ and a quiescent state at $t=9,500\,M$. The field-line colors encode the relative strength of the toroidal component. The enlarged region in panel (a) isolates the upper-left polarity inversion, where tightly wound toroidal field lines change sign; panel (b) contains no comparably extended structure. Panels (c) and (d) show the matching GRRT intensity maps $I_{\nu}^{1/4}$ for the flaring and quiescent states, both normalized to panel (c). In panel (c), Position A and Position B are marked by a yellow triangle and a blue star, while panel (a) marks their 3D emitting locations with yellow and blue spheres.}
 	\label{Fig: f2}
\end{figure*}

\begin{figure}
    \centering
    \includegraphics[width=.6\linewidth]{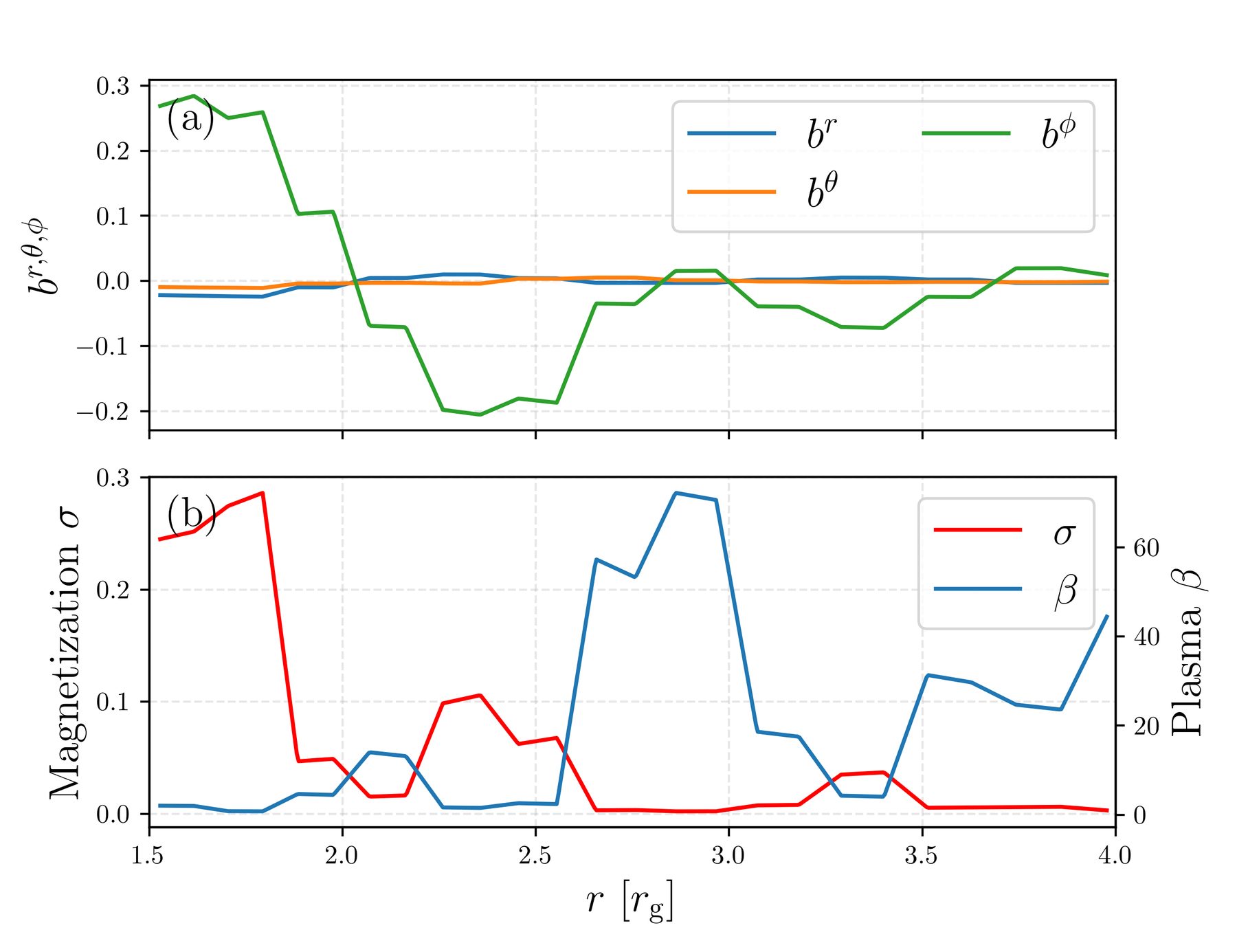}
    \caption{One-dimensional radial cut through the zoomed region of Figure~\ref{Fig: f2}(a). The slice is taken at $\theta$ and $\phi$ values close to Position B in Figure~\ref{Fig: f2}(a). Panel (a) plots the magnetic-field components and shows the toroidal-field reversal across the large-scale polarity inversion explicitly. Panel (b) gives the matching radial profiles of the magnetization $\sigma$ (red) and plasma $\beta$ (blue).}
    \label{Fig: 1D_cut}
\end{figure}

\begin{figure*}
\centering

\includegraphics[width=.49\linewidth]{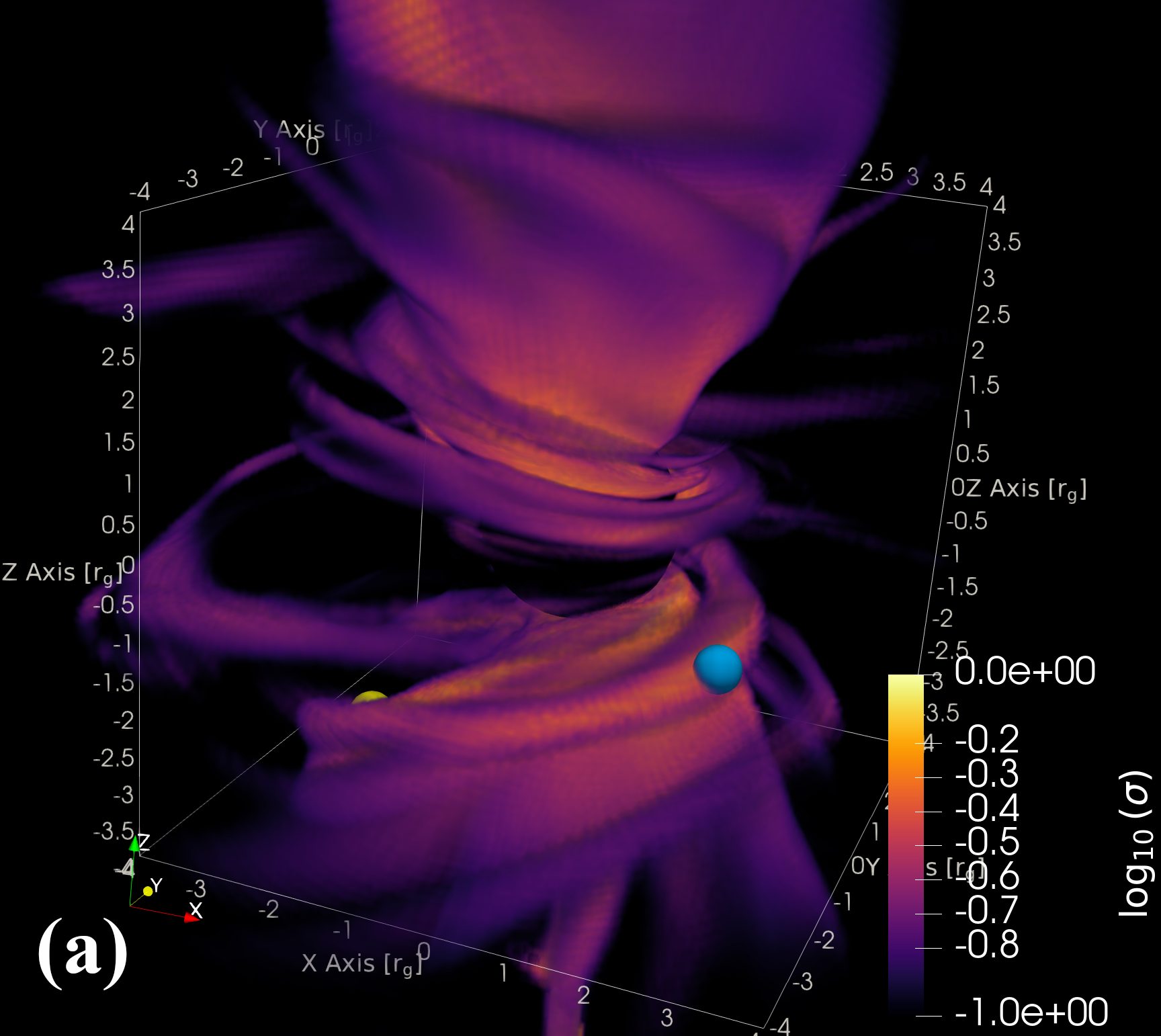}
\includegraphics[width=.49\linewidth]{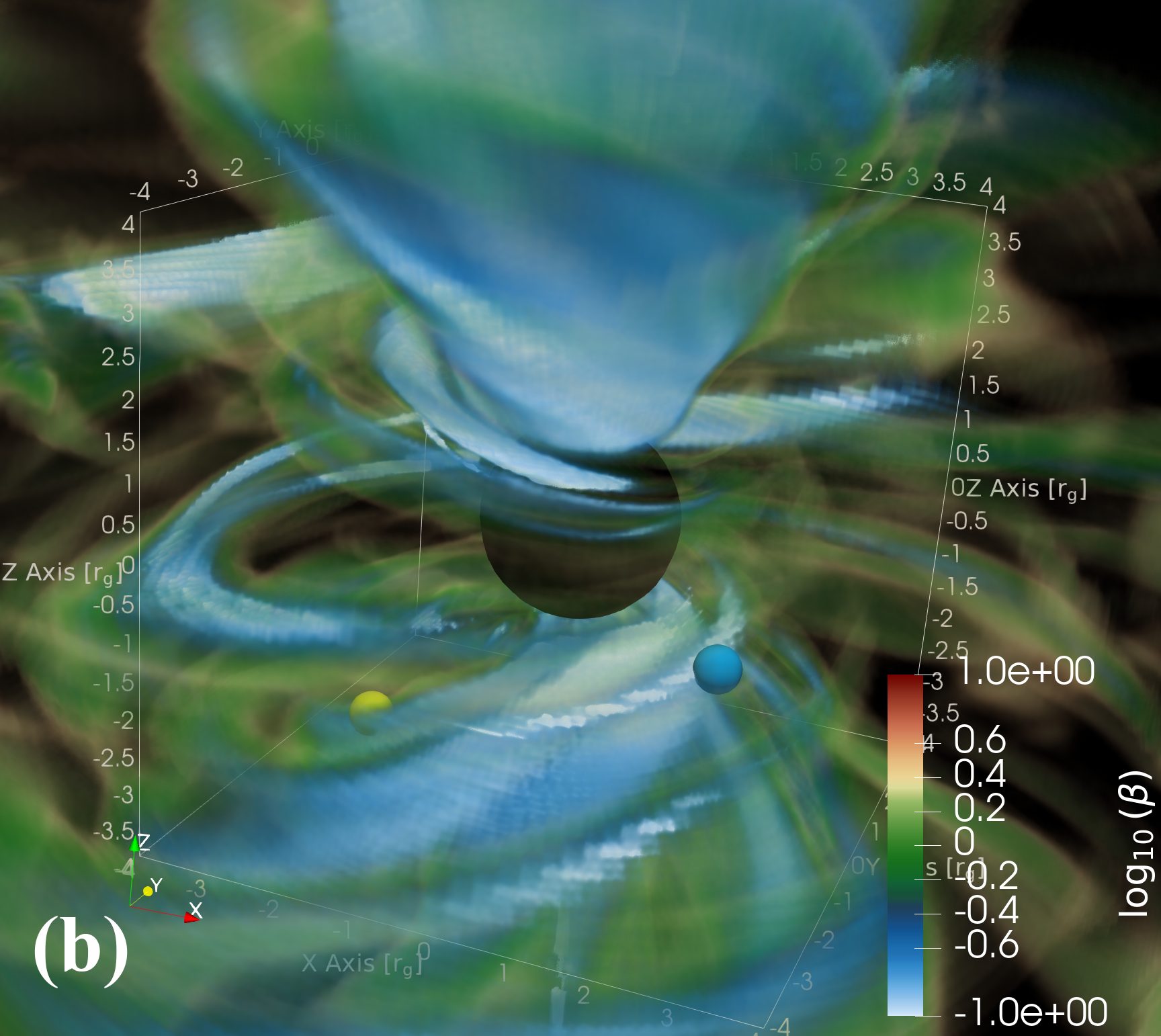}
\includegraphics[width=.49\linewidth]{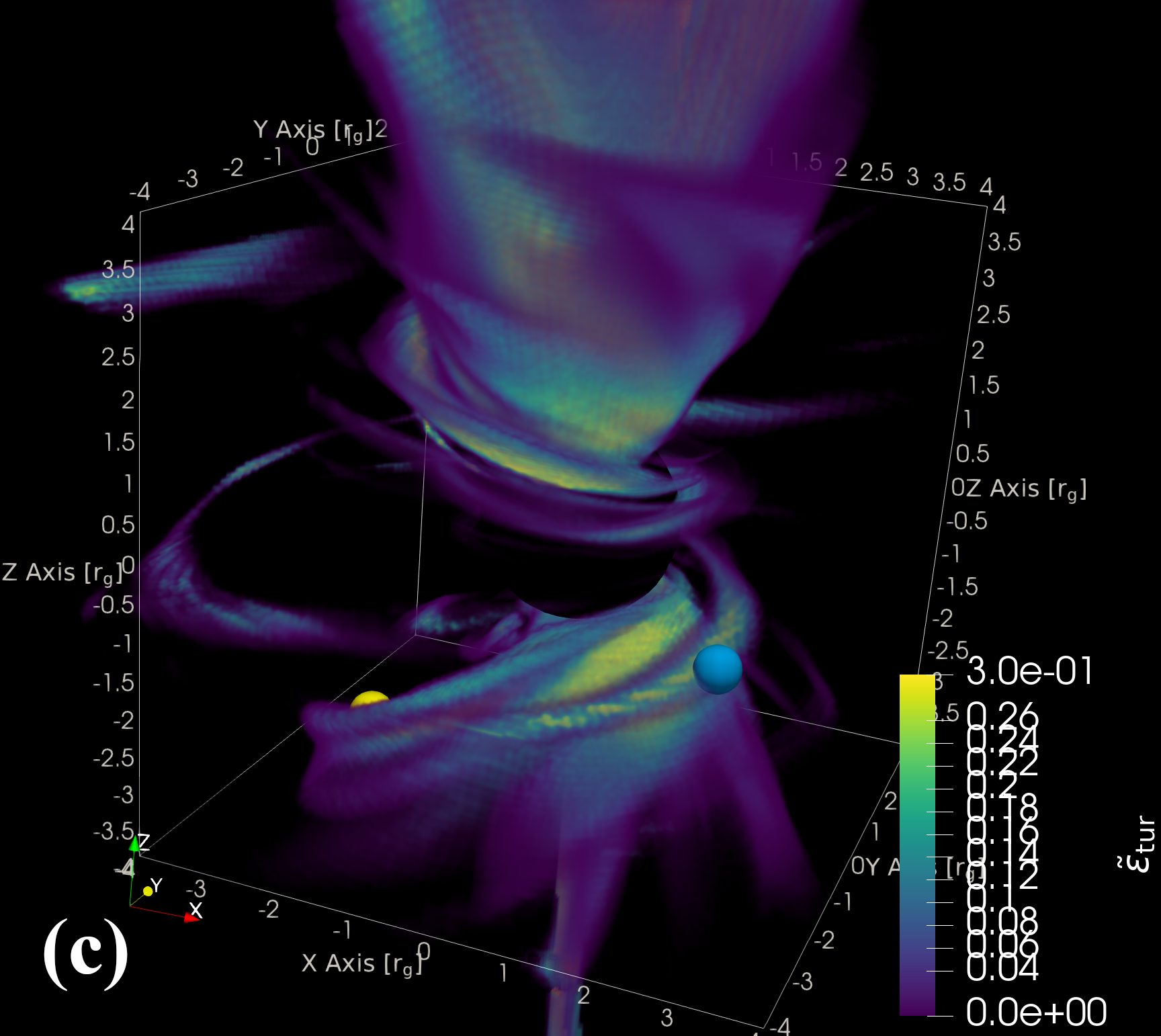}
\includegraphics[width=.49\linewidth]{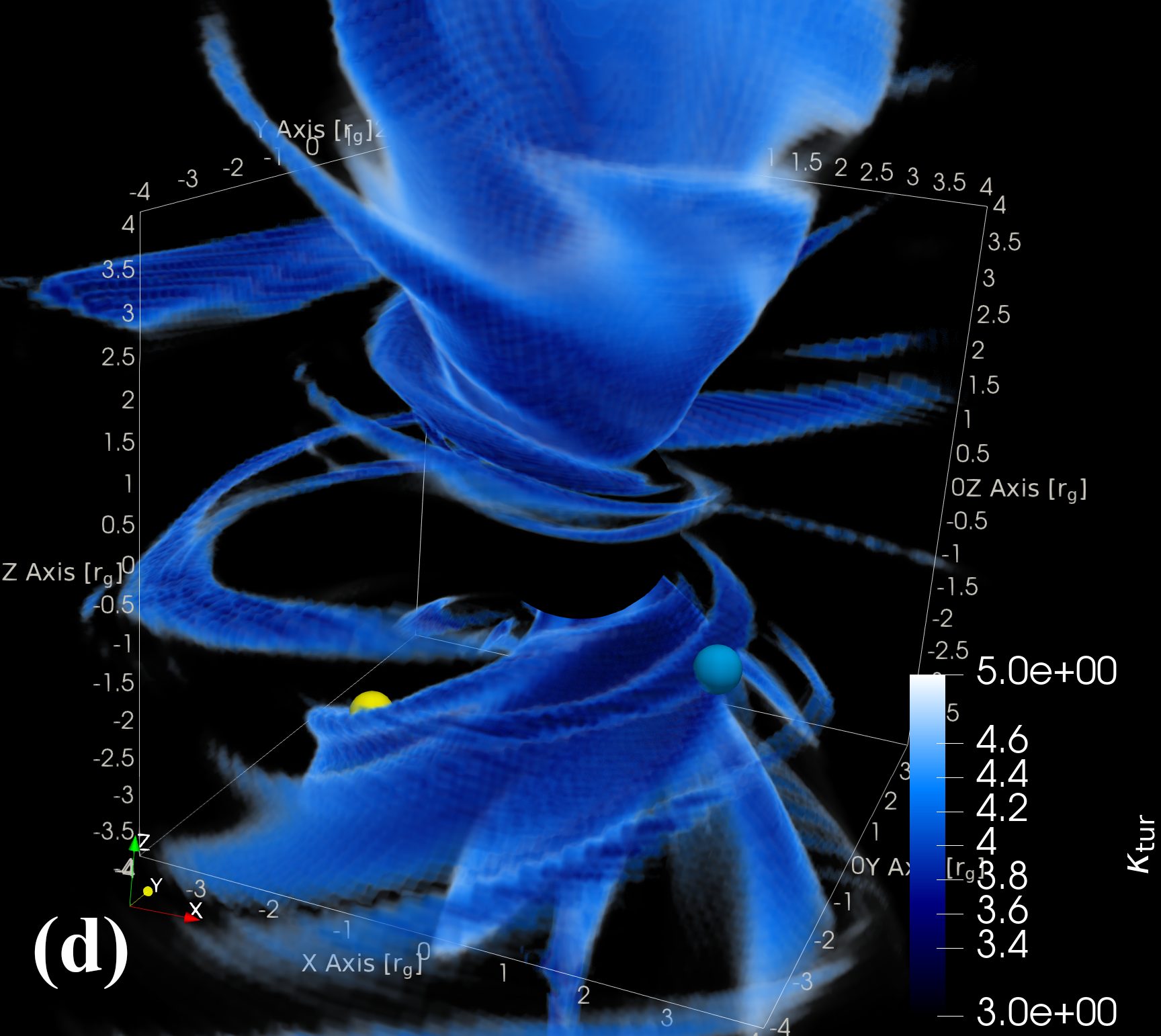}
 	\caption{Local plasma conditions and non-thermal-electron diagnostics near the reconnection site. Panel (a) shows the magnetization $\sigma$, panel (b) shows the plasma beta $\beta$, panel (c) shows the non-thermal acceleration efficiency from the turbulence-based model, and panel (d) shows the resulting non-thermal index $\kappa$. The yellow and blue spheres are the same markers used in Figure~\ref{fig: f1}(b) and identify the polarity-inversion region.}
 	\label{Fig: beta&sigma}
\end{figure*}

\begin{figure*}
    \centering
    \includegraphics[height=.35\linewidth]{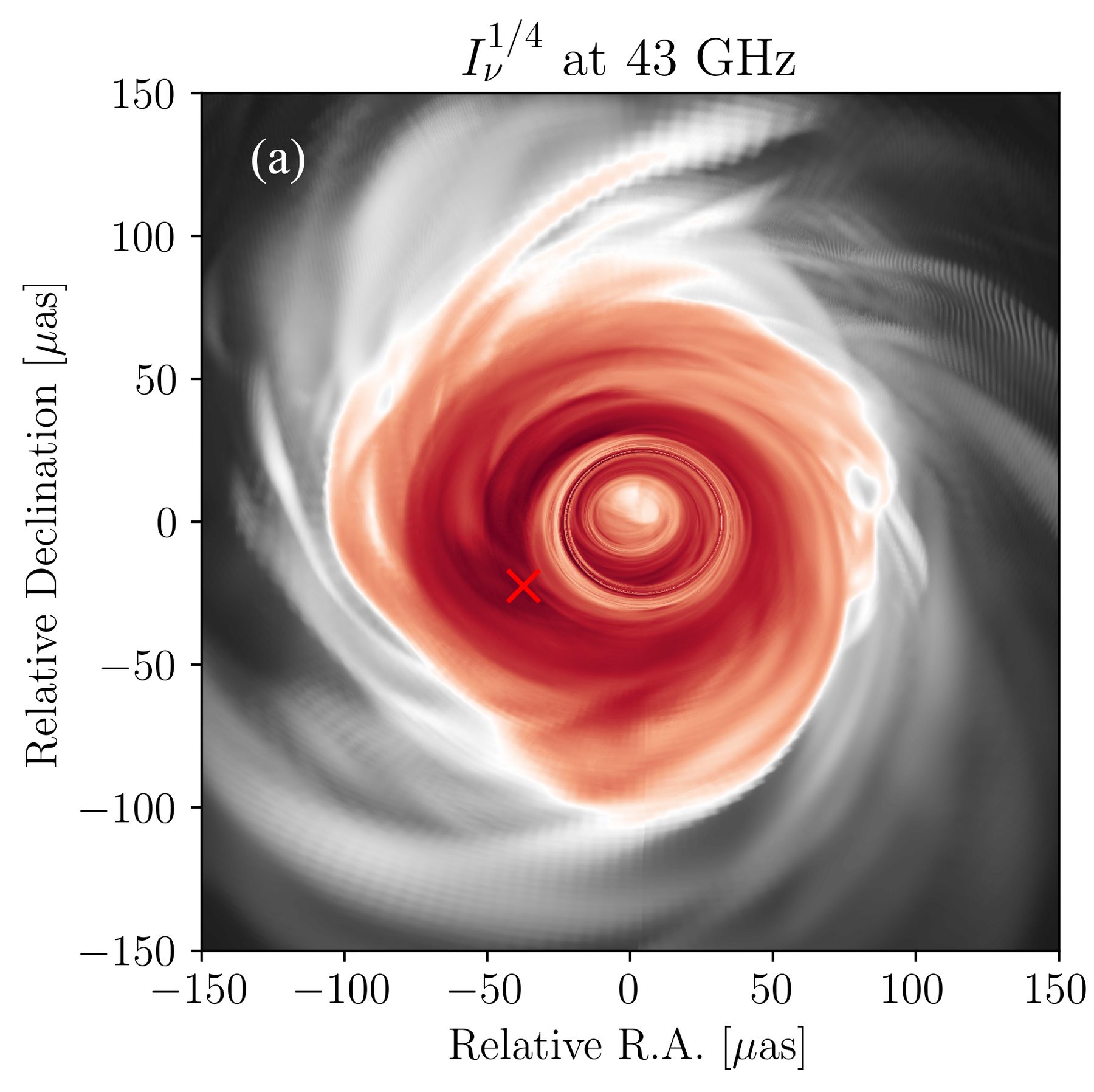}
    \includegraphics[height=.35\linewidth]{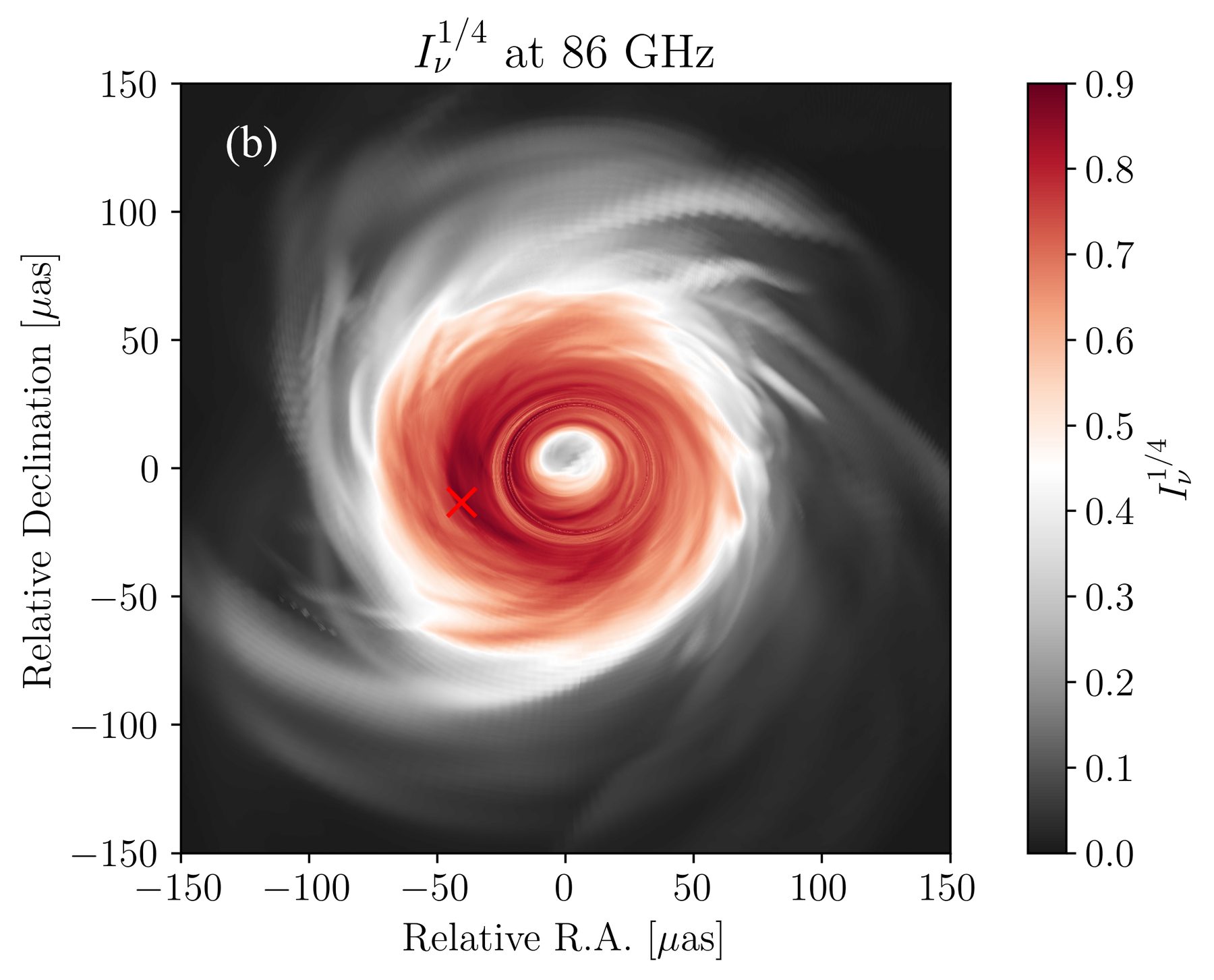}
    \includegraphics[width=.8\linewidth]{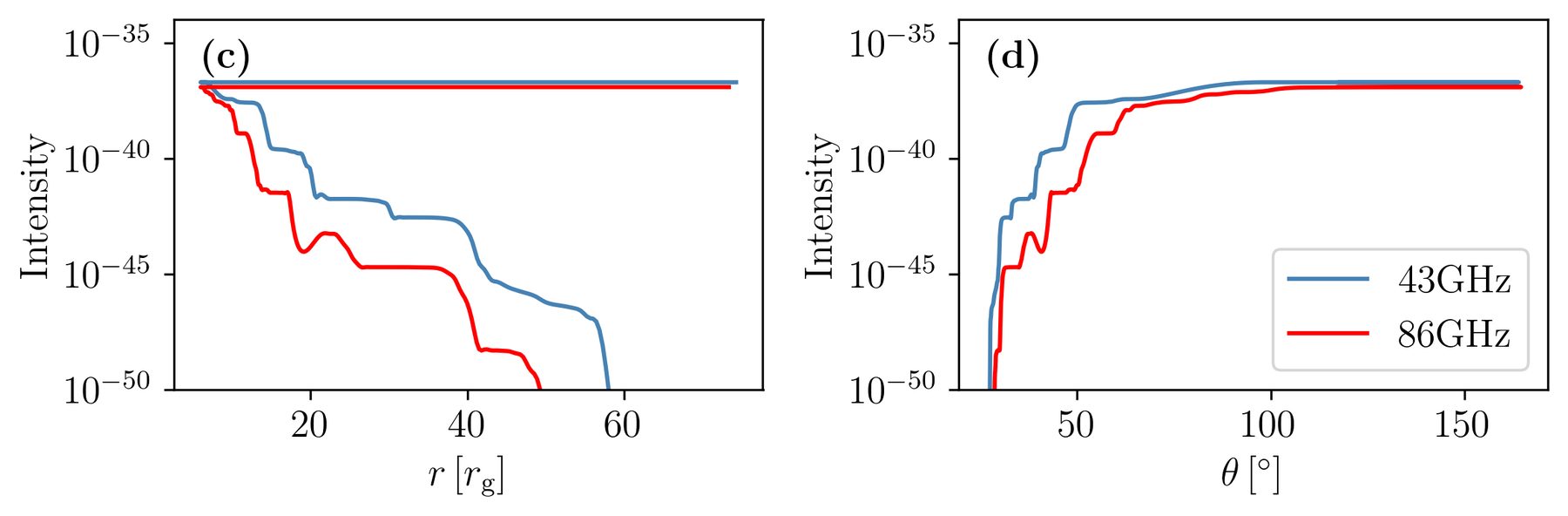}
    \caption{Outward drift of the peak-emission region in the radiative-transfer calculation. Panels (a) and (b) show GRRT snapshots at $t=8,380$ and $8,420\,GM/c^3$, corresponding to the 43 and 86 GHz flare peaks. A red cross marks the brightest pixel in each image. Panels (c) and (d) trace the intensity accumulated along the corresponding geodesics. Blue and red curves denote the 43 and 86 GHz rays, respectively. Panel (c) shows the radial dependence, and panel (d) shows the polar-angle dependence.}
    \label{fig: f3}
\end{figure*}

\begin{figure}
    \centering
    \includegraphics[width=.6\linewidth]{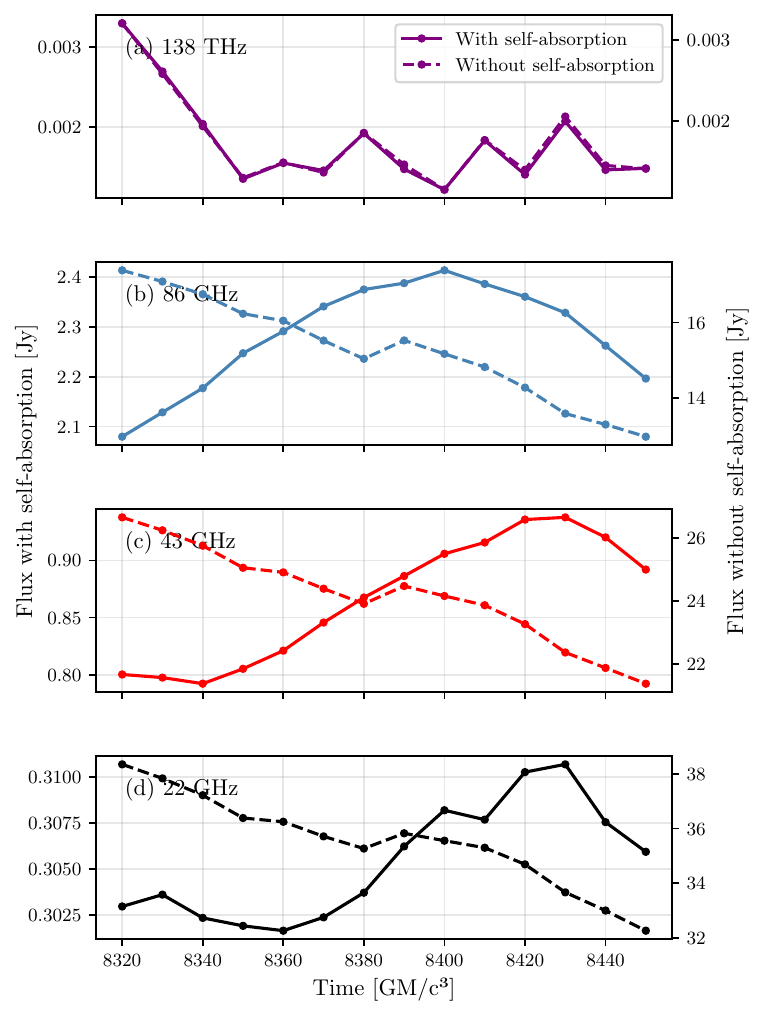}
    \caption{Common origin of the multi-frequency flares and the lag caused by self-absorption. Panels (a)--(d) show the light curves at 138 THz, 86 GHz, 43 GHz, and 22 GHz in purple, blue, red, and black. The solid curves include plasma self-absorption in the GRRT calculation, whereas the dashed curves omit it.}
    \label{fig: f4}
\end{figure}

\subsection{Physical Origin of the NIR Flares}\label{Sec: NIR_origin}

In the multi-loop GRMHD models, polarity inversions are accompanied by comparatively ordered magnetic fields, enhanced density, and elevated electron temperature in the jet sheath. These properties make them natural candidates for the flare-producing regions in Sgr~A$^*$ \cite{Nathanail2020, 2015MNRAS.446L..61P}.

Figure~\ref{fig: f1}(d) shows that the modeled NIR light curve reaches flare peaks nearly two orders of magnitude above the quiescent level. The amplitude, duration, and peak brightness remain close to the values reported by \cite{Abuter2020}. To identify the emitting region, I examine the 134 THz emissivity in 3D volume renderings with magnetic-field lines, together with the corresponding NIR GRRT images, for a flaring state ($t=8,320\,GM/c^3$, total flux 4 mJy) and a quiescent state ($t=9,500\,GM/c^3$, total flux 0.6 mJy) shown in Figure~\ref{Fig: f2}. Panels (a) and (b) show that the flaring snapshot contains an extended spiral emissivity feature inside the accretion flow. In the enlarged upper-left region of panel (a), this bright structure coincides with strongly twisted magnetic field lines. The two sides of the spiral have opposite polarity, identifying the feature as a polarity-inversion layer. The bright emission is therefore linked directly to magnetic-energy release in a large-scale polarity inversion. By contrast, the quiescent state shows neither the strongly twisted field geometry nor the extended spiral emissivity pattern, either in the 3D rendering or in the GRRT image. Small polarity reversals are still present, but they contribute little radiation.

Panel (c) contains an extended bright feature, shown in reddish colors, that is absent from the quiescent image. To determine where this radiation originates, I track how the intensity accumulates along individual geodesics through the GRMHD domain. During the GRRT post-processing, both the ray trajectory and the local intensity are recorded, as described in Appendix~\ref{Sec: ray_path}. This makes it possible to identify the part of the flow where each photon packet acquires most of its flux while crossing the polarity inversion. In panel (c), two representative points inside the brightest region are marked by a gold triangle (Position A) and a blue star (Position B). Figure~\ref{Fig: ray_path} then shows where the corresponding rays accumulate most of their intensity. Once the rays have left the flaring zone, their intensity changes only weakly before reaching the observer.
For Position A, the dominant contribution comes from approximately $r\approx3.8$, $\theta\approx138^\circ$, and $\phi\approx212^\circ$, whereas Position B is dominated by emission from about $r\approx3.0$, $\theta\approx115^\circ$, and $\phi\approx321^\circ$. In the 3D rendering of panel (a), those two locations are marked by the same gold and blue spheres and coincide with distinct emissivity peaks. As these structures rotate toward the observer, whose azimuth is zero, Doppler boosting amplifies their radiation and produces the flare. The blue sphere lies directly on the reconnection site and corresponds to strong NIR emission in the GRRT image. This mapping shows that the brightest NIR features are generated by polarity-inversion events.

As Figure~\ref{Fig: f2}(a) illustrates, large-scale polarity inversions generate filamentary emissivity structures. To characterize the local plasma there, I extract the one-dimensional profile across the current sheet shown in Figure~\ref{Fig: 1D_cut}. That cut provides direct evidence for magnetic reconnection.

Figure~\ref{Fig: 1D_cut}(a) shows that the dominant toroidal component ($b^\phi$) reverses sign completely, identifying it as the main magnetic-energy reservoir dissipated in the event. By contrast, the poloidal component ($b^\theta$) keeps the same sign and therefore acts only as a weak guide field. This agrees with the global result that the ratio of reconnecting to guide field remains well below 0.1, indicating weak-guide-field reconnection. Figure~\ref{Fig: 1D_cut}(b) further shows that the upstream plasma in the reconnection layer has $\sigma\approx0.28$ and $\beta\approx2$.
These plasma conditions fall exactly in the regime where the adopted sub-grid models predict efficient non-thermal acceleration. The full 3D distributions of $\sigma$ and $\beta$ are shown in Figure~\ref{Fig: beta&sigma}(a) and (b). In the GRRT calculation, the non-thermal acceleration efficiency $\tilde{\epsilon}$ varies across the domain and is plotted in Figure~\ref{Fig: beta&sigma}(c); it injects non-thermal electrons directly into these localized structures. The resulting non-thermal radiation is therefore concentrated at the polarity inversions and contributes directly to the observed NIR flares. Figure~\ref{Fig: beta&sigma}(d) shows that most of the volume remains close to thermal ($\kappa>5$), whereas the polarity-inversion layer reaches low $\kappa$, indicating the strong non-thermal component responsible for the enhanced emission.

\subsection{Energetic Requirements for the NIR Synchrotron Flares}\label{sec: energy}

In this section, I show that the magnetic-reconnection regions identified in the GRMHD simulation contain enough energy to accelerate electrons to the energies required for the NIR flares. The synchrotron frequency characteristic of an electron with Lorentz factor $\gamma_{\rm e}$ radiating in a magnetic field $B$ is
\begin{equation}
\label{eq:nu_c}
\nu_c = \frac{3 e B \gamma_e^2}{4 \pi m_e c}.
\end{equation}
To estimate the electron energy needed for NIR emission, we adopt a representative flare wavelength of $\lambda\approx2.2\,\mu{\rm m}$, corresponding to an observed frequency of $\nu_{\rm obs}\approx1.38\times10^{14}$ Hz. In the reconnection zones of our simulation, the magnetic field is typically $B\gtrsim80\,{\rm G}$. Solving for the Lorentz factor then gives
\begin{equation}
\label{eq:gamma_e_req}
\gamma_e \approx \left( \frac{4\pi m_e c \, \nu_{\text{obs}}}{3 e B} \right)^{1/2} \approx 640.
\end{equation}
This estimate shows that the observed NIR radiation requires highly relativistic electrons, with Lorentz factors of order $\gamma_e\sim10^3$.

It is important to distinguish the ion-dominated bulk-flow magnetization $\sigma_i$ from the electron magnetization $\sigma_e$, because the latter is the quantity more directly tied to particle acceleration. In the GRMHD simulation, the measured magnetization refers to the bulk fluid, whose inertia is dominated by protons. The value $\sigma\gtrsim0.5$ measured in the reconnection sites therefore corresponds to the ion magnetization ($\sigma_i$):
\begin{equation}
\label{eq:sigma_i}
\sigma_i = \frac{B^2}{4\pi \rho c^2} \approx \frac{B^2}{4\pi n_p m_p c^2} \gtrsim 0.5.
\end{equation}
Direct magnetic energization of electrons, however, is governed by the electron magnetization $\sigma_e$, defined using the electron rest-mass energy density:
\begin{equation}
\label{eq:sigma_e}
\sigma_e = \frac{B^2}{4\pi n_e m_e c^2}.
\end{equation}

For a quasi-neutral electron-ion plasma with $n_e\approx n_p$, the two magnetizations differ only by the proton-to-electron mass ratio:
\begin{equation}
\label{eq:sigma_relation}
\sigma_e = \left( \frac{m_p}{m_e} \right) \sigma_i \approx 1836 \, \sigma_i.
\end{equation}
Using the simulated value $\sigma_i\gtrsim0.5$, we estimate the corresponding electron magnetization in the reconnection region as
\begin{equation}
\label{eq:sigma_e_val}
\sigma_e \approx 1836 \times 0.5 \approx 918.
\end{equation}
The resulting local electron magnetization is therefore $\sigma_e\sim10^3$, which measures the magnetic energy available per electron. That scale is comparable to the kinetic energy needed for the emitting particles, $\gamma_e\approx640$. In other words, the relation $\gamma_e\sim\sigma_e$ is satisfied. This supports the conclusion that the magnetic-reconnection events present in the GRMHD simulation can accelerate electrons to the Lorentz factors required by the observed NIR flare. Additional processes, such as stochastic acceleration within a turbulent reconnection layer, will likely modify the final particle distribution, but this basic energy-budget estimate already supports the physical plausibility of the proposed mechanism.

\subsection{Time Delays and Radio-Band Flares}\label{sec: 22GHz_flare}

One well-known property of Sgr~A$^*$ flares is that the radio peaks usually occur after the NIR and X-ray peaks \cite[e.g.,][]{witzel_rapid_2021}. Earlier interpretations associated that delay with adiabatically expanding hot spots \cite{Yusef-Zadeh2006, Xi2024, 2024ApJ...971...52M}, although the physical nature of those transient structures remained uncertain. The previous subsection showed that, in the present model, the NIR flares are produced by polarity-inversion events and that the millimeter flares appear later.

At millimeter wavelengths, plasma self-absorption becomes important. A polarity inversion therefore produces an evolving emitting structure whose visibility depends on the local opacity. To determine the origin of the lag, I examine the 43 and 86 GHz flares at $t=8420$ and $8380\,GM/c^3$, which follow the NIR event discussed in Sec.~\ref{Sec: NIR_origin}. Panels (a) and (b) show the corresponding GRRT images, while panels (c) and (d) trace the intensity accumulated along the geodesic that passes through the brightest pixel, marked by a red cross. Although the projected peaks are nearly coincident at the two frequencies ($38.2\,\rm \mu as$ at 86 GHz and $38.9\,\rm \mu as$ at 43 GHz), the radiation originates from slightly different parts of the flow. Panels (c) and (d) show that the emitting geometry changes with frequency. At both 43 and 86 GHz, most of the radiation still comes from the same compact region inside $r\lesssim10\,r_{\rm g}$. Even so, the dominant 43 GHz contribution is displaced slightly outward and toward higher latitude relative to the 86 GHz emission site. That outward shift is the most plausible source of the lag. Radiative-transfer effects reinforce the trend, because the lower-frequency photosphere lies farther out in denser and more opaque plasma. Observationally, a modest delay of about 20 minutes has also been reported between 230 GHz and NIR flares \cite{witzel_rapid_2021}. In general, however, theory does not predict a large lag there because the plasma should already be mostly optically thin at those frequencies. When such a lag is observed, it may therefore indicate additional structure either in the flare mechanism itself or in the emitting region \cite{Jiang2024}.

Once this outward shift of the millimeter emitting region is recognized, the connection between the NIR and millimeter flares becomes straightforward. The millimeter events occur after the NIR flare but follow the same broad temporal evolution, which strongly suggests a common physical origin. Because self-absorption matters at millimeter wavelengths, Figure~\ref{fig: f4} compares light curves computed with and without it. When self-absorption is removed, all radio bands peak simultaneously with the NIR flare, directly tying them to the same underlying event, namely the polarity inversion shown in Figure~\ref{Fig: f2}. The lags appear only after self-absorption is restored, because the lower-frequency radiation remains hidden until the polarity-inversion structure has moved outward into gas that is less dense and less opaque. The delayed radio peaks therefore arise naturally, while the shared origin of the NIR and millimeter flares remains unchanged. Taken together, these results show that a single polarity-inversion event can generate both the NIR and radio flares, with the observed delays set by the combined effects of evolving geometry and frequency-dependent self-absorption.

\section{Summary and Implications}

I carried out 3D two-temperature GRMHD simulations of magnetized accretion flows around a rotating black hole threaded by multiple magnetic loops and combined them with GRRT calculations that use hybrid thermal and non-thermal electron distributions. The analysis focused on the origin of the multi-band flares in Sgr~A$^*$ and on the delays between those bands. The main conclusions are:

\begin{enumerate}
    \item Non-thermal electrons are necessary to reproduce the observed NIR light curves of Sgr~A$^*$. In a SANE flow with a multi-loop magnetic topology, large-scale polarity inversions can produce NIR flares whose amplitudes and variability are comparable to the observations.
    
    \item Filamentary structures appear in the GRRT images during the NIR flares. By tracking both the ray paths and the accumulated intensity along them, I show that the brightest parts of the image originate in large-scale polarity inversions close to the black hole.
    
    \item The GRRT light curves show clear multi-frequency delays. The millimeter-radio flares reach maximum after the NIR flares, and the lag can be as large as roughly 50 minutes, well within the observationally reported range.

    \item The radio flares are produced by the same large-scale polarity-inversion events that generate the NIR flares. Their delayed appearance is caused by frequency-dependent self-absorption at millimeter wavelengths: the emitting structure must move outward before the lower-frequency radiation becomes visible.
   
\end{enumerate}

The results support a picture in which large-scale polarity inversions in SANE accretion flows can account for the NIR flares of Sgr~A$^*$ while remaining broadly consistent with the current data. In that sense, the scenario provides a credible alternative to the MAD-based flare models that are more often discussed for Sgr~A$^*$ \cite[e.g.,][]{Scepi2022, 2020MNRAS.497.4999D, 2021MNRAS.502.2023P, 2023arXiv230816740N, 2024arXiv240410982A}. It also reproduces several additional observational traits of the Galactic center. In particular, the models do not produce a persistent relativistic jet (see Appendix~\ref{sec: jet_nojet} for a fuller discussion of the accretion state and jet behavior), they naturally generate delays between the NIR and 43 GHz flares, and they provide a reasonable match to the observed SED.

The multi-loop configuration still has clear limitations. It does not maintain frequent, long-lived, large-scale polarity inversions, and earlier work suggests that it more closely resembles a transitional state \cite{2023MNRAS.522.2307J, Jiang_2024}. Understanding how Sgr~A$^*$ sustains persistent flaring therefore remains an open problem. Although the adopted resolution is high enough to capture the formation and large-scale evolution of magnetic polarity inversions, it is not sufficient to resolve the smallest turbulent structures or the asymptotic plasmoid-dominated reconnection regime fully. Ultra-high-resolution GRMHD calculations with GPU-accelerated codes \cite{Ripperda2021} can reach a regime in which reconnection becomes nearly grid independent, but such simulations require far more computational resources than are available here. The present global 3D study therefore concentrates on the large-scale polarity-inversion process rather than on the detailed microphysics of the reconnection layer.

Because the resolution remains limited, the disk dynamo is not fully resolved and therefore does not continuously regenerate alternating polarities \cite{DelZanna2022, Mattia2020, Mattia2022}. I also do not resolve the tearing instability needed for plasmoid formation at the reconnection site \cite[e.g.,][]{2023MNRAS.522.2307J, Ripperda2021}, which may reduce the number of non-thermal electrons. In addition, a polarity inversion becomes an observable flare only if two conditions are satisfied: the event must be intrinsically energetic enough, and the emitting plasma must move toward the observer rapidly enough for Doppler boosting to amplify the signal. Equally energetic events moving away from the line of sight would instead be de-boosted and appear faint. These geometric selection effects therefore shape both the apparent flare rate and the measured flare amplitude.

The present GRRT calculation also neglects Compton scattering, so the model cannot yet address the X-ray counterparts. Synchrotron self-Compton emission may contribute significantly to the observed X-ray flares \cite[e.g.,][]{2012A&A...537A..52E, 2016MNRAS.461..552D}, but that channel is not included here. In addition, the current ray-tracing method is limited when one tries to extract truly local properties of the flaring region; a Monte Carlo treatment that follows photons from a reconnection event to infinity would improve this aspect. These missing ingredients will be explored in future higher-resolution simulations that can also address the weaker X-ray flares.


\newpage
{
\section*{Supplementary Appendix}  
\setcounter{section}{0}
\renewcommand{\thesection}{\thechapter.\Alph{section}}
\renewcommand{\theHsection}{appendix.\thechapter.\Alph{section}}
\renewcommand{\theHsubsection}{appendix.\thechapter.\Alph{section}.\arabic{subsection}}

\section{Code Configuration}
\subsection{GRMHD Setup with {\rm KHARMA}} \label{Sec: kharma}
All simulations in this appendix are evolved with KHARMA\footnote{\url https://github.com/AFD-Illinois/kharma} \cite{2024arXiv240801361P}, the GPU-oriented GRMHD code built on top of {\tt iharm3D} \cite{2021JOSS....6.3336P}. Numerically, both codes descend from the conservative ideal-GRMHD scheme introduced in {\tt HARM} \cite{2003ApJ...589..444G}. In the Eulerian form used here, the governing equations are
\begin{equation}
    \begin{aligned}
        \partial_t(\sqrt{-g}\rho u^t)&=-\partial_i(\sqrt{-g}\rho u^i), \\
        \partial_t(\sqrt{-g}T^t_{\nu})&=-\partial_i(\sqrt{-g} T^i_{\nu}) + \sqrt{-g}T^\kappa_\lambda\Gamma^\lambda_{\nu\kappa}, \\
        \partial_t(\sqrt{-g}B^i)&=-\partial_j\left[\sqrt{-g}(b^ju^i - b^i u^j)\right], \\
        \partial_i(\sqrt{-g}B^i)&=0,
    \end{aligned}
\end{equation}
Here $\rho$ is the rest-mass density, $u^\mu$ is the fluid four-velocity, $\Gamma^\alpha_{\beta\gamma}$ are the connection coefficients, $B^i$ and $b^\mu$ are the three- and four-magnetic-field components, $T^\mu_\nu$ is the stress-energy tensor, and $g$ is the metric determinant \cite[see][]{2003ApJ...589..444G}. On the implementation side, KHARMA combines the Parthenon mesh infrastructure \cite{2022arXiv220212309G} with the Kokkos performance-portability layer \cite{9485033}. Numerical floors are applied to stabilize the evolution in low-density, highly magnetized regions, especially near the horizon and along the rotation axis \cite{Rezzolla-Zanotti2013}. Concretely, I set $\rho_{\rm fl}=10^{-6}r^{-3/2}$ and $p_{\rm fl}=10^{-8}r^{-5/2}$, and cap the magnetization and Lorentz factor at $\sigma_{\rm max}=50$ and $\gamma_{\rm max}=15$. Magnetic-flux conservation is maintained with Face-Centered Constrained Transport (Face-CT) \cite{2024arXiv240801361P}; more specifically, I adopt the three-dimensional gs05$_c$ variant based on the upwind $E_z^c$ method of \cite{2005JCoPh.205..509G} as generalized by \cite{2008JCoPh.227.4123G}.

\begin{figure}
\centering
 	\includegraphics[height=.6\linewidth]{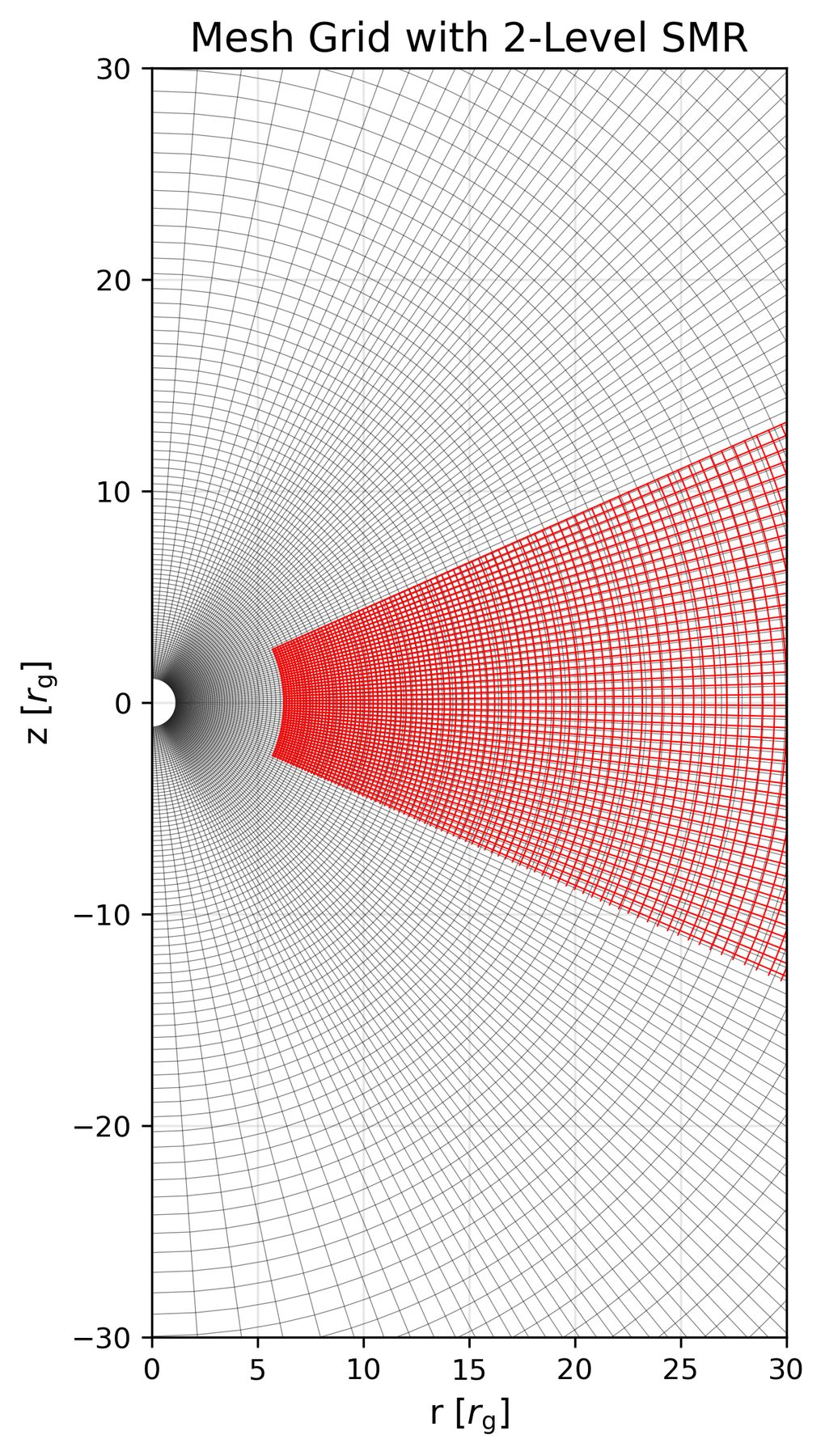}
        \includegraphics[height=.6\linewidth]{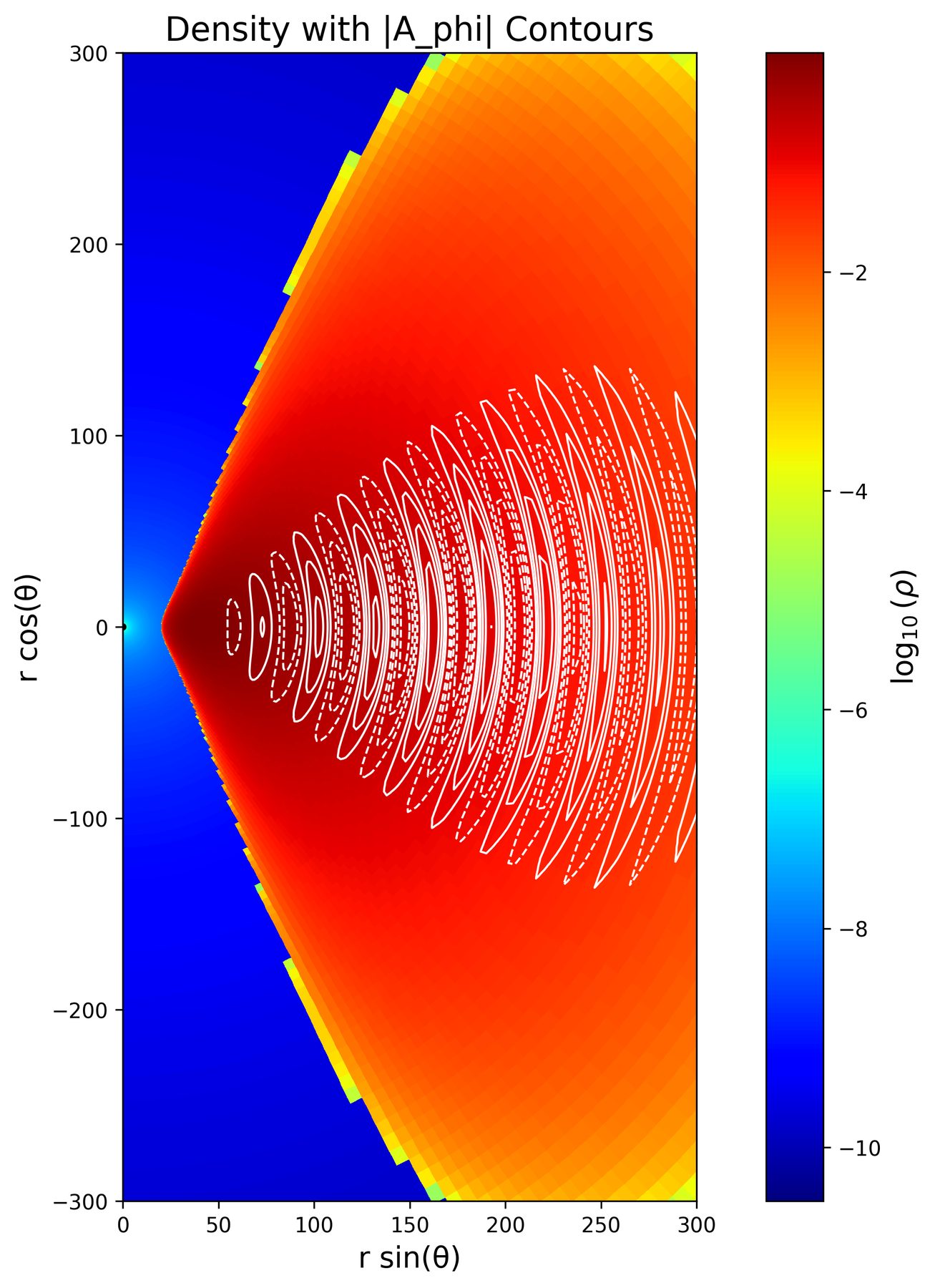}
 	\caption{Grid layout and seed magnetic-field structure used in the simulation. The left panel shows the two-level SMR hierarchy, with the coarse mesh in black and the refined patch in red. The right panel visualizes the initial field through contours of the vector potential $A_\phi$; solid and dashed lines denote loops of opposite magnetic polarity.}
 	\label{Fig: mesh}
\end{figure}

\subsection{Resolution and Initial Magnetic Geometry} \label{sec:res_check}

The mesh starts from $192\times144\times128$ zones and is supplemented by two static-refinement levels, so the finest effective resolution becomes $384\times288\times256$. Refinement is applied only inside a selected region of Wide-pole Kerr-Schild (WKS) coordinates; see \cite{2024ApJ...977..200C} together with \cite{McKinney2006, 2012MNRAS.423.3083M, 2017MNRAS.467.3604R} for related coordinate choices. Radially, that high-resolution region extends from $r_{\rm min}=6.24\,r_{\rm g}$ to $r_{\rm max}=281\,r_{\rm g}$, while in polar angle it covers $\theta_{\rm min}=66^\circ$ to $\theta_{\rm max}=114^\circ$. Figure~\ref{Fig: mesh} marks this refined domain in red.
 
The magnetic seed field shown in the right panel of Figure~\ref{Fig: mesh} is a radial multi-loop configuration in which neighboring loops reverse polarity from one ring to the next.

\subsection{Cooling Model} \label{sec: cooling}

Radiative cooling can reshape the particle distribution by limiting the highest Lorentz factor and altering the effective power-law slope. Here I adopt the simple approximation introduced by \cite{Scepi2022}:

The synchrotron cooling timescale $t_{\rm sync}$ is
\begin{equation}
    t_{\rm sync} = \frac{3m_{\rm e}c}{4\sigma_{T}U_{\rm B}\gamma\beta^2},
\end{equation}
where $\sigma_{\rm T}$ is the Thomson cross-section, $U_{\rm B}$ is the magnetic energy density, $\gamma$ is the Lorentz factor, and $\beta\equiv v/c\sim1$. The photon escape time is approximated by the light-crossing time $t_{\rm esc}=r_{\rm g}/c$. Cooling therefore becomes important near
\begin{equation}
    \gamma_{\rm break} \approx 3.9\times 10^{3}\left(\frac{|b|}{100\,\rm Gauss}\right)^{-2},
\end{equation}
which corresponds to a characteristic break frequency of
\begin{equation}
    \nu_{\rm break}\approx 2.5\times 10^{15}\left(\frac{|b|}{100\,\rm Gauss} \right)^{-1}\,\rm Hz.
\end{equation}
This corresponds to strong cooling above approximately $10^{15}$ Hz. In practice, following \cite{Scepi2022}, I divide the emissivity and absorptivity from Eq.~\ref{Eq: emi&alpha-c4} by $\sqrt{\nu/\nu_{\rm break}}$ whenever $\nu>\nu_{\rm break}$. Including this prescription suppresses the X-ray flux substantially and produces a more realistic high-frequency SED.

\section{Supplementary Material}
\subsection{Spectral Energy Distribution} \label{sec: sed}

\begin{figure*}
    \centering
    \includegraphics[width=.7\linewidth]{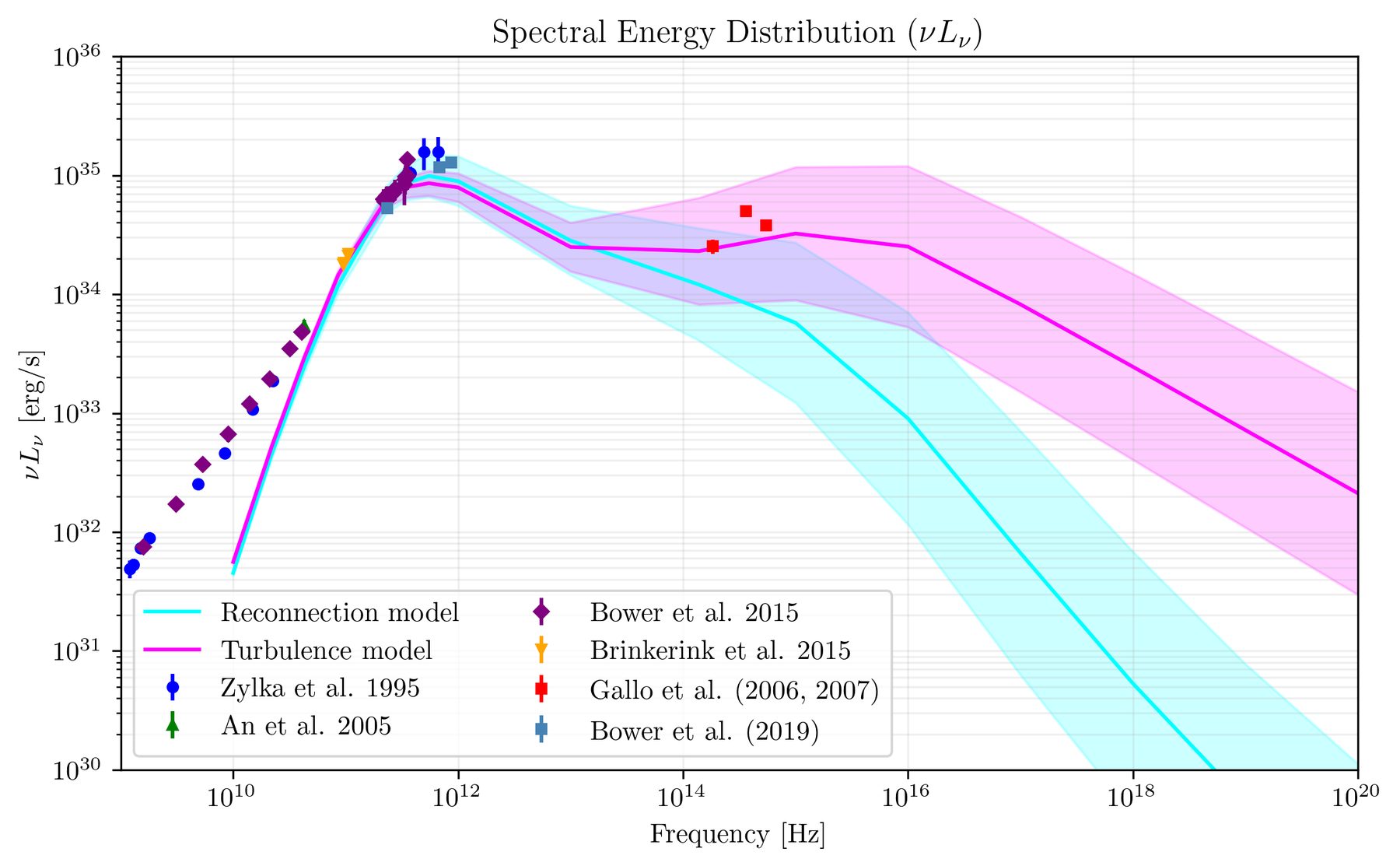}
    \caption{Broadband data from \cite{1995A&A...297...83Z, 2005ApJ...634L..49A, 2015ApJ...802...69B, 2015ASPC..499..167B, 2019ApJ...881L...2B, 2017MNRAS.466.4121C, 2006MNRAS.370.1351G, 2007ApJ...670..600G} plotted together with the model spectra. The pink and cyan curves show the time-averaged SEDs obtained with the turbulence-driven and reconnection-driven sub-grid prescriptions over $t=8,000$--$11,000\,[GM/c^3]$, and the shaded envelopes mark the associated temporal scatter.}
    \label{fig: SED}
\end{figure*}

Figure~\ref{fig: SED} places the simulated spectral energy distributions beside the available multi-wavelength measurements. The millimeter and sub-millimeter fluxes come from \cite{1995A&A...297...83Z, 2005ApJ...634L..49A, 2015ApJ...802...69B, 2015ASPC..499..167B, 2019ApJ...881L...2B}, while the infrared points are taken from the compilation of \cite{2017MNRAS.466.4121C}, which is itself based on \cite{2006MNRAS.370.1351G, 2007ApJ...670..600G}.

The calculation uses two alternative prescriptions for the electron thermodynamics and non-thermal component (see Sec.~\ref{sec:2T-c4}). One is tied to turbulent dissipation and the other to magnetic reconnection. Because the simulation does not resolve every eddy and current sheet, it is not yet possible to assign those microphysical channels to each fluid element in a fully self-consistent way. Instead, I follow the pragmatic strategy used in earlier flare studies \cite[e.g.,][]{Fromm2022, Yang2024} and examine the observational consequences of the two prescriptions separately.

The comparison in Figure~\ref{fig: SED} shows that the two models remain almost indistinguishable through the millimeter band, where thermal synchrotron emission from the bulk flow dominates. Their predictions begin to separate only toward the infrared and above, once the non-thermal electrons become important. In that regime, the turbulence-based prescription more often falls inside the observed infrared range and can generate NIR peaks of roughly 10--25 mJy, comparable to the flares reported by \cite{GRAVITY2018b}. For that reason, I keep the turbulence-based model as the fiducial choice in the main discussion.

The low-frequency radio emission below 43 GHz is still underestimated. In GRMHD jet solutions, adiabatic expansion cools the outflow and lowers the electron temperature, whereas the observed flat radio spectrum points to a jet that stays closer to isothermal. The multi-loop initial field used here also weakens the large-scale jet because repeated polarity reversals impede the buildup of a persistent outflow, which further worsens the radio deficit.

\subsection{Ray Paths and Intensity in the GRRT Calculation}\label{Sec: ray_path}

\begin{figure}
\centering
 	\includegraphics[height=.3\linewidth]{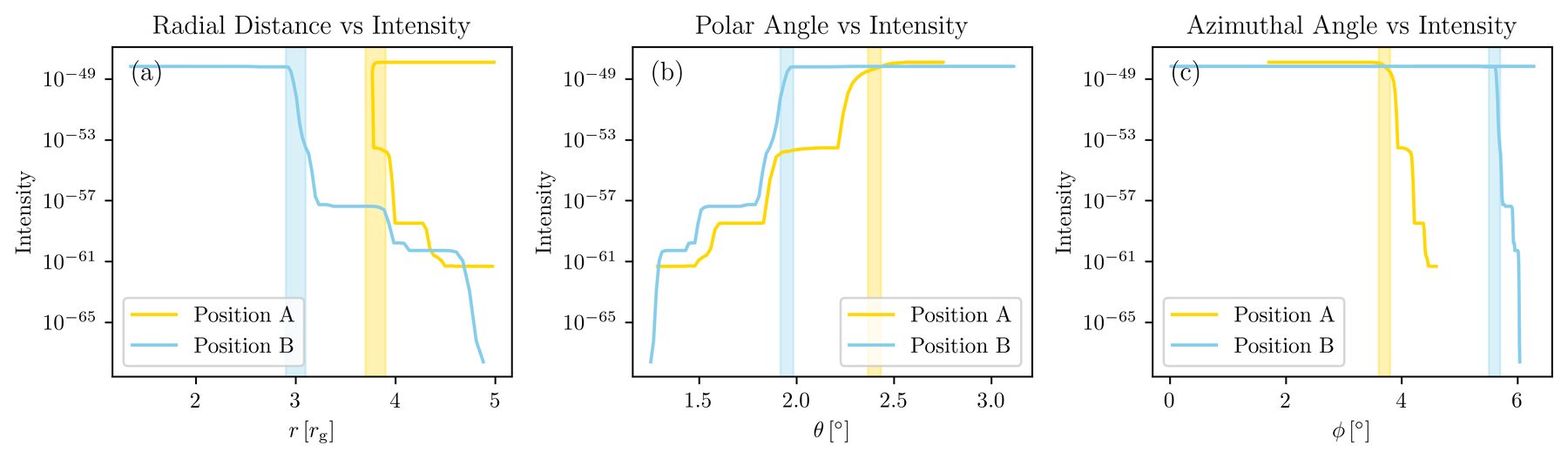}
 	\includegraphics[height=.4\linewidth]{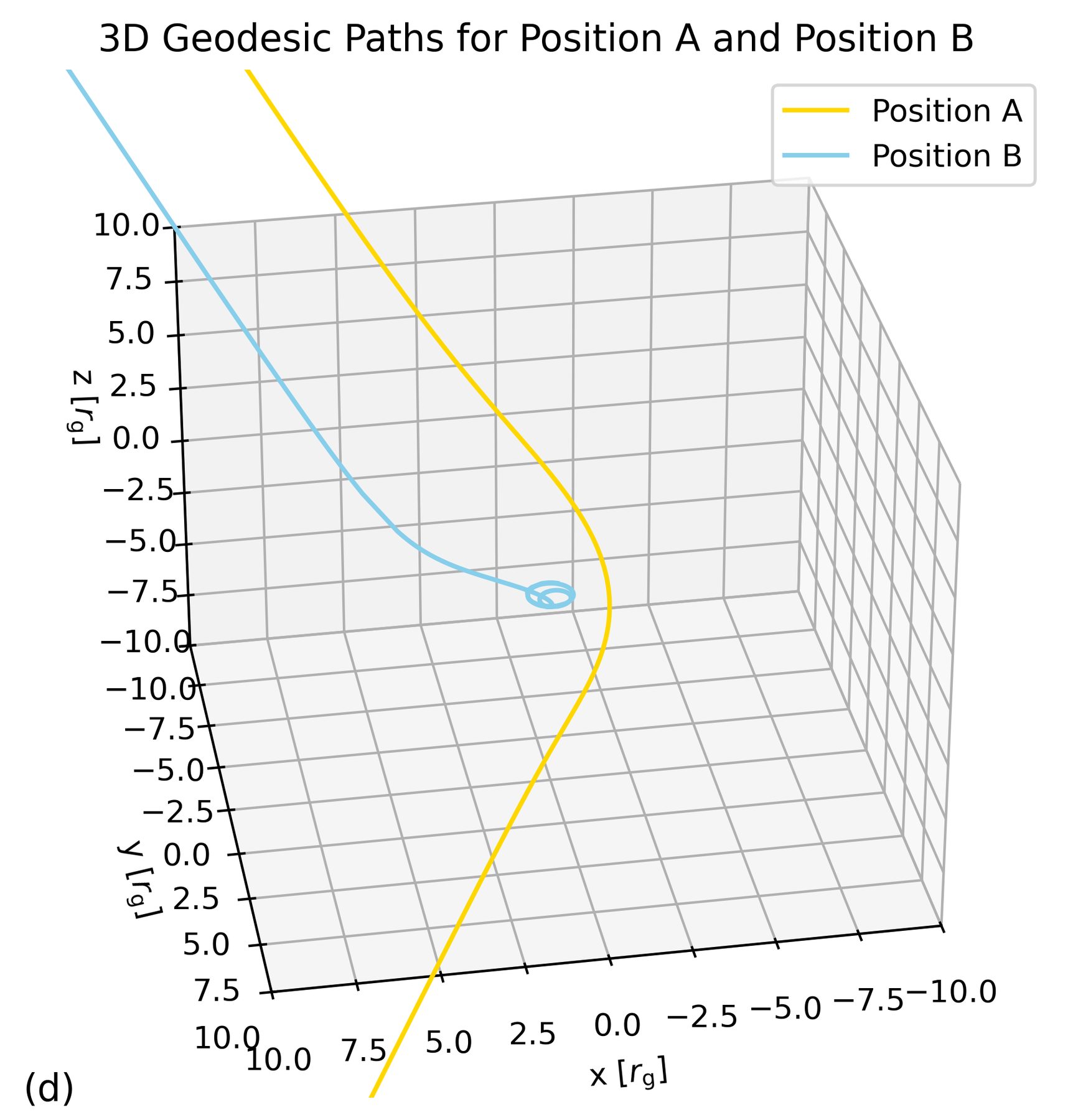}
 	\caption{Ray trajectories and intensity buildup. Panels (a)--(c) trace how the specific intensity evolves along the same two rays when projected onto the $r$, $\theta$, and $\phi$ coordinates. The yellow and blue curves correspond to the two image-plane locations highlighted in panel (c) of Figure~\ref{Fig: f2}. Panel (d) shows the full three-dimensional geodesics of those rays and indicates where the intensity growth seen in panels (a)--(c) occurs in space.}
 	\label{Fig: ray_path}
\end{figure}

To determine where each image pixel acquires its flux in the GRRT calculation, I extract both the null geodesic and the accumulated intensity along that path. Figure~\ref{Fig: ray_path} displays this intensity history in spherical coordinates, making it possible to identify the region from which most of the radiation is received and therefore to localize the flare site.

The yellow and blue curves correspond to the two rays passing through Positions A and B in panel (c) of Figure~\ref{Fig: f2}. The shaded interval marks the part of the trajectory where the intensity increases most strongly, which identifies the coordinates of the dominant emitting region. The same site can then be located directly in the 3D volume-rendered view shown in Figure~\ref{Fig: f2}.

Panel (d) of Figure~\ref{Fig: ray_path} gives the same two trajectories in full three dimensions.

\subsection{Discrete Cross-Correlation Function} \label{Sec: CCF}

The cross-correlation function (CCF) is a standard tool for measuring the time lag between two time-dependent signals \cite{1997ASSL..218..163A}. In practice, one usually uses the discrete cross-correlation function (DCCF) rather than the continuous CCF. For two light curves $F(t)$ and $G(t)$, the lag between a pair of points $F(t_j)$ and $G(t_k)$ is
\begin{equation}
  \Delta t_{jk} = t_j-t_k.  
\end{equation}
After defining the lag sequence $\tau$, all pairs whose separations satisfy $\tau_i\leq \Delta t_{jk}\leq \tau_{i+1}$ are grouped together, and each pair is assigned the correlation coefficient
\begin{equation}
    U_{jk} = \frac{(F(t_j)-\overline{F})(G(t_k) - \overline{G})}{\sigma_F\sigma_G},
\end{equation}
where $\overline{F}$ and $\overline{G}$ are the averaged values of the two light curves, and $\sigma_F$ and $\sigma_G$ are their standard deviations. Thus, the DCCF within the given bin of time lag $\tau_i$ is obtained as
\begin{equation}
    DCCF(\tau_i) = \frac{1}{n_i}\sum_{jk} U_{jk},
\end{equation}
Here $n_i$ is the number of point pairs that fall inside the lag bin $\tau_i$. I estimate the DCCF uncertainty from the standard deviation within that bin:
\begin{equation}
    \sigma_{DCCF}(\tau_i) = \frac{1}{\sqrt{n_i}}\sqrt{\sum_{jk}\left(U_{jk}-DCCF(\tau_i)\right)^2}.
\end{equation}

\subsection{Accretion Stages in the Multi-Loop Model and Convergence Check} \label{sec: jet_nojet}

\begin{figure}
    \centering
    \includegraphics[width=\linewidth]{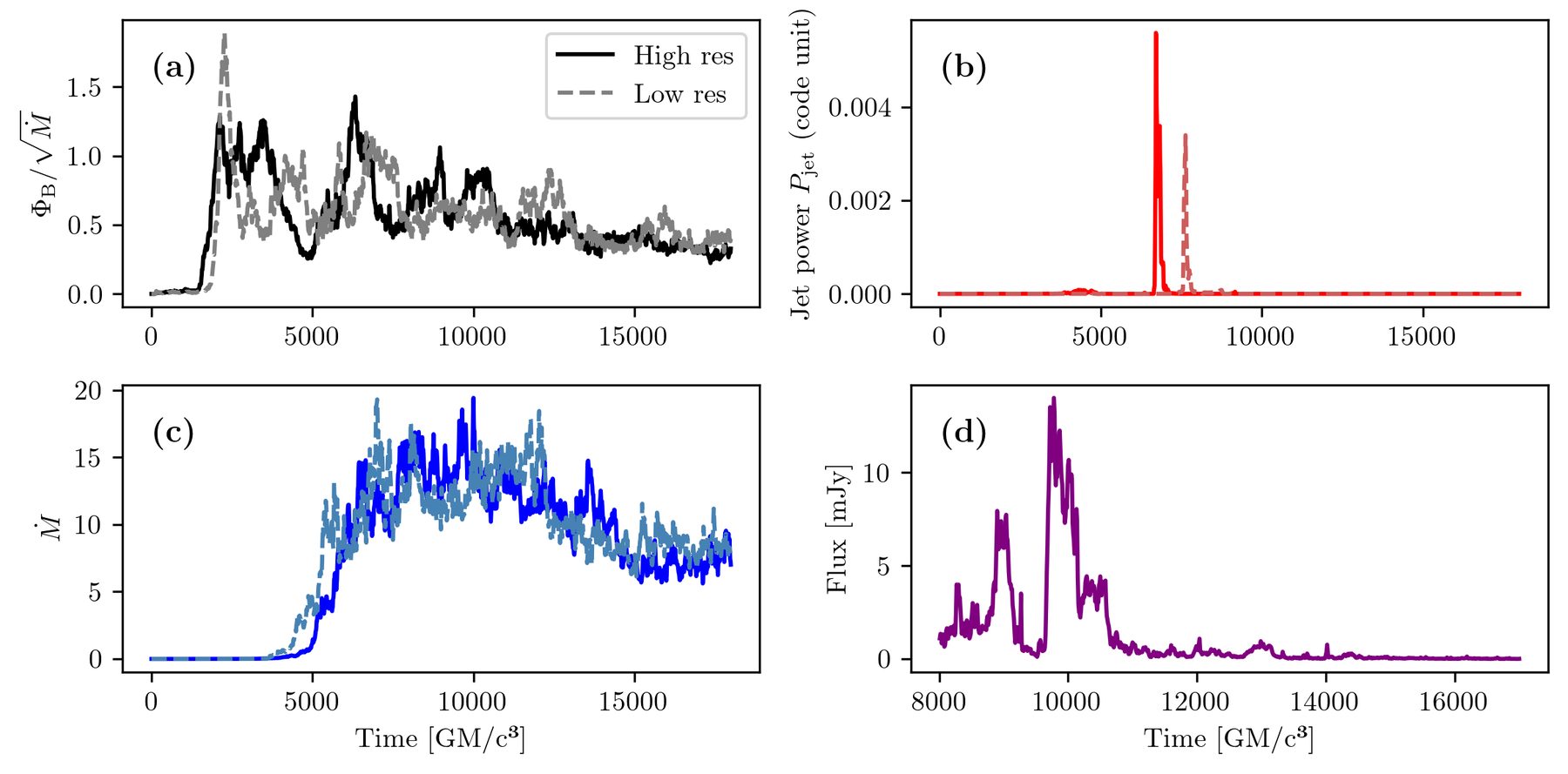}
    \caption{Panels (a)--(c) show the time evolution of the normalized magnetic flux (black), jet power (red), and accretion rate (blue). Solid lines correspond to the fiducial simulation, and dashed lines show the lower-resolution comparison run. Panel (d) gives the associated NIR light curve over $t=8,000$ to $18,000\,GM/c^3$ (purple).}
    \label{fig: Mdot-c4}
\end{figure}
Figure~\ref{fig: Mdot-c4} summarizes the evolution of the magnetic flux, jet power, and accretion rate. In panels (a)--(c), I use the horizon-based definitions of $\dot M$ and $\Phi_{\rm B}$ from \cite{Porth2019} together with the jet-power definition of \cite{Nathanail2020}. Specifically, $\Phi_{\rm B}=\frac{1}{2}\int_0^{2\pi}\int_0^\pi |B^r|\sqrt{-g}\,d\theta\,d\phi$ and $\dot M=\int_0^{2\pi}\int_0^\pi \rho u^r\sqrt{-g}\,d\theta\,d\phi$ are measured at the event horizon, while the jet power is computed at $r=50\,r_{\rm g}$ as $P_{\rm jet}=\int_0^{2\pi}\int_{\theta_{\rm jet}}(-T_t^r-\rho u^r)\sqrt{-g}\,d\theta\,d\phi$, with $\theta_{\rm jet}$ spanning the funnel region defined by $\sigma\geq1$. The joint evolution of $\dot M$ and $\Phi_{\rm B}$ indicates that the system passes through distinct accretion stages. The simulation is followed beyond $t=18,000\,GM/c^3$. After about $8,000\,GM/c^3$, the inner flow inside roughly $50\,r_{\rm g}$ reaches a quasi-steady state. Before $t=12,000\,GM/c^3$, repeated polarity reversals drive stronger fluctuations in both $\dot M$ and $\Phi_{\rm B}$, and those same inversions generate the NIR flares shown in panel (d). At later times, the available magnetic energy declines, the inversions cease, and the NIR variability relaxes back toward a quiescent level. Because frequent reconnection during the inversion phase suppresses jet formation, the jet power in panel (b) stays close to zero for most of the run and no sustained jet develops.

Panels (a)--(c) also compare the SMR run with a run without refinement, using resolutions of $384\times288\times256$ and $192\times144\times128$, respectively. The close agreement in their time histories indicates that the simulation is numerically well converged at the level relevant for this analysis.

\subsection{Effect of $\sigma_{\rm cut}$}
\label{sec: sigma_cut}
\begin{figure}
\centering
 	\includegraphics[width=.8\linewidth]{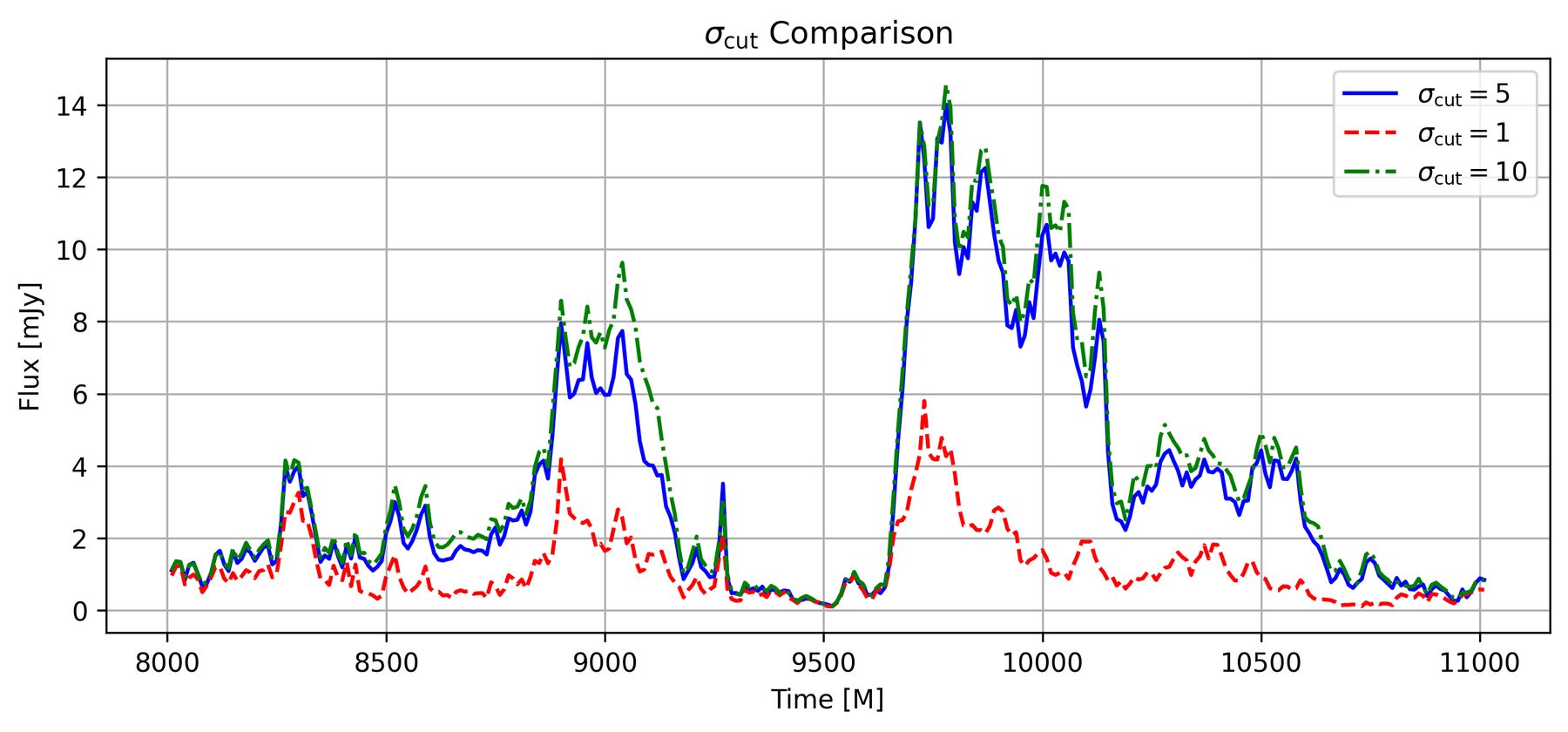}
 	\caption{NIR light curves computed with different values of $\sigma_{\rm cut}$. The solid blue curve is the fiducial choice ($\sigma_{\rm cut}=5$), the dashed red curve shows $\sigma_{\rm cut}=1$, and the dot-dashed green curve shows $\sigma_{\rm cut}=10$.}
 	\label{Fig: sigma_cut_comp}
\end{figure}

In the GRRT post-processing, highly magnetized cells are excluded through the cutoff parameter $\sigma_{\rm cut}$, so the NIR light curves can depend on that choice. Here I test the sensitivity by comparing the fiducial value $\sigma_{\rm cut}=5$ with two alternatives, $\sigma_{\rm cut}=1$ and $\sigma_{\rm cut}=10$, as shown in Figure~\ref{Fig: sigma_cut_comp}.

The main effect appears at the low-cutoff end: using $\sigma_{\rm cut}=1$ noticeably suppresses the light curve. Raising the cutoff to $\sigma_{\rm cut}=10$, by contrast, produces a result that is almost identical to the fiducial case. This indicates that $\sigma_{\rm cut}=5$ is already large enough for the present application and that increasing it further has little practical effect on the GRRT output.

\begin{figure}
\centering
 	\includegraphics[width=\linewidth]{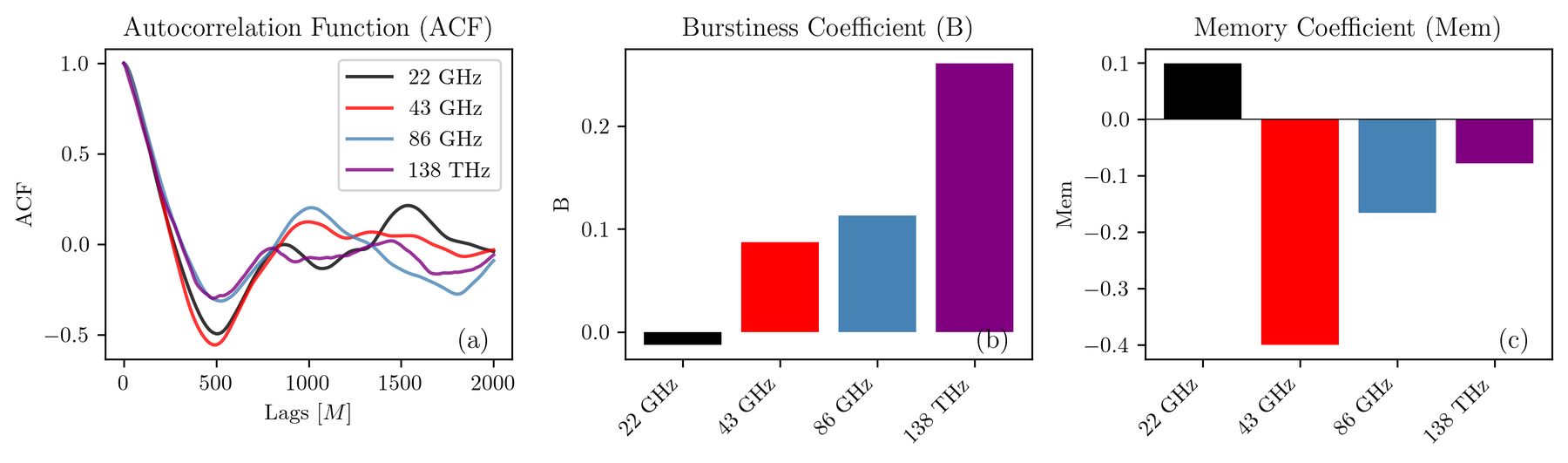}
 	\includegraphics[width=.45\linewidth]{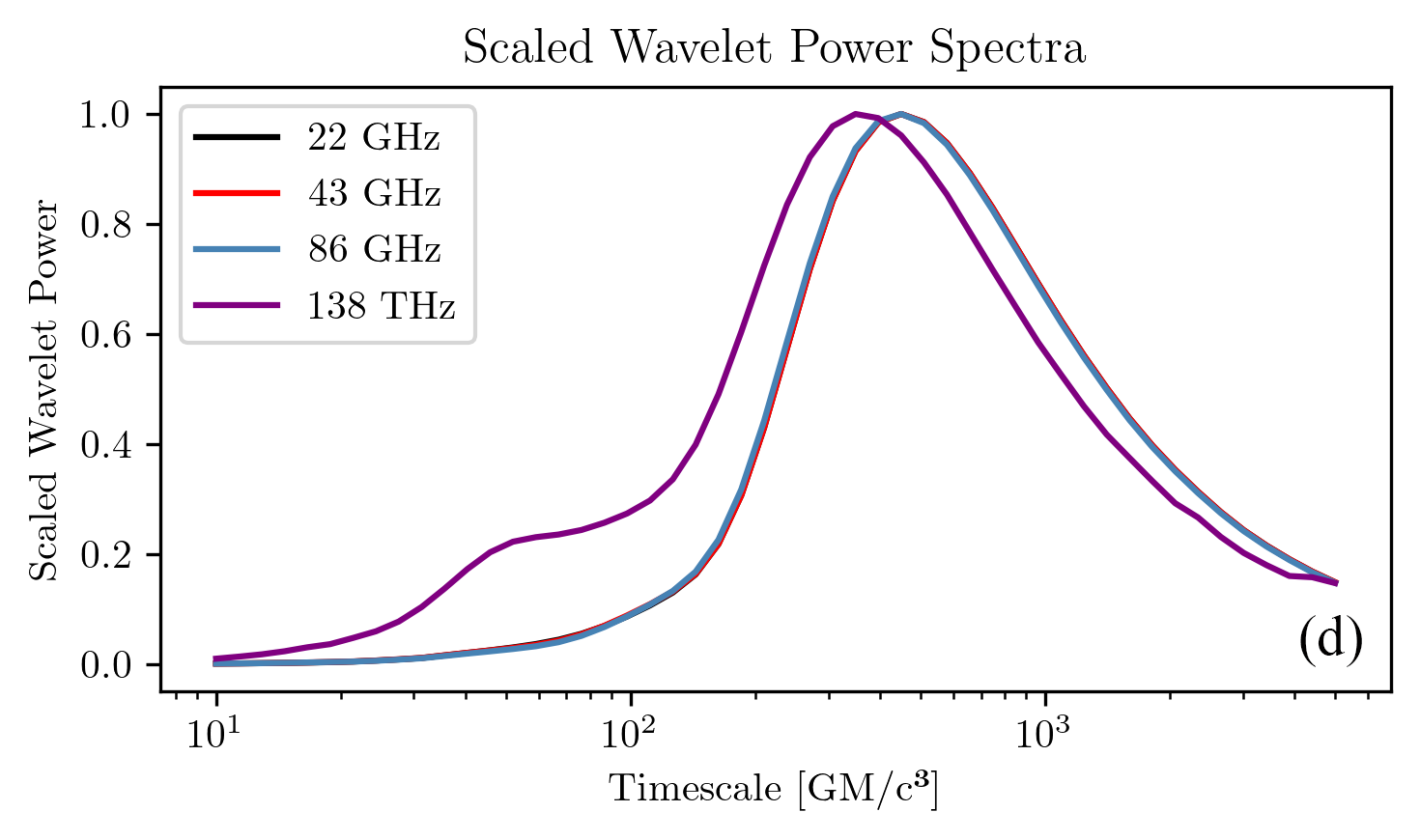}
 	\caption{Statistical diagnostics of the multi-frequency light curves. Panel (a) shows the autocorrelation functions at 22 GHz (black), 43 GHz (red), 86 GHz (blue), and 138 THz (purple); the same colors are reused in panels (b)--(d). Panels (b) and (c) show the burstiness and memory coefficients, while panel (d) gives the scaled wavelet power spectra.}
 	\label{Fig: ACF}
\end{figure}

\subsection{Additional Statistical Analysis of the Light Curves}
\label{sec: statistics}

This subsection examines the light curves with several additional statistical diagnostics in order to isolate their characteristic timescales and variability patterns.

The multi-frequency light curves in Figure~\ref{fig: f1} already suggest that different bands are governed by different parts of the flow. That contrast becomes clearer in Figure~\ref{Fig: ACF}. In panel (a), the autocorrelation function (ACF; see \cite{1976tsaf.conf.....B, 2019A&A...627A.103V} for the definition) drops rapidly at 138 THz, indicating short-timescale variability. The radio bands (22--86 GHz), by contrast, retain correlation over much longer lags and therefore appear more periodic. Panel (b) shows that the burstiness coefficient \cite{2019PhRvE.100b2307J} increases with frequency, implying that the high-frequency emission is more clustered in time and more stochastic. Panel (c) shows the memory coefficient \cite{2019PhRvE.100b2307J}. It is weakly positive at 22 GHz, becomes negative at intermediate frequencies, and then shifts back toward less negative values at 138 THz. Taken together, these trends point to shorter-lived events with smaller separations at high frequency, where compact turbulent regions dominate, while the lower-frequency variability is produced by larger and more slowly evolving structures.

Panel (d) of Figure~\ref{Fig: ACF} expresses the same contrast in the wavelet domain. The three radio-band spectra nearly coincide, indicating a common variability pattern with a dominant timescale near $\sim450\,GM/c^3$. The 138 THz curve is different: most of its power is concentrated near $\sim348\,GM/c^3$, and a weaker component remains around $\sim80\,GM/c^3$. This suggests that the NIR outbursts are either shorter in duration or more closely spaced in time, most likely because they originate from rapidly orbiting gas in the inner 3--$5\,r_{\rm g}$ at roughly $\sim0.3c$ (see panel (a) of Figure~\ref{Fig: f2}). In other words, the high-frequency end of the spectrum encodes the narrow spike-like structures seen in the NIR flare light curves.
}

\chapter[Tilted MAD retrograde precession]{Tilted MAD retrograde precession}
\section*{Retrograde Precession from Magnetic Torques in Tilted Black-Hole Accretion Flows}
A shorter version of this work was published in ApJ, volume 995, article 112 (2025), by Jiang H.-X., Mizuno Y., Lai D., Dihingia I. K., and Fromm C. M. \cite{2025ApJ...995..112J}.

\section{Physical Motivation and Context} \label{sec:intro-c5}
Spin-misaligned accretion should arise in many black-hole systems of interest, from X-ray binaries to active galactic nuclei \cite{2002ApJ...565L..75E, 2024ApJ...967L...9Z, 2011ApJ...742...85G, 2011MNRAS.416..941S, 2020A&A...643A..31K}. Whenever the angular momentum of the supplied gas differs from that of the hole, the inner flow settles into a tilted disk-jet configuration rather than an aligned one \cite{2010ApJ...719L..79F, 2022Sci...375..874P}. Frame dragging then induces Lense-Thirring precession and continuously redistributes angular momentum throughout the tilted flow \cite{1918PhyZ...19...33T, Narayan2003}.

How often such tilts occur depends on both how the system formed and how it is fed later on \cite{2021ARA&A..59..117R, 2020A&A...635A..97B}. Gas captured from winds, disrupted stars, or post-merger inflows need not share the spin axis of the central hole, so warped and tilted disks are expected to be common \cite{2005ApJ...623..347F, 2007ApJ...668..417F, Liska2018, 2021MNRAS.507..983L, Chatterjee2023}. This picture has been invoked for microquasars such as SS 433, GRO J1655-40, and GRS 1758-258 \cite{1984ARAA..22..507M, 2014MNRAS.437.2554M, 2015A&A...584A.122L, 2002ApJ...578L.129S}, and it is likewise relevant to tidal-disruption events and post-merger AGN, where the impact geometry is often effectively random \cite{2019MNRAS.487.4965Z, Tchekhovskoy2014, 2016MNRAS.455.1946F, 2024Natur.630..325P, 2012PhRvL.108f1302S, Andalman2022, 2005ApJ...618..618K, 2006ApJ...638..120C, 2007MNRAS.379..135C}. Thin disks can diffuse into Bardeen-Petterson alignment \cite{1996MNRAS.282..291S, 1975ApJ...195L..65B, 2019MNRAS.487..550L, 2021MNRAS.507..983L}, whereas thick disks communicate warps as bending waves and may preserve a global LT mode in the SANE regime \cite{2005ApJ...623..347F, 2007ApJ...668..417F, 2008ApJ...687..757F, Liska2018, 2019ApJ...878...51W, 2019MNRAS.487.4965Z}. That coherence requires the sound-crossing time to remain shorter than the precession time \cite{2024arXiv240410052F}; once MRI-driven spreading enlarges the torus, the precession frequency tends to drop \cite{Liska2018}.

The MAD state introduces another torque channel. When large-scale magnetic flux accumulates near the horizon, the jet and ordered field can torque the disk toward the black-hole equator and compete directly with LT precession \cite{Chatterjee2023, McKinney2013, 2017MNRAS.464.2660P, 2021MNRAS.504.6076R}. That makes the persistence of precession uncertain in strongly magnetized systems. Indeed, \cite{Chatterjee2023} found no compelling long-lived precession in very extended tilted MAD tori \cite{McKinney2013, Chatterjee2023}. The question is particularly relevant for M~87$^*$, whose accretion flow is widely interpreted as MAD \cite{2019ApJ...875L...5E, Yuan2022} even though VLBI observations suggest jet precession \cite{Cui2023}. In this chapter I examine whether tilted MAD disks can continue to precess across several magnetic setups. The central result is that once the accumulated flux becomes strong enough, the magnetic torque can flip the sign of the net precession rate and drive retrograde motion of both the disk and the jet.

Section~6.2 presents the simulation results, Section~6.3 summarizes the main conclusions, and the Appendix collects the numerical setup and related supplementary material.
\section{Results}\label{sec2}

\begin{figure*}
\centering 	
\includegraphics[width=\linewidth]{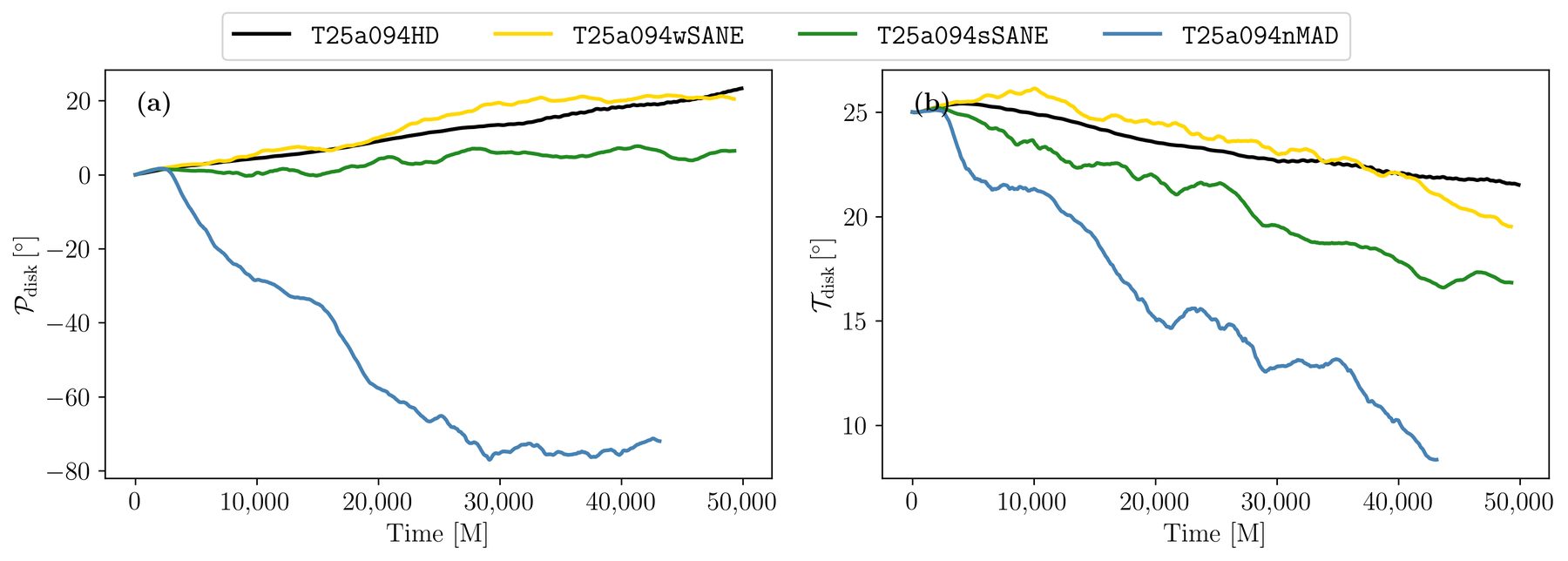}
\caption{Histories of the disk precession and tilt angles for four representative tilted-disk calculations. The black, yellow, green, and blue curves correspond to {\tt T25a094HD}, {\tt T25a094wSANE}, {\tt T25a094sSANE}, and {\tt T25a094nMAD}; the model nomenclature is listed in Table~\ref{table:models}.}
 	\label{fig: precession_exmple}
\end{figure*}

\begin{figure*}
\centering 	
\includegraphics[width=.49\linewidth]{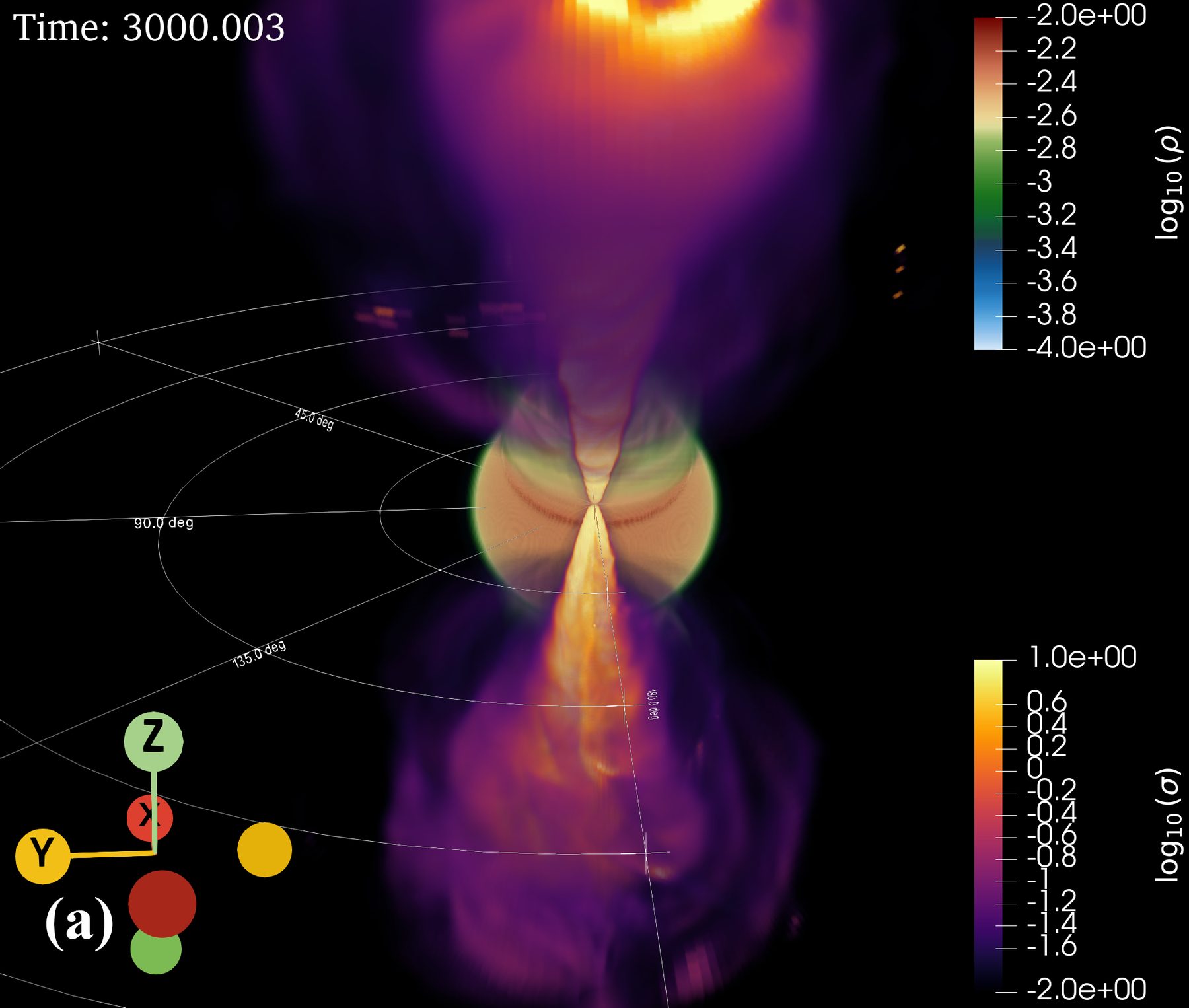}
\includegraphics[width=.49\linewidth]{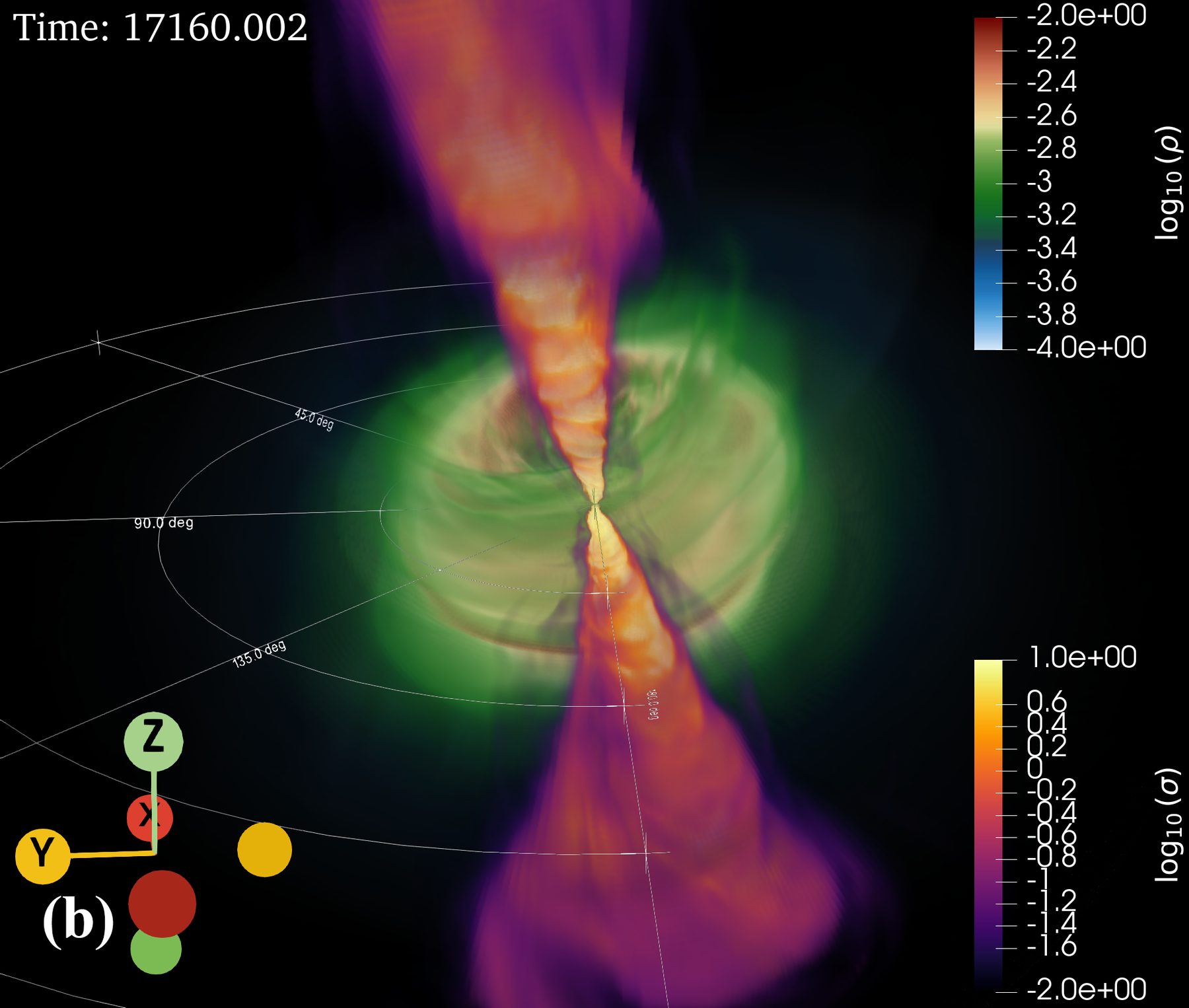}
 	\caption{Volume renderings of the strongly magnetized run {\tt T25a094nMAD}. The jet is colored by magnetization $\sigma$ and the disk by density $\rho$. Panels (a) and (b) show snapshots at $t=3{,}000\,M$ and $t=17{,}160\,M$, and their relative reorientation demonstrates retrograde precession of the coupled jet-disk structure with respect to the black-hole spin.}
    \label{fig: volume_rendering}
\end{figure*}

\begin{figure*}
\centering 	
\includegraphics[width=\linewidth]{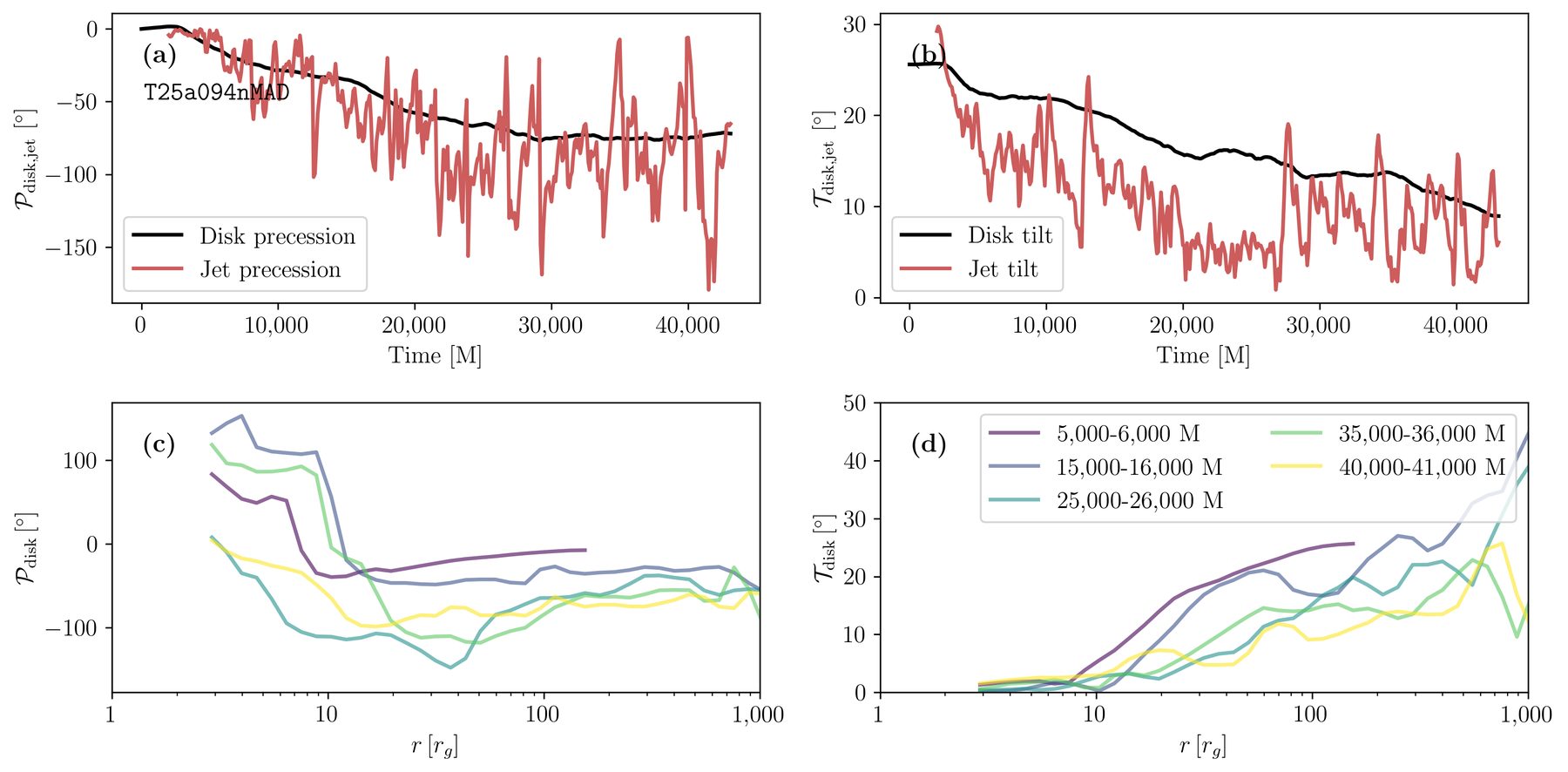}
\caption{Orientation diagnostics for the strongly magnetized model {\tt T25a094nMAD}. Panels (a) and (b) plot the time evolution of the disk (black) and jet (red) precession and tilt angles, while panels (c) and (d) give radial profiles of the disk precession and tilt measured over several averaging windows.}
 	\label{fig: T25a094nMAD}
\end{figure*}

This section analyzes a suite of tilted-disk GR(M)HD calculations carried out with KHARMA, the GPU implementation that succeeds {\tt iharm3D} \cite{2021JOSS....6.3336P}. By varying both the magnetic-field configuration and the hole spin $a$, the model set isolates the competition between LT frame dragging and magnetically induced torques. Details of the numerical setup and grid choices are deferred to Appendix~6.A and 6.B.

\subsection{Precession and Tilt in the GRMHD Simulations}

\begin{figure*}
\centering 	
\includegraphics[height=.41\linewidth]{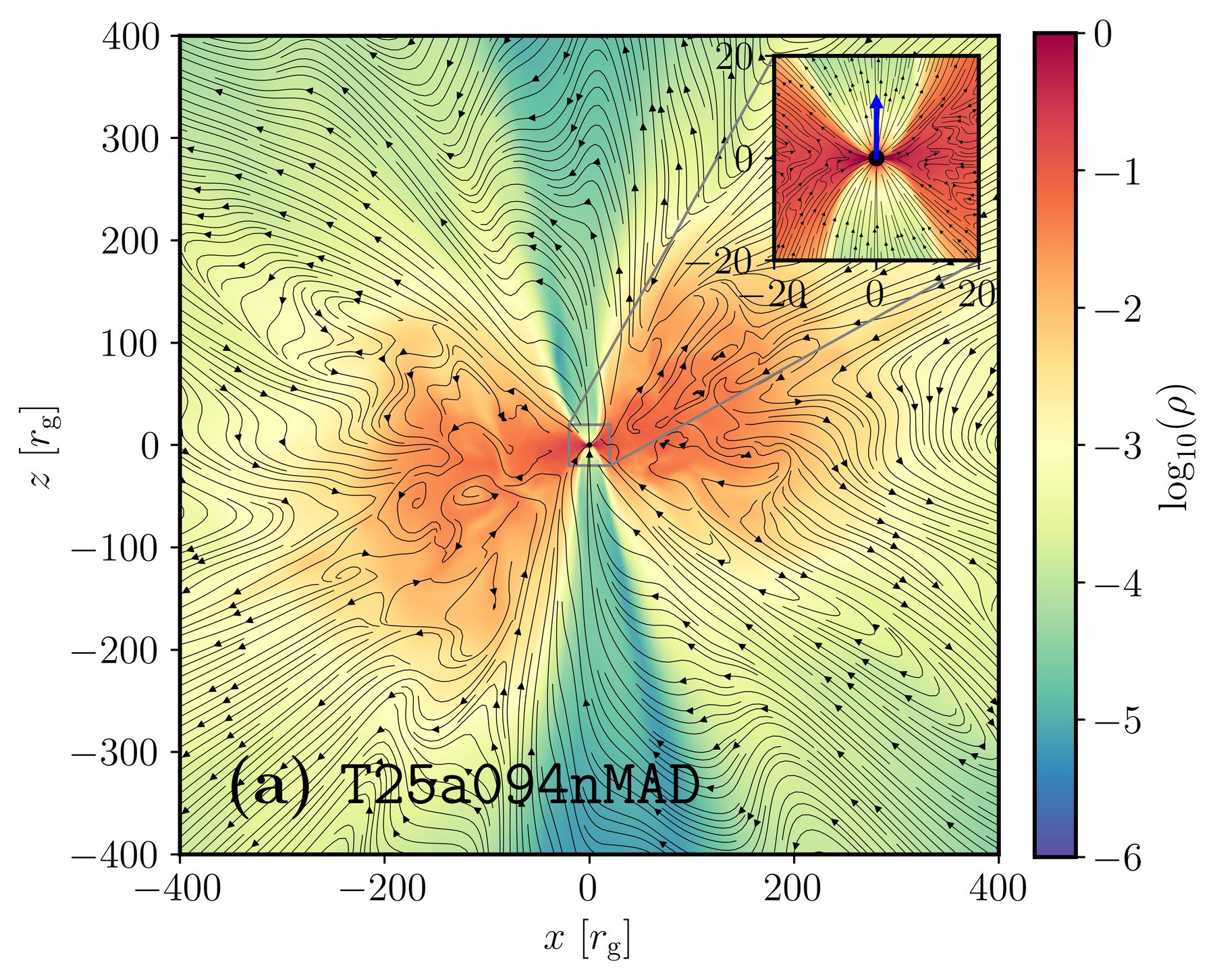}
\includegraphics[height=.41\linewidth]{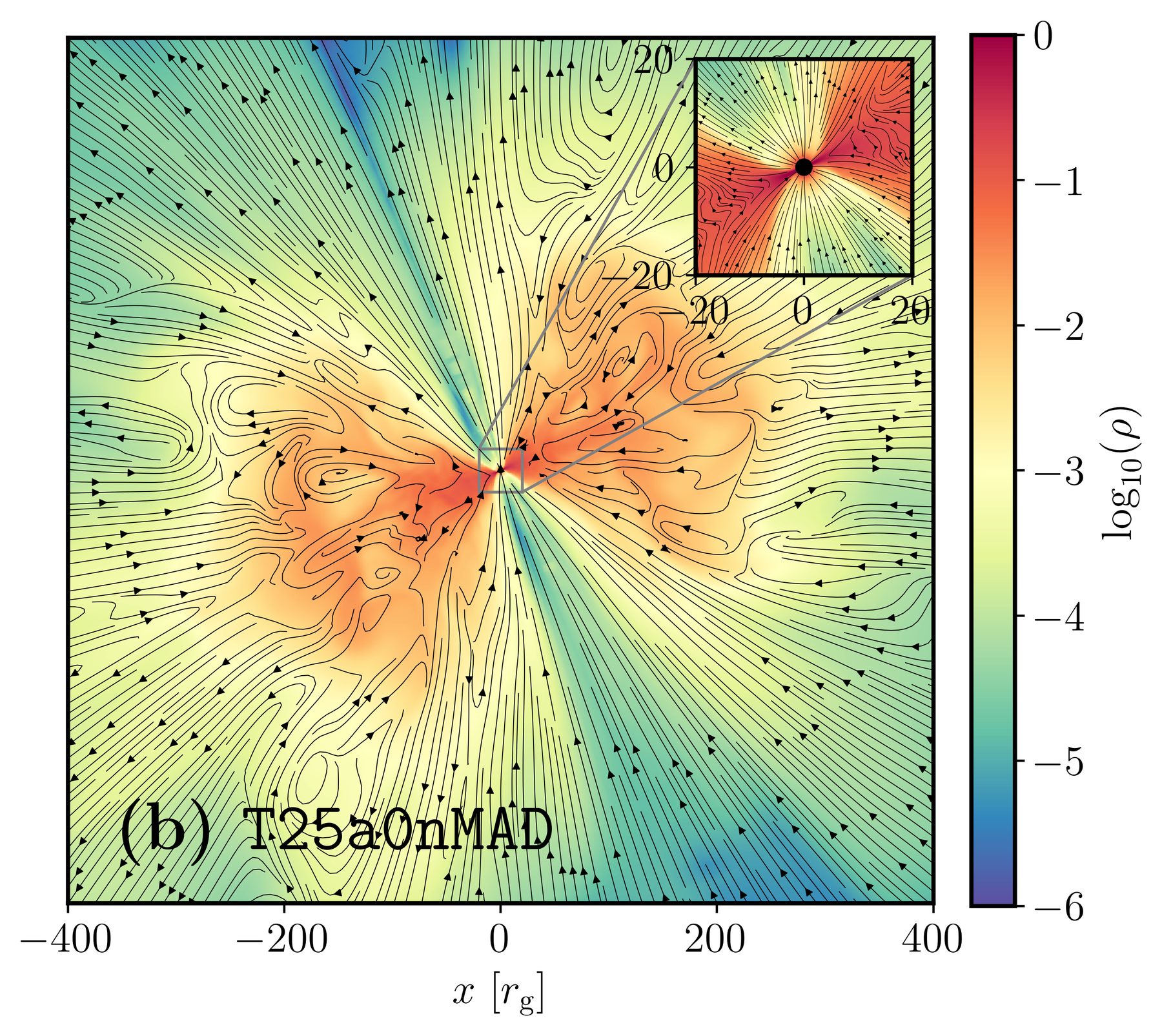}
 	\caption{Time-averaged logarithmic density slices in the $x$-$z$ plane for the rotating model {\tt T25a094nMAD} and the non-rotating control {\tt T25a0nMAD}. Panels (a) and (b) correspond to $a=0.9375$ and $a=0$, with the overplotted streamlines showing the averaged poloidal field. The averages cover $t=14{,}000\,M$ to $15{,}000\,M$, and the inset in panel (a) marks the spin direction with a blue arrow.}
    \label{fig: stream}
\end{figure*}

Figure~\ref{fig: precession_exmple} shows the evolution of the disk precession angle $\mathcal{P}_{\rm disk}$ and tilt angle $\mathcal{T}_{\rm disk}$ for four representative runs. The angle definitions follow \cite{Liska2018} and are summarized in the Appendix\footnote{The tilt angle is the angle between the disk angular-momentum vector and the positive $z$-axis, while the precession angle is the position angle of that vector. Following \cite{Liska2018}, the jet orientation is measured by isolating the magnetically dominated jet and applying the same construction.}.

The hydrodynamic control (black) and the two SANE models (yellow and green) all exhibit ordinary prograde LT precession, meaning that the disk turns in the same sense as the black-hole spin. In {\tt T25a094HD}, the precession remains comparatively steady because the torus expands only modestly. The two SANE cases, {\tt T25a094wSANE} and {\tt T25a094sSANE}, differ only in field strength and therefore provide a clean magnetic comparison. The weaker-field run stays close to the hydrodynamic behavior, showing that magnetic stresses remain secondary there. The stronger SANE case, by contrast, aligns more rapidly and precesses more slowly, consistent with stronger magneto-spin coupling. In both SANE runs the precession signal fades substantially at late times ($30{,}000$ and $40{,}000\,M$).

In the MAD case the behavior changes qualitatively. Once the horizon-threading flux reaches the usual MAD level ($\Phi_{\rm B}/\sqrt{\dot M}\sim15$ for prograde holes; \cite{tchekhovskoy_efficient_2011}), the blue curve in Figure~\ref{fig: precession_exmple}(a) crosses through zero and becomes retrograde. Early in the evolution, before the jet is established, model {\tt T25a094nMAD} still resembles the hydrodynamic reference and follows LT-driven prograde motion. After strong magnetic flux piles up and the jet forms, the sign flips and the disk precesses opposite to the hole spin. The simplest interpretation is that the magnetic torque has overtaken the LT torque. The same change of sense is visible in the 3D renderings of Figure~\ref{fig: volume_rendering}: between the two snapshots the tilted jet-disk system clearly rotates opposite to $\hat{z}$. At the same time, the tilt declines in every magnetized run because magneto-spin alignment steadily draws the flow closer to the spin axis \cite{McKinney2013, Chatterjee2023}. A gradual tilt decay is also present in the HD model because the initial tilted torus is not an exact equilibrium.

The jet broadly mirrors the disk orientation, but its response is noisier. In Figures~\ref{fig: T25a094nMAD}(a) and (b), the red curves show that the jet reacts more sharply than the disk to the flux eruptions characteristic of the MAD state. That behavior is expected because the jet is threaded by a more coherent large-scale field than the dense disk body, so magneto-spin alignment acts on it more efficiently and reduces the jet tilt faster.

The transition in {\tt T25a094nMAD} therefore marks the point where magnetic forcing overtakes LT forcing. The strongly magnetized SANE case supports the same picture in a weaker form: its slower precession rate cannot be explained by disk expansion alone and already points to a magnetic reduction of the ordinary LT signal.

The radial behavior makes the competition clearer. Since $\tau_{\rm LT}\propto r^{-3}$, LT forcing is concentrated toward the inner flow. Figures~\ref{fig: T25a094nMAD}(c) and (d) show radial profiles of the disk precession and tilt in {\tt T25a094nMAD}, obtained by dividing the flow into annuli eight cells wide and integrating the angular momentum in each shell separately. Near the hole, at $r\lesssim10\,r_{\rm g}$, frame dragging twists the disk toward positive $\phi$. As time passes, however, the contrast between inner and outer radii diminishes, so the profile gradually smooths even after the global precession has almost stalled by $t\sim30{,}000\,M$. The tilt profile shows a similar trend: the inner disk aligns first, again consistent with strong magnetic alignment. The end result is a significantly warped and twisted disk shaped jointly by LT precession, magnetic torques, and spin alignment, as illustrated by Figure~\ref{fig: stream}(a).

\subsection{Origin of the Magnetic Torque}

A key question is how the field manages to reverse the precession sense even though the initial tilt is only moderate and the disk and spin axes are already fairly close. Figure~\ref{fig: stream} provides the essential clue. It compares the time-averaged density and poloidal field structure in the rotating model {\tt T25a094nMAD} with the otherwise identical non-rotating control {\tt T25a0nMAD}, averaged over $t=14{,}000$--$15{,}000\,M$. When the hole does not rotate, the poloidal field stays roughly radial. When $a=0.9375$, however, the rotating spacetime drags the initially tilted field toward the spin axis, strengthens the toroidal component, and reorganizes the global magnetic geometry. The field therefore aligns faster than the dense disk body does, leaving a finite offset between the magnetic symmetry axis and the disk angular momentum for an extended period.

\cite{Lai2003} proposed a simple model in which a misaligned disk threaded by a vertical magnetic field $B_z$ aligned with the black-hole spin axis $\hat{z}$ experiences a magnetic torque that drives retrograde precession. Such a torque exists whenever a conducting disk is embedded in an inclined external magnetic field (see \cite{1999ApJ...524.1030L}). When the external field $\mathbf{B}=B_z\hat{z}$ is misaligned with the disk axis $\hat{l}$, it projects a radial component onto the disk plane, $B_r=B_z\sin\tilde{\beta}\sin\phi$, where $\tilde{\beta}$ is the angle between $\hat{z}$ and $\hat{l}$ and $\phi$ is the azimuthal angle around the disk. Meanwhile, the component perpendicular to the disk, $B_z\cos\tilde{\beta}\simeq B_z$ for small $\tilde{\beta}$, induces an azimuthal screening surface current $K_\phi=(c/2\pi)B_z\tan\theta$, where $\tan\theta=B_r^{\rm (ind)}/B_z$ and $B_r^{\rm (ind)}$ is the induced radial field on the upper disk surface. The interaction between $K_\phi$ and $B_r$ then produces a $\phi$-dependent vertical force, which yields the magnetic torque per unit area \cite{Lai2003}
\begin{equation}
    \mathbf{T_{\rm mag}}=-\frac{1}{4\pi}r B_z^2\tan{\theta}\hat{z}\times\hat{l}.
\end{equation}
This torque in turn produces a local retrograde precession,
\begin{equation}
    \Omega_{\rm mag}(r) = -\frac{B_z^2\tan\theta}{4\pi\Sigma r \Omega(r)},
\end{equation}
where $\Sigma$ is the disk surface density and $\Omega(r)$ is the disk angular frequency. Comparing this with the LT precession frequency, where $G=c=1$ and $S$ denotes the BH angular momentum,
\begin{equation}
    \Omega_{\rm LT}(r)=\frac{2S}{r^3}=\frac{2M^2a}{r^3},
\end{equation}
we obtain
\begin{equation}
    \frac{\Omega_{\rm mag}(r)}{\Omega_{\rm LT}(r)}\simeq - \frac{B_z^2\tan\theta/8\pi}{a\rho c c_{\rm s} (r_{\rm g}/r)^2}, \label{Eq: ratio}
\end{equation} 
where we have used $\Sigma\sim\rho H\sim\rho c_{\rm s}/\Omega$, with $c_{\rm s}$ the sound speed in the disk. Whenever the ratio $|\Omega_{\rm mag}/\Omega_{\rm LT}|$ exceeds unity, the magnetic torque dominates over the LT torque.

\begin{figure*}
\centering 	
\includegraphics[width=\linewidth]{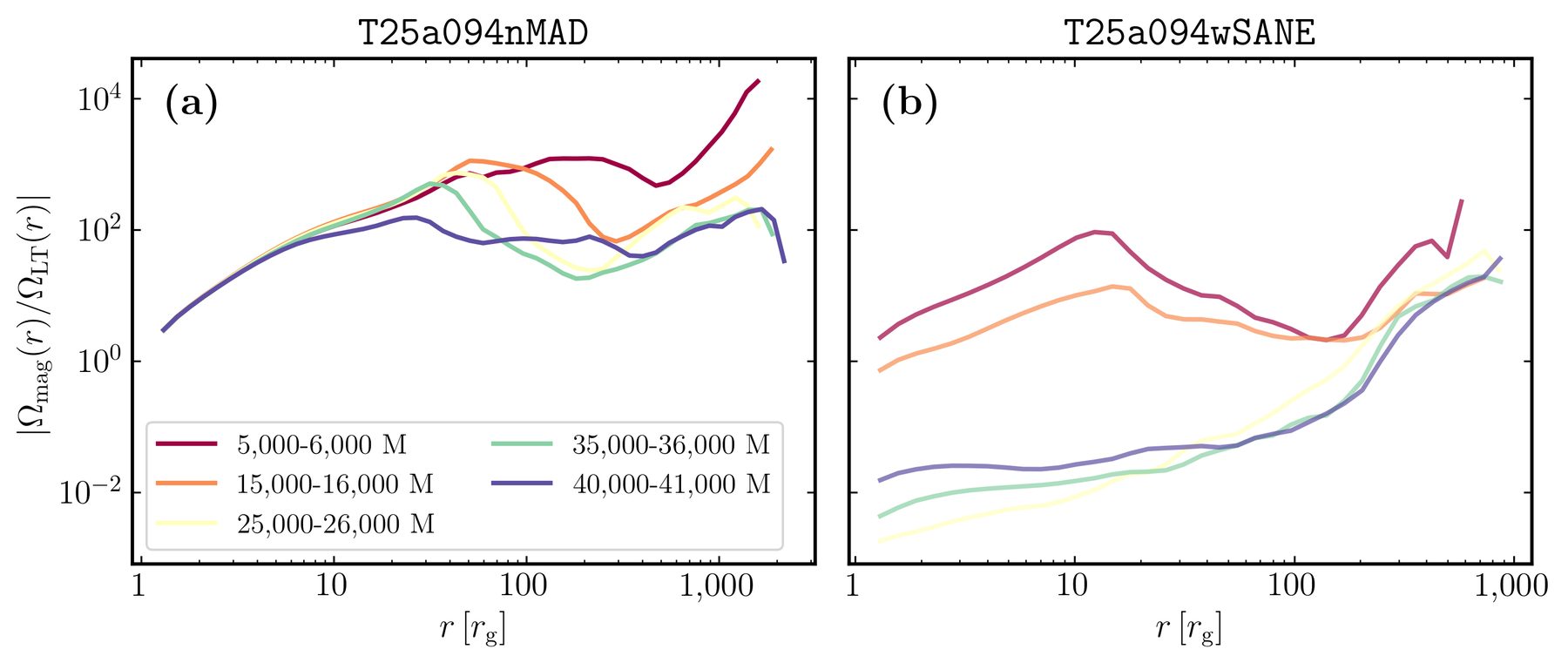}
\caption{Radial profiles of the ratio $|\Omega_{\rm mag}/\Omega_{\rm LT}|$ for two representative simulations: panel (a) shows {\tt T25a094nMAD}, and panel (b) shows {\tt T25a094wSANE}. Colors indicate different averaging windows, and each curve is built from shell averages over annuli that are 8 cells wide.}
 	\label{fig: Omega_ratio}
\end{figure*}

Since LT forcing decays as $r^{-3}$, coherent global precession requires angular-momentum transport that can communicate the inner torque through the broader disk \cite{Liska2018, 2024arXiv240410052F}. To measure the magnetic contribution directly, I evaluate the ratio $|\Omega_{\rm mag}/\Omega_{\rm LT}|$ from the simulations themselves. Figure~\ref{fig: Omega_ratio} highlights the contrast between the representative SANE model {\tt T25a094wSANE} and the MAD model {\tt T25a094nMAD}. In the weakly magnetized SANE run, the ratio stays around $\sim10^{-2}$--$10^{0}$ inside $r\lesssim100\,r_{\rm g}$ at late times ($t\gtrsim20{,}000\,M$), which means LT torques still control the global precession sense. In the MAD run, the same ratio rises to order ten, showing that the magnetic term has become dominant. That large excess naturally explains why the entire flow switches to retrograde precession.

\section{Discussion and Conclusions}

Taken together, these simulations show that thick tilted disks need not remain prograde once magnetic stresses become strong. In the weakly magnetized runs, LT forcing still sets the precession sense. After the flow enters the MAD state, however, the ordered poloidal field supplies a larger counter-torque and the motion reverses. The calculations also make clear that the precession rate evolves with time: as the torus spreads radially, both the prograde and retrograde solutions slow down and can eventually fade, consistent with the disk-expansion behavior reported by \cite{Liska2018}.

The astrophysical implication is significant. In these runs the disk reaches only about $200\,r_{\rm g}$, still smaller than the larger tori often used in some MAD calculations \cite{2022ApJ...930L..16E}, yet even here a purely LT interpretation cannot reproduce the precession rate inferred for the M~87 jet \cite{Cui2023}. If M~87$^*$ is indeed a MAD source \cite[e.g.,][]{2019ApJ...875L...5E, Yuan2022}, then magnetically driven retrograde precession offers a more plausible explanation than standard prograde LT precession alone. Jet orientation in such systems is therefore controlled jointly by frame dragging and magnetic structure.

Distinguishing those two senses of precession observationally remains difficult but important. Side-on jets often hide the direction of motion, yet a few diagnostics may still work. The nearly circular jet motion already seen in V404 Cygni \cite{2019Natur.569..374M} demonstrates that the precession sense can, in principle, be measured; if the spin direction were also known, one could determine whether the motion is prograde or retrograde directly. For M~87-like systems, secular variations in jet width or morphology may provide a comparable handle. Measurements of that kind will be essential for separating magnetic retrograde precession from ordinary LT-driven motion, and future facilities such as the ngEHT and ngVLA should make such tests more practical.


\newpage
{
\section*{Appendix and Supplementary Material}  
\setcounter{section}{0}
\renewcommand{\thesection}{\thechapter.\Alph{section}}
\renewcommand{\theHsection}{appendix.\thechapter.\Alph{section}}
\renewcommand{\theHsubsection}{appendix.\thechapter.\Alph{section}.\arabic{subsection}}
\section{GRMHD Code and Model Setup}\label{sec: numerical_method}

All simulations in this chapter are performed with KHARMA\footnote{\url https://github.com/AFD-Illinois/kharma} \cite{2024arXiv240801361P}, the GPU-based successor to {\tt iharm3D} \cite{2021JOSS....6.3336P}. Both codes descend from HARM \cite{2003ApJ...589..444G} and evolve ideal MHD in curved spacetime. In Eulerian form, the GRMHD system is
\begin{equation}
    \begin{aligned}
        \partial_t(\sqrt{-g}\rho u^t)&=-\partial_i(\sqrt{-g}\rho u^i),\\
        \partial_t(\sqrt{-g}T^t_{\nu})&=-\partial_i(\sqrt{-g} T^i_{\nu}) + \sqrt{-g}T^\kappa_\lambda\Gamma^\lambda_{\nu\kappa},\\
        \partial_t(\sqrt{-g}B^i)&=-\partial_j\left[\sqrt{-g}(b^ju^i - b^i u^j)\right],\\
        \frac{1}{\sqrt{-g}}\partial_i(\sqrt{-g}B^i)&=0,
    \end{aligned}
\end{equation}
where $\rho$ is the rest-mass density, $u^\mu$ is the fluid four-velocity, $\Gamma$ denotes the Christoffel symbols, $B^i$ and $b$ are the three- and four-magnetic fields, and $g$ is the metric determinant \cite[see][]{2003ApJ...589..444G}. As in most modern GRMHD calculations, I use Kerr-Schild coordinates to avoid coordinate pathologies near the horizon \cite[e.g.,][]{Porth2017, 2022ApJS..259...64W}, together with an exponential radial coordinate that concentrates resolution toward small radii \cite{2003ApJ...589..444G}.

The initial condition is a Fishbone-Moncrief equilibrium torus \cite{1976ApJ...207..962F} threaded by a single poloidal loop. Exact equilibrium would hold only for a spin-aligned torus, so I introduce the desired misalignment gently by tilting the disk by $25^\circ$ about the $y$-axis \cite{Liska2018, 2019ApJ...878...51W}. The torus is chosen to be comparatively extended, with outer radius $r_{\rm out}\sim200\,r_{\rm g}$, where $r_{\rm g}=GM/c^2$. Some transient inflow from imperfect equilibrium is unavoidable, but it remains much weaker than the subsequent magnetically driven evolution; the HD control accretes at roughly an order of magnitude below the MHD cases (Figure~\ref{fig: Mdot-c5}a). Across the suite, the Kerr spin is either 0 or 0.9375 so that the role of rotation can be isolated cleanly. I adopt an ideal-gas equation of state with $\Gamma_{\rm g}=5/3$, and for the strongest MAD case ({\tt T25a094nMAD}) I also compute a higher-resolution companion run marked with the suffix H.
\begin{table}[h]
\caption{Main parameters of the GR(M)HD simulations analyzed in this chapter.}
\centering
\begin{tabular}{lllllllll}
\hline
Models                                & a       & $\mathcal{T}_0\,(^\circ)$ & ${\tt lin\_frac}$ & ${\tt smoothness}$ & $r_{\rm d} \,[{r_{\rm g}}]$ & $A_\phi$               & $\beta_{\rm min}$ & Resolution              \\ \hline
{\tt T25a094HD}   & 0.9375  & 25         & 0.8            & 0.03             & 500                     & -                      & -                 & $144\times72\times96$   \\
{\tt T25a094wSANE}& 0.9375  & 25         & 0.8            & 0.03             & 1000                     & $A_{\phi}^{\tt wSANE}$ & 100               & $288\times128\times128$ \\
{\tt T25a094sSANE}& 0.9375  & 25         & 0.8            & 0.03             & 1000                     & $A_{\phi}^{\tt sSANE}$ & 100               & $288\times128\times128$ \\
{\tt T25a094nMAD} & 0.9375  & 25         & 0.9            & 0.02             & 2500                     & $A_{\phi}^{\tt MAD}$  & 100               & $288\times128\times128$ \\
{\tt T25a094nMADH} & 0.9375 & 25         & 0.9            & 0.02             & 2500                     & $A_{\phi}^{\tt MAD}$  & 100               & $392\times256\times192$ \\
{\tt T25a0nMAD}    & 0       & 25         & 0.8            & 0.03             & 1000                     & $A_{\phi}^{\tt MAD}$  & 100               & $288\times128\times128$ \\
 \hline
\end{tabular}\label{table:models}
\end{table}

Table~\ref{table:models} summarizes the parameter choices for all simulations discussed in this chapter. The naming scheme is intentionally compact, with each label encoding the essential setup choices directly. Each label follows the pattern
\begin{equation*}
\texttt{T[Tilt Angle]a[Spin] [Magnetic Configuration]}
\end{equation*}

\begin{itemize}
    \item \textbf{Tilt angle:} The leading \texttt{T} gives the initial tilt angle in degrees. For example, \texttt{T25} denotes a tilt of $25^\circ$.
    \item \textbf{BH spin:} The factor \texttt{a[Value]} specifies the dimensionless BH spin. For instance, \texttt{a094} means $a=0.9375$, whereas \texttt{a-094} would indicate $a=-0.9375$.
    \item \textbf{Magnetic configuration:} The suffix encodes the magnetic setup, for example \texttt{nMAD} for a normal-strength MAD configuration, \texttt{sSANE} for a strong-field SANE setup, and \texttt{wSANE} for a weak-field SANE setup.
\end{itemize}

As one explicit example, the label \texttt{T25a094nMAD} denotes a torus with an initial tilt of $25^\circ$ (\texttt{T25}), a black-hole spin of $a=0.94$ (\texttt{a094}), and a normal-strength MAD magnetic configuration (\texttt{nMAD}).

The seed field is specified through the vector potential. To obtain a purely poloidal magnetic field, I initialize only the toroidal component and set the remaining components to zero. Three choices are considered for $A_\phi$: weak SANE (wSANE), strong SANE (sSANE), and MAD (nMAD). They are
\begin{equation}
\begin{aligned}
    A_{\phi}^{\rm \tt wSANE} &\propto {\rm max}[\rho-0.2,0],\\
    A_{\phi}^{\rm \tt sSANE} &\propto {\rm max}[(\rho-0.5)^3r^3,0],\\
    A_{\phi}^{\rm \tt MAD} &\propto {\rm max}\left[(r/r_{\rm in})^3\exp{(-r/400)}\rho-0.2,0\right].
\end{aligned}
\label{Eq: Aphi}
\end{equation}
The overall field normalization is set through the minimum plasma beta, $\beta_{\rm min}$, with $\beta\equiv p_{\rm gas}/p_{\rm mag}$. To seed the MRI, the internal energy is perturbed randomly according to $u\rightarrow u+\delta u$ with $|\delta u/u|\leq u_{\rm jitter}$ \cite{2022ApJS..259...64W}. In all runs I use $u_{\rm jitter}=0.1$.

\section{Coordinate System and Numerical Resolution}\label{sec: resolution}

\begin{figure*}
    \centering
    \includegraphics[width=\linewidth]{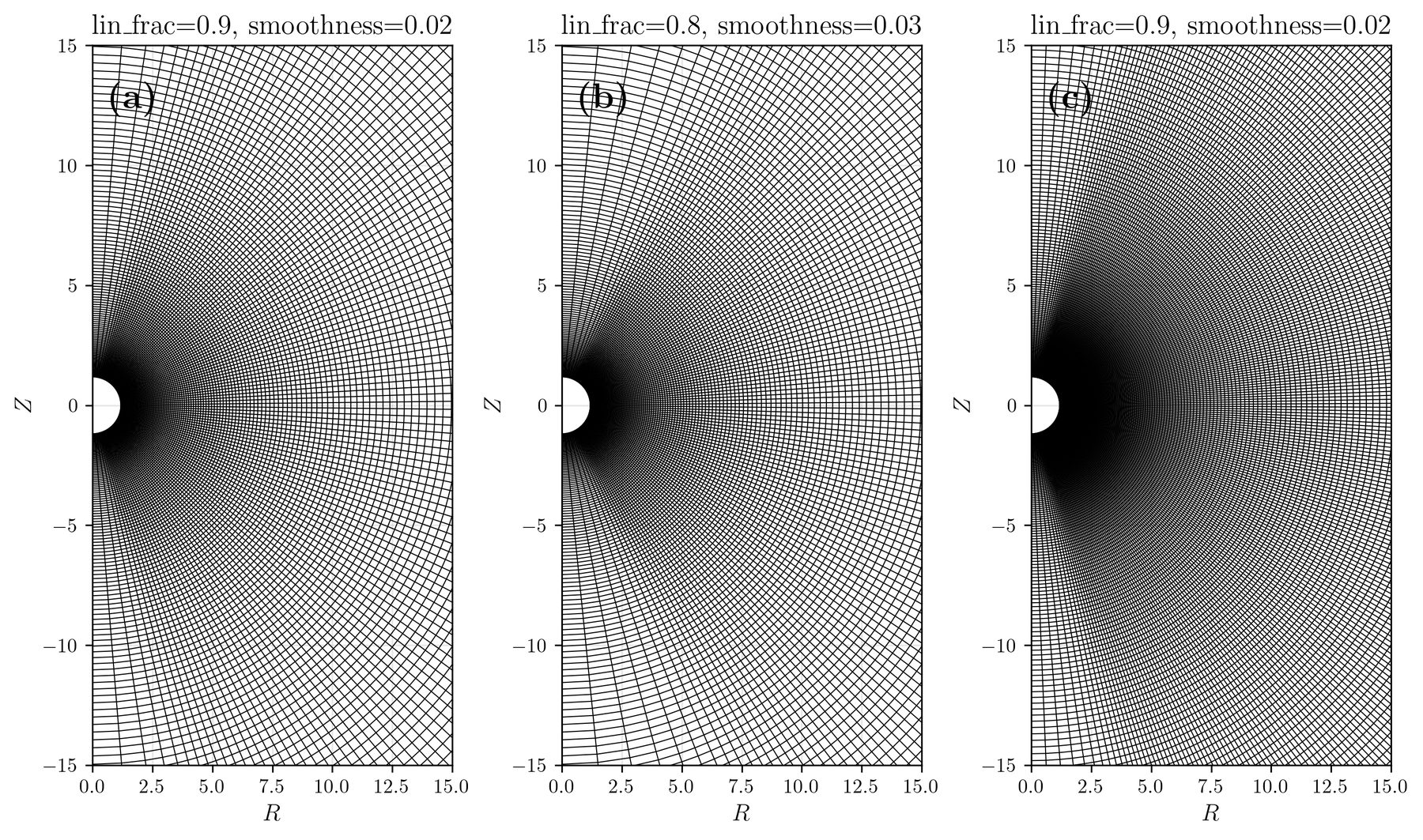}
	    \caption{Mesh configurations used in the simulations. Panels (a) and (c) correspond to models \texttt{T25a094nMAD} and \texttt{T25a094nMADH}, for which \texttt{lin\_frac}=0.9 and \texttt{smoothness}=0.02. Panel (b) shows the grid adopted for the other MHD runs, where \texttt{lin\_frac}=0.8 and \texttt{smoothness}=0.03.}
    \label{fig: mesh}
\end{figure*}

\begin{figure*}
    \centering
    \includegraphics[width=\linewidth]{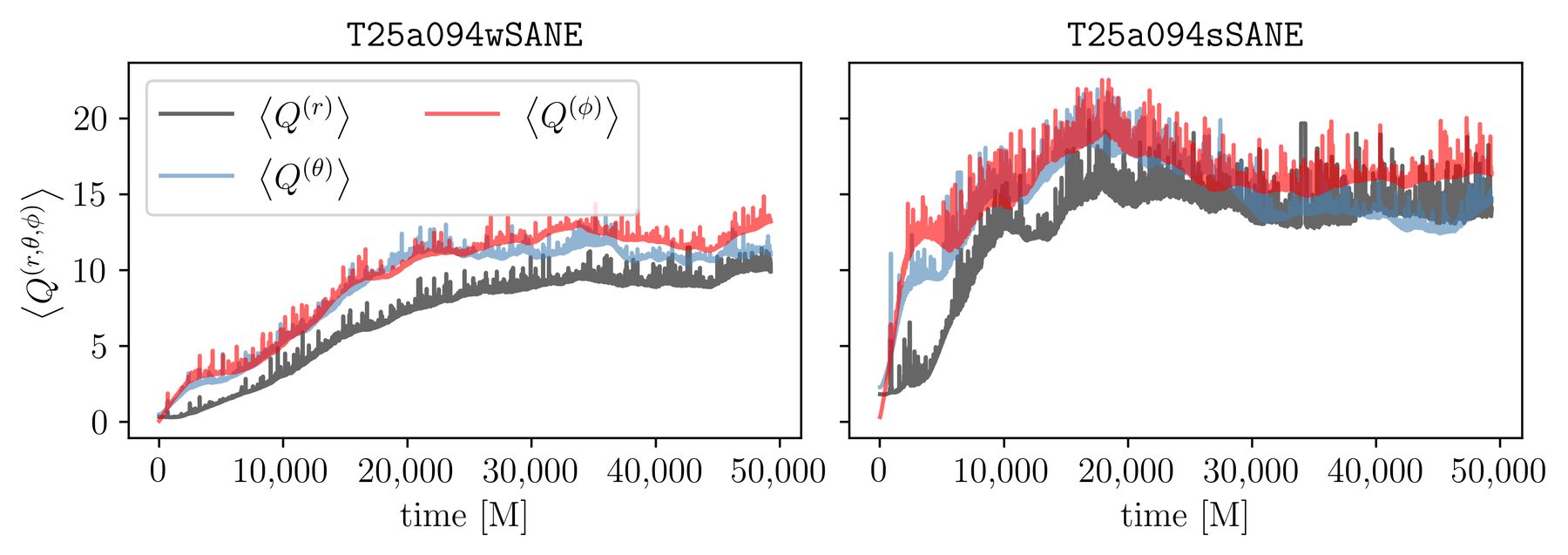}
	    \caption{Time evolution of the density-averaged MRI quality factors $\langle Q^{(r,\theta,\phi)} \rangle$ for the two SANE simulations. Panel (a) corresponds to model {\tt T25a094wSANE}, and panel (b) corresponds to model {\tt T25a094sSANE}. The black, blue, and red curves represent $\langle Q^{(r)}\rangle$, $\langle Q^{(\theta)} \rangle$, and $\langle Q^{(\phi)} \rangle$, respectively.}    \label{fig: Q_factor}
\end{figure*}

Following \cite{2024ApJ...977..200C}, I use uniform wide-pole Kerr-Schild coordinates. This grid keeps most of the disk well resolved while relaxing the resolution near the poles, which roughly doubles the allowed time step. Because the imposed tilt is only moderate, that polar coarsening does not materially alter the precession behavior of interest. Figure~\ref{fig: mesh} shows the resulting grid layouts for the different parameter choices.

The default MHD resolution is $288 \times 128 \times 128$, consistent with recent KHARMA studies \cite{2025ApJS..277...16D, 2022ApJ...930L..16E}. Figure~\ref{fig: Q_factor} shows that this mesh resolves the MRI at the level needed here. The purely hydrodynamic control run {\tt T25a094HD} is instead evolved at $144 \times 72 \times 96$ because its precession does not depend on magnetic turbulence and a lower-cost mesh is sufficient.

Most GRMHD runs use an outer boundary at $r_{\rm d}=1{,}000\,r_{\rm g}$, while the more extended MAD configuration is evolved out to $r_{\rm d}=2{,}500\,r_{\rm g}$ so that the jet remains well inside the numerical domain. The GRHD run uses a smaller boundary, $r_{\rm d}=500\,r_{\rm g}$, because its outflows remain weak and the more compact domain provides higher effective resolution. At the poles I adopt a transmitting boundary condition to let matter leave the grid without introducing unnecessary dissipation.

For the SANE models, Figure~\ref{fig: Q_factor} plots the density-weighted MRI quality factors $\langle Q^{(r,\theta,\phi)}\rangle$. These are computed from the simulation output and averaged over $r<150\,r_{\rm g}$ following \cite{Porth2019}. The resulting $Q^{(r)}$ and $Q^{(\theta)}$ values satisfy the MRI-resolution criteria discussed by \cite{Sorathia2012}, and the weaker SANE case also matches the values reported by \cite{2025ApJS..277...16D}. The key point for this chapter is that the precession reversal is governed by the large-scale ordered poloidal field, not by small-scale fully developed turbulence. Our grid is therefore adequate for the magnetic-precession problem even though it is coarser than the meshes used in some dedicated turbulence studies such as \cite{Liska2018}.

\section{Procedure for Measuring Disk and Jet Tilt and Precession} \label{sec: measurement}

In this work, the disk and jet tilt and precession angles are evaluated with the same framework adopted in earlier studies \cite{2005ApJ...623..347F, Liska2018, 2019ApJ...878...51W}. We summarize that procedure below for completeness.

Following \cite{2005ApJ...623..347F}, the disk angular-momentum vector is written as
\begin{equation}
    (J_{\rm disk})_{\rho} = \frac{\epsilon_{\mu\nu\sigma\rho}L^{\mu\nu}S^{\sigma}}{2\sqrt{-S^\alpha S_\alpha}},
\end{equation}
where the total angular momentum is given by
\begin{equation}
    L^{\mu\nu} = \int\left(x^\mu T^{\nu 0} - x^\nu T^{\mu 0}\right) d^3x,
\end{equation}
and the four-momentum $S^\sigma$ is defined as
\begin{equation}
    S^\sigma=\int T^{\sigma0}d^3x.
\end{equation}
The fluid contribution to the stress-energy tensor is $T^{\mu\nu}=\rho h u^\mu u^\nu + p g^{\mu\nu}$, where $h$ denotes the enthalpy and $p$ the gas pressure. Adopting the Cartesian-coordinate construction of \cite{Liska2018}, the disk tilt angle $\mathcal{T}_{\rm disk}$ is then evaluated as
\begin{equation}
    \mathcal{T}_{\rm disk}=\cos^{-1}\left(\frac{J_{\rm disk}^z}{|\mathbf{J_{\rm disk}}|}\right), \label{Eq: tilt}
\end{equation}
while the disk precession angle $\mathcal{P}_{\rm disk}$ is
\begin{equation}
    \mathcal{P}_{\rm disk} = \tan^{-1}\left(J_{\rm disk}^y, J_{\rm disk}^x\right). \label{Eq: precession}
\end{equation}

Our definition of the jet orientation also follows \cite{Liska2018}. We identify the jet region with the condition $r p_{\rm mag}/\rho>0.5$, where $p_{\rm mag}$ denotes the magnetic pressure. Using the upper jet as an example, its Cartesian centroid is measured through
\begin{equation}
    x^i_{\rm jet}=\frac{\int p_{\rm mag}^{\rm up} x^i d^3x}{\int p_{\rm mag}^{\rm up}d^3x},
\end{equation}
where $p_{\rm mag}^{\rm up}$ is the magnetic pressure in the upper jet and $i=1,2,3$. The jet tilt and precession angles are then extracted in the same manner as for the disk by applying Eqs.~\ref{Eq: tilt} and \ref{Eq: precession}.

\section{Torque Estimates in a Simplified Model} \label{sec: Torque}

We next summarize the estimate of the magnetic torque acting on the disk following \cite{Lai2003}. For an untilted disk threaded by a poloidal magnetic field, and using the notation of \cite{Lai2003}, the surface current in the disk can be written as
\begin{equation}
    K_{\phi}=\int J_{\phi} dz = \frac{c}{2\pi}B_{\rm R}^+, \label{Eq: Kphi} 
\end{equation}
where $J_\phi$ is the current density, $B_{\rm R}^+\equiv B_Z\tan\theta$ denotes the magnetic field on the upper disk surface, and the pitch angle of the poloidal magnetic field is $\theta=\tan^{-1}|B_{\rm R}^+/B_Z|$. For an untilted disk, the magnetic force takes the form $F_{\rm mag}=K_\phi\times B_Z$, which is purely radial and therefore cannot generate a precessional torque.

For the tilted-disk case, we assume that the spinning black hole tends to align the magnetic field with the black-hole spin axis while the disk itself stays misaligned. The disk tilt angle is denoted by $\tilde{\beta}$. We introduce a coordinate system whose $z$-axis is aligned with the disk angular momentum. The corresponding unit vector is $\hat{l}$, whereas $\hat{Z}$ denotes the unit vector along the $Z$-axis of the untilted frame. In this tilted frame, the vertical magnetic field $\mathbf{B_Z}$ can be decomposed into two parts:
\begin{equation}
    \mathbf{B_{Z}} = B_Z \cos{\tilde{\beta}} {\hat{l}} + B_Z \sin{\tilde{\beta}}\sin{\phi} {\hat{r}}.
    \label{Eq: Bz_decompose}
\end{equation}
Accordingly, the component of the magnetic force perpendicular to the disk becomes
\begin{equation}
    F_z=-\frac{1}{c}K_\phi B_Z \sin{\tilde{\beta}}\sin{\phi}.
\end{equation}
For a small tilt angle, Eq.~\ref{Eq: Kphi} remains a good approximation. After integrating over the azimuthal direction $\phi$, the torque per unit area becomes
\begin{equation}
    \left<\mathbf{T_{\rm prec}}\right>=-\frac{1}{2c}r K_\phi B_Z {\hat{Z}}\times{\hat{l}} 
    =-\frac{1}{4\pi}rB_Z^2\tan{\theta}{\hat{Z}}\times{\hat{l}}.
\end{equation}
The total torque acting on the disk, $\mathbf{T_{\rm tot}}$, is therefore
\begin{equation}
    \mathbf{T_{\rm tot}}=\int_{r_{\rm in}}^{r_{\rm out}}2\pi rdr\left<\mathbf{T_{\rm prec}}\right>=-\int_{r_{\rm in}}^{r_{\rm out}}dr \frac{1}{2}r^2 B_Z^2\tan{\theta} {\hat{Z}}\times{\hat{l}}. \label{Eq: mag_torq}
\end{equation}
Hence, the angular frequency associated with magnetically driven precession is
\begin{equation}
    \mathbf{\Omega_{\rm prec}} = -\frac{1}{L_{\rm disk}}\int_{r_{\rm in}}^{r_{\rm out}}dr \frac{1}{2}r^2B_Z^2\tan{\theta} {\hat{Z}},
\end{equation}
where $L_{\rm disk} = \int_{r_{\rm in}}^{r_{\rm out}}\Sigma r^2 \Omega_{\rm d} 2\pi r dr$ is the disk angular momentum, $\Sigma$ is the surface density, and $\Omega_{\rm d}$ is the disk angular frequency.

In a realistic accretion flow, the magnetic field is unlikely to realize such an idealized alignment. Even so, black-hole rotation generally causes the poloidal-field direction, including the jet orientation, to differ from the disk orientation by a finite amount, which is sufficient for the simple model considered here.

Previous work has repeatedly shown that tilted disks can undergo prograde LT precession \cite[e.g.,][]{2024Natur.630..325P}. The discussion above indicates that, in weakly magnetized disks, the magnetic torque mainly acts to slow the LT precession rate. The LT torque per unit area can be written as \cite{McKinney2013}
\begin{equation}
T_{\rm LT} = \sin\tilde{\beta} \Omega_{\rm LT} L_{\rm disk}, 
\end{equation}
with
\begin{equation}
\Omega_{\rm LT}=\frac{1}{L_{\rm disk}} \int_{r_{\rm in}}^{r_{\rm out}}\frac{2M^2 a}{r^3}r^2\Omega_{\rm d}\Sigma 2\pi r dr.
\end{equation}

\section{Disk Size, Accretion Rate, and Magnetic Flux} \label{sec: Mdot}

\begin{figure}
    \centering
    \includegraphics[width=\linewidth]{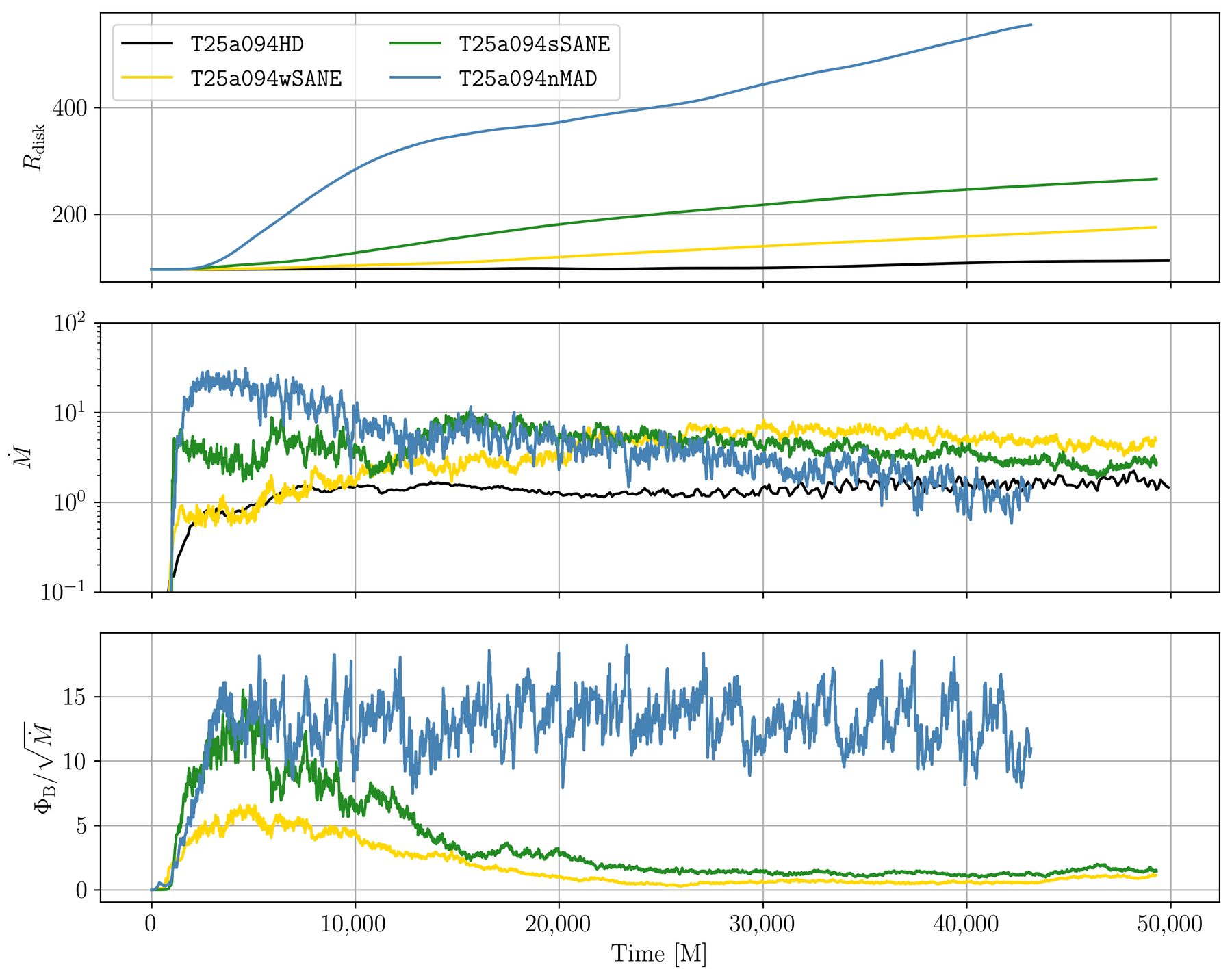}
	    \caption{Time evolution of three bulk diagnostics for four representative tilted-disk models. Panel (a) gives the characteristic disk size, panel (b) the accretion rate, and panel (c) the normalized magnetic-flux rate. Black, yellow, green, and blue curves correspond to {\tt T25a094HD}, {\tt T25a094wSANE}, {\tt T25a094sSANE}, and {\tt T25a094nMAD}, respectively.}    \label{fig: Mdot-c5}
\end{figure}

Figure~\ref{fig: Mdot-c5} summarizes three global quantities: the mean disk radius $R_{\rm disk}$, the accretion rate $\dot M$, and the event-horizon magnetic flux normalized by $\sqrt{\dot M}$. I use the definition of $R_{\rm disk}$ from \cite{Liska2018}, while $\dot M$ and $\Phi_{\rm B}$ follow \cite{Porth2019}; explicitly,
\begin{equation}
    \begin{aligned}
        R_{\rm disk} &= \frac{\int\int_0^{\pi}\int_0^{2\pi} r \rho\sqrt{-g}drd\theta d\phi}{\int\int_0^{\pi}\int_0^{2\pi} \rho\sqrt{-g}drd\theta d\phi},\\
        \dot{M} &= \int_0^{2\pi} \int_0^{\pi} \rho u^r \sqrt{-g} \, d\theta \, d\phi,\\
        \Phi_{\rm B} &= \frac{1}{2} \int_0^{2\pi} \int_{0}^{\pi} \left|B^r\right| \sqrt{-g} \, d\theta \, d\phi,
    \end{aligned}
\end{equation}
where, when evaluating $R_{\rm disk}$, we retain only the region with $\rho>10^{-5}$, and $g$ is the metric determinant.

Panel (a) of Figure~\ref{fig: Mdot-c5} shows that the strongly magnetized MAD model {\tt T25a094nMAD} grows radially much faster than either SANE run, while the hydrodynamic model {\tt T25a094HD} changes only modestly in size. The difference is produced by the much stronger outflows launched in the MAD case. In the hydrodynamic run, by contrast, most of the torus mass stays close to its initial location.

Because the torus-atmosphere transition is smooth, the non-magnetized model {\tt T25a094HD} still shows a small residual inflow from its outer edge, as seen in panel (b). That weak accretion remains far below the values reached in the MHD models and does not affect the overall interpretation. In {\tt T25a094nMAD}, however, the pronounced radial spreading of the torus lowers the disk density by more than an order of magnitude. By the end of the run, its accretion rate therefore falls below that of the hydrodynamic case.

The two SANE models both remain at normalized magnetic fluxes below unity in code units. The strongly magnetized run {\tt T25a094nMAD}, on the other hand, reaches $\Phi_{\rm B}/\sqrt{\dot M}\sim15$, which places it safely in the MAD regime \cite{Begelman2022, tchekhovskoy_efficient_2011}.

\section{Comparison of High and Low Resolutions}
\label{sec: res_comp}

\begin{figure}
    \centering
    \includegraphics[width=\linewidth]{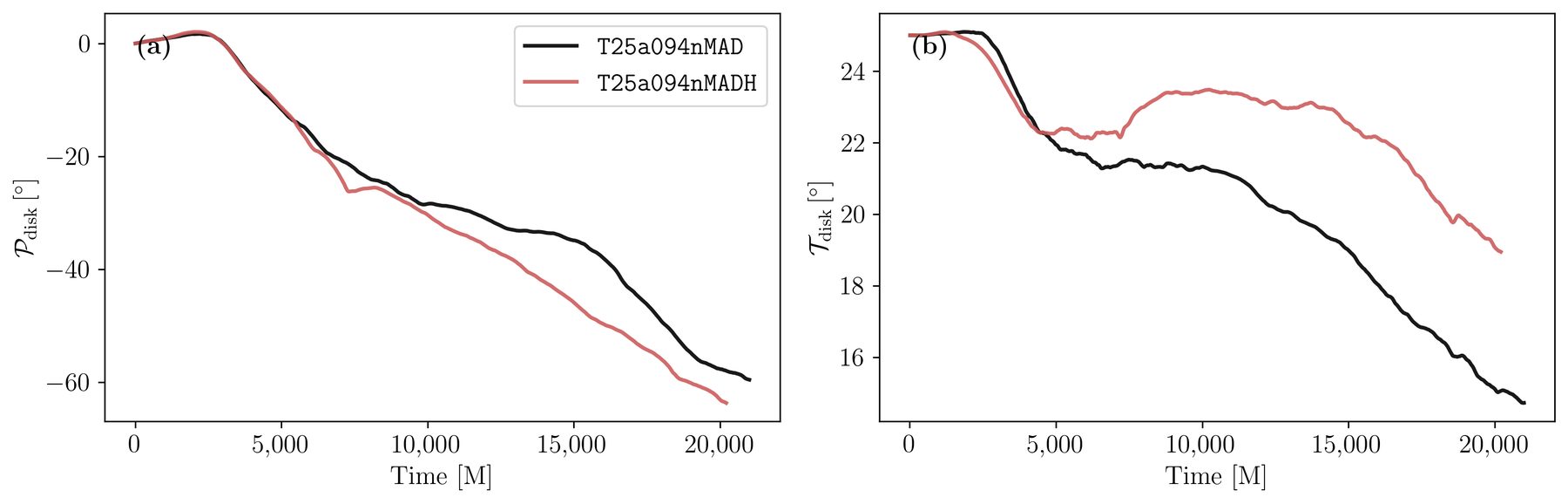}
	    \caption{Resolution check for the MAD model {\tt T25a094nMAD}. Shown are the histories of the disk precession angle $\mathcal{P}_{\rm disk}$ and tilt angle $\mathcal{T}_{\rm disk}$ from the baseline grid and from the refined calculation {\tt T25a094nMADH}.}    \label{fig: res_comp}
\end{figure}

Figure~\ref{fig: res_comp} compares the fiducial MAD model {\tt T25a094nMAD} with the higher-resolution realization {\tt T25a094nMADH}. The precession histories from the two meshes are almost identical. The tilt evolution is likewise the same in broad terms, with only a mildly slower alignment on the refined grid. The retrograde-precession signal is therefore not a numerical artifact, and the standard grid already reproduces the essential behavior discussed in this chapter.
}

\part{Accretion Signatures and Shadows in Complex Spacetimes}

\chapter{Imaging Rotating Loop-Quantum Black Holes at Horizon Scales: Semi-Analytical Tests with Sagittarius A$^*$ and M\,87$^*$}
\markboth{第7章 Imaging Rotating LQBHs at Horizon Scales}{第7章 Imaging Rotating LQBHs at Horizon Scales}

This chapter is based on an earlier article by Jiang H.-X., Liu C., Dihingia I. K., Mizuno Y., Xu H., Zhu T., and Wu Q., published in 2024 in JCAP, volume 2024, article 059 \cite{2024JCAP...01..059J}.


\section{Motivation and Observational Context} \label{secintro}

General relativity remains extraordinarily successful in strong-gravity astrophysics, with support ranging from Solar-System tests \cite{Will2014kxa} and binary pulsars \cite{Hulse1974eb} to gravitational-wave detections \cite{LIGOScientific2016aoc} and horizon-scale imaging \cite{EventHorizonTelescope2019dse}. That success, however, does not solve its deepest conceptual problems. Singularities still arise, a full quantum completion is still missing, and the dark sector remains unexplained \cite{Ng2003jk, Adler2010wf}. For that reason, one expects the classical spacetime description to fail precisely where curvature becomes largest, namely in the early universe \cite{Borde1993xh, Borde2001nh} and in black-hole interiors \cite{Hawking1973uf}.

This motivates asking whether quantum-gravity corrections can leave any observable imprint outside the horizon. Loop quantum gravity is a useful framework for that question because singularity resolution is a recurring outcome in both cosmological and black-hole settings. In symmetry-reduced polymerized constructions, one obtains a regular self-dual black-hole metric whose corrections are governed mainly by the area gap and the Barbero-Immirzi parameter \cite{Modesto2008im, Yang2023gas, Modesto2009ve, Sahu2015dea}. More broadly, many LQG-inspired black-hole models have now been explored, and the replacement of the central singularity by a regular core is among the most robust qualitative results \cite{Ashtekar2018cay, Ashtekar2018lag, Bojowald2018xxu, ABP19, ADL20, Perez17, Han2023wxg, Rovelli18, BMM18, Ashtekar20, Gan2020dkb}.

The practical issue is whether those corrections are large enough to affect data. That possibility has motivated studies spanning conceptual tests, lensing calculations, and astrophysical imaging constraints \cite[e.g.,][]{Sahu2015dea, Alesci2011wn, Chen2011zzi, Dasgupta2012nk, Hossenfelder2012tc, Barrau2014yka, Cruz2015bcj, add1, add2, Cruz2020emz, Santos2021wsw, Liu2020ola, Zhu2020tcf, Virbhadra2022iiy, Yan2022fkr, Tu2023xab, Papanikolaou2023crz}. Rotating LQBH shadows were already explored by \cite{Liu2020ola}; the same self-dual geometry has also been tested with gravitational lensing, solar-radio deflection, Solar-System constraints, and stellar orbits around Sgr\,A$^*$ \cite{Sahu2015dea, Zhu2020tcf, Yan2023vdg}. Even so, horizon-scale radiative calculations for rotating LQBHs remain comparatively limited.

The EHT sharpened that question by delivering horizon-scale images of M\,87$^*$ and Sgr\,A$^*$. In M\,87$^*$, the measured ring size and its near-circular morphology are consistent with a Kerr black hole of mass $(6.5\pm 0.7)\times 10^9\,M_{\odot}$ \cite{EventHorizonTelescope2019dse, Akiyama2019brx, Akiyama2019eap, Akiyama2019bqs, 2019ApJ...875L...5E, Akiyama2019sww}. The Sgr\,A$^*$ campaign extended the same kind of test to the Galactic center \cite{2022ApJ...930L..12E, 2022ApJ...930L..13E, 2022ApJ...930L..14E, 2022ApJ...930L..15E, 2022ApJ...930L..16E, EventHorizonTelescope2022xqj}, although rapid intrinsic variability and the presence of NIR/X-ray flares complicate the interpretation \cite{2022ApJ...930L..15E, Genzel2003, Mossoux2020}. Current analyses do not demand departures from GR \cite{EventHorizonTelescope2022xqj}, but neither do they exclude every non-Kerr alternative \cite{Mizuno2018lxz, EventHorizonTelescope2021dqv}. Horizon-scale images therefore remain among the most direct observables for testing LQBH spacetimes \cite{Akiyama2019eap, EventHorizonTelescope2022xqj, Mizuno2018lxz, EventHorizonTelescope2021dqv, Tsukamoto2014tja, Tsukamoto2017fxq, Chael2018oym, Kumar2023jgh, Roder2023oqa, Psaltis2020}.

The analysis below uses semi-analytical GRRT calculations at $230\,\rm GHz$ to generate images of rotating LQBHs and compare them with current EHT constraints for Sgr\,A$^*$ and M\,87$^*$. The purpose is to determine how the loop-quantum correction modifies image morphology, polarization structure, and the parameter bounds inferred from existing observations.

The chapter proceeds as follows. Section 7.2 reviews the rotating LQBH spacetime, Section 7.3 describes the GRRT setup and the resulting image signatures, Section 7.4 compares those models with EHT constraints, and Section 7.5 summarizes the implications.

\section{Rotating LQBH Geometry: Brief Summary}\label{secLQBH}

The radiative calculations below use the rotating self-dual loop-quantum black-hole metric derived in symmetry-reduced LQG models of homogeneous spacetimes. The geometry is geodesically complete and replaces the classical singularity with a regular interior region. In Boyer-Lindquist coordinates it can be written as \cite{Liu2020ola},
	\begin{eqnarray}
	d s^2 &=& \frac{\mathscr{H}}{\Sigma} \left[\frac{\Delta}{\Sigma}(dt-a\sin^2\theta d\phi)^2-\frac{\Sigma}{\Delta}dr^2-\Sigma d\theta^2-\frac{\sin^2\theta}{\Sigma}(a dt-(k^2+a^2)d\phi)^2 \right],  \label{mmm} \label{metric-c6}
	\end{eqnarray}
with
\begin{eqnarray}
\Delta(r)&=& \frac{(r-r_+)(r-r_-)r^2}{(r+r_*)^2}+a^2, \label{Delta}\\
\Sigma(r)&=&k^2(r)+a^2\cos^2\theta,\\
k^2(r)&=& \frac{r^4+a^2_0}{(r+r_*)^2},
\end{eqnarray}
Here $a$ denotes the specific angular momentum of the hole. The prefactor $\mathscr{H}$ remains finite and satisfies $\lim\limits_{a \to 0} \mathscr{H}=r^2+a_0^2/r^2$, so $\mathscr{H}/\Sigma$ rescales the line element without changing the intrinsic null structure. In the non-rotating limit the metric reduces to the static self-dual solution discussed by \cite{Zhu2020tcf}. There the outer and inner horizons lie at $r_+=2 G_\text{LQG} M/(1+P)^2$ and $r_{-} = 2G_\text{LQG} M P^2/(1+P)^2$, while the intermediate scale is $r_{*}= \sqrt{r_+ r_-} = 2G_\text{LQG} MP/(1+P)^2$. The effective gravitational constant in the loop-quantum spacetime is $G_\text{LQG}$, $M$ is the ADM mass, and the correction strength is encoded by
\begin{eqnarray}
P \equiv \frac{\sqrt{1+\epsilon^2}-1}{\sqrt{1+\epsilon^2}+1},   \label{P_epsilon-c6}
\end{eqnarray} 
where $\epsilon=\gamma \delta \ll 1$ is the product of the Immirzi parameter $\gamma$ and the polymeric parameter $\delta$. The quantity $a_{0}$ is
\begin{eqnarray}
a_0 = \frac{A_{\rm min}}{8\pi},
\end{eqnarray}
where $A_{\rm min}$ is the minimum area gap of LQG. Since $A_{\min} \simeq 4 \pi \gamma \sqrt{3} l_{\rm Pl}^2$ \cite{Sahu2015dea}, the associated scale is Planckian and irrelevant for the horizon-scale phenomenology studied here. I therefore set $a_0=0$ and retain only the loop-quantum correction that operates over macroscopic radii. In this notation,
\begin{eqnarray}
G_\text{LQG}=G_{\rm N}\frac{(1+P)^2}{(1-P)^2}.
\end{eqnarray}  
The horizon locations follow from solving $\Delta(r)=0$. The larger solution defines the outer horizon $r_{h+}$ and the smaller one gives the inner horizon $r_{h-}$. Setting $P=0$ immediately restores the Kerr expression, $r_{h\pm}=M \pm \sqrt{M^2-a^2}$.

For the numerical comparison with Kerr, I rewrite the metric using the Newtonian constant $G_{\rm N}$ so that every model is expressed in a common normalization. The dimensionless spin must then be rescaled by the same factor $(1+P)^2/(1-P)^2$. This differs from the choice made by \cite{Afrin2022ztr}, who left $G_\text{LQG}$ fixed; that normalization difference is the main reason their quantitative bounds do not match those reported here.

The Immirzi parameter itself is not unique across the literature \cite[see][]{BenAchour2014qca, Frodden2012dq, Achour2014eqa, Han2014xna, Carlip2014bfa, Taveras2008yf}. Some constructions even allow complex values \cite{Frodden2012dq, BenAchour2014qca, Carlip2014bfa, Meissner2004ju}, whereas others promote $\gamma$ to a dynamical field \cite{Taveras2008yf}. Throughout this chapter I adopt the entropy-motivated value $\gamma = 0.2375$ \cite{Meissner2004ju}. 

\begin{figure}
    \centering
    \includegraphics[width=0.8\linewidth]{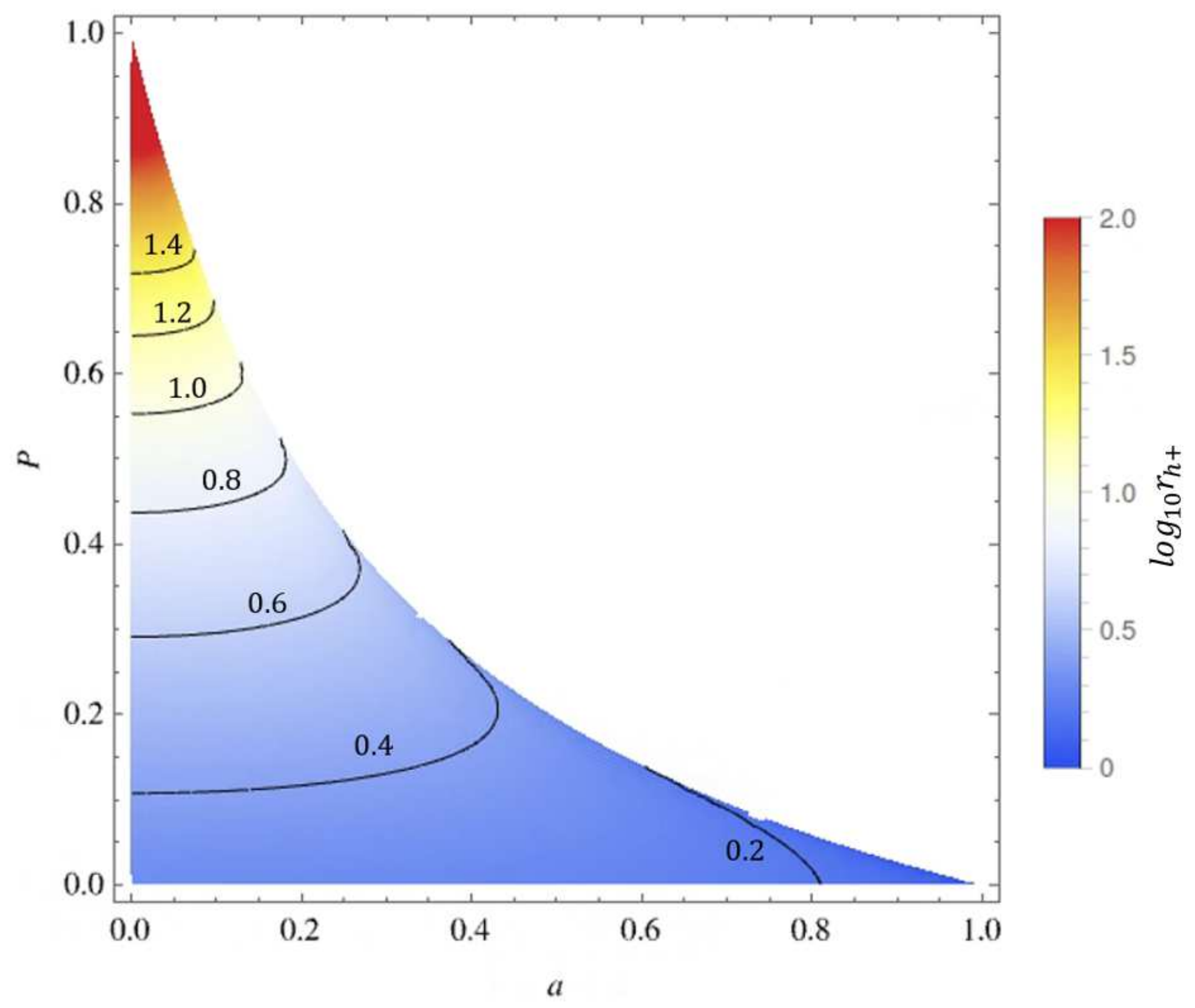}
    \caption{Logarithmic contour map of the outer-horizon radius $r_{h+}$ in units of $G_{\rm N} M/c^2$ across the $(P,a)$ plane.
    }  
    \label{horizon_contour}
\end{figure}

Figure~\ref{horizon_contour} illustrates how the outer horizon varies across the $(P,a)$ plane. At fixed spin, increasing $P$ initially enlarges $r_{h+}$, but the growth weakens and may reverse for larger polymeric corrections. A single value of $r_{h+}$ can therefore correspond to more than one value of $P$ at fixed $a$.

\section{GRRT Images of LQBH Shadows}

The appearance of a black-hole shadow is set jointly by the metric and by the radiating plasma crossed by the photon trajectories. For that reason, confronting synthetic images with the EHT measurements \cite{2022ApJ...930L..12E, 2022ApJ...930L..13E, 2022ApJ...930L..14E, 2022ApJ...930L..15E, 2022ApJ...930L..16E, EventHorizonTelescope2022xqj} provides a direct way to bound the polymeric parameter $P$. Analytic shadow calculations \cite{Liu2020ola, Afrin2022ztr} identify only the formal edge of the shadow, whereas full images also include emission from the photon ring and the surrounding accretion flow. Earlier studies have already shown that both the spacetime and the plasma model can move the bright ring appreciably \cite{Ozel2022, Younsi2023}. Here I therefore generate semi-analytical GRRT images in rotating LQBH spacetimes and compare them with the present EHT limits.

\subsection{Flow Prescription and Viewing Geometry}

For the ray-tracing calculations, I adopt a semi-analytical optically thin RIAF model \cite{Yuan2003, Pu2018ute} embedded in the LQBH spacetime, and I solve the polarized transfer with {\tt ipole}\footnote{\url{https://github.com/AFD-Illinois/ipole}} \cite{Moscibrodzka2017lcu, Noble2007zx}. This subsection summarizes that plasma prescription together with the camera configuration.
The virtual camera sits at $1\,000\,r_{\rm g}$ and covers a field of view of $200\times200\,\rm \mu as$. Photon geodesics are traced across a $1\,000\times1\,000$ image plane at $230\,\rm GHz$. Unless otherwise noted, the fiducial source is Sgr\,A$^*$ with mass $4\times10^6\,\rm M_{\odot}$ and distance $8\,\rm kpc$ \cite{2022ApJ...930L..12E}. Because the analysis stays at $230\,\rm GHz$, where non-thermal electrons are expected to contribute only weakly, I compute the synchrotron signal using a purely thermal electron population.

The semi-analytical plasma model is mainly controlled by the electron density $n_{\rm e}$, the electron-temperature profile $T_{\rm e}$, and the scale height $H$; the velocity field then sets part of the polarization morphology. Since current EHT fits tend to favor MAD-like prescriptions \cite{2022ApJ...930L..16E}, I adopt the phenomenological model of \cite{Chen2021lvo}. Inside the ISCO the gas is assumed to free-fall, while outside it rotates sub-Keplerianly. For each spacetime the ISCO radius $R_{\rm ISCO}$ is calculated numerically from the geodesic equations in Appendix~\ref{sec:geodesic} and passed to {\tt ipole}. The free-fall angular frequency is $\omega_{\rm FF} = -g_{t\phi}/g_{\phi\phi}$, whereas the Keplerian angular frequency is $\omega_{\rm K}=\sqrt{-g_{tt,r}/g_{\phi\phi,r}}$. We write the sub-Keplerian angular velocity as
 \begin{equation}
     \omega_{\rm SubK}=\omega_{\rm K} + (1 - K)(\omega_{\rm FF} - \omega_{\rm K}), \label{EqsubKepler}
 \end{equation}
where $K$ measures how close the flow is to Keplerian rotation. The corresponding free-fall radial velocity is $u^r_{\rm FF} = \sqrt{|-(1 + g_{tt})/g_{rr}|}$, so the sub-Keplerian inflow velocity becomes $u^r_{\rm SubK}=(1-K)\,u^r_{\rm FF}$. 
 
 In the parameter survey I use the usual Sgr\,A$^*$ normalization, with $n_{\rm e}$ measured in units of $3\times10^7\,\rm cm^{-3}$ and $T_{\rm e}$ in units of $3\times10^{11}\,\rm K$. For reasonable $H$ and inclination $i$, this prescription yields total fluxes of order $1$--$3\,\rm Jy$. The fiducial model follows \cite{Chen2021lvo} and adopts $K=0.5$ with $H=0.3$. 

 \subsection{Lensing-Ring Expansion in LQBHs}
 
The horizon analysis above showed that $r_{h+}$ does not vary with $P$ in a uniform way across all spins. Broadly speaking, the horizon grows with increasing $P$ at low spin but can shrink once the hole rotates rapidly. The image-plane ring size and photon sphere behave more simply: they tend to expand as the loop-quantum correction becomes stronger.
 
\begin{figure}
     \centering
     \includegraphics[height=0.35\linewidth]{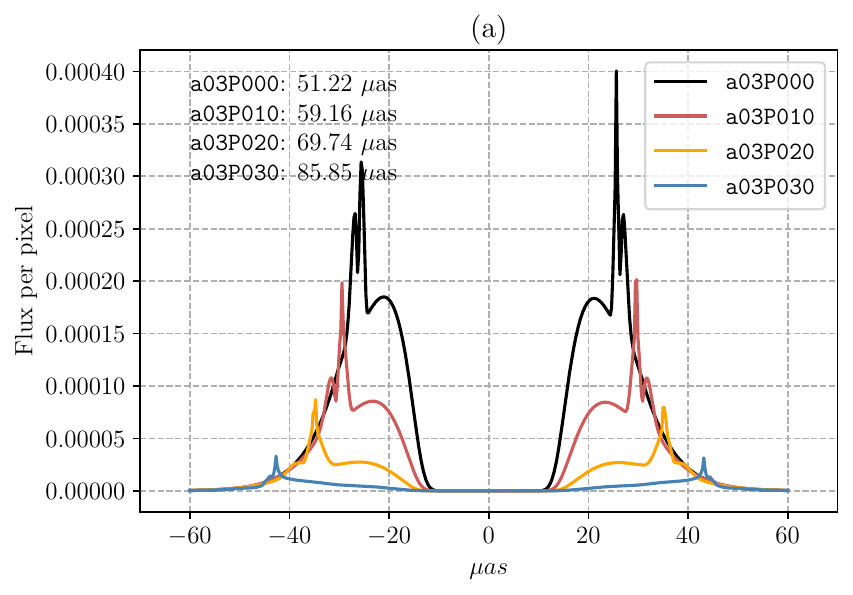}
      \includegraphics[height=0.35\linewidth]{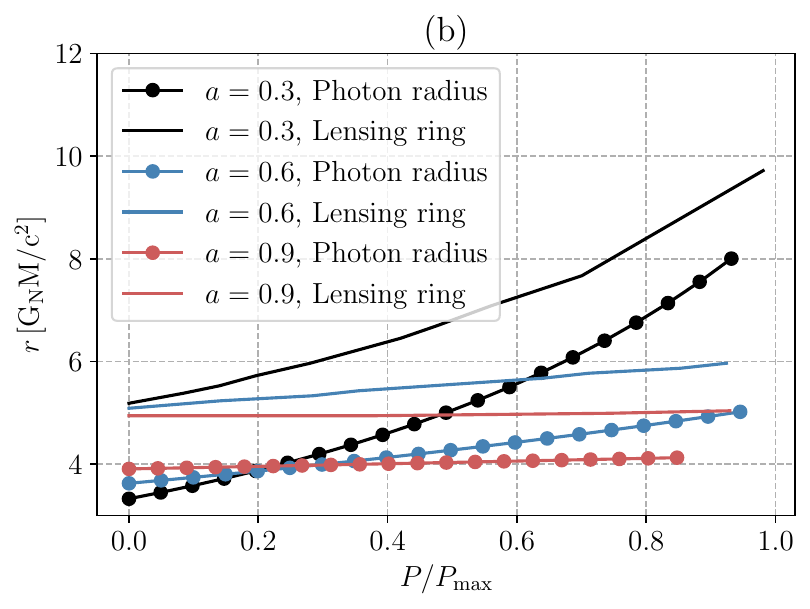}
     \caption{Panel (a) shows one-dimensional intensity cuts for the $a=0.3$ sequence with $P=0$ (black), $0.1$ (red), $0.2$ (orange), and $0.3$ (blue). Panel (b) compares the evolution of the photon-sphere radius and the image-plane ring diameter as $P$ increases. Solid black, blue, and red curves correspond to $a=0.3$, $0.6$, and $0.9$, while the dotted curves give the lensing-ring radii.
     }
     \label{figdsh_expansion}
 \end{figure}

To suppress strong spin-driven asymmetries, we fix the inclination angle to $1^\circ$, so the images remain close to circular symmetry.
I measure the shadow size with the cross-cut method of \cite{Bronzwaer2020}\footnote{If the image-plane intensity is $I(X,Y)$, the corresponding one-dimensional cut is $I(X,0)$.}. The ring diameter $d_{\rm ring}$ is defined as the separation between the two lensing-ring peaks in that profile. Figure~\ref{figdsh_expansion}a shows the cuts for the $a=0.3$ sequence, where the $P=0.3$ model already stands out by producing both a wider ring and a more symmetric profile. 
Applying the same measurement to GRRT images with $a=0.3$, $0.6$, and $0.9$ yields Figure~\ref{figdsh_expansion}b, which compares the unstable-photon-orbit radius $R_{\rm photon}$ with the image-plane photon-ring radius. Because the largest admissible polymeric correction depends on spin, the horizontal axis is expressed as $P/P_{\rm max}$ so that all sequences can be shown together. As emphasized by \cite{Ozel2022, Younsi2023}, the lensing ring in a realistic image is not identical to the formal photon sphere, so I plot both quantities explicitly \cite{Akiyama2019eap, EventHorizonTelescope2022xqj, Younsi2023}. Unlike the outer horizon, both radii increase monotonically with $P$. 

 \subsection{Polarization Signatures from LQBHs}
 
\begin{figure}
    \centering
    \includegraphics[height=0.33\linewidth]{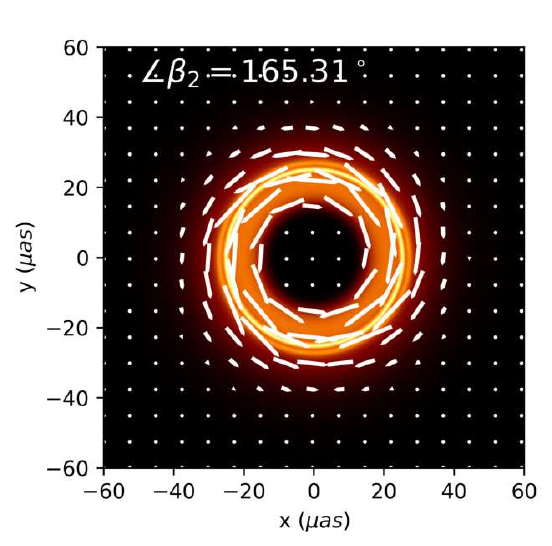}
    \includegraphics[height=0.33\linewidth]{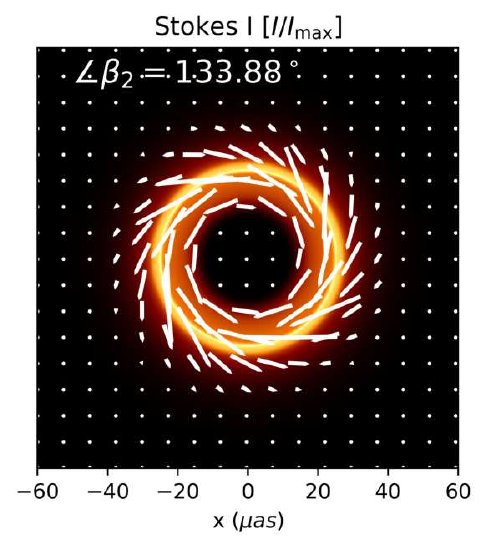}
    \includegraphics[height=0.33\linewidth]{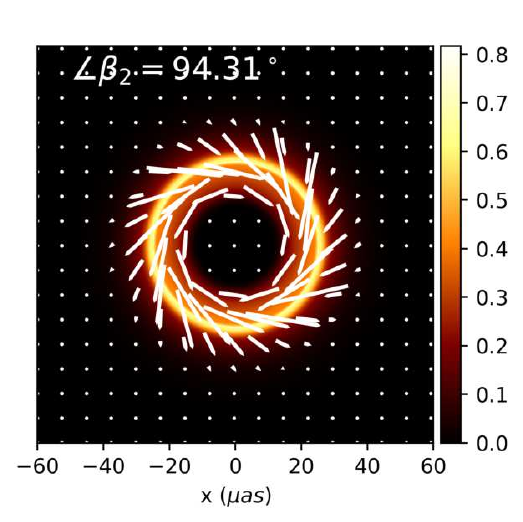}
    \caption{Normalized total-intensity images ($I/I_{\rm max}$) and polarization patterns for three plasma prescriptions around a Kerr hole with $a=0.6$: free fall (left), sub-Keplerian rotation (middle), and Keplerian rotation (right).}
    \label{figpolarization}
\end{figure}

The EHT polarimetric images of M\,87$^*$ already show that the field geometry near the horizon can be read directly from the image-plane polarization pattern \cite{Akiyama2021a, Akiyama2021b}. Comparable data are not yet available for Sgr\,A$^*$, but it is still informative to compare the polarization patterns predicted by Kerr and LQBH spacetimes using Sgr\,A$^*$ as the fiducial source. In what follows I keep the same fiducial plasma prescription and again fix the inclination to $1^\circ$ so that the loop-quantum effect can be isolated more cleanly.

Throughout this chapter, we follow the International Astronomical Union (IAU) convention for the Stokes parameters $(\mathcal{I, Q, U ,V})$ \cite{Smirnov2011vp}. To characterize the polarization pattern, we use the complex linear polarization introduced in \cite{Akiyama2021a, Akiyama2021b},
 \begin{equation}
     \mathcal{P} = \mathcal{Q} + i\mathcal{U}.
 \end{equation}
The electric-vector position angle (EVPA) is defined as 
\begin{equation}
    {\rm EVPA} \equiv {\rm arg}(\mathcal{P}).
\end{equation}
To quantify the EVPA pattern across the image, we expand the complex linear polarization $\mathcal{P}$ into azimuthal modes with complex coefficients $\beta_{\rm m}$ \cite{Akiyama2021b, Palumbo2020flt, Chael2023pwp}. The coefficient $\beta_{\rm m}$ is defined by
 \begin{equation}
     \beta_{\rm m}=\frac{1}{I_{\rm ann}}\int_{r_{\rm min}}^{r_{\rm max}}\int_0^{2\pi} 
     \mathcal{P}(r,\phi)\exp{(-im\phi)}rd\phi dr,
 \end{equation}
where $r_{\rm max}$ and $r_{\rm min}$ bracket the bright ring. Following \cite{2022ApJ...930L..15E, Palumbo2020flt}, I locate the ring center from 360 azimuthal cuts by requiring the peak emission to occur at the same radius in every direction. The ring width is then the azimuthal average of the full width measured from that center, using only pixels with $I/I_{\rm max}>0.05$. This suppresses small flow-dependent fluctuations while preserving the dominant bright ring. Because both Sgr\,A$^*$ and M\,87$^*$ are viewed at relatively low inclination, the rotationally symmetric $\beta_2$ mode dominates here, so I use its phase $\angle \beta_2$ as a compact measure of the large-scale EVPA structure.  
\begin{figure}
     \centering
     \includegraphics[width=0.47\linewidth]{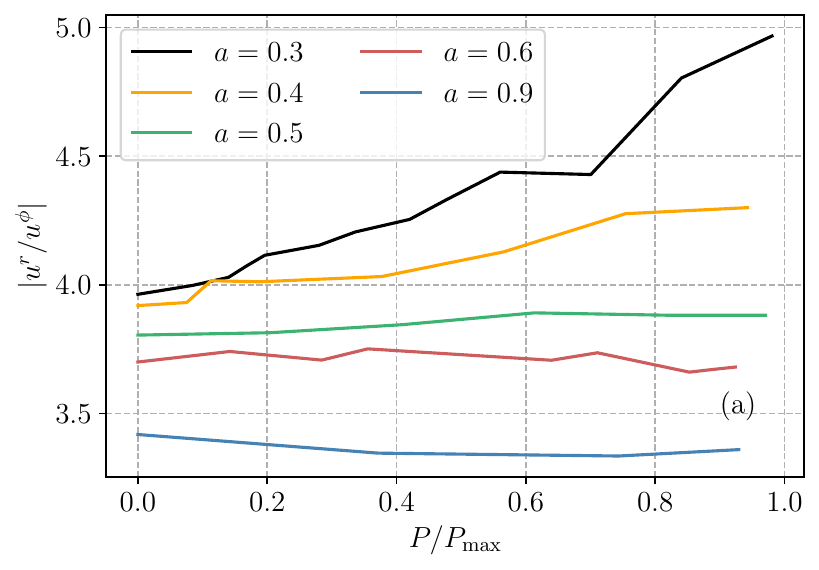}
     \includegraphics[width=0.47\linewidth]{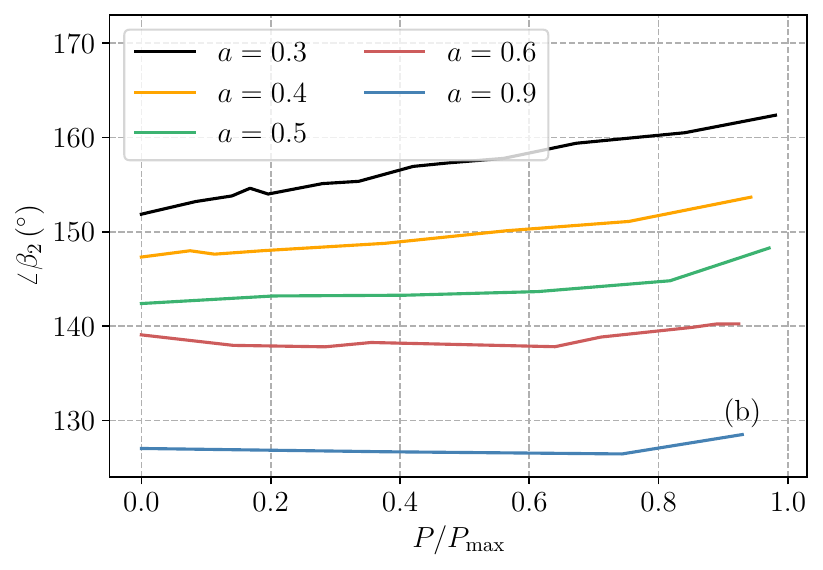}
     \caption{Panel (a) plots the ratio of radial to azimuthal motion evaluated at the lensing ring. Panel (b) shows the corresponding phase of the $\beta_2$ mode, $\angle \beta_2$, as a function of $P$. The black, red, and blue curves correspond to $a=0.3$, $0.6$, and $0.9$, respectively.}
     \label{figur_uph}
 \end{figure}
The polarization pattern depends on both the magnetic geometry and the gas kinematics. Assuming a vertical magnetic field, Figure~\ref{figpolarization} compares free-fall, sub-Keplerian, and Keplerian flows around a Kerr hole with $a=0.6$. The main difference is an EVPA rotation as the flow prescription changes, consistent with the trend emphasized by \cite{Akiyama2021b}. These three cases correspond to $K=0$, $0.5$, and $1$ in Eq.~\ref{EqsubKepler}, and thus to different values of $|u^r/u^\phi|$:
 \begin{equation}
     |u^r/u^\phi| = \frac{(1-K)u^r_{\rm FF}}{\left[\omega_{\rm K}+(1+K)(\omega_{\rm FF}-\omega_{\rm K})\right]u^t},
 \end{equation}
where $u^t=\sqrt{|-(1+ {(u^r)}^2 g^{rr})/\mathcal{C}|}$ and $\mathcal{C}$ denotes the Carter constant. 
Because most of the flux comes from near the lensing ring, the weighting in $\beta_2$ makes that region especially important. Evaluated there, the three plasma prescriptions yield $|u^r/u^\phi| = 36.63$, $3.70$, and $0$, with corresponding phases $\angle \beta_2=165.31^\circ$, $133.88^\circ$, and $94.31^\circ$. 

To isolate the metric effect, I also computed a free-fall LQBH model with the same spin, $a=0.6$, but a relatively large polymeric correction, $P=0.16$. In that case $\angle \beta_2=165.10^\circ$, essentially identical to the Kerr value. Within this setup, the plasma kinematics therefore control the polarimetric morphology more strongly than the spacetime change itself. More radial motion raises $\angle \beta_2$ and makes the pattern more circular, while stronger azimuthal motion lowers $\angle \beta_2$ and produces a more radial-looking EVPA structure.
 
Figure~\ref{figur_uph}b shows the dependence of $\angle\beta_2$ on $P$ for $a=0.3$, $0.6$, and $0.9$. For $a\lesssim 0.5$, the phase rises monotonically with the polymeric correction, indicating that the dominant emitting region becomes less nearly circular in its motion. 
For sub-Keplerian flows, $|u^r/u^\phi|$ typically grows with radius. Figure~\ref{figur_uph}a evaluates that ratio at the lensing ring. Along the $a=0.3$ sequence (black), the radial component increases steadily with $P$, naturally explaining the larger $\angle\beta_2$ because the free-fall case in Figure~\ref{figpolarization} already gives the largest phase. At $a=0.9$, the trend reverses: the motion becomes more circular as $P$ increases, and $\angle\beta_2$ correspondingly decreases slightly.

Overall, $\angle \beta_2$ remains a useful spin diagnostic, but once $a\gtrsim0.5$ its direct sensitivity to the polymeric correction becomes rather weak.

\section{Constraints from EHT Shadow Measurements}\label{secEHT_constraint}

The EHT measurements of Sgr\,A$^*$ already provide a direct handle on the allowed parameter space \cite{2022ApJ...930L..12E, 2022ApJ...930L..13E, 2022ApJ...930L..14E, 2022ApJ...930L..15E, 2022ApJ...930L..16E, EventHorizonTelescope2022xqj}. For the present purpose, the most useful observable is the angular ring diameter, $d_{\rm ring}=51.8\pm2.3\,\rm \mu as$ \cite{2022ApJ...930L..12E}. Because an LQBH with $P\ne0$ does not share the same horizon and photon-ring scales as a Kerr hole at fixed spin, the observed diameter can be turned into a bound on $P$. Earlier analytic work \cite{Afrin2022ztr} quoted $P\lesssim 0.0423$, but that estimate used a different normalization of the gravitational constant. Here I instead adopt the self-consistent normalization described above and use {\tt ipole} images built from the same MAD-like plasma prescription as \cite{Chen2021lvo}. The sampled parameter grid is listed in Appendix~\ref{Sec:survey}. 

\subsection{Survey of the Parameter Space}
\begin{figure*}
    \centering
    \includegraphics[height=0.48\linewidth]{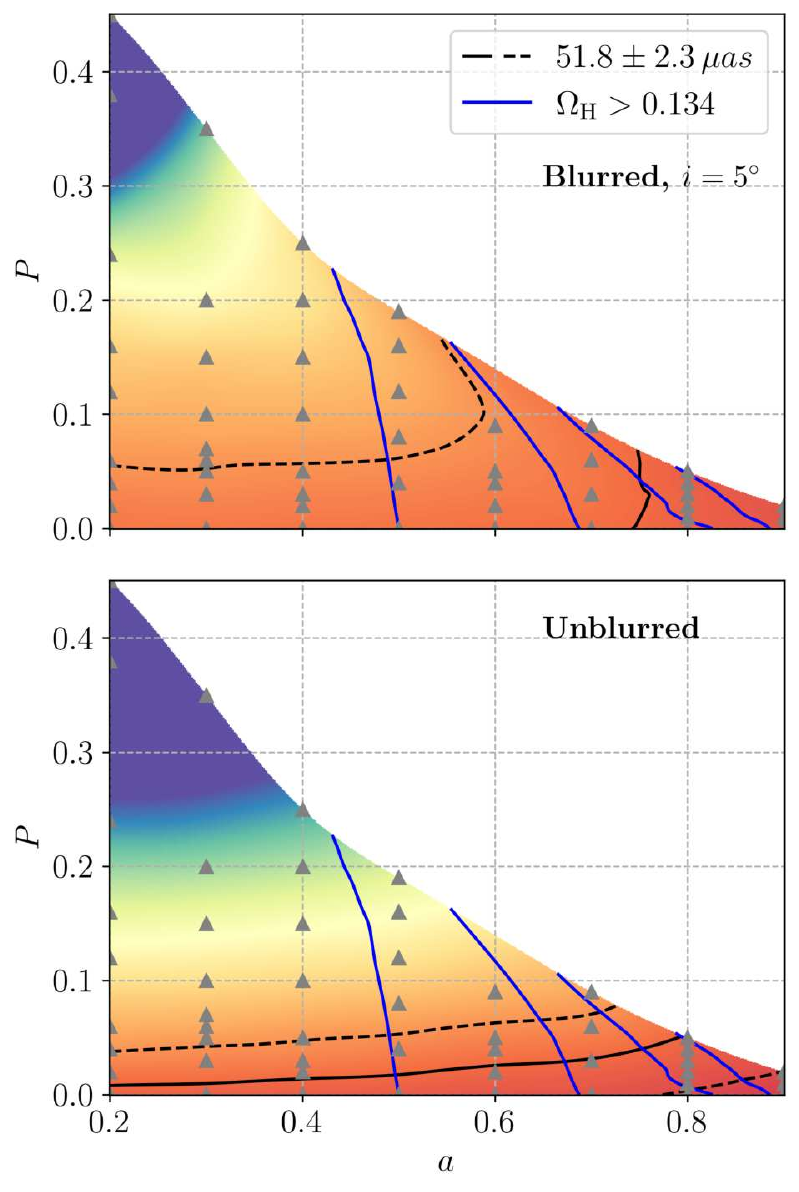}
    \includegraphics[height=0.48\linewidth]{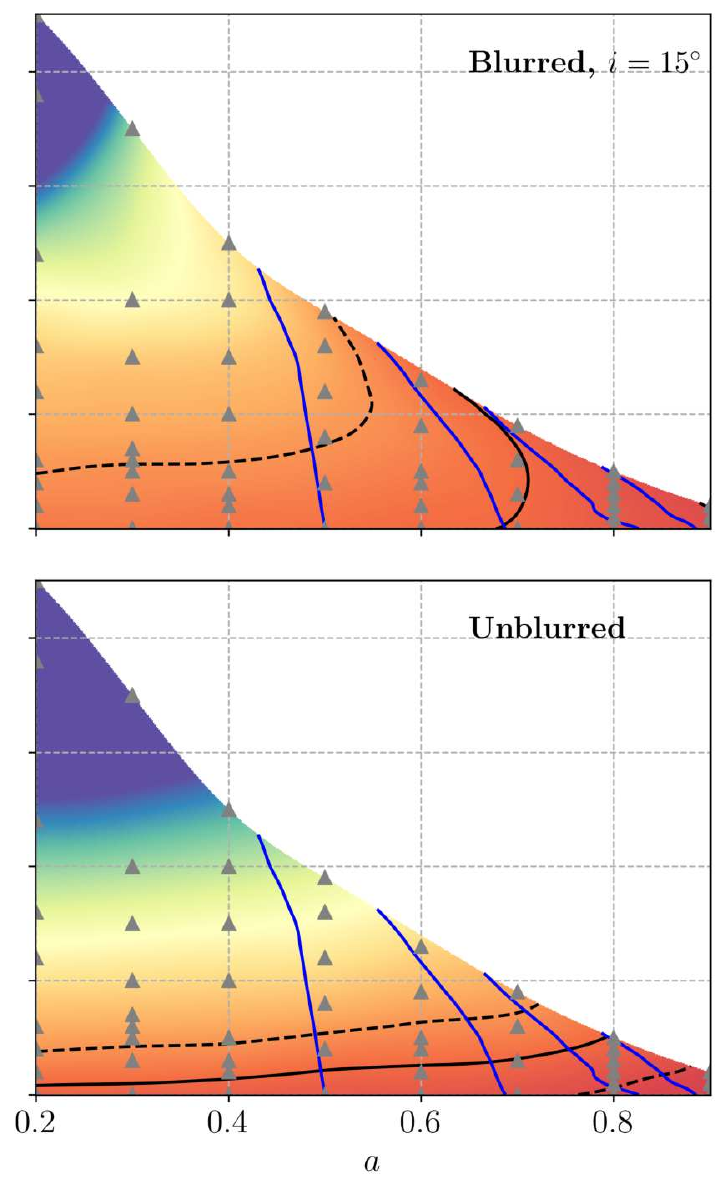}
    \includegraphics[height=0.48\linewidth]{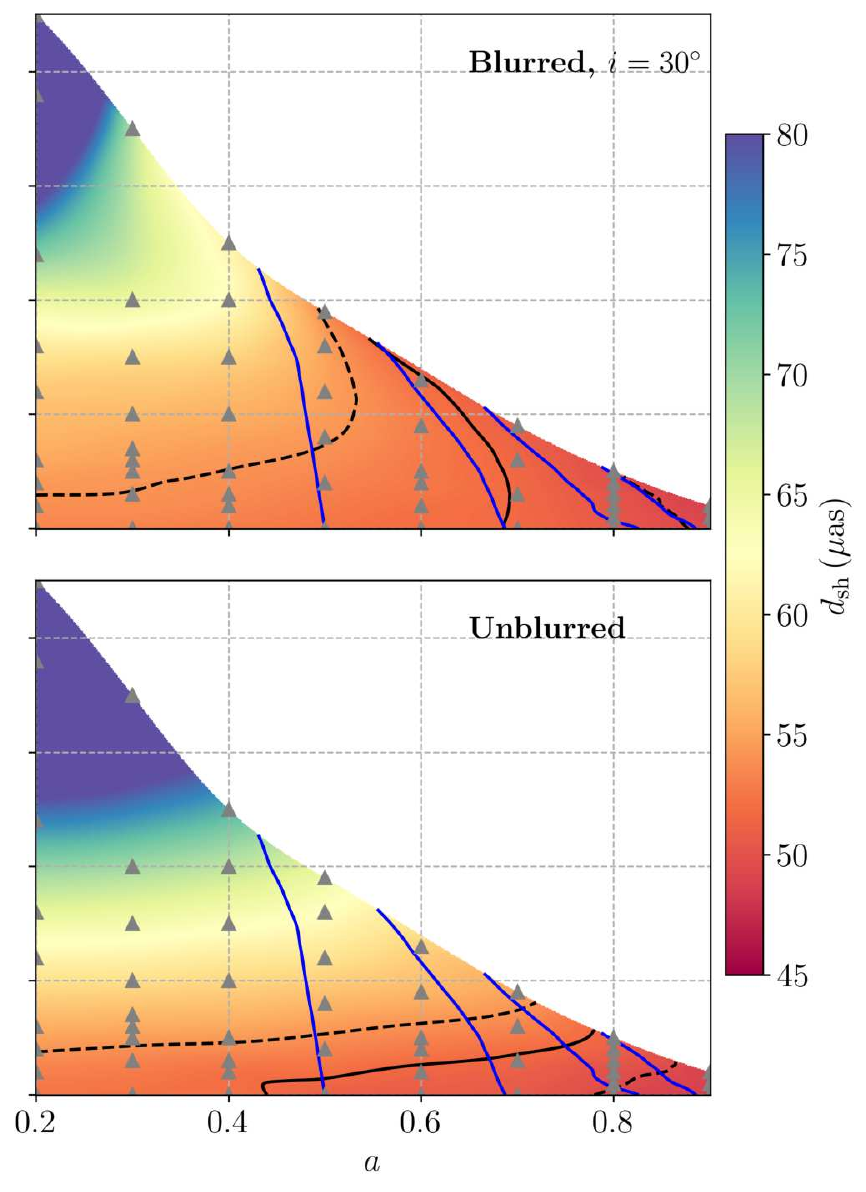}
    
    \caption{Interpolated ring diameters across the $(a,P)$ plane for the LQBH models. The upper row uses blurred images, the lower row uses the original GRRT images, and the three columns correspond to viewing inclinations of $5^\circ$, $15^\circ$, and $30^\circ$ from left to right.}     \label{figdsh}
\end{figure*}

\begin{figure*}
    \centering
    \includegraphics[width=\linewidth]{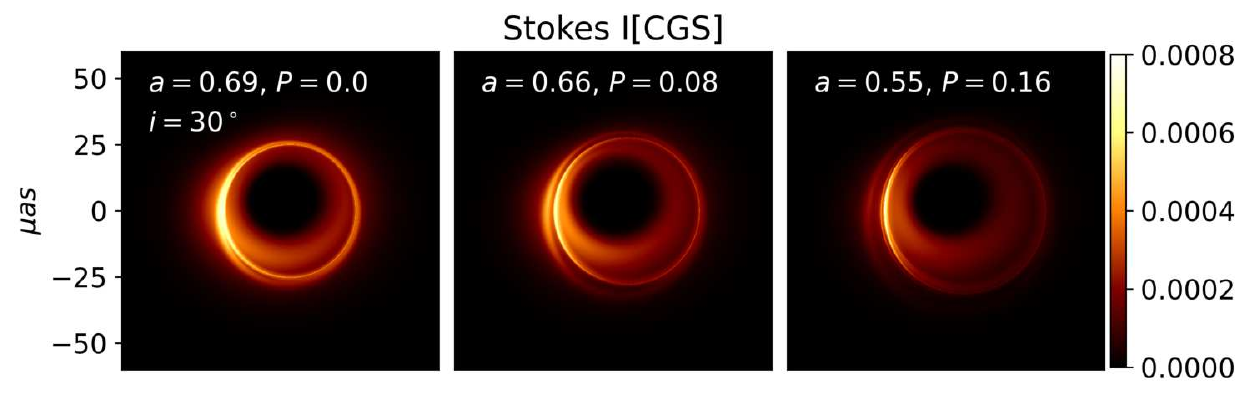}
    \includegraphics[width=\linewidth]{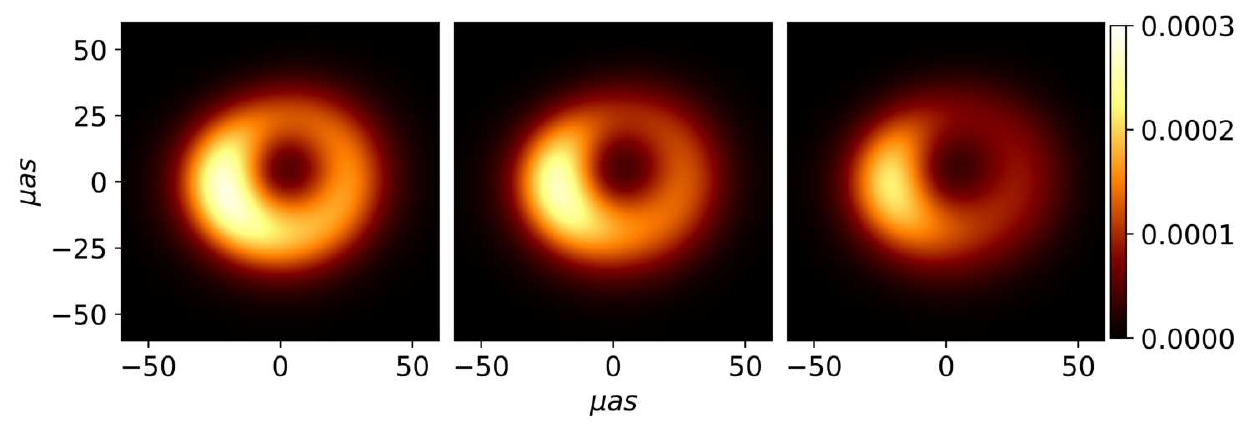}
    \caption{Example shadow images for three models evaluated at $i=30^\circ$: left, $(a,P)=(0.69,0.0)$; middle, $(0.66,0.08)$; right, $(0.55,0.16)$. The upper row shows the original GRRT images, and the lower row shows the same models after EHT-like blurring.}     \label{figobs_par}
\end{figure*}

The polymeric correction changes the image scale in a straightforward way: increasing $P$ generally pushes the photon ring outward and enlarges the apparent shadow. I extract $d_{\rm ring}$ from the GRRT images using the same procedure described above. However, as emphasized by \cite{EventHorizonTelescope2022xqj}, the measured shadow diameter $\hat{d}_{\rm sh}$ is related to the intrinsic one by a calibration factor, $\hat{d}_{\rm sh}=\alpha_{\rm c} d_{\rm sh}$, and the same type of observational bias also affects the measured ring diameter. Figure~\ref{figdsh} shows that finite angular resolution changes the inferred size substantially. For that reason, the most self-consistent comparison is obtained only after the theoretical images are blurred to EHT resolution. I therefore analyze both the raw GRRT outputs and versions convolved with a Gaussian beam of FWHM $20\,\rm \mu as$ \cite{EventHorizonTelescope2022xqj, Chael2018oym}.

Figure~\ref{figdsh} shows the interpolated ring diameter across parameter space. The gray triangles mark the models that were explicitly simulated, the red solid contour traces the measured EHT diameter, and the dashed red contours indicate the corresponding $1\sigma$ interval. The blue contours identify models whose angular momentum exceeds that of a Kerr black hole with $a=0.5$. Moving from left to right increases the inclination from $5^\circ$ to $30^\circ$, while the upper and lower rows compare blurred and unblurred images. Because the current EHT uncertainty remains fairly large, a wide region of parameter space is still allowed. The preferred spin changes only modestly as the inclination rises from $5^\circ$ to $30^\circ$, by roughly $\sim0.1$, whereas the difference between the blurred and unblurred panels is much larger. The inference is therefore limited more strongly by finite image resolution than by moderate changes in inclination.

The parameter regions inferred from the unblurred images are noticeably tighter than those obtained from the blurred ones. That difference is not surprising, because the unblurred comparison matches idealized theoretical images directly to finite-resolution data and is therefore not fully self-consistent. If one nevertheless uses the original images, the allowed range is roughly limited to $P\sim0.1$ at the $1\sigma$ level. Once EHT-like blurring is included, the upper bound becomes much weaker. Shadow size alone therefore does not yet provide a stringent limit, although the blurred comparison still suggests that values near $P\lesssim0.2$ remain plausible for Sgr\,A$^*$.

The bound on $P$ for Sgr\,A$^*$ can be sharpened by incorporating information about the black-hole angular momentum. The angular frequency at the event horizon is
\begin{equation}
    \Omega_{\rm H} = \left(-\frac{g_{t\phi}^{\rm BL}}{g_{\phi\phi}^{\rm BL}}\right)_{r_{\rm H}},
\end{equation}
where $r_{\rm H}=r_{h+}$ is the outer-horizon radius. 
Previous EHT modeling favors a relatively large spin for Sgr\,A$^*$ \cite{2022ApJ...930L..16E}. If I impose the conservative lower bound $a\geq0.5$, then the corresponding Kerr horizon frequency cannot drop below $\Omega_{\rm H}=0.134$. I therefore require $\Omega_{\rm H}\geq0.134$ when interpreting Figure~\ref{figdsh}; the allowed values of $P$ are then determined by the overlap between the blue and red contours. In the blurred panels, the measured ring diameter is controlled mainly by the size of the outer horizon. For $a>0.5$, that horizon changes only weakly with $P$ (Figure~\ref{horizon_contour}), so the relevant contour is nearly vertical. Although the formal photon-ring radius continues to increase with $P$, current EHT resolution is not sufficient to distinguish those differences cleanly. Under present observational conditions, Sgr\,A$^*$ is therefore consistent with $P\lesssim 0.2$ provided that $a\gtrsim0.5$.

I also repeated the survey for thinner flows with $H=0.1$ and $0.05$ in addition to the fiducial $H=0.3$. In both the blurred and unblurred analyses, the inferred limits on $a$ and $P$ shift by only about $\sim0.01$. The qualitative conclusion for Sgr\,A$^*$ is therefore unchanged: the present data remain consistent with $P\lesssim 0.2$.

\subsection{Degeneracies After Image Blurring}

To illustrate the degeneracy introduced by image blurring, I select three points from the $d_{\rm ring}=51.8\,\mu{\rm as}$ contour in the upper-right panel of Figure~\ref{figdsh}, corresponding to $i=30^\circ$. These points have nearly identical diameters after blurring even though their parameters differ: $(a,P)=(0.69,0.00)$, $(0.66,0.08)$, and $(0.55,0.16)$. Figure~\ref{figobs_par} shows the resulting images. Before blurring, the photon ring expands noticeably as $P$ increases, with lensing-ring diameters of $51.45$, $54.25$, and $60.66\,\rm \mu as$. After blurring, however, the models become visually similar. Their outer-horizon radii are $1.73\,r_{\rm g}$, $1.79\,r_{\rm g}$, and $1.87\,r_{\rm g}$, corresponding to blurred ring diameters of $51.45$, $51.05$, and $51.05\,\rm \mu as$. Breaking this spin-polymeric-function degeneracy will therefore require better angular resolution, most likely from future EHT or ngEHT observations \cite{Tiede2022grp, Johnson2023ynn}. Figure~\ref{figobs_par} also shows that the largest-$P$ model is slightly dimmer than the other two. 
Comparing the blurred images directly, the brightest pixel in the right panel, which corresponds to the LQBH model with $a=0.55$ and $P=0.16$, reaches only about $77\%$ of the brightest pixel in the left panel, namely the Kerr case with $a=0.69$ and $P=0$. 

The phase $\angle\beta_2$ provides a second possible discriminator. For the three original images it takes the values $125^\circ$, $123^\circ$, and $124^\circ$, while after blurring the corresponding phases are $129^\circ$, $126^\circ$, and $123^\circ$. These differences are too small to distinguish with current polarimetric data \cite{Akiyama2021b}. Polarization of Sgr\,A$^*$ alone is therefore unlikely to determine $a$ and $P$ uniquely at present, and tighter bounds on $P$ will probably require higher-resolution observations such as those anticipated from ngEHT.

\subsection{EHT Limits on the Polymeric Function for M\,87$^*$}
\begin{figure}
    \centering
    \includegraphics[width=\linewidth]{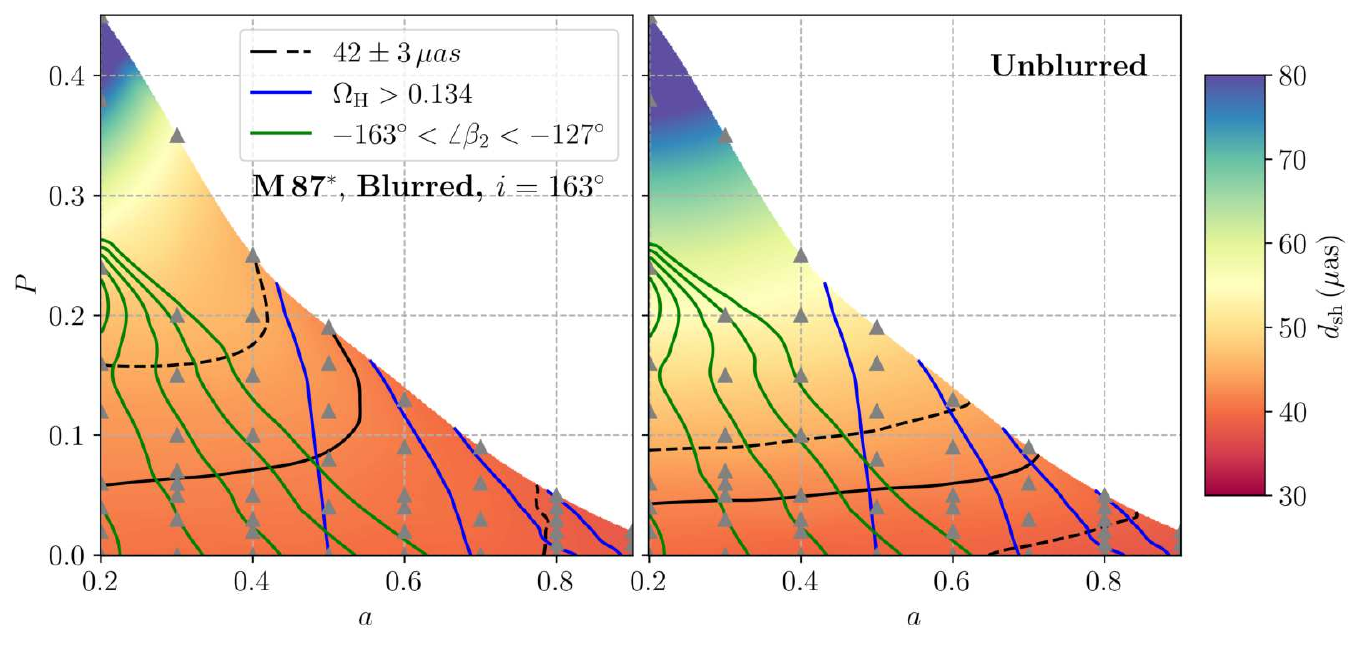}
    \caption{Same presentation as Figure~\ref{figdsh}, now applied to M\,87$^*$. The green contours indicate the region satisfying $-163^\circ < \angle\beta_2<-127^\circ$.}
    \label{fig:M87}
\end{figure}

The other EHT-imaged source relevant here is M\,87$^*$ \cite{EventHorizonTelescope2019dse, Akiyama2019brx, Akiyama2019eap, Akiyama2019bqs, 2019ApJ...875L...5E, Akiyama2019sww}. The observed ring diameter is $42\pm3\,\mu{\rm as}$, and the source properties are approximately $D=16.8\pm0.8\,\rm Mpc$ and $M_{\bullet}=(6.5\pm0.7)\times10^9\,M_{\odot}$. I use those values in the GRRT calculations together with the same plasma setup as \cite{Chen2021lvo}, namely $n_{\rm e} = 1.25\times 10^5\,\rm cm^{-3}$. Because M\,87 hosts a bright, highly collimated jet, the viewing angle is inferred to be $i \sim 163\pm 2^\circ$ \cite{Walker2018}, and I adopt that inclination throughout the analysis. I also use $M_{\bullet}=6.5\times10^9\,M_{\odot}$ and $16.7\,\rm Mpc$ as the fiducial black-hole mass and source distance \cite{2019ApJ...875L...5E}. The spin of M\,87$^*$ remains uncertain: some analyses prefer values near $a=0.9\pm0.05$ \cite{Tamburini2019vrf}, whereas others allow a much broader range extending from about $0.5$ to nearly $1$ \cite{Fromm2022, Cruz-Osorio2021cob, Dokuchaev2019bbf}. I therefore impose only the conservative prior $a\gtrsim0.5$. Polarization adds an additional constraint, because the EHT favors a vertical-field EVPA structure with $-163^\circ < \angle\beta_2<-127^\circ$ \cite{Akiyama2021b}. Figure~\ref{fig:M87} therefore combines the EHT limits on $\Omega_{\rm H}$, $\angle\beta_2$, and $d_{\rm ring}$. The green contours show the allowed phase range, the blue contours identify models with larger horizon frequency than a Kerr black hole at $a=0.5$, and the black contours show the measured ring diameter with its $1\sigma$ interval. Panels A and B correspond to blurred and unblurred GRRT images, respectively.

Figure~\ref{fig:M87} shows that image blurring has only a minor effect on $\angle\beta_2$, whereas it changes the inferred ring diameter much more strongly. As discussed above, the blurred image is dominated mainly by the inner shadow and therefore tracks the event-horizon scale more closely \cite{Chael2021rjo}. The unblurred analysis would suggest a tighter limit, roughly $P\sim0.12$, while the blurred images allow values up to about $P\sim0.25$. As in the Sgr\,A$^*$ analysis, I regard the blurred comparison as the more realistic one. When these results are combined with the $\angle\beta_2$ and $\Omega_{\rm H}$ constraints, the allowed parameter space in Figure~\ref{fig:M87}a is reduced to the overlap of the green and blue contours, implying $P\lesssim0.07$ with $0.5\lesssim a\lesssim0.7$.

This comparison makes clear that $\beta_2$ carries essential information for both M\,87$^*$ and Sgr\,A$^*$. Once comparable polarimetric constraints become available for Sgr\,A$^*$, the allowed loop-quantum parameter range should tighten further. Beyond size and polarization, jet-power measurements offer another promising diagnostic \cite{2019ApJ...875L...5E, 2022ApJ...930L..16E}, but using that observable requires explicit GRMHD calculations in different black-hole spacetimes. I leave that extension to future work.

Because the mass of M\,87$^*$ is still uncertain, Appendix~\ref{diff_mass} repeats the analysis for different assumed masses. In practice, the upper bound on $P$ changes only modestly, although smaller black-hole masses shift the preferred region toward somewhat larger polymeric corrections. In the extreme case $M_\bullet=5.8\times 10^9\,M_\odot$, no Kerr model satisfies the adopted ring-diameter constraint.

\section{Conclusions and Discussion}

Using semi-analytical GRRT calculations spanning several spins and polymeric corrections, I investigated how the parameter $P$ changes black-hole shadow images in the rotating LQBH spacetime. The main results can be summarized as follows.

\begin{enumerate}
    \item Within the semi-analytical RIAF model implemented in {\tt ipole}, increasing $P$ makes the image ring both brighter and larger than in the Kerr case at fixed spin. Both the image-plane lensing ring and the photon-sphere scale increase monotonically with the polymeric correction, whereas the event horizon responds non-monotonically by first growing and then slightly shrinking as $P$ approaches its maximum allowed value.
    \item The polarimetric behavior can be summarized by the phase of the $\beta_2$ mode. For $a\lesssim0.5$, that phase rises as $P$ increases, while for $a\gtrsim0.6$ it decreases slightly. This behavior follows the corresponding trend in $|u^r/u^\phi|$ at the lensing ring, showing that the image polarization is controlled primarily by the local flow kinematics there.
    \item For Sgr\,A$^*$, comparison with the observed ring diameter implies $P\lesssim0.2$ if the source spin satisfies $a\geq0.5$, consistent with \cite{2022ApJ...930L..16E}. Ring diameter alone does not determine the model uniquely, so additional observables such as polarization and jet power will be needed. A fully self-consistent treatment of those effects will require 3D GRMHD simulations of MAD flows.
    \item At current EHT resolution, the image size remains degenerate between spin and polymeric correction. The ring diameter by itself therefore cannot distinguish Kerr and LQBH models. This degeneracy is likely to be reduced only when substantially sharper horizon-scale imaging becomes available, for example from ngEHT. 
    \item For M\,87$^*$, the additional information carried by $\angle\beta_2$ yields a stronger bound, namely $P\lesssim0.07$ in the range $0.5\lesssim a\lesssim0.7$. That result also suggests that future polarimetric measurements of Sgr\,A$^*$ should tighten the allowed LQG parameter space further.

\end{enumerate}

Several caveats remain. First, the metric studied here is only one particular realization of loop-quantum gravity. Different LQG quantization prescriptions can produce black holes with different spacetime properties \cite{Modesto2008im, Sahu2015dea, Ashtekar2018cay, Ashtekar2018lag, Bojowald2018xxu, Gan2020dkb}. The present analysis applies specifically to the rotating geometry of \cite{Liu2020ola}, obtained from the self-dual spherical solution of \cite{Modesto2008im} through the Newman-Janis construction. In this model, $P$ measures the amplitude of the loop-quantum correction, and the spacetime preserves the self-dual structure associated with the transformation $r\rightarrow a_0/r$ \cite{Modesto2008im, Sahu2015dea, Ashtekar2018cay}. Within that framework, comparison with Sgr\,A$^*$ and M\,87$^*$ provides no evidence for a departure from Kerr, so the present observational result should be interpreted as an upper bound of about $P\lesssim0.2$.

Second, the calculations in this chapter are semi-analytical GRRT models rather than self-consistent GRMHD simulations. They therefore omit higher-order plasma variability such as hot spots and intrinsic light-curve fluctuations. Addressing those effects will require 2D and 3D GRMHD simulations coupled directly to GRRT. More realistic source modeling will also need two-temperature accretion prescriptions and a broader set of magnetic-field configurations \cite[e.g.,][]{2023MNRAS.522.2307J, Dihingia2023, Mizuno2021}. I leave that broader program to future work. 


{
\section*{Supplementary Appendices}  
\setcounter{section}{0}
\renewcommand{\thesection}{\thechapter.\Alph{section}}
\renewcommand{\theHsection}{appendix.\thechapter.\Alph{section}}
\renewcommand{\theHsubsection}{appendix.\thechapter.\Alph{section}.\arabic{subsection}}

\section{Geodesic Structure of the LQBH Spacetime}\label{sec:geodesic}
	
This appendix summarizes the geodesic relations needed for the rotating LQBH spacetime. Because both massive-particle trajectories and photon paths follow from the same underlying geodesic structure, I begin with the Hamilton-Jacobi equation,
	\begin{eqnarray}
	\frac{\partial S}{\partial \lambda}=-\frac{1}{2}g^{\mu\nu}\frac{\partial S}{\partial x^\mu}\frac{\partial S}{\partial x^\nu},
	\end{eqnarray}
where $\lambda$ is the affine parameter and $S$ is the Jacobi action. A separable ansatz may be written as
	\begin{equation}
	S=\frac{1}{2}m^2\lambda-Et+L\phi+S_r(r)+S_\theta(\theta),
	\end{equation}
	where $m$ is the particle mass, so $m^2= 1$ for timelike geodesics and $m^2=0$ for null ones. The conserved quantities $E$ and $L$ denote the energy and the component of angular momentum parallel to the rotation axis, while $S_r$ and $S_\theta$ depend only on $r$ and $\theta$.
	
	Substituting the ansatz into the Hamilton-Jacobi equation gives
	\begin{eqnarray}
	S_r(r)&=&\int^r\frac{\sqrt{R(r)}}{\Delta}dr,\\
	S_\theta(\theta)&=&\int^\theta\sqrt{\Theta(\theta)}d\theta,
	\end{eqnarray}
	where
	\begin{eqnarray}
    R(r) &=&[X(r)E-aL]^2-\Delta(r)[\mathcal{C}+ m^2 r^2+(L-aE)^2],\\
	\Theta(\theta)&=& \mathcal{C}+(a^2E^2-m^2a^2-L^2\csc^2\theta)\cos^2\theta,
	\end{eqnarray}
	where $\mathcal{C}$ is the Carter constant. In these expressions, $\Delta(r)$ is the same function introduced in Eq.~(\ref{Delta}), and $X(r)\equiv k+a^2$.
	
Because the shadow is determined by null geodesics, we next vary the Jacobi action to obtain the explicit photon equations of motion,
	\begin{eqnarray}
	\Sigma\frac{dt}{d\lambda} &=&a(L-aE\sin^2\theta) 
	+\frac{r^2+a^2}{\Delta}[(r^2+a^2)E -aL], \label{YYY} \\
	\Sigma\frac{d\phi}{d\lambda} &=&\frac{L}{\sin^2\theta}-aE+\frac{a}{\Delta}[(R^2+a^2)E-aL],\\
	\Sigma\frac{dr}{d\lambda} &=&\pm\sqrt{R(r)},\\
	\Sigma\frac{d\theta}{d\lambda}&=&\pm\sqrt{\Theta(\theta)}.\label{XXX}
	\end{eqnarray}
where the signs $\pm$ are chosen according to the directions of the radial and angular motion. The photon trajectories can then be characterized by the two usual impact parameters,
	\begin{eqnarray}
	\xi=\frac{L}{E},\qquad \eta=\frac{\mathcal{C}}{E^2}.
	\end{eqnarray}
To determine the geometric boundary of the shadow, one solves for the unstable circular photon orbit, which satisfies
	\begin{eqnarray}
	R(r)=0,\qquad \frac{dR(r)}{dr}=0,\qquad
\frac{d^2R(r)}{dr^2}>0\label{ZZZ}
	\end{eqnarray}
The corresponding shadow shape is therefore controlled jointly by the spin $a$ and the polymeric function $P$ through the allowed impact parameters $\xi$ and $\eta$.
		
Solving Eq.~(\ref{ZZZ}) for spherical photon motion yields the corresponding impact parameters,
\begin{align}  \label{AAA}
&\xi=\frac{X_\text{ps}\Delta'_\text{ps}-2\Delta_\text{ps}X'_\text{ps}}{a\Delta'_\text{ps}},\nonumber\\
&\eta=\frac{4a^2X'^2_\text{ps}\Delta_\text{ps}-\left[\left(X_\text{ps}-a^2\right)   \Delta'_\text{ps}-2X'_\text{ps}\Delta_\text{ps} \right]^2}{a^2\Delta'^2_\text{ps}},
	\end{align}
	where
	
\begin{eqnarray} \label{BBB}
	X(r,M,a,P)&=&a^2+\frac{r^4+a_0}{(2 M P+r)^2},\nonumber\\
\Delta(r,M,a,P)&=&a^2+\frac{r^2 (2 M-r) \left(2 M P^2-r\right)}{(2 M P+r)^2}.
	\end{eqnarray}

The expressions in Eqs.~(\ref{AAA}) and~(\ref{BBB}) reduce to the standard Kerr result when $P=0$ and $a_0=0$. Although $\xi$ and $\eta$ separately diverge as $a\to 0$, the combination $\xi^2+\eta$ remains finite and reproduces the spherical static LQBH limit \cite{Liu:2020ola}.

\section{Parameter Survey Used for the GRRT Models} \label{Sec:survey}
The semi-analytical GRRT survey covers the spin--polymeric-function combinations listed in Table~\ref{list}. For every pair in that grid, I also examined viewing inclinations from $5^\circ$ to $30^\circ$.
\begin{table}[]
\centering
\begin{tabular}{@{}ll@{}}
\toprule
a   & P    \\ \midrule
0.2 & 0.00 \\
0.2 & 0.02 \\
0.2 & 0.04 \\
0.2 & 0.06 \\
0.2 & 0.12 \\
0.2 & 0.16 \\
0.2 & 0.24 \\
0.2 & 0.38 \\
0.2 & 0.45 \\
0.3 & 0.00 \\
0.3 & 0.03 \\
0.3 & 0.05 \\
0.3 & 0.06 \\
0.3 & 0.07 \\
0.3 & 0.10 \\
0.3 & 0.15 \\
0.3 & 0.20 \\
0.3 & 0.35 \\
0.4 & 0.00 \\
0.4 & 0.02 \\
0.4 & 0.03 \\
0.4 & 0.05 \\
0.4 & 0.10 \\
0.4 & 0.20 \\
0.4 & 0.25 \\ \bottomrule
\end{tabular}
\begin{tabular}{@{}ll@{}}
\toprule
a   & P     \\ \midrule
0.5 & 0.00  \\
0.5 & 0.04  \\
0.5 & 0.08  \\
0.5 & 0.12  \\
0.5 & 0.16  \\
0.5 & 0.19  \\
0.6 & 0.00  \\
0.6 & 0.02  \\
0.6 & 0.04  \\
0.6 & 0.05  \\
0.6 & 0.09  \\
0.6 & 0.14  \\
0.7 & 0.00  \\
0.7 & 0.03  \\
0.7 & 0.06  \\
0.7 & 0.09  \\
0.8 & 0.00  \\
0.8 & 0.01  \\
0.8 & 0.02  \\
0.8 & 0.03  \\
0.8 & 0.04  \\
0.8 & 0.055 \\
0.9 & 0.00  \\
0.9 & 0.01  \\
0.9 & 0.02  \\ \bottomrule
\end{tabular}
\caption{Black-hole parameter combinations used in the semi-analytical GRRT calculations.}
\label{list}
\end{table}

\section{Impact of the Mass Uncertainty of M\,87$^*$}
\label{diff_mass}
The mass of M\,87$^*$ remains uncertain because stellar-dynamical constraints are difficult to obtain at that distance. The EHT analysis inferred $(6.5\pm0.7) \times 10^9\,M_{\odot}$ \cite{EventHorizonTelescope2019dse}, and changing that mass shifts the predicted ring diameter even when $a$ and $P$ are kept fixed. I therefore repeated the parameter survey for two alternative masses and summarize the outcome in Figure~\ref{fig:M87pm}. The phase $\angle \beta_2$ depends only weakly on the assumed mass, and $\Omega_{\rm H}$ is unchanged because it is mass independent. The primary change appears in the ring-diameter contour. For larger black-hole masses, the upper bound remains near $P\sim0.07$. When the mass drops below $6.5\times10^9\,M_\odot$, the allowed region shifts toward somewhat larger $P$. In the extreme case $M_\bullet=5.8\times10^9\,M_{\odot}$, the Kerr limit $P=0$ no longer satisfies the adopted $1\sigma$ ring-diameter constraint. Accurate mass measurements of M\,87$^*$ are therefore important for constraining the LQG parameter.
\begin{figure}
    \centering
    \includegraphics[height=0.7\linewidth]{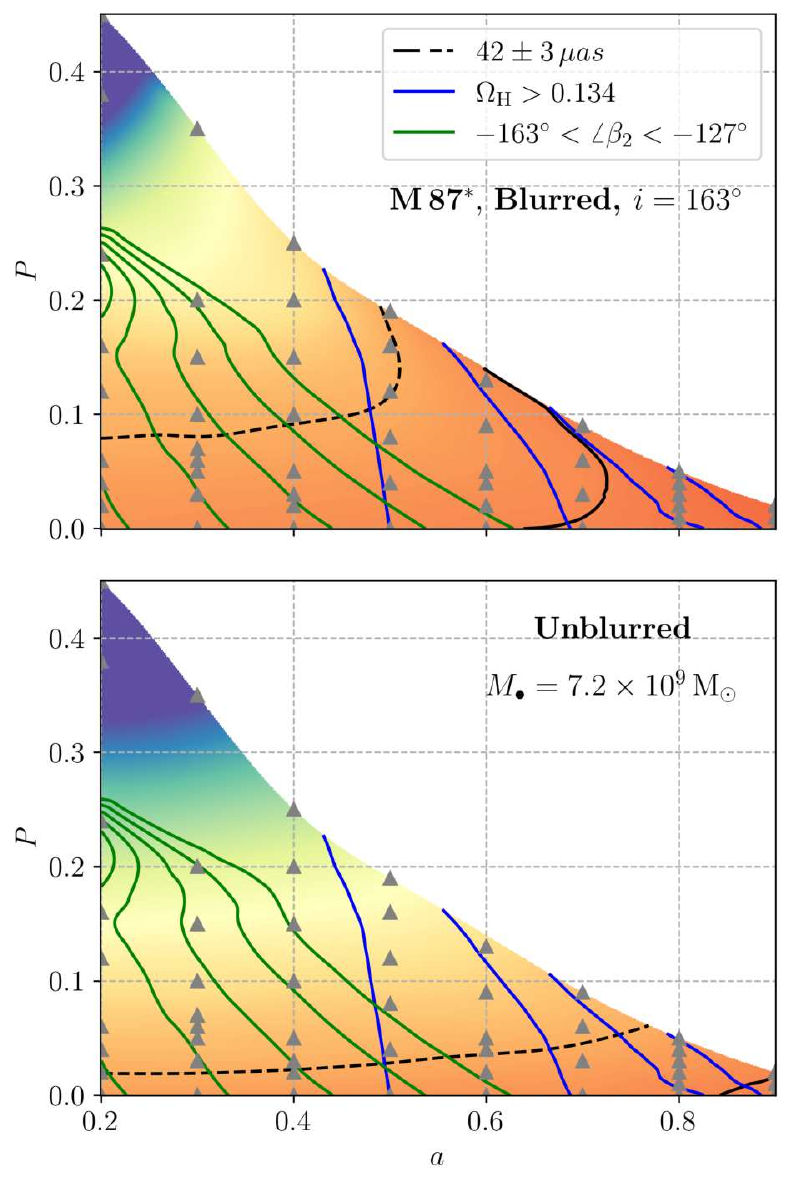}
    \includegraphics[height=0.7\linewidth]{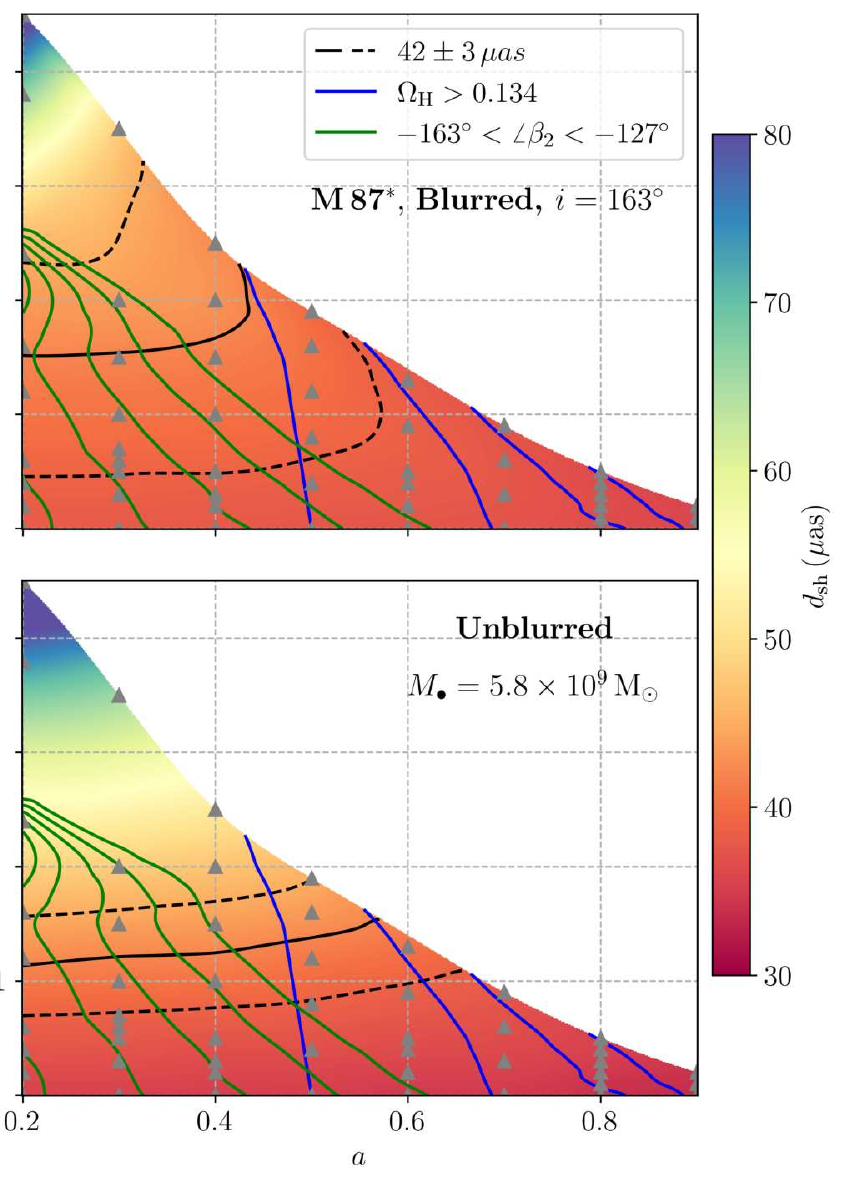}
    \caption{Same layout as Figure~\ref{fig:M87}, but evaluated for two alternative black-hole masses: $M_\bullet=7.2\times10^9\,M_{\odot}$ (left) and $M_\bullet=5.8\times10^9\,M_{\odot}$ (right).
    }
    \label{fig:M87pm}
\end{figure}
}

\chapter{Blandford-Znajek Jet Launching in Loop-Quantum Black Holes}
\markboth{第8章 BZ Jet Launching in LQBHs}{第8章 BZ Jet Launching in LQBHs}

This chapter is based on an earlier article by Jiang H.-X., Dihingia I. K., Liu C., Mizuno Y., and Zhu T., published in JCAP in 2024 as article 101 \cite{Jiang_2024}.


\section{Physical Introduction} \label{sec:intro-c7}
Horizon-scale observations of supermassive black holes now provide one of the clearest probes of strong-field gravity. Although general relativity continues to pass all established experimental tests, the EHT images of M\,87$^*$ and Sgr\,A$^*$ directly probe both the spacetime and the emitting plasma on event-horizon scales \citep{EventHorizonTelescope2019dse, 2022ApJ...930L..12E}. Those images do not yet rule out every alternative to Kerr \citep[e.g.,][]{2023arXiv231204288J, Psaltis2020, Ozel2022, Younsi2023, EHT2022f, Afrin2022ztr, 2023CQGra..40p5007V}, but they already place meaningful bounds on possible deviations \citep{EventHorizonTelescope2021dqv, EHT2022f}. That makes it natural to ask not only how non-Kerr spacetimes affect shadow size, but also how they modify jet launching.

Relativistic jets are observed from black holes across the full mass spectrum, from X-ray binaries and gamma-ray bursts to radio-loud AGN \citep[e.g.,][]{1992ApJ...397L...5M, 1999ApJ...519L..17S, 2006csxs.book..381F, 2019ARA&A..57..467B}. M87 is especially useful because its nearby jet is resolved well enough to show limb brightening, large-scale collimation, and possible precession signatures \citep{2018A&A...616A.188K, Lu2023, Cui2023}. These observables encode how the outflow is launched and how spacetime rotation couples to the magnetic field. The conventional interpretation is the Blandford-Znajek mechanism \citep{Blandford1977}, in which magnetic flux threading the horizon extracts black-hole rotational energy. Since that framework is usually discussed in Kerr spacetime, an obvious question is whether the same channel remains comparably efficient in alternative metrics. Analytical studies outside GR already exist \cite[e.g.,][]{Pei2016, Banerjee2021, 2023arXiv230706878C}.

Numerical GRMHD calculations have started to answer that question directly in non-Kerr backgrounds, including string-inspired black-hole metrics \citep{Mizuno2018lxz, Roder2023oqa, 2023arXiv231020040C}. For example, \cite{2023arXiv231020040C} found that rotating stringy black holes can launch jets more powerful than Kerr holes with matched spin. That result suggests that the BZ mechanism may remain robust even when the geometry departs substantially from Kerr.

Loop quantum gravity provides another well-motivated alternative setting, with singularity resolution arising naturally for both black holes and the early universe \citep{2023arXiv231204288J, Liu:2020ola}. In our earlier parameter survey \citep{2023arXiv231204288J}, we identified the region of LQBH parameter space still allowed by present EHT constraints. That region favored moderately large to high spin ($a\gtrsim0.5$) together with polymeric corrections roughly limited to $P\lesssim0.15$ for both Sgr\,A$^*$ and M\,87$^*$. Starting from that range, I now ask how the BZ mechanism behaves in the same LQBH metric. The purpose is to determine how loop-quantum corrections reshape the flow, alter the near-horizon magnetic field, and modify the resulting jet properties in GRMHD and GRRT calculations.

Section 8.2 summarizes the numerical setup, Section 8.3 presents the main results, and Section 8.4 concludes.

\section{Numerical Method}

I use {\tt BHAC}\footnote{\url{https://bhac.science/}} \citep{Porth2017, Olivares2019} to evolve a suite of two-temperature GRMHD models of magnetized tori in the LQBH spacetime; details of the metric are collected in Appendix~\ref{sec:LQBH}. In the adopted units, $G_{\rm N}M=c=1$, and the factor $1/\sqrt{4\pi}$ is absorbed into the magnetic field\footnote{Because the effective gravitational coupling in the LQBH metric depends on the polymeric parameter $P$, I distinguish the loop-quantum and Newtonian constants as $G_{\rm LQG}$ and $G_{\rm N}$; see Appendix~\ref{sec:LQBH}.}. The system solved by {\tt BHAC} is
\begin{equation}
	\nabla_{\mu}\left(\rho u^{\mu}\right)=0,\,\,\,
	\nabla_{\mu}T^{\mu\nu}=0,\,\,\,
	\mathrm{and}\,\,\,
	\nabla_{\mu}^*F^{\mu\nu}=0,
\end{equation} 
Here $\rho$ is the rest-mass density, $u^{\rm \mu}$ the fluid four-velocity, $T^{\rm \mu\nu}$ the stress-energy tensor, and $^*F^{\rm \mu\nu}$ the dual Faraday tensor. Full details of the numerical scheme are given by \citet{Porth2017, Olivares2019}. 

Each model begins from a hydrostatic torus constructed following \cite{Font2002}. As in \cite{Mizuno2018lxz}, I keep the inner radius fixed at $r_{\rm{in}} = 10.3 \,r_{\rm g}$ in every run, where $r_{\rm g}\equiv G_{\rm N}M/c^2$. The specific angular momentum is then adjusted so that the total torus mass is held fixed while $a$ and $P$ are varied. This gives a controlled comparison across spacetimes with different spin and polymeric correction. The radial domain extends to $r=2\,500\,r_{\rm g}$. The 2D calculations use Kerr-Schild-like coordinates (Appendix~\ref{sec:LQBH}) on a $1\,024\times 512$ mesh, while the 3D runs employ static mesh refinement with an effective resolution of $384\times192\times192$. All models use an ideal-gas equation of state with fixed adiabatic index $\Tilde{\Gamma} = 4/3$. 

To seed the MRI in every run, I place a weak poloidal magnetic field inside the torus. In practice, I prescribe only the toroidal component of the initial vector potential,
\begin{equation}
    A_{\rm \phi} \propto \rho/\rho_{\rm max}-0.25.
    \label{Eq: SANE}
\end{equation}
with all other components set to zero. The initial field amplitude is fixed by the minimum plasma beta, $\beta_{\rm min}=100$, which keeps the field relatively weak and places the flow in the SANE regime.

The 3D GRMHD snapshots are then post-processed with {\tt RAPTOR}\footnote{\url{https://github.com/jordydavelaar/raptor}} to generate GRRT images. Its spacetime module was extended following the framework developed in our earlier semi-analytical LQBH imaging study \citep{2023arXiv231204288J}. Non-thermal electrons are included through the kappa eDF described in Appendix~\ref{sec:nonthermal}. The GRRT images are produced with a $1\,000\times1\,000$ camera covering roughly $1.5\times 1.5\,\rm mas^2$ ($400\times400\,r_{\rm g}^2$). I adopt M\,87$^*$ as the fiducial source, with mass $6.5\times10^9\,M_{\odot}$ and distance $16.8\,\rm Mpc$ \citep{EventHorizonTelescope2019dse}. The simulation mass unit is chosen so that the total $230\,\rm GHz$ flux equals $1\,\rm Jy$.

\begin{figure*}
    \centering    
    \includegraphics[height=0.41\linewidth]{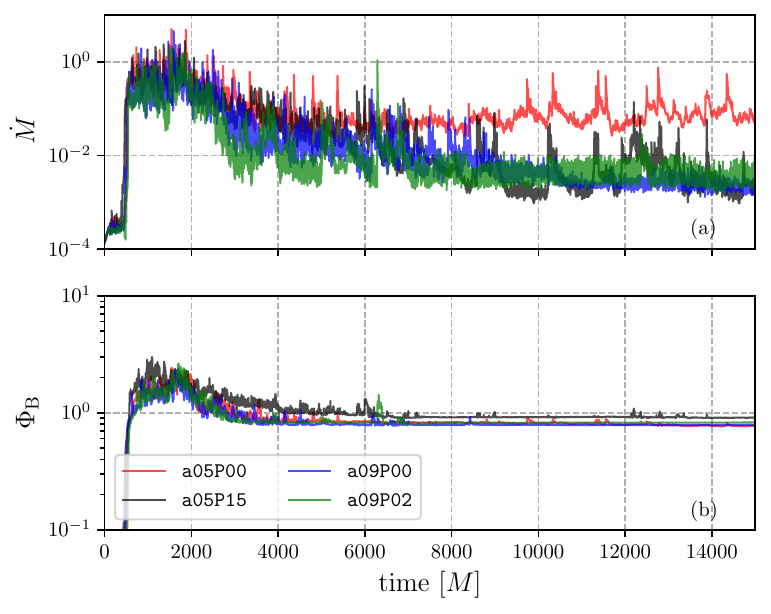}
    \includegraphics[height=0.41\linewidth]{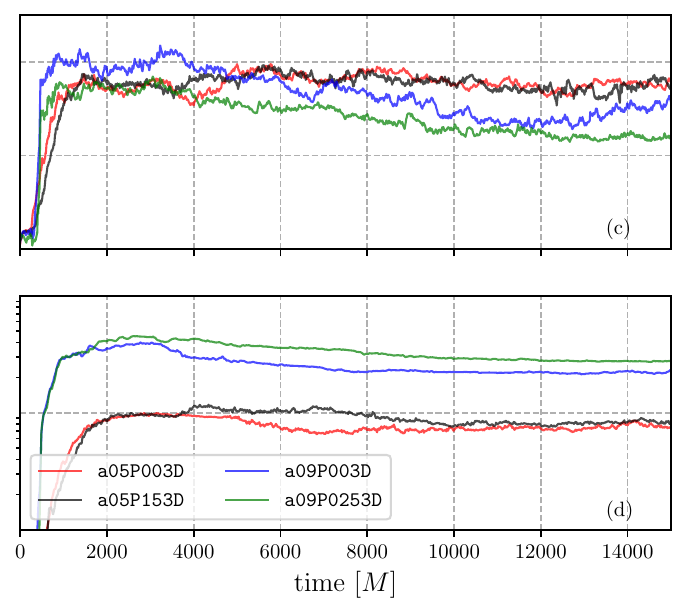}
    \caption{Time histories of the mass-accretion rate and magnetic flux for the LQBH model suite. Panels (a) and (b) show the 2D runs {\tt a05P00}, {\tt a05P15}, {\tt a09P00}, and {\tt a09P02}; panels (c) and (d) show the corresponding 3D runs {\tt a05P003D}, {\tt a05P153D}, {\tt a09P003D}, and {\tt a09P0253D}.}     \label{fig: Mdot-c7}
\end{figure*}

\section{Results}

\subsection{Accretion Flows Around an LQBH}

\begin{figure*}[t]
    \centering
    \includegraphics[height=0.41\linewidth]{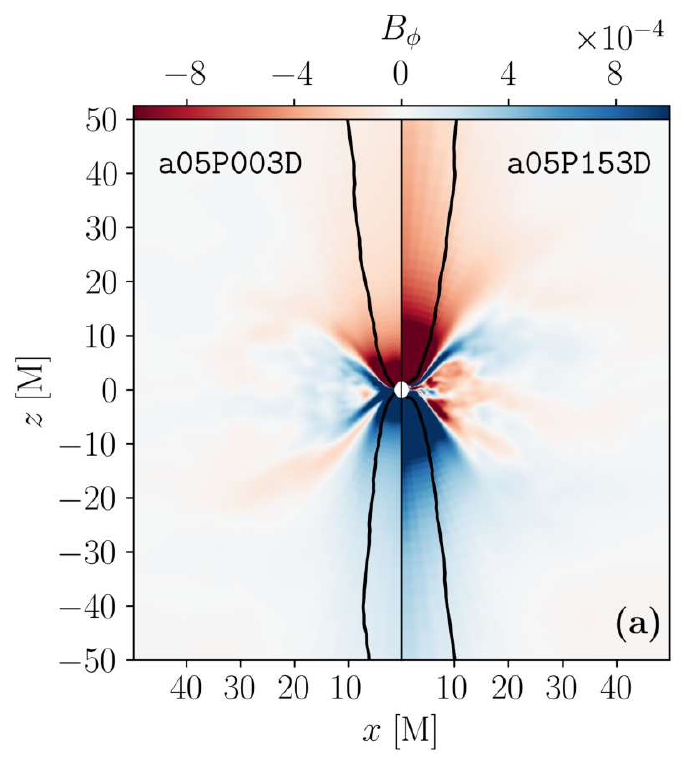}
    \includegraphics[height=0.41\linewidth]{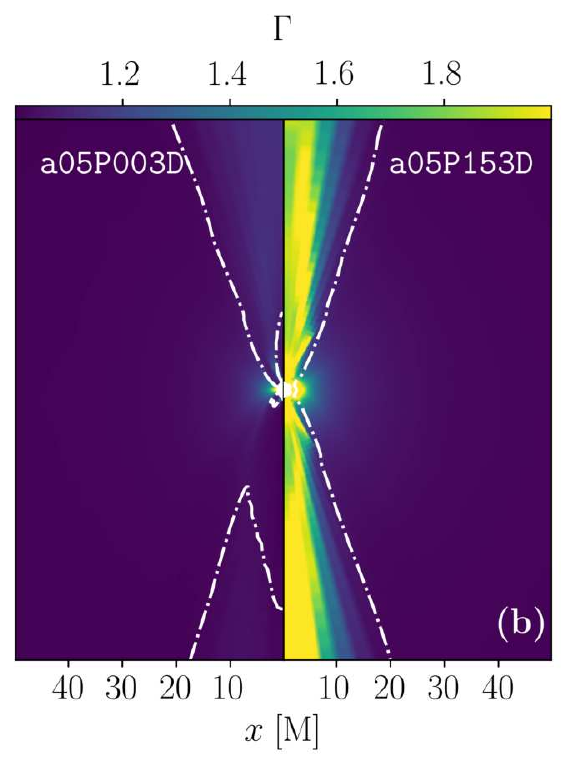}
    \includegraphics[height=0.41\linewidth]{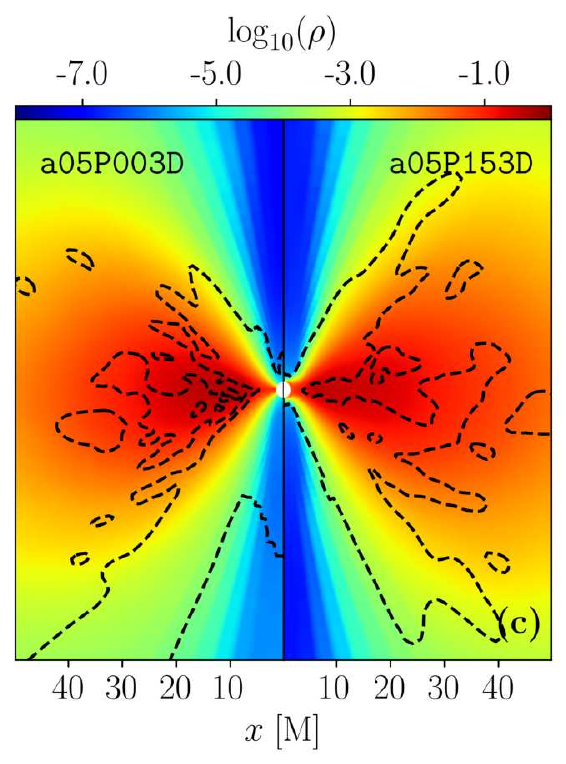}
    \caption{Comparison of 3D GRMHD flow structures for a Kerr BH and an LQBH. In each panel, the Kerr solution with $a=0.5$ is shown on the left, and the LQBH solution with $a=0.5$ and $P=0.15$ is shown on the right. Panels (a), (b), and (c) display the toroidal magnetic field $B_{\rm \phi}$, the Lorentz factor $\Gamma$, and the rest-mass density $\rho$, respectively. Overlaid contours mark $\sigma=1$ (black), $-hu_t=1$ (white dash-dotted), and $u^r=0$ (black dashed).}
    \label{fig: LQBH_compare}
\end{figure*}

The LQBH models span the portion of parameter space allowed by the two quantities that most directly affect the dynamics, the spin $a$ and the polymeric correction $P$. The adopted $(a,P)$ combinations are listed in Table~\ref{table: LQBH}, and the grid stops short of the region where naked singularities would appear \citep{Liu:2020ola}. Each run is identified by its $(a,P)$ pair, with the suffix {\tt 3D} reserved for the fully three-dimensional models. Below I concentrate on the sequences with $a=0.5$ and $a=0.9$. Figure~\ref{fig: Mdot-c7} collects the corresponding histories of the accretion rate $\dot{M}$ and magnetic flux $\Phi_{\rm B}$. At the horizon, the mass inflow and dimensionless magnetic-flux rate \citep{Porth2019} are
\begin{equation}
\begin{aligned}
    &\dot M = \int_0^{2\pi}\int_0^{\pi} \rho u^r \sqrt{-g}d\theta d\phi, \\
    &\Phi_{\rm B} = \frac{1}{2}\int_0^{2\pi}\int_{0}^{\pi}\left|B^r\right|\sqrt{-g}d\theta d\phi. \label{Eq: B-flux-c7}
\end{aligned}
\end{equation}

The histories in Figure~\ref{fig: Mdot-c7}(a,b) indicate that every run settles into a quasi-steady state by roughly $t\gtrsim6,000\,M$, both in $\dot{M}$ and in $\Phi_{\rm B}$. Over that interval, the accretion rate decreases as either $a$ or $P$ is increased. For the $a=0.5$ sequence, the 2D models drop from $\langle \dot{M} \rangle =0.062$ at $P=0$ to $\langle \dot{M} \rangle =0.005$ at $P=0.15$, and the same trend remains present in 3D. The $a=0.9$ models behave similarly, with the LQBH cases accreting more slowly than the Kerr reference. A notable difference appears in 3D, however, because the magnetic flux grows with both $a$ and $P$, likely owing to the richer field structure supported by a fully three-dimensional dynamo \citep[e.g.,][]{2023arXiv231100034J}.

To understand the origin of that trend, I compare the time-averaged 3D flow structure in representative Kerr and LQBH models. Figure~\ref{fig: LQBH_compare} displays the toroidal magnetic field, Lorentz factor, and density for the Kerr model {\tt a5P003D} on the left half of each panel and the LQBH model {\tt a5P153D} on the right. In the latter, $P=0.15$ lies already close to the critical value $P_{\rm max}=0.1957$. The contours in panels (a), (b), and (c) mark $\sigma=b^2/\rho=1$, $-hu_t=1$, and $u^r=0$, respectively. Figure~\ref{fig: LQBH_compare}(a) shows that the LQBH run develops a stronger toroidal field and a somewhat broader $\sigma>1$ funnel than the Kerr reference. Since frame dragging winds up the magnetic field in rotating spacetimes \citep[e.g.,][]{2023MNRAS.522.2307J}, this is naturally consistent with the stronger rotational effects of the LQBH geometry and with the lower accretion rate at larger $P$.
The same comparison indicates that {\tt a05P153D} also attains a larger jet Lorentz factor than {\tt a05P003D}, especially near the equatorial region close to the horizon ($r<10\,r_{\rm g}$). I interpret that as a consequence of the faster plasma rotation permitted by the LQBH spacetime. The $-hu_t=1$ contour additionally shows that the gravitationally bound region covers a smaller area in the LQBH case. Panel (c) reveals that the Kerr model keeps slightly more density inside the funnel, which implies a weaker degree of magnetization there. The LQBH run also displays a thicker inflow channel, reflecting the fact that the geodesic structure differs from Kerr. Finally, the $u^r=0$ contour places the outflow surface of case {\tt a05P15} closer to the horizon than in the Kerr comparison run.

Taken together, these diagnostics indicate that the loop-quantum correction reinforces the funnel magnetic field, suppresses the inflow, and favors a more energetic jet. The next subsection links that behavior to the larger horizon angular frequency of the LQBH.

\subsection{Jet Launching in the LQBH Model}\label{sec:enhanced_spin}

The Blandford-Znajek mechanism extracts rotational energy from a spinning black hole and converts it into a relativistic jet \citep{Blandford1977}. To examine how that process is altered in the LQBH spacetime, I first evaluate the angular frequency at the event horizon,
\begin{equation}
    \Omega_{\rm H} = \left(-\frac{g_{t\phi}^{\rm BL}}{g_{\phi\phi}^{\rm BL}}\right)_{r_{\rm H}},
\end{equation}
where $r_{\rm H}$ is the outer-horizon radius. For the LQBH model, this expression reduces to
\begin{equation}
    \Omega_{\rm H}^{\rm LQBH} = \\
    \Omega^{\text{Kerr}}_{\text{H}}\Bigg(1+\frac{2(a^2+a^4+2\sqrt{1-a^2}-2)}{a^2 (1-a^2)}P + O(P^2)\Bigg). \label{Eq:Omega_H} 
\end{equation}
For Kerr, the corresponding horizon frequency is $\Omega_{\rm H}^{\rm Kerr} = a/2r_{\rm H}^{\rm Kerr}$ \citep{Tchekhovskoy2010}. Equation~(\ref{Eq:Omega_H}) therefore makes explicit that the LQG correction increases the horizon angular frequency. At fixed spin, a positive polymeric parameter $P$ causes the LQBH horizon to rotate faster.
 
\cite{Tchekhovskoy2010} showed that the jet power measured in GRMHD simulations scales with the black-hole angular frequency. In that analysis, the Kerr jet power was related to the magnetic flux and horizon frequency through
\begin{equation}
    P^{\rm BZ}\approx k \Phi_{\rm tot}^2\Omega_{\rm H}^2,\label{Eq:BZpower}
\end{equation}
where $\Phi_{\rm tot}$ denotes the total magnetic flux threading the jet. 
\begin{figure*}
    \centering
    \includegraphics[height=0.33\linewidth]{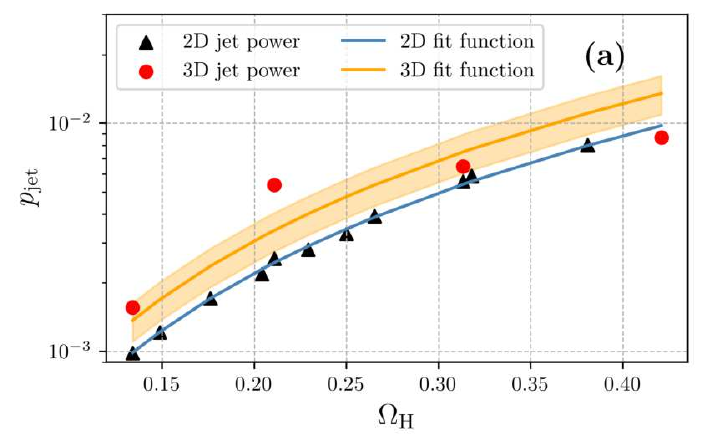}
    \includegraphics[height=0.33\linewidth]{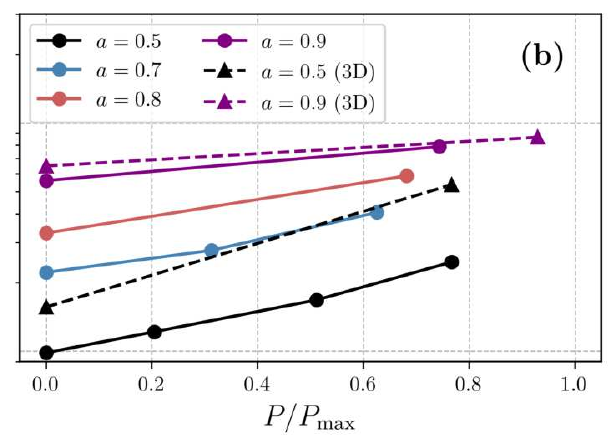}
    \caption{Panel (a) shows the normalized, time-averaged ($t=13,000$--$15,000\,M$) BZ jet power $p_{\rm jet}$ for the simulated models. The solid curves are fits based on Eq.~\ref{Eq:BZpower}, and the shaded bands indicate the uncertainty in the fitted coefficient $k$. Panel (b) gives the jet power as a function of $P$ at fixed black-hole spin $a$, with solid and dashed curves denoting the 2D and 3D runs.} \label{fig:BZ_fit}
\end{figure*}

To measure the jet energetics, I first define the jet boundary through the contour $\sigma=1$. I then evaluate the energy flux through spherical shells at $50\,r_{\rm g}$ within the funnel region satisfying $\sigma\geq1$ \citep{Nathanail2020, 2022ApJ...930L..16E}. The resulting jet power is
\begin{equation}
    P_{\rm jet}=\int_0^{2\pi}\int_{\theta_{\rm jet}}(-T_t^r-\rho u^r)\sqrt{-g}d\theta d\phi,
\end{equation}
where the angular domain $\theta_{\rm jet}$ covers the funnel region for which $\sigma\geq1$. 

Figure~\ref{fig:BZ_fit}(a) makes the rotation dependence clearer by plotting the normalized jet power, $p_{\rm jet}=P_{\rm jet}/\Phi^2_{\rm jet}$, measured from the LQBH simulations. Black triangles represent the 2D models and red circles the 3D models, with each point averaged over $t=13,000$--$15,000\,M$. I fit Eq.~\ref{Eq:BZpower} separately to the 2D and 3D subsets, which gives the blue and orange curves. Because the jet is largely Poynting dominated ($P_{\rm jet}/P_{\rm em}\sim1$), I use the approximation $P_{\rm jet}\approx P_{\rm BZ}$. The best-fit coefficients are $k=0.055\pm0.0006$ in 2D and $k=0.076\pm0.0147$ in 3D. The larger uncertainty in the 3D fit mainly comes from the smaller number of models. Both fitted values differ from the axisymmetric force-free result of \cite{Tchekhovskoy2010} and from the original analytic estimate of \cite{Blandford1977}, which is unsurprising because the present GRMHD models include plasma inertia and hot-jet thermodynamics.

Figure~\ref{fig:BZ_fit}(b) isolates the influence of the polymeric correction by plotting the jet power versus $P$ at fixed spin. Since Eq.~\ref{Eq:Omega_H} shows that the horizon angular frequency increases with $P$, the BZ scaling predicts more powerful jets for larger polymeric corrections. That expectation is borne out by the simulations at every spin considered here. For example, the 3D model {\tt a05P153D}, with $P/P_{\rm max}\sim0.78$, launches a jet about four times more powerful than {\tt a05P003D}.

All GRMHD models in this chapter lie in the SANE regime. Observational modeling of Sgr\,A$^*$ and M\,87$^*$ nevertheless suggests that their real accretion flows may not be purely SANE \cite{2019ApJ...875L...5E, 2022ApJ...930L..16E}. In a MAD state, the dynamically dominant magnetic field could partly obscure the spacetime imprint of the LQBH. The present calculations are therefore designed to isolate the direct influence of the LQG correction itself, leaving the more complicated plasma and magnetic structure of MAD flows to future work.

\subsection{Jet-Power Variations Across LQBH Parameters}

\begin{figure}
    \centering
    \includegraphics[width=\linewidth]{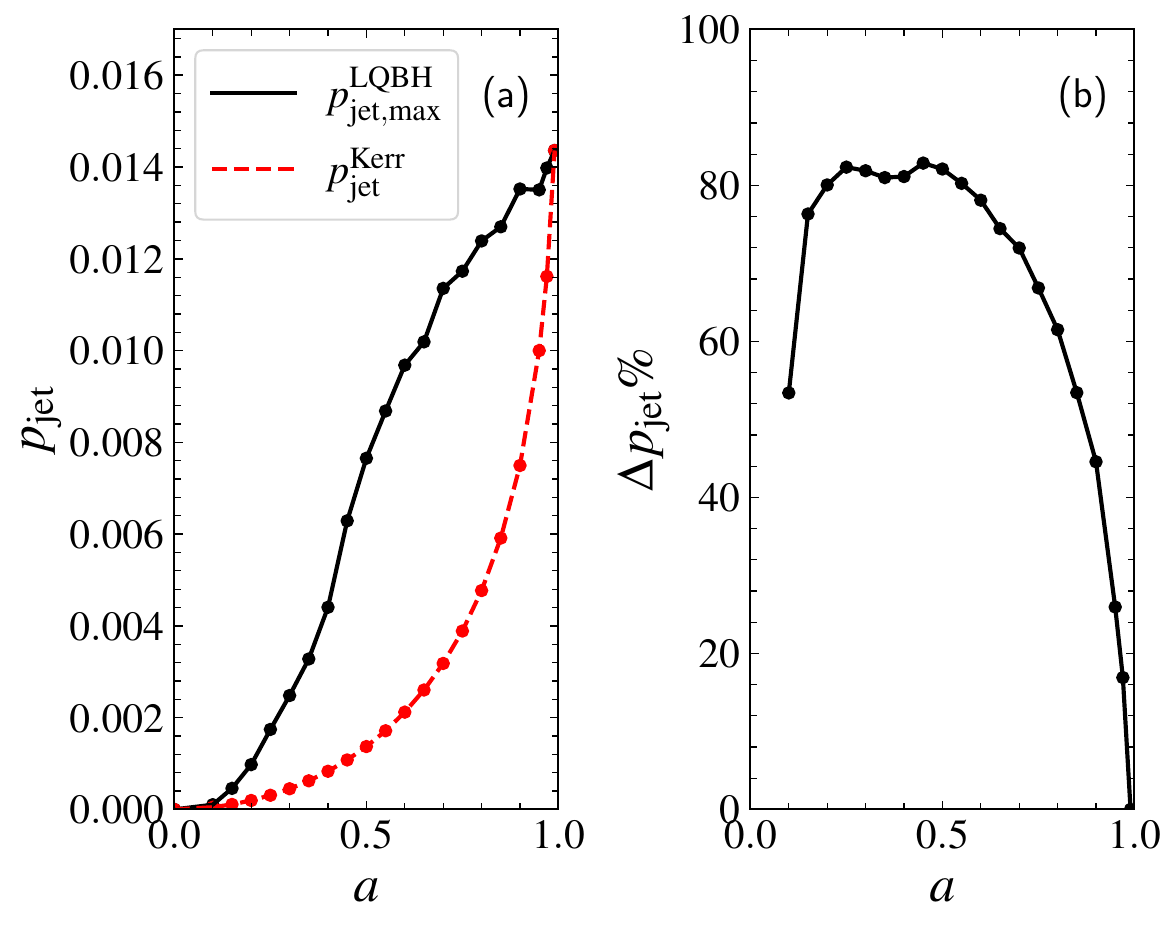}
    \caption{Panel (a): maximum LQBH jet power, $p^{\rm LQBH}_{\rm jet,max}$, shown by the black solid curve together with the Kerr value $p^{\rm Kerr}_{\rm jet}$ shown by the red dashed curve, both plotted against spin. Panel (b): corresponding fractional enhancement of the LQBH jet power, $\Delta p_{\rm jet}$.}     \label{fig:Pjetminmax}
\end{figure}

\begin{figure*}
    \centering    
    \includegraphics[height=0.42\linewidth]{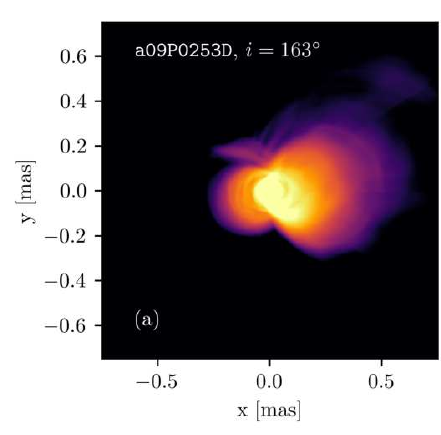}
    \includegraphics[height=0.42\linewidth]{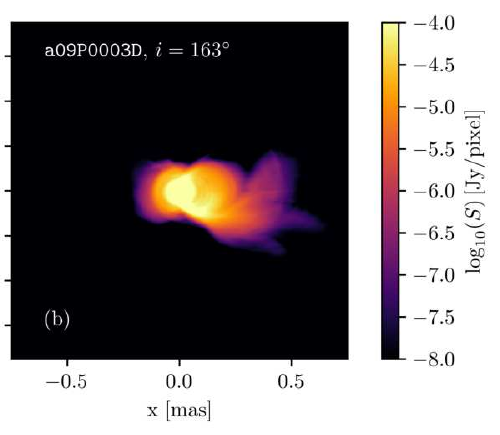}
    \caption{Time-averaged $86\,\rm GHz$ GRRT jet images for models {\tt a09P0253D} and {\tt a09P003D}, shown in panels (a) and (b), respectively. In both panels, the averaging interval is $14,000$ to $15,000\,M$.}     \label{fig:Jet}
\end{figure*}

The previous subsection established that the LQBH geometry can strengthen the BZ jet relative to Kerr. Because the largest allowed polymeric correction depends on spin, the maximum possible enhancement is itself spin dependent. Figure~\ref{fig:Pjetminmax}(a) therefore plots the largest normalized LQBH jet power, $p^{\rm LQBH}_{\rm jet,max}=P_{\rm jet}|_{\rm max}/\Phi_{\rm jet}^2$, as a function of $a$, while the red dashed line shows the Kerr reference with $P=0$. These values are computed from Eq.~\ref{Eq:BZpower} using the coefficient obtained from the 3D fits. Panel (b) presents the same result as a fractional excess over Kerr at fixed spin, $\Delta p_{\rm jet}=(p^{\rm LQBH}_{\rm jet,max}-p^{\rm Kerr}_{\rm jet})\times100/p^{\rm LQBH}_{\rm jet,max}$. In both spacetimes the jet power increases with spin, but the LQBH trend rises more steeply. The two cases become similar again at very low and nearly maximal spin, which means that a nonzero polymeric correction is most advantageous at intermediate spin and does not surpass an extremal Kerr hole. The figure also makes clear that different combinations of $(a,P)$ can yield the same jet power.

\subsection{More Extended Jet Images in the LQBH Case}
The stronger BZ outflow inferred above should also be visible in the synthetic images. At the same spin, Figure~\ref{fig: LQBH_compare} shows that the LQBH funnel is both faster and more strongly magnetized than the Kerr funnel. The additional magnetic energy can feed particle acceleration through turbulence \citep{Meringolo2023} and reconnection \citep{Ball2018}, thereby enhancing the non-thermal electron population. In the GRRT calculation, I represent that component with a kappa eDF added to the thermal Maxwell-Juttner background, as described in Appendix~\ref{sec:nonthermal}. The resulting $86\,\rm GHz$ images in Figure~\ref{fig:Jet} illustrate the effect directly: relative to {\tt a09P003D}, the stronger-LQG model {\tt a09P0253D} produces a jet that is brighter, broader, and more prominently edge brightened.

\section{Summary and Discussion}

This chapter used coupled GRMHD and GRRT calculations to examine how loop-quantum corrections alter accretion and jet launching in the LQBH spacetime. The model suite includes both 2D and 3D evolutions and covers several values of the spin $a$ and polymeric function $P$. The main conclusions are:

\begin{enumerate}
    \item The loop-quantum correction acts most strongly in the near-horizon region. Increasing $P$ lowers the accretion rate, but it raises both the jet Lorentz factor and the BZ power. I interpret this combination as the result of stronger magnetic support produced by more efficient field amplification in the faster-rotating LQBH geometry.
    
    \item The variation of jet power across parameter space is captured well by the standard BZ scaling written in terms of the horizon angular frequency. The fitted coefficient $k$ differs somewhat from the value reported by \cite{Tchekhovskoy2010}, which is expected because the present GRMHD models include plasma inertia and thermodynamics that are absent in an axisymmetric force-free treatment. The 3D fits also yield a somewhat larger $k$ than the 2D runs, showing that non-axisymmetric structure already matters.
    
    \item The strongest jet-power contrast between Kerr and LQBH spacetimes appears at intermediate spin, where the LQBH jet can approach roughly twice the Kerr value. In the limits $a\rightarrow0$ and $a\rightarrow1$, however, the two spacetimes produce nearly indistinguishable jet powers.
    
    \item Stronger frame dragging in the LQBH spacetime amplifies the funnel magnetic field more efficiently than in Kerr, and the corresponding non-thermal jet emission becomes brighter. In the $86\,\rm GHz$ GRRT images, this appears as a jet that is both more extended and easier to detect.
\end{enumerate}

Several caveats should be kept in mind. Every simulation in this chapter is SANE, even though Sgr\,A$^*$ and M\,87$^*$ may in reality be closer to MAD flows \citep[e.g.,][]{2019ApJ...875L...5E, 2022ApJ...930L..16E, Yuan2022, Cruz-Osorio2021}. I also chose the simplest single-loop SANE initial condition so that the effect of the LQG correction on the flow and jet could be isolated as cleanly as possible; consequently, higher-order phenomena such as hot spots and intrinsic light-curve variability were intentionally excluded. A more realistic treatment will require two-temperature, radiatively cooling flows together with a broader range of magnetic-field topologies \citep[e.g.,][]{Mizuno2021, Dihingia2023, 2023MNRAS.522.2307J}. Those extensions are deferred to future work.


\newpage
{
\section*{Supplementary Appendix}  
\setcounter{section}{0}
\renewcommand{\thesection}{\thechapter.\Alph{section}}
\renewcommand{\theHsection}{appendix.\thechapter.\Alph{section}}
\renewcommand{\theHsubsection}{appendix.\thechapter.\Alph{section}.\arabic{subsection}}

\section{LQBH Spacetime}\label{sec:LQBH}

The appendix uses the self-dual LQBH metric obtained from a symmetry-reduced LQG treatment of homogeneous geometries. In that construction, geodesic completeness is preserved and the central singularity is replaced by a regular core. Written in Boyer-Lindquist coordinates, the metric takes the form \cite{2023arXiv231204288J, Liu:2020ola},
 \begin{equation}
     d s^2 = \frac{\Delta}{\Sigma}(dt-a\sin^2\theta d\phi)^2-\frac{\Sigma}{\Delta}dr^2-\Sigma d\theta^2 -\frac{\sin^2\theta}{\Sigma}(a dt-(k^2+a^2)d\phi)^2,  
      \label{metric-c7}
 \end{equation}
 with
\begin{eqnarray}
\Delta= \frac{(r-r_+)(r-r_-)r^2}{(r+r_*)^2}+a^2,
\Sigma=k^2(r)+a^2\cos^2\theta,\
k^2= \frac{r^4+a^2_0}{(r+r_*)^2},
\end{eqnarray}
where $a$ is the specific angular momentum of the black hole. In the associated spherical self-dual limit, the two horizons are located at $r_+=2 G_\text{LQG} M/(1+P)^2$ and $r_{-} = 2G_\text{LQG} M P^2/(1+P)^2$, with $r_{*}= \sqrt{r_+ r_-} = 2G_\text{LQG} MP/(1+P)^2$. Here $G_\text{LQG}$ is the effective gravitational constant of the LQBH geometry, $M$ denotes the ADM (Arnowitt-Deser-Misner) mass, and the loop-quantum correction enters through the polymeric function
\begin{eqnarray}
P \equiv \frac{\sqrt{1+\epsilon^2}-1}{\sqrt{1+\epsilon^2}+1},   \label{P_epsilon-c7}
\end{eqnarray} 
where $\epsilon$ is the product of the Immirzi parameter $\gamma$ and the polymeric parameter $\delta$, so $\epsilon=\gamma \delta \ll 1$. The quantity $a_{0}$ is
\begin{eqnarray}
a_0 = \frac{A_{\rm min}}{8\pi},
\end{eqnarray}
where $A_{\rm min}$ is the minimum area gap in LQG. Since $A_{\min} \simeq 4 \pi \gamma \sqrt{3} l_{\rm Pl}^2$ \citep{Sahu2015dea}, the scale $a_0$ is Planckian and should be irrelevant for the phenomenology considered here. I therefore set $a_0=0$ and focus on loop-quantum effects that act from a few to thousands of Schwarzschild radii. The effective gravitational constant of the rotating LQBH is related to the Newtonian one by 
\begin{eqnarray}
G_\text{LQG}=G_{\rm N}\frac{(1+P)^2}{(1-P)^2}.
\end{eqnarray}  
Finally, the rotating-LQBH horizons are determined by $\Delta(r_{h\pm})=0$. Closed expressions for $r_{h\pm}$ were given in \cite{Liu:2020ola}, and the familiar Kerr result $r_{h\pm}=M \pm \sqrt{M^2-a^2}$ is recovered in the limit $P=0$.

For the numerical work, I re-express the metric using the Newtonian gravitational constant so that Kerr and LQBH models share the same normalization. The dimensionless spin must therefore be rescaled by the factor $(1+P)^2/(1-P)^2$. By contrast, \cite{Afrin2022ztr} kept the LQBH gravitational constant rather than converting back to $G_{\rm N}$, so their normalization is different from the one adopted here.

The Immirzi parameter $\gamma$ is not unique, and its preferred value depends on the theoretical framework \citep[see][and references therein]{BenAchour2014qca, Frodden:2012dq, Carlip2014bfa, Taveras:2008yf}. Some treatments allow complex values of $\gamma$ \citep{Frodden:2012dq, BenAchour2014qca, Carlip2014bfa}, while others promote it to a scalar field determined dynamically \citep{Taveras:2008yf}. Here I adopt the standard value $\gamma = 0.2375$ motivated by black-hole entropy arguments \citep{Meissner:2004ju}. 

To remove the coordinate singularity at the horizon, we transform to Kerr-Schild coordinates. The Boyer-Lindquist-like to Kerr-Schild-like transformation is
\begin{eqnarray}
dr_\text{KS}&=&dr, \\
d\theta_\text{KS}&=&d\theta, \\
dt_\text{KS}&=&dt - \frac{2M\mathbf{A} r}{\Delta}dr_\text{KS}, \\
d\phi_\text{KS} &=& d\phi -\frac{a}{\Delta}dr_{KS}.
\end{eqnarray}
After this transformation, the metric in the new coordinates becomes \citep{Kocherlakota2023}
\begin{eqnarray}
ds^2 = g^{\rm KS}_{tt}dt^2 + 2g^{\rm KS}_{tr}drdt + 2g^{\rm KS}_{t\phi}dtd\phi \nonumber
+ g^{\rm KS}_{rr}dr^2 + g^{\rm KS}_{r\phi}drd\phi + g^{\rm KS}_{\theta\theta}d\theta^2 +g^{\rm KS}_{\phi\phi}d\phi^2,
\end{eqnarray}
with
\begin{eqnarray}
g^{\rm KS}_{tt}&=&-\frac{\Delta-a^2 \sin^2\theta}{\Sigma},  \\
g^{\rm KS}_{tr}&=&\frac{1}{\Sigma}[\mathbf{A} (2Mr+a^2 \sin^2\theta)-a^2\sin^2\theta], \\
g^{\rm KS}_{t\phi}&=&-\frac{a \sin^2\theta (k^2+a^2-\Delta)}{\Sigma},\\
g^{\rm KS}_{rr}&=&\frac{\mathbf{A}}{\Sigma}[\mathbf{A}(\Delta+4Mr+a^2\sin^2\theta)-2a^2\sin^2\theta], \\
g^{\rm KS}_{r\phi}&=&-\frac{a\sin^2\theta}{\Sigma}[(r^2+a^2)\mathbf{A}^2+2Mr\mathbf{A}-a^2 \sin^2\theta],\\
g^{\rm KS}_{\theta\theta}&=&\Sigma, \\
g^{\rm KS}_{\phi\phi}&=&-\sin^2\theta\Bigg[\frac{a^2 \sin^2\theta\Delta-(k^2+a^2)^2}{\Sigma}\Bigg],
\end{eqnarray}
where
\begin{eqnarray}
\mathbf{A}=\frac{k^2+a^2}{r^2+a^2}.
\end{eqnarray}

\section{LQBH Parameters}\label{sec:LQBH_parameter}

Table~\ref{table: LQBH} summarizes the GRMHD model set together with the spin and polymeric parameter of each run. The first two blocks correspond to axisymmetric calculations, whereas the final block collects the fully three-dimensional models labeled with the suffix {\tt 3D}.

\begin{table}[t]
\centering
\caption{Spin parameters ($a$) and polymeric-function values ($P$) adopted in the GRMHD simulations.}
\small
\setlength{\tabcolsep}{4pt}
\begin{tabular}{lcccccc}
\hline
Case & {\tt a5P00} & {\tt a5P04} & {\tt a5P10} & {\tt a5P15} & {\tt a7P00} & {\tt a7P03} \\
\hline
$a$ & 0.5 & 0.5 & 0.5 & 0.5 & 0.7 & 0.7 \\
$P$ & 0.00 & 0.04 & 0.10 & 0.15 & 0.00 & 0.03 \\
\hline
Case & {\tt a7P06} & {\tt a8P00} & {\tt a8P04} & {\tt a9P00} & {\tt a9P02} & -- \\
$a$ & 0.7 & 0.8 & 0.8 & 0.9 & 0.9 & -- \\
$P$ & 0.06 & 0.00 & 0.04 & 0.00 & 0.02 & -- \\
\hline
\end{tabular}

\vspace{6pt}
\begin{tabular}{lcccc}
\hline
Case & {\tt a5P003D} & {\tt a5P153D} & {\tt a9P003D} & {\tt a9P0253D} \\
\hline
$a$ & 0.5 & 0.5 & 0.9 & 0.9 \\
$P$ & 0.00 & 0.15 & 0.00 & 0.025 \\
\hline
\end{tabular}
\label{table: LQBH}
\end{table}

\section{Non-thermal Electron Distribution Function}
\label{sec:nonthermal}
In the GRRT post-processing, the synchrotron signal is built from both thermal and non-thermal electrons. The thermal emissivity and absorption coefficients are taken directly from the electron-temperature data in the two-temperature GRMHD snapshots \cite{Mizuno2021, 2023MNRAS.522.2307J}. The non-thermal component is described with a kappa eDF,
\begin{equation}
    \frac{dn_{\rm e}}{d\gamma_{\rm e}}=\frac{N}{4\pi}\gamma_{\rm e}\sqrt{\gamma_{\rm e}^2-1}\left(1+\frac{\gamma_{\rm e}-1}{\kappa w}\right)^{-(\kappa+1)},
\end{equation}
where $n_{\rm e}$ is the electron number density, $\gamma_{\rm e}$ is the electron Lorentz factor, $\kappa$ sets the slope of the non-thermal tail, and $N$ is a normalization constant; further details are given in \cite{Pandya2016}. To model unresolved particle acceleration, we adopt a sub-grid prescription calibrated against Particle-In-Cell (PIC) simulations of special-relativistic turbulence \citep{Meringolo2023}:
\begin{equation}
        \kappa(\beta,\sigma)=2.8+\frac{0.2}{\sqrt{\sigma}} + 1.6\sigma^{-6/10}\tanh{\left(2.25\beta\sigma^{1/3}\right)}. \label{Eq: kappa_tur-c7}
\end{equation}
}

\chapter{Electromagnetic Tracers of Supermassive Black-Hole Binaries}
\markboth{第9章 EM Tracers of SMBBH Binaries}{第9章 EM Tracers of SMBBH Binaries}

This chapter examines how the dynamics of a high-mass-ratio supermassive black-hole binary are encoded in observable electromagnetic signals.

\section{Astrophysical and Multi-messenger Context}
In hierarchical models of galaxy growth, repeated mergers should naturally leave behind supermassive black-hole binaries (SMBBHs) \citep{1980Natur.287..307B, 2010A&ARv..18..279V}. Their gravitational-wave emission peaks in the nanohertz band, precisely where pulsar timing arrays search for a stochastic background \citep{2019A&ARv..27....5B, 2025arXiv250500797K}. The most recent PTA releases already provide strong evidence for such a background \citep{2023ApJ...951L...8A, 2026arXiv260109481A, 2023A&A...678A..48E, 2023A&A...678A..49E, 2023A&A...678A..50E}.

Electromagnetic information remains essential because PTA measurements alone do not pinpoint individual sources, and future missions such as LISA will benefit greatly from independent sky localization \citep{2017arXiv170200786A, 2019BAAS...51g..77T}. For that reason, candidate binaries have long been sought through repeated flares, quasi-periodic variability, and jet modulation, with OJ~287 remaining the best-known example \citep{1988ApJ...325..628S, 1996A&A...315L..13S, 1996ApJ...460..207L, 2008Natur.452..851V, 2016ApJ...819L..37V, 2018ApJ...866...11D, 2020ApJ...894L...1L}. VLBI polarimetry adds a complementary probe by revealing twisted inner jets, strong magnetization, and rapid polarization changes on horizon scales \citep{2022ApJ...932...72Z, 2026A&A...705A..23G}. Nearly edge-on binaries are especially interesting because the foreground hole can lens radiation from the background minidisk or jet, producing short and nearly time-symmetric flares \citep[e.g.,][]{2018MNRAS.474.2975D, 2020MNRAS.495.4061H, 2022PhRvD.105j3010D, 2022PhRvL.128s1101D, 2024PhRvD.109j3014K}. In favorable geometries, the flare profile itself may preserve information about source size and orbital configuration.

Connecting those candidate signals to the underlying binary dynamics requires strong-field GRMHD calculations that evolve both the accreting gas and the jet-launching region around the two holes. Purely Newtonian circumbinary-disk studies \citep[e.g.,][]{2024ApJ...970..156D, 2020ApJ...889..114M} omit several ingredients that matter here, including LT precession, magnetic-flux buildup near the horizons, and relativistic jet launching. Fully self-consistent Einstein-plus-GRMHD evolutions remain prohibitively expensive for the long integrations and high resolutions needed to study sustained accretion and emission \citep[e.g.,][]{2012PhRvL.109v1102F, 2015CQGra..32q5009E, 2024PhRvD.109j3024F, 2025PhRvD.112d3004M}. A practical compromise is to evolve GRMHD on an approximate but explicitly time-dependent binary spacetime. Earlier PN near-zone metrics already coupled orbital motion to a 2.5PN background, but they required excising the central cavity because the approximation breaks down near the horizons \citep{2012ApJ...755...51N, 2021PhRvD.104d4041C}. The superposed Kerr-Schild construction avoids that limitation, preserves horizon-scale resolution, and tracks numerical-relativity results well over long integrations \citep{2021ApJ...913...16L, 2024arXiv240313308C}. Recent studies using the same framework have already followed inclined passages through MAD disks and OJ~287-like systems, finding enhanced accretion, stronger outflows, and orbit-driven precession \citep{2024ApJ...967...70R, 2025ApJ...993L..22R}.

\begin{table*}
\centering
\caption{Binary models considered in this chapter. Every binary run adopts $q=0.1$, and the listed inclination is measured relative to the initial torus midplane.}
\label{tab:sim_parameters}
\begin{tabular}{lccccc}
\hline \hline
Run & $a_1$ & $a_2$ & Inclination ($i$) & Eccentricity ($e$) & Description \\
\hline
1 & 0 & 0.9375 & $90^\circ$ & 0   & Vertical impact (transient shock) \\
2 & 0 & 0.9375 & $0^\circ$  & 0   & Coplanar (persistently embedded) \\
3 & 0.9375 & 0.9375 & $90^\circ$ & 0.3 & High-spin, eccentric, precessing \\
4 & 0 & \textemdash & \textemdash & \textemdash & Single BH \\
\hline
\end{tabular}
\end{table*}

\begin{figure*}
    \centering
    \includegraphics[width=0.85\linewidth]{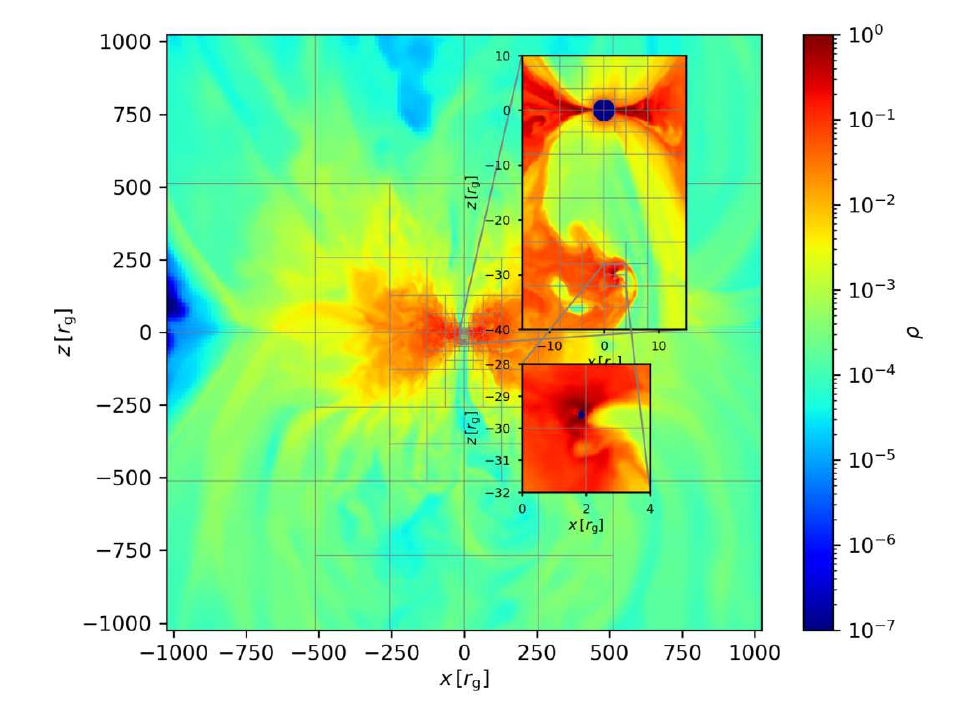}
    \caption{Representative edge-on density slice from Run~1 at $t=26{,}410\,M$, with the refinement-block hierarchy drawn on top of the flow.
}
    \label{fig: meshblock}
\end{figure*}

Among the accretion states relevant to this problem, magnetically arrested disks are especially important because they trap large magnetic flux near the horizon and therefore launch efficient Blandford-Znajek jets \citep{Blandford1977, tchekhovskoy_efficient_2011}. MAD models already reproduce several properties inferred for M\,87$^*$, including its variability and polarization structure \citep{Narayan2012, 2015MNRAS.447...49S, Akiyama2019eap, 2019ApJ...875L...5E, Akiyama2021a, 2025A&A...704A..91E}. What remains much less explored is how an M\,87$^*$-like MAD would appear if a secondary black hole either orbited within the disk or repeatedly crossed it. To address that question, I perform controlled global 3D GRMHD simulations in which a companion of moderate mass ratio, $q\lesssim 10^{-1}$, perturbs a MAD torus on an approximate but explicitly time-dependent BBH spacetime, and I then derive broadband observables through GRRT post-processing.

Section 9.2 describes the numerical setup, the following sections present the main results, and the Appendix collects the technical details.

\section{Numerical Method and Setup}
\label{sec:numerical}

I evolve a MAD torus with global 3D GRMHD while a secondary black hole perturbs the flow on a time-dependent binary background. Throughout this chapter I adopt units with $G=c=M_1=1$, where $M_1$ denotes the primary mass. The companion mass is fixed at $M_2=qM_1=0.1$, and the corresponding primary gravitational radius is $r_{\rm g}\equiv GM_1/c^2$. Table~\ref{tab:sim_parameters} lists the simulated models. All binary runs therefore share the same mass ratio, $q=0.1$, and the same orbital separation, $d=30\,r_{\rm g}$, with $r_{\rm g}$ always defined using the primary. Following \cite{2024ApJ...967...70R}, I place the primary at the coordinate origin and allow spin-orbit coupling to tilt the primary spin, disk, and jet self-consistently.

\subsection{GRMHD Evolution on a Time-dependent Superposed Kerr-Schild Background}
\label{subsec:method_grmhd}

Following \citep{2021ApJ...913...16L, 2024ApJ...967...70R}, I represent the binary spacetime with a superposed Kerr-Schild metric. The orbital trajectories inserted into that metric are supplied by high-order PN evolution \citep{2014LRR....17....2B, 2015JHEP...09..219L}. Appendices~\ref{app:sks} and~\ref{app:pn} summarize the relevant technical details. 

The GRMHD evolution is carried out with the \texttt{DynGRMHD} module of \texttt{AthenaK} \citep{Stone2024} and an ideal-gas equation of state with $\Gamma=5/3$. To resolve both the global torus and the horizon-scale environment of the companion, I combine static mesh refinement (SMR) with adaptive mesh refinement (AMR). The domain uses 7 SMR levels together with 3 additional AMR levels focused on the near-horizon region. The finest AMR mesh tracks both holes with a black-hole-following criterion that keeps the local resolution nearly uniform and reaches a minimum cell size of $\Delta x_{\rm min}\equiv 0.05\,r_{\rm g}$. Figure~\ref{fig: meshblock} shows a representative density slice from Run~1 together with the meshblock layout; each block contains $40^3$ cells. Inside $1\,GM_{1,2}/c^2$ of either hole, the plasma is reset to floor values, which produces the evacuated blue regions in the zoomed views.

During the GRMHD run I impose a magnetization ceiling of $\sigma_{\rm ceiling}=100$. The initial gas distribution is a Fishbone-Moncrief hydrostatic torus \citep{1976ApJ...207..962F} with $r_{\rm min}=20\,r_{\rm g}$ and $r_{\rm max}=40\,r_{\rm g}$. That torus is threaded by a poloidal magnetic field chosen so that the subsequent evolution reaches the MAD state. The initial vector potential is prescribed by
\begin{equation}
    A_{\rm \phi}\propto (\rho - 0.01)(r/r_{\rm in})^3\sin^3\theta \exp{(-r/400)}
    \sin(2\pi(r-r_{\rm in})),
\end{equation}
The field is normalized so that the minimum plasma beta is $\beta_{\rm min}=100$, where $\beta\equiv p_{\rm gas}/p_{\rm mag}$. To avoid introducing the companion impulsively, I first evolve the torus around the primary alone and only then ramp the secondary in by increasing its mass smoothly from zero to $0.1\,M$ over $t=10,000$--$11,000\,M$.

\subsection{GRRT Post-processing}
\label{subsec:method_grrt}

To generate mock observables, I post-process selected GRMHD snapshots with \texttt{BHOSS} \citep{2012A&A...545A..13Y, 2020IAUS..342....9Y}. The code is modified so that null geodesics are integrated in Cartesian coordinates on the superposed Kerr-Schild spacetime. All transfer calculations use the fast-light approximation. For the thermal synchrotron emissivities and absorptivities, I adopt the $R$-$\beta$ electron-temperature prescription \citep{Moscibrodzka2016} with fiducial parameters $(R_{\rm low},R_{\rm high})=(1,10)$, motivated by comparisons with two-temperature GRMHD models \citep{2023MNRAS.522.2307J, 2024A&A...688A..82J}. Unless stated otherwise, the viewing inclination is fixed to the edge-on choice $\theta_{\rm obs}=90^\circ$, and I sample azimuths $\phi_{\rm obs}=0^\circ$ and $90^\circ$ to isolate line-of-sight configurations with strong and weak lensing. M\,87$^*$ is used as the fiducial source, with $M_{\rm BH}=6.5\times10^9\,M_\odot$ and $d_{\rm BH}=16.8\,{\rm Mpc}$. Cells with $\sigma>5$ are excluded from the transfer calculation. The electrons are taken to follow a thermal Maxwell-J\"uttner distribution, while non-thermal particles are deferred to future work. 

The density normalization is chosen so that the single-BH control model (Run~4) yields about $0.5\,{\rm Jy}$ at $230\,{\rm GHz}$ for M\,87$^*$. For Runs~1, 2, and 4, I adopt $\rho_{\rm unit}=1.5\times10^{-20}\,{\rm g\,cm^{-3}}$ using snapshots from the quasi-steady interval $t=25{,}000$--$30{,}000\,M$. For Run~3, I instead use $\rho_{\rm unit}=2\times10^{-21}\,{\rm g\,cm^{-3}}$ so that the total $230\,{\rm GHz}$ flux stays comparable. Every \texttt{BHOSS} image uses a $40\,r_{\rm g}\times 40\,r_{\rm g}$ field of view and a fixed $1000\times 1000$ camera, which fully resolves the photon ring of the secondary.

\section{Simulation Results and Observable Signatures}

\begin{figure*}
\centering 	
\includegraphics[width=\linewidth]{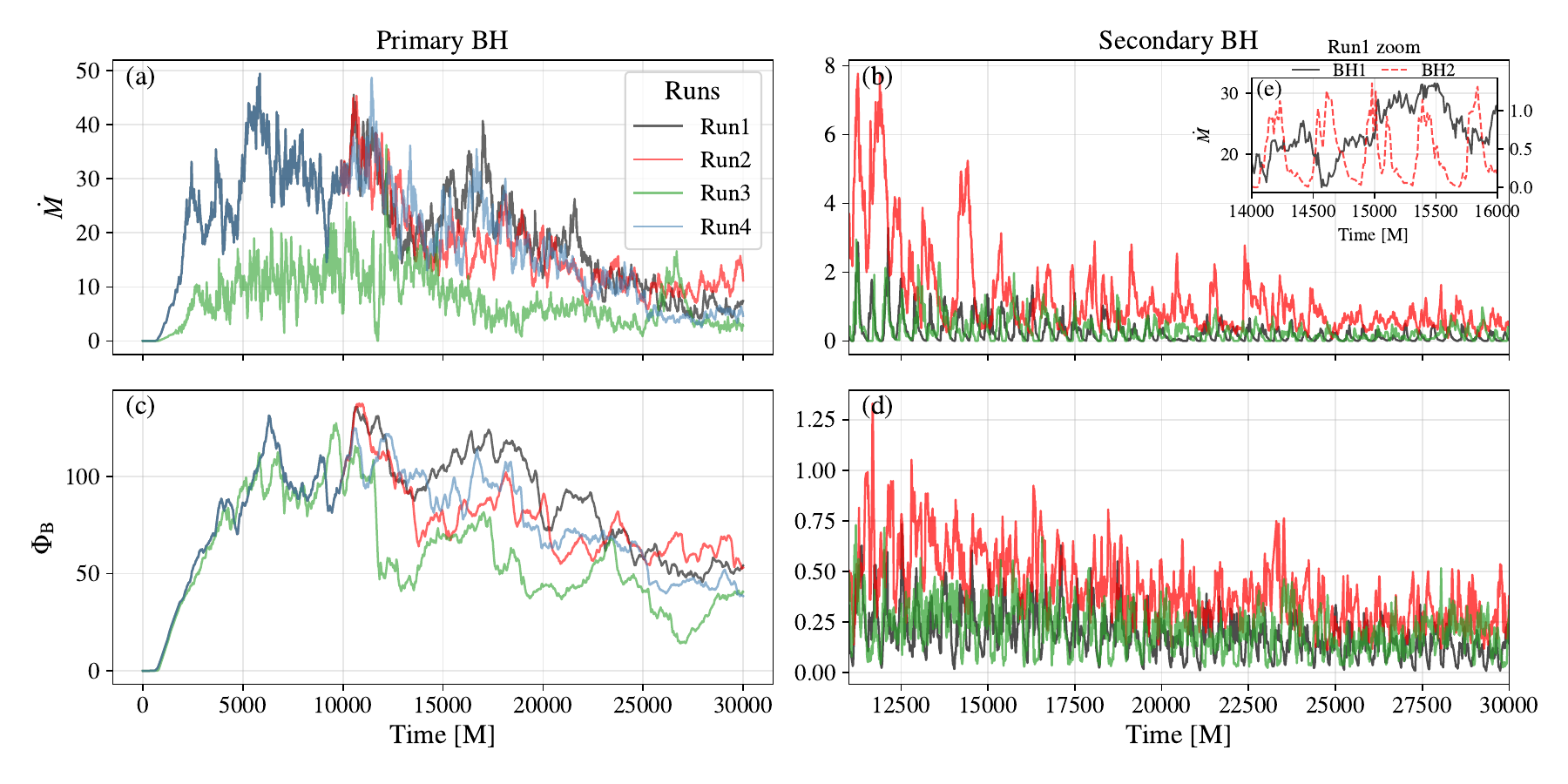}
\caption{Accretion-rate and horizon-flux histories for the two holes. Panel (a) plots $\dot{M}_{\rm BH1}$ for Runs 1--4 (black, red, green, and blue), panel (b) plots $\dot{M}_{\rm BH2}$ for Runs 1--3, panel (c) shows the magnetic flux through the primary horizon, $\Phi_{\rm B,BH1}$, and panel (d) shows the corresponding flux through the secondary horizon, $\Phi_{\rm B,BH2}$. Panel (e) enlarges the Run~1 interval $t \in [1.4, 1.6]\times10^4\,M$ and compares $\dot{M}_{\rm BH1}$ (black solid; left axis) with $\dot{M}_{\rm BH2}$ (red dashed; right axis).}
\label{fig: mdot_run12}
\end{figure*}

\begin{figure}
\centering 	
\includegraphics[width=\linewidth]{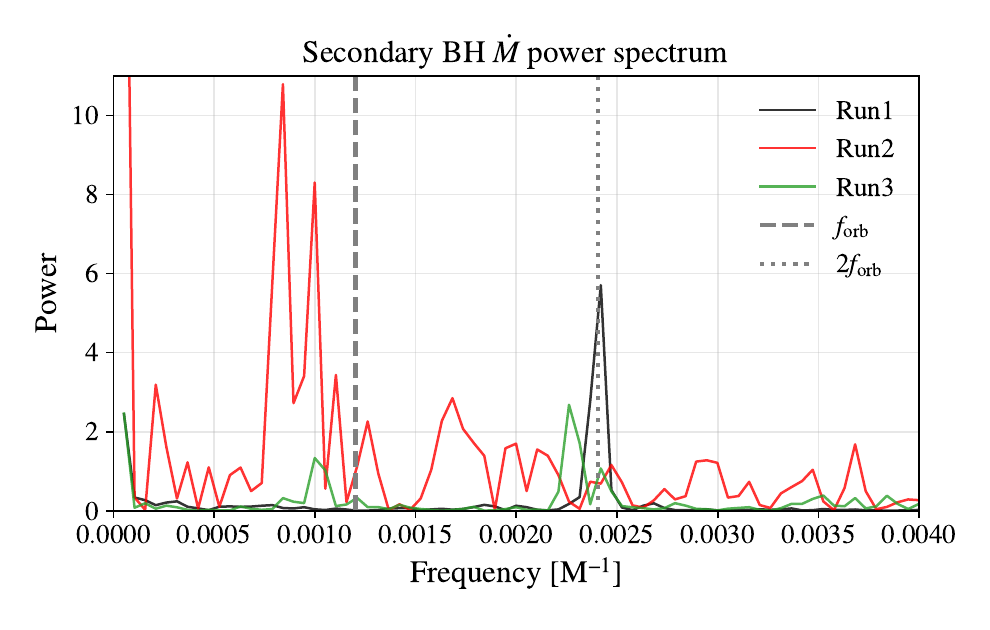}
\caption{Frequency-space view of the secondary accretion history, $\dot{M}_{\rm BH2}$, for Run~1 (black), Run~2 (red), and Run~3 (green). The gray dashed and dotted reference lines indicate the orbital frequency $f_{\rm orb}$ and its first harmonic $2 f_{\rm orb}$.
}
\label{fig: bh2_spectrum}
\end{figure}

\begin{figure*}
\centering 	
\includegraphics[width=\linewidth]{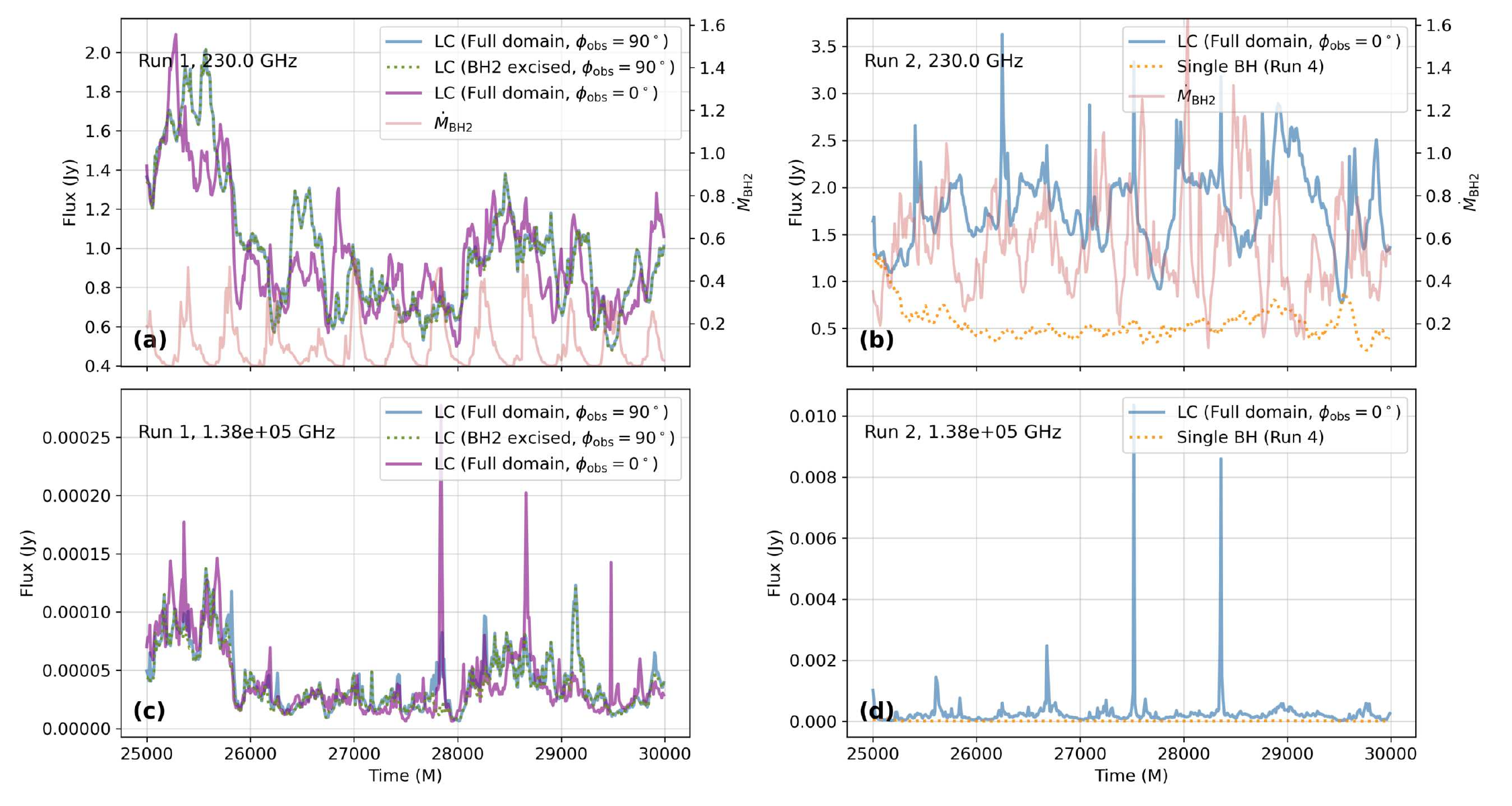}
\caption{GRRT thermal-synchrotron light curves compared with the secondary accretion history. Panel (a) gives the Run~1 $230\,\mathrm{GHz}$ signal for $\phi_{\rm obs}=90^\circ$ (blue solid), the corresponding curve after excising plasma inside $r<10\,r_{\rm g}$ around the secondary (green dotted), and the $\phi_{\rm obs}=0^\circ$ view (purple solid); the pink curve on the right axis is $\dot{M}_{\rm BH2}$. Panel (b) shows the same frequency for Run~2, comparing the binary model at $\phi_{\rm obs}=0^\circ$ (blue solid) with the single-hole control Run~4 (orange dotted), again with $\dot{M}_{\rm BH2}$ in pink. Panels (c) and (d) repeat the comparison in the NIR at $1.38\times10^{5}\,\mathrm{GHz}$.
}
\label{fig: lc_BH2}
\end{figure*}

\begin{figure*}
\centering 	
\includegraphics[height=.33\linewidth]{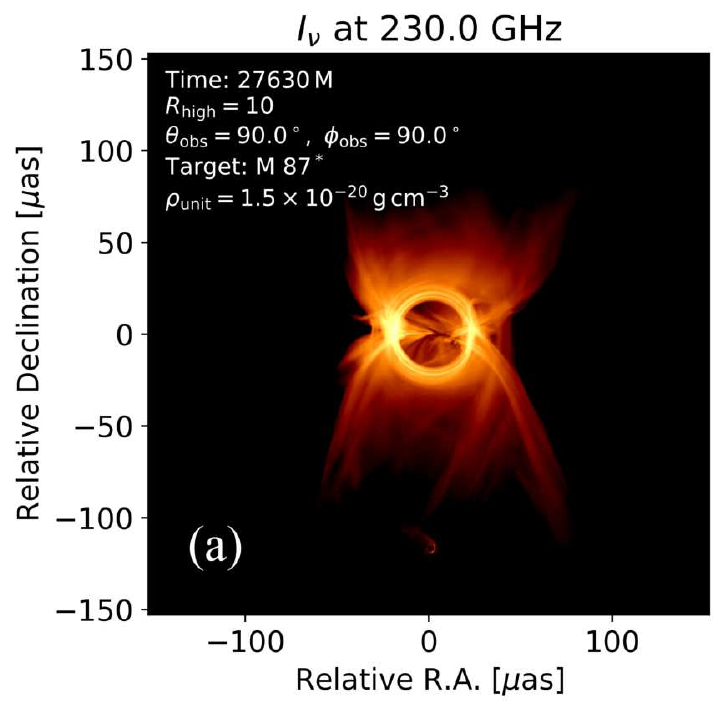}
\includegraphics[height=.33\linewidth]{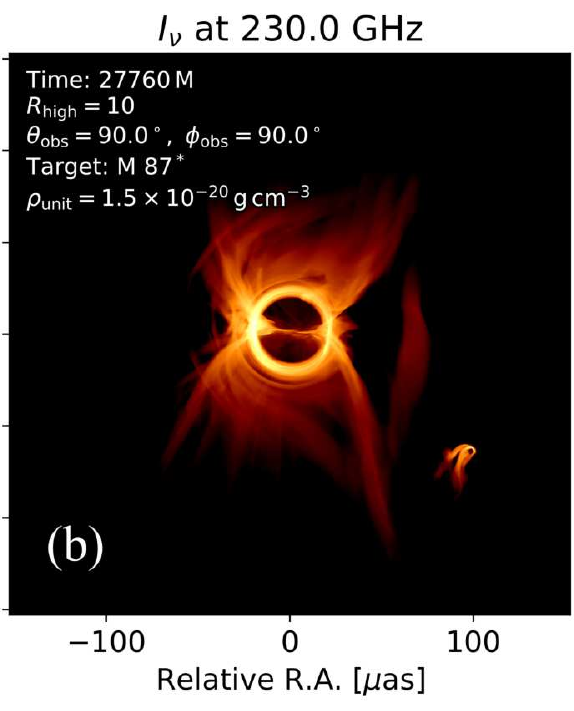}
\includegraphics[height=.33\linewidth]{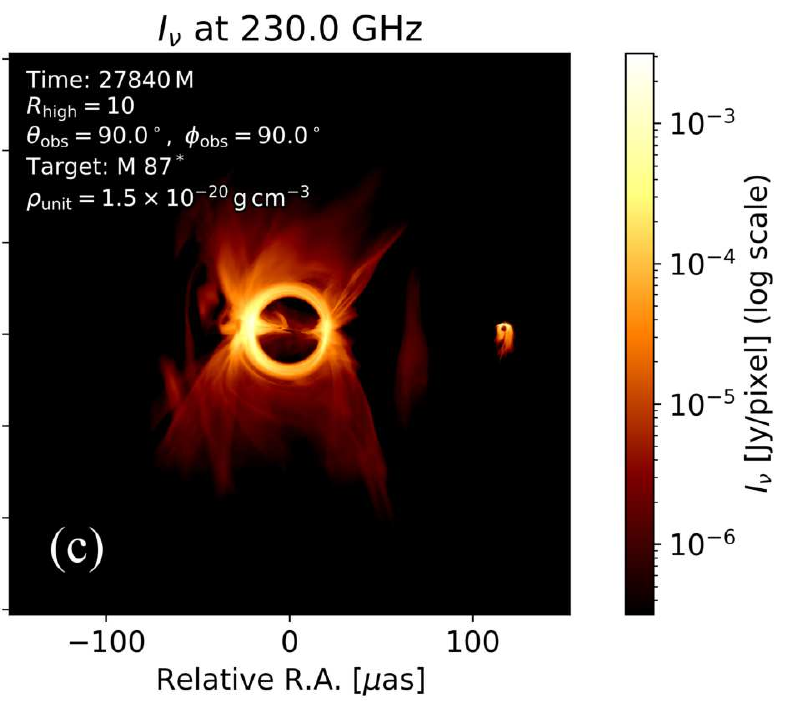}
\includegraphics[height=.33\linewidth]{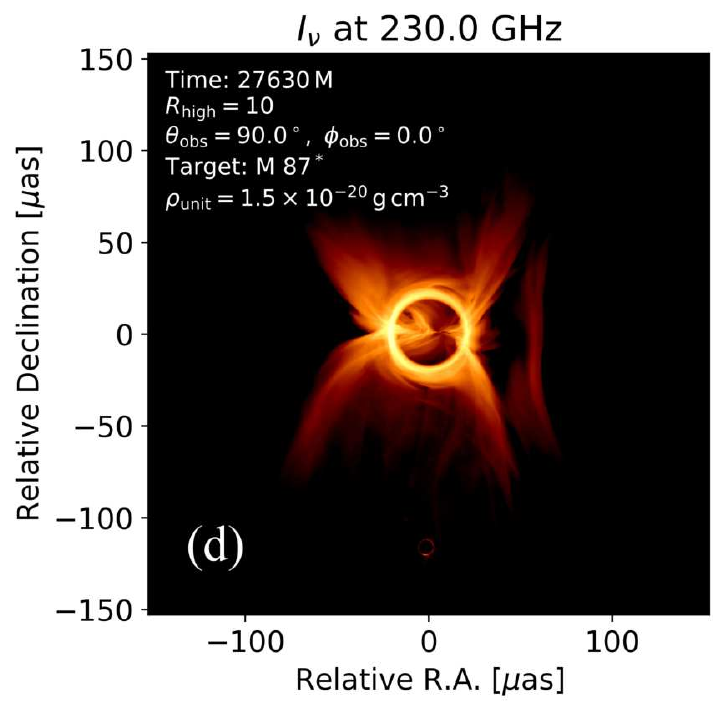}
\includegraphics[height=.33\linewidth]{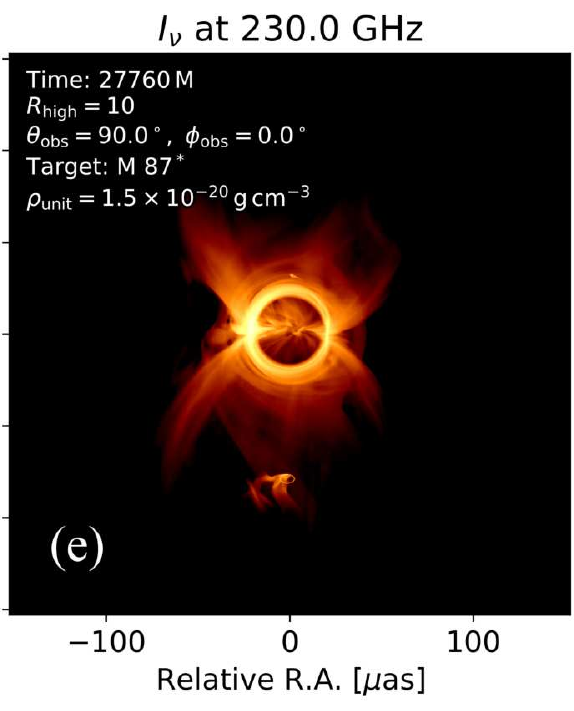}
\includegraphics[height=.33\linewidth]{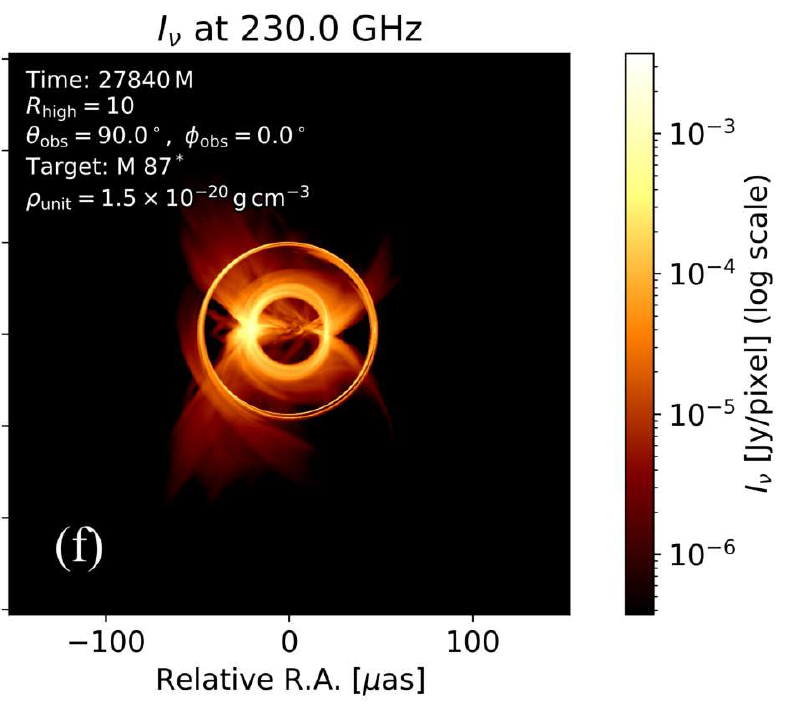}
\includegraphics[height=.33\linewidth]{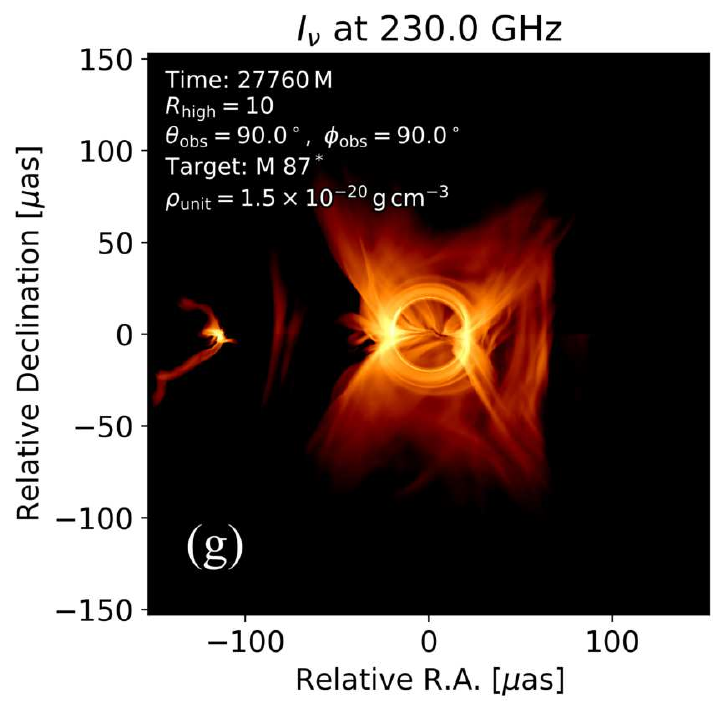}
\includegraphics[height=.33\linewidth]{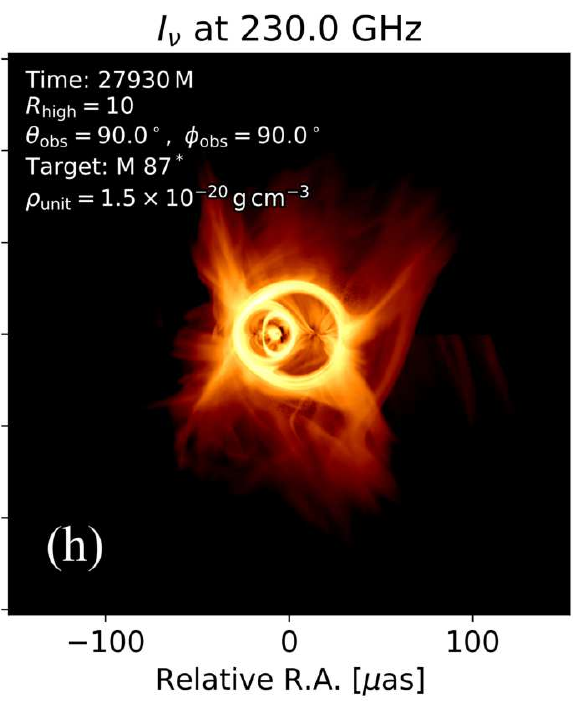}
\includegraphics[height=.33\linewidth]{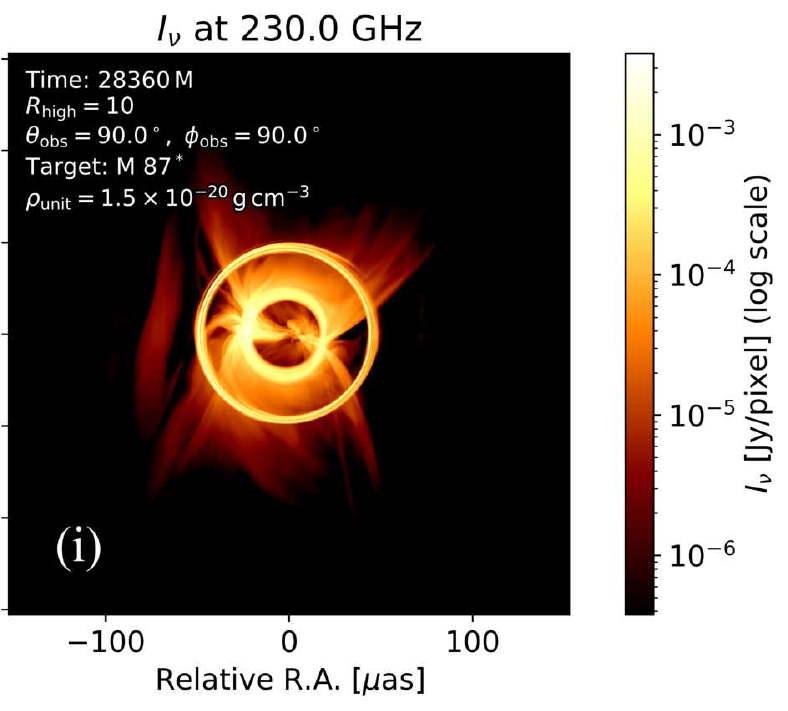}
\caption{Synthetic $230\,\mathrm{GHz}$ horizon-scale total-intensity maps ($I_\nu$). Panels (a)--(c) in the top row show three representative snapshots from Run~1 for $\phi_{\rm obs}=90^\circ$. Movie: \href{https://youtu.be/Z201nzWcnss}{https://youtu.be/Z201nzWcnss}. Panels (d)--(f) in the middle row show the same run at the same times but for $\phi_{\rm obs}=0^\circ$. Movie: \href{https://youtu.be/R_2J4vxRqgg}{https://youtu.be/R\_2J4vxRqgg}. Panels (g)--(i) in the bottom row display three representative Run~2 snapshots for $\phi_{\rm obs}=90^\circ$. Movie: \href{https://youtu.be/1NRrb8LMbWI}{https://youtu.be/1NRrb8LMbWI}. The color scale is logarithmic in $\rm Jy\,pixel^{-1}$, and each image is annotated with the snapshot time and the viewing geometry.}
\label{fig: BH_shadow}
\end{figure*}

The observable diagnostics of a $q=0.1$ companion are extracted from four GRMHD models plus their GRRT post-processing, summarized in Table~\ref{tab:sim_parameters}. Runs~1 and 2 were chosen as clean reference cases with a non-spinning primary ($a_1=0$), so they omit the strong LT-driven reorientation seen in \cite{2024ApJ...967...70R} and in our Run~3. In Run~1, the secondary follows a $90^\circ$ orbit and repeatedly pierces the torus. In Run~2, the same companion stays embedded in the flow on a coplanar orbit ($i=0^\circ$). Run~3 is deliberately more complex: both holes spin rapidly ($a_1=a_2=0.9375$), and the secondary starts on an eccentric orbit ($e=0.3$) tilted by $90^\circ$, which drives strong orbital precession and reorients the primary as well. Run~4 provides the single-hole baseline.

\subsection{MHD Evolution During the BH-Disk Interaction} \label{sec: BH-disk interaction}

Runs~1 and 2 both use a Schwarzschild primary, so the secondary remains on an orbit that stays nearly circular and does not reorient strongly. Figure~\ref{fig: mdot_run12} gathers the accretion-rate and magnetic-flux diagnostics measured on the two horizons. For each hole, the surface integral is evaluated in local Kerr-Schild coordinates including the orbital Lorentz boost. The mass inflow rate is
\begin{equation}
    \dot{M} = -\int_{S} \rho\, u^r \sqrt{-g}\, d\theta\, d\phi,
\end{equation}
where $\rho$ denotes the rest-mass density, $u$ is the radial component of the four-velocity, $\partial r/\partial x^i$ is transformed consistently into the lab frame, and $\sqrt{-g}\,d\theta\,d\phi$ is the invariant area element. The magnetic flux through the horizon is evaluated as
\begin{equation}
    \Phi_{\rm B} = \frac{1}{2} \int_{S} \left| b^r u^0 - b^0 u^r \right| 
    \sqrt{-g}\, d\theta\, d\phi,
\end{equation}
where $b^\mu$ is the fluid-frame four-magnetic field. It is related to the coordinate magnetic field $B^i$ through $b^0=u_iB^i$ and $b^i=(B^i+b^0u^i)/u^0$, so the integrand is equivalently the magnitude of the dual Faraday-tensor component $|{}^*F^{rt}|$. 

In Run~1, the companion leaves the primary's long-term average accretion rate close to the single-hole control Run~4, as panel (a) of Figure~\ref{fig: mdot_run12} shows. What changes more noticeably is the periodic modulation produced by repeated disk crossings. The zoomed view in panel (e) makes the sequence plain: each crossing triggers a small rise in $\dot{M}$, after which the flow relaxes before the next passage. Run~2 shows a different pattern. After the companion is introduced, the primary accretion rate initially increases, but the enhancement later disappears and becomes a mild deficit by $t=15,000\,M$. Early in the run, the embedded secondary perturbs the torus strongly enough to drive shocks and temporarily feed additional gas inward, much as in Run~1. At later times, however, it excavates its orbital path and weakens the local gas supply. The resulting depletion resembles the low-density cavities commonly found in circumbinary-disk calculations \citep[e.g.][]{2020ApJ...889..114M, 2019ApJ...871...84M, 2024ApJ...970..156D}. Run~3 departs farther from these Schwarzschild-primary cases because the rapidly spinning primary exhibits both lower accretion rates and weaker horizon-threading flux.

Panels (c) and (d) show the same diagnostics for the secondary. In Run~1 (black), both $\dot{M}$ and $\Phi_{\rm B}$ spike at each torus crossing. The Run~2 curves (red) are much less orderly and instead resemble ordinary MRI variability. Run~3 also begins at $90^\circ$ inclination, but the orbit later reorients strongly (Appendix~\ref{sec:orientation}) and alternates between more vertical and more coplanar episodes. Its bursts in $\dot{M}_{\rm BH2}$ are therefore not as clocklike as in Run~1, though they remain far more organized than the almost stochastic signal in Run~2.

Figure~\ref{fig: bh2_spectrum} recasts the same behavior in frequency space. Both Runs~1 and 2 are quasiperiodic, but for different physical reasons. In Run~1, the companion intersects the torus twice per orbit, so the dominant spectral feature appears at $2 f_{\rm orb}$. In Run~2, the strongest peak instead falls slightly below $f_{\rm orb}$. That shift is expected because a MAD flow is not axisymmetric: co-rotating magnetic-flux bundles and related structures move through the disk, so the embedded companion samples them at a beat frequency $f_{\rm obs}\approx f_{\rm orb}-f_{\rm flow}$. Run~3 shows one strong feature near $f_{\rm orb}$ and another near $2f_{\rm orb}$, consistent with a trajectory that alternates between embedded and disk-crossing phases.

\begin{figure*}
\centering 	
\includegraphics[width=.9\linewidth]{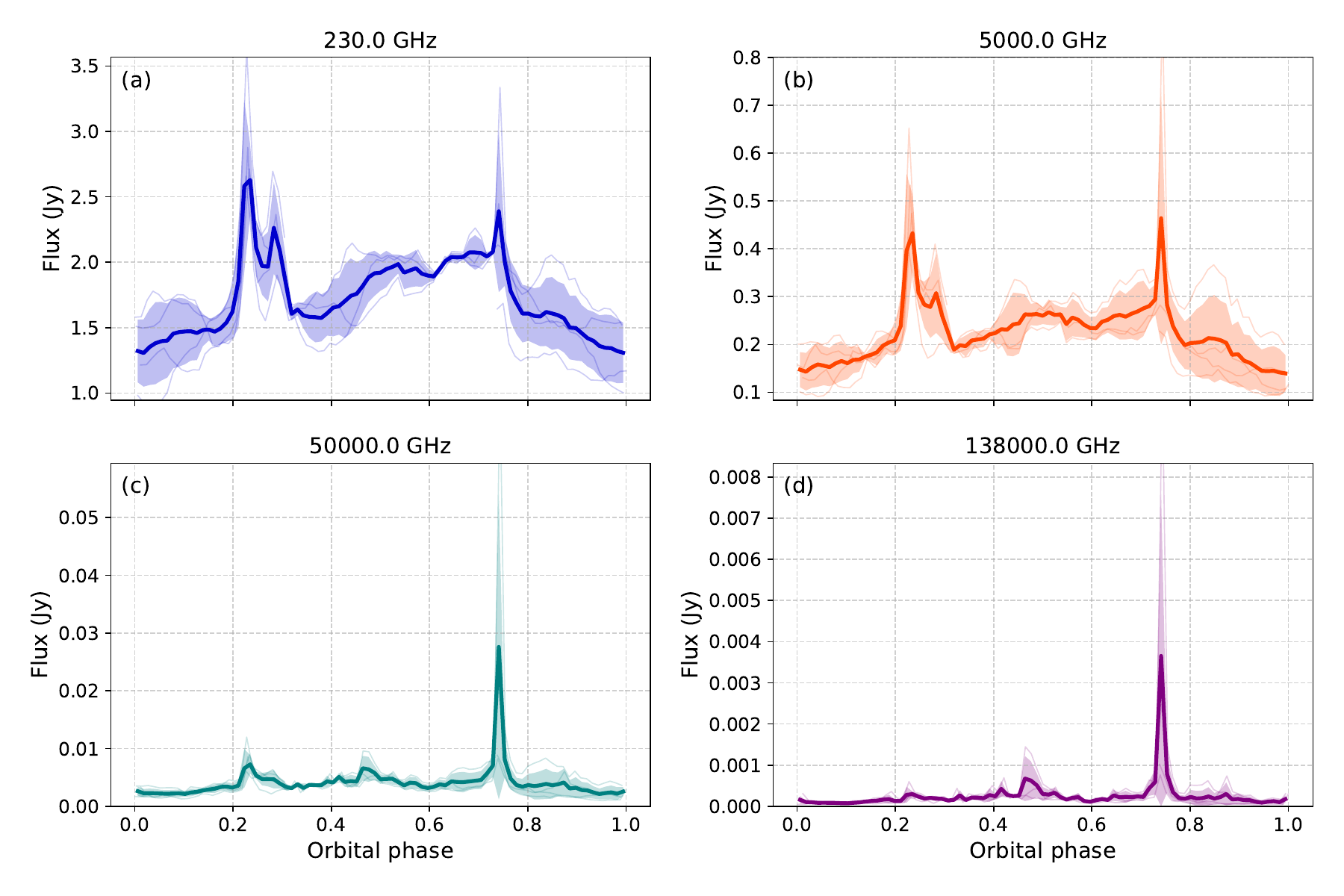}
\caption{Run~2 light curves folded by orbital phase at four observing frequencies: panel (a) $230\,\mathrm{GHz}$ (dark blue), panel (b) $5000\,\mathrm{GHz}$ (orange), panel (c) $5\times10^{4}\,\mathrm{GHz}$ (teal), and panel (d) $1.38\times10^{5}\,\mathrm{GHz}$ (purple). In every panel, the thick curve is the phase-averaged profile, the shaded region marks $\pm1\sigma$, and the faint lines show the orbit-to-orbit scatter.
}
\label{fig: folded_lc}
\end{figure*}

\begin{figure*}
\centering 	
\includegraphics[height=.33\linewidth]{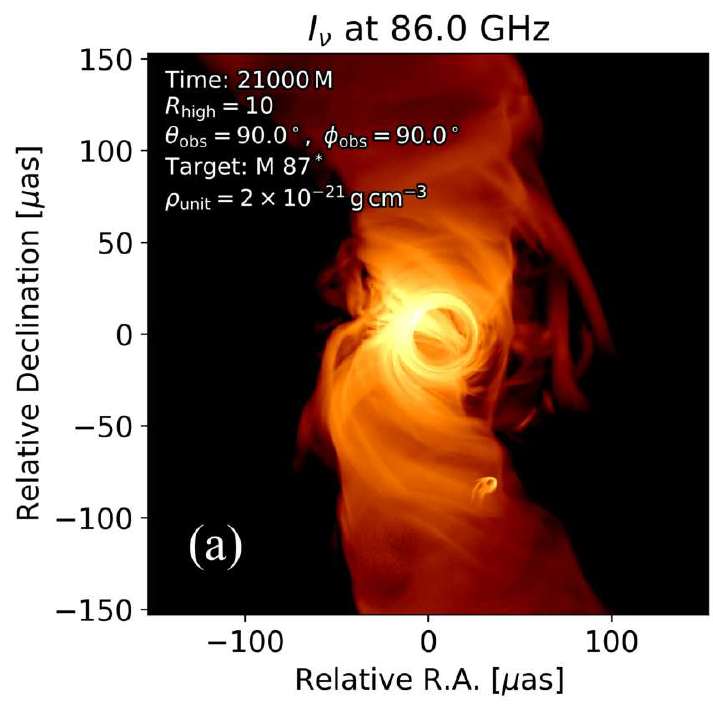}
\includegraphics[height=.33\linewidth]{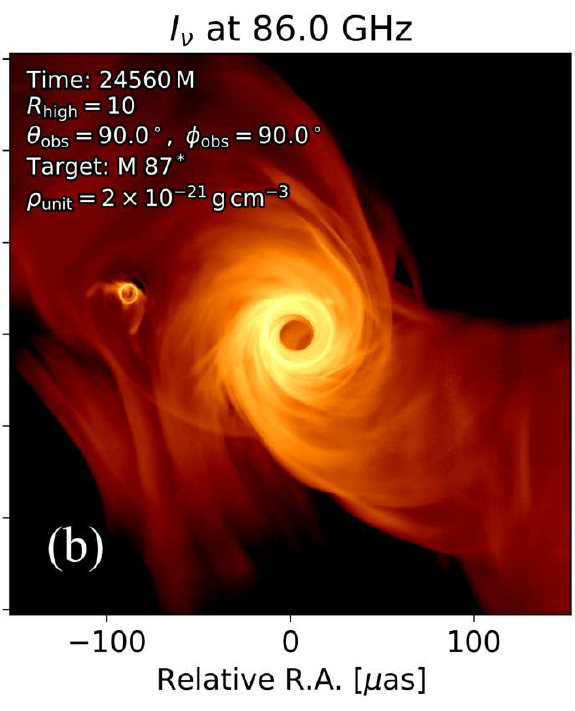}
\includegraphics[height=.33\linewidth]{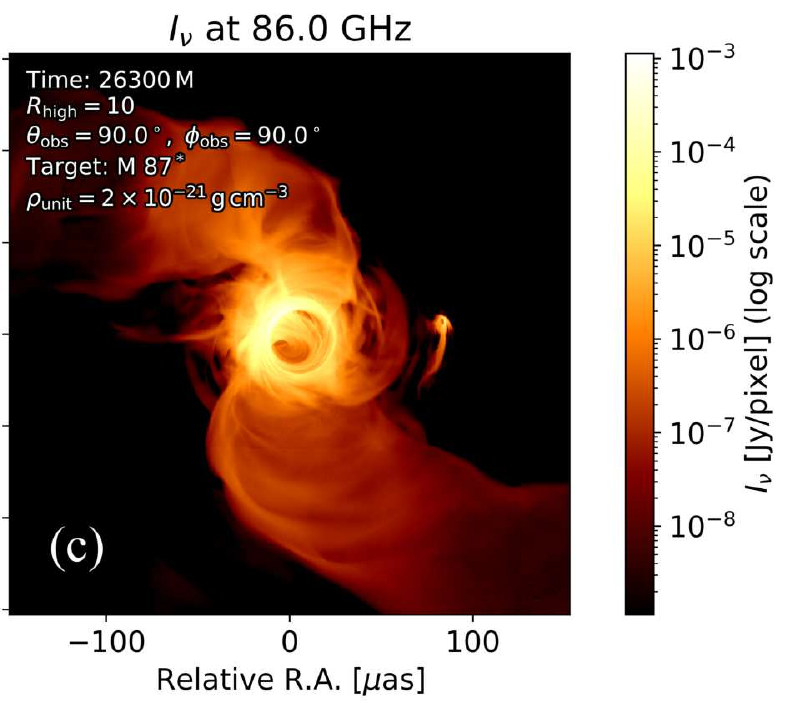}
\caption{Synthetic horizon-scale $86\,\mathrm{GHz}$ intensity maps ($I_\nu$) from Run~3 at three representative epochs: panel (a) $t=21,000\,M$, panel (b) $t=24,560\,M$, and panel (c) $t=26,300\,M$. The color scale is logarithmic in $\mathrm{Jy\,pixel^{-1}}$, and each frame records its snapshot time and viewing geometry. Movie: \href{https://youtu.be/dOQ8ppdLvWU}{https://youtu.be/dOQ8ppdLvWU}.}
\label{fig: precessing_jet}
\end{figure*}

\subsection{Observable Signatures of Runs 1 and 2}
\label{sec:obs_signiture}

I convert the BBH dynamics into observable quantities by post-processing the GRMHD snapshots with GRRT. The transfer calculation uses the same time-dependent BBH spacetime as the dynamical evolution. Figure~\ref{fig: lc_BH2} compares the resulting thermal-synchrotron light curves for Runs~1, 2, and 4 with the accretion history onto the secondary, $\dot{M}_{\rm BH2}$. 

The clearest example that luminosity does not simply mirror instantaneous accretion is provided by the vertical-impact configuration in panels (a) and (c) of Figure~\ref{fig: lc_BH2}. I therefore contrast two observing azimuths, $\phi_{\rm obs}=90^\circ$ and $\phi_{\rm obs}=0^\circ$, in order to disentangle intrinsic plasma variability from lensing. When $\phi_{\rm obs}=90^\circ$ (blue curves), the sightline is normal to the orbital plane and strong magnification is largely suppressed. When $\phi_{\rm obs}=0^\circ$ (purple curves), the observer, primary, and secondary approach near alignment twice during each orbit, so geometric amplification becomes important. Under the perpendicular view, the repeated spikes in $\dot{M}_{\rm BH2}$ produced by torus crossings do not translate into equally regular flares at $230\,\rm GHz$ or $138\,\rm THz$; instead, the curves are dominated by the irregular variability of the MAD background. To gauge how much of the signal comes from local shocks, I remove the gas inside $r<10\,r_{\rm g}$ around the secondary and plot the corresponding light curves as green dotted lines in panels (a) and (c). Only at a few epochs, such as $t=27,840\,M$, does that excision erase a narrow NIR spike, indicating that shocks can occasionally control the flare. Most of the time, however, the change is small, so shocks are not the main driver of the variability.

The behavior is different when the observer looks nearly along the binary axis. In that orientation, the curves develop brief, high-contrast spikes, most noticeably in the NIR, and panel c contains three prominent examples superposed on the broader MAD fluctuations. Similar short brightenings also occur in the coplanar configuration, shown in panels (b) and (d), at both sub-millimeter and NIR frequencies. These events are most naturally interpreted as self-lensing episodes that occur when the observer and the two holes are almost colinear. The magnification is largest when the primary passes into the foreground, because it is the stronger lens and can amplify the compact emitting region around the secondary. The observed light curve is then a sharp lensing flare riding on top of a much broader turbulence-dominated background.

Run~2 differs from Run~1 because it does not develop the same strong localized shocks (Appendix~\ref{sec: shock_3D}). Its spike-like brightenings are therefore more plausibly attributed to geometry than to shock dissipation. Favorable alignments temporarily magnify the compact emission, and the secondary may also stay hotter and denser than the gas near the primary because it preferentially accretes from the circumbinary flow \citep{2023MNRAS.518.5059S}. If that picture is correct, then lensing flares should be more common and easier to identify in Run~2 than the less frequent shock-powered events found in Run~1.

\subsection{Synthetic Horizon-Scale Views of the BBH System}

Figure~\ref{fig: BH_shadow} shows an illustrative orbital sequence of synthetic horizon-scale images. In the top row, panels (a)--(c), the secondary approaches the torus and then penetrates it. Because the binary has a small mass ratio, $q=0.1$, the secondary shadow is far smaller than the primary one, but the companion still appears as a compact luminous source. After entering denser gas, it excites a bow shock followed by a wake. As discussed in Appendix~\ref{sec: shock_3D}, that structure is produced when the companion gravitationally focuses the surrounding plasma, thereby compressing and heating it. Even so, the secondary remains subdominant to the primary in the sub-millimeter images, consistent with the light curves in Figure~\ref{fig: lc_BH2}a,c. 

Panels (d)--(f) display the same run from the azimuth $\phi_{\rm obs}=0^\circ$. At several orbital phases, that line of sight brings the observer and the two black holes close to colinearity, which strongly enhances the lensing signal. One such example occurs at $t=27,840\,M$, where magnification of the compact binary emission produces a conspicuous transient brightening. This is the same episode that appears as the bright NIR flare in Figure~\ref{fig: lc_BH2}c.

The lower row, panels (g)--(i), gives the corresponding image sequence for Run~2. Because the orbit is coplanar, the secondary remains inside the torus and alternates between foreground and background locations relative to the primary. It lies in front in panel h, whereas by panel i it has passed behind the primary and brightened. Each swap in foreground/background ordering creates another favorable lensing geometry, distorts the image morphology, and intermittently enhances the compact emission.

The phase-folded light curves in Figure~\ref{fig: folded_lc} make this geometric behavior easier to interpret. At 230 and $5000\,\rm GHz$, the primary flow supplies most of the radiation. When the secondary crosses the observer's sightline to the primary near phase $\simeq0.2$--$0.3$, the magnification rises quickly and the secondary lenses the primary photon ring, yielding the double-peaked brightening in panels (a) and (b). The same configuration is visible in Figure~\ref{fig: BH_shadow}h, where the primary ring is strongly deformed. At higher frequencies the situation reverses: the hotter plasma around the secondary dominates while the primary contribution weakens. The main peak therefore moves to the phase near $\sim0.72$, when the secondary lies behind the primary and the primary lenses the compact secondary source into the Einstein-ring-like structure seen in Figure~\ref{fig: BH_shadow}i.

Panel g of Figure~\ref{fig: BH_shadow} further hints that the secondary may be capable of launching a mini-jet of its own. As the companion travels through the disk, that outflow becomes stretched and twisted and may in turn react back on the ambient gas. Previous hydrodynamic studies show that sufficiently strong outflows can disrupt the gravitational wake and reduce, or even reverse, the effective drag force \citep{2020MNRAS.494.2327L}; other work suggests that interacting outflows in binaries can exert a positive net torque and promote orbital expansion \citep{2022ApJ...932..108W}. By the same logic, a time-dependent mini-jet from the secondary could alter the disk torques and partially counteract inward migration, although quantifying that effect lies beyond the scope of this thesis.

\begin{figure*}
\centering 	
\includegraphics[width=.7\linewidth]{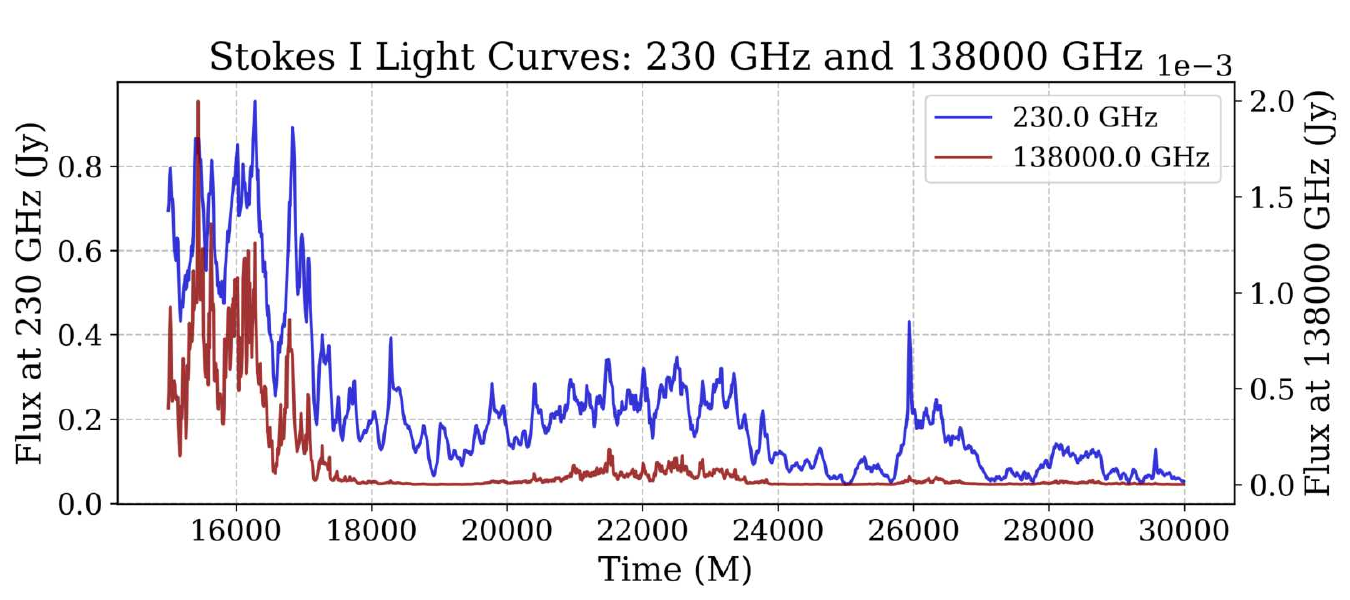}
\caption{Stokes-$I$ light curves for Run~3 at $230\,\mathrm{GHz}$ (blue; left axis) and $1.38\times10^{5}\,\mathrm{GHz}$ (brown; right axis).
}
\label{fig: lc_run3}
\end{figure*}

\subsection{Morphology and Observables of the Precessing Binary}
\label{sec:run4_results}
Run~3 occupies the most strongly time-dependent end of the model set. As found in \cite{2024ApJ...967...70R}, spin-orbit coupling gradually changes the direction of the primary spin axis, and that evolving orientation warps the inner flow while redirecting the jet. In this configuration both holes spin rapidly, and the binary orbit is simultaneously eccentric and highly inclined.

Compared with Runs~1 and 2, Run~3 is distinguished by a primary whose spin is large enough for frame dragging to impose strong LT torques on the secondary orbit. At the same time, the binary coupling also drives secular changes in the primary spin itself (Appendix~\ref{sec:orientation}). The jet therefore does not merely precess as a rigid structure; it twists while its launch direction changes. Figure~\ref{fig: precessing_jet} shows this directly in three synthetic $86\,{\rm GHz}$ snapshots: the nozzle wanders, the jet bends, and the brightness pattern evolves asymmetrically from epoch to epoch. Such behavior offers a natural route to the twisted VLBI jet morphologies reported for systems such as OJ~287 \citep{2022ApJ...932...72Z, 2026A&A...705A..23G}. 

The corresponding Run~3 light curves are plotted in Figure~\ref{fig: lc_run3} at $230\,{\rm GHz}$ and $138\,{\rm THz}$. Although the inner flow and jet orientation vary strongly, the thermal synchrotron output does not show the clean quasi-periodic flares seen in the simpler Runs~1 and 2. Instead, stochastic fluctuations dominate most of the signal. A recognizable self-lensing event appears only near $t\approx26{,}000\,M$, when the secondary moves behind the primary and approaches line-of-sight conjunction. Shock-driven brightenings are still present at other times, as in Run~1, but in Run~3 they are typically too weak to stand out as isolated flares.

\section{Discussion and Key Findings}

This chapter combines global 3D GRMHD calculations with GRRT post-processing to determine how an embedded SMBBH in a MAD flow may appear observationally. The binary spacetime is approximate, but it still retains the horizon-scale plasma effects most relevant for a moderate mass ratio, $q=0.1$, including magnetic-flux transport and jet launching. The main conclusions are:
\begin{enumerate}
    \item Strong modulation in the accretion flow does not by itself guarantee an equally regular flare pattern. This is clearest in the vertical-impact case of Run~1, where the observed light curve depends strongly on viewing geometry. In most orientations the variability is dominated by stochastic MAD turbulence, and obvious periodic signatures appear only when lensing is especially favorable or a shock is unusually strong.
    
    \item The orbital geometry leaves a direct imprint on the observables. Coplanar systems such as Run~2 repeatedly generate self-lensing flares. At $230\,\rm GHz$, the primary emission dominates and is amplified by the secondary, whereas in the NIR the hotter and more compact plasma around the secondary becomes dominant and is instead lensed by the primary into sharp Einstein-ring-like events. Inclined encounters such as Run~1 can drive strong bow shocks, but those shocks become obvious flares only for certain sightlines.
    
    \item Spin-orbit coupling can generate precessing and twisted jets. In the high-spin eccentric configuration of Run~3, the companion exerts enough torque to tilt and reorient the primary spin axis. The jet then develops a wobbling base together with a twisted large-scale morphology. This provides a physical explanation for helical structures like those inferred from VLBI observations of systems such as OJ~287, without requiring an ad hoc precessing nozzle.
    
    \item The secondary may launch a dynamically important mini-jet. Even while embedded in the primary accretion flow, the companion can drive its own relativistic outflow. That outflow interacts with the nearby plasma and could alter both the effective dynamical friction and the migration torque, although a dedicated calculation of this feedback is left for future work.
\end{enumerate}

The broader lesson is that identifying SMBBHs cannot be reduced to a simple search for periodic light curves. Strong-field lensing, shock dissipation, and relativistic jet precession together create a far richer set of signatures than periodicity alone. Distinguishing those binary-specific signals from the intrinsic variability of single-black-hole AGN will therefore require both horizon-scale VLBI measurements, including EHT observations, and dedicated time-domain monitoring.

This analysis also has several important limitations. First, the transfer calculation includes only thermal synchrotron radiation. That approximation is suitable from the sub-millimeter to the NIR, but it neglects inverse Compton scattering. Since the strongest shocks in the vertical-impact configuration are plausible high-energy emitters, IC physics will be needed to predict the corresponding X-ray and gamma-ray counterparts. Second, the gas is modeled as a geometrically thick, radiatively inefficient flow. Systems such as OJ~287, with $\dot{M}\sim0.1$--$0.01\,\dot{M}_{\rm Edd}$ \citep{2019ApJ...882...88V}, are often interpreted instead in terms of a geometrically thin, optically thick disk, where both shock propagation and thermal structure may be very different. Third, most of the discussion concentrates on the edge-on orientation ($i=90^\circ$), because that viewing angle maximizes lensing and obscuration. A realistic estimate of detectability across the full SMBBH population will therefore require a dedicated inclination survey.

\newpage
{
\section*{Supplementary Appendix}  
\setcounter{section}{0}
\renewcommand{\thesection}{\thechapter.\Alph{section}}
\renewcommand{\theHsection}{appendix.\thechapter.\Alph{section}}
\renewcommand{\theHsubsection}{appendix.\thechapter.\Alph{section}.\arabic{subsection}}

\section{Technical Setup and Additional Prescriptions}
\label{app:sim_setup}

\subsection{PN Orbit and Spin Evolution}
\label{app:pn}
The BBH trajectories are computed with post-Newtonian equations of motion that include both conservative contributions and radiation reaction, following \citep{2014LRR....17....2B, 2015JHEP...09..219L}:
\begin{equation}
    \dot{\mathbf{r}} = \mathbf{v}, \quad \dot{\mathbf{S}}_A = \boldsymbol{\Omega}_A \times \mathbf{S}_A, \quad A \in \{1, 2\}.
\end{equation}
In this notation, $\mathbf{r}$ is the relative separation, $\mathbf{v}$ is the orbital velocity, and $\mathbf{S}_A$ is the spin of black hole $A$. The additional kinematic variables are
\begin{equation}
    r = |\mathbf{r}|, \quad \mathbf{n} = \frac{\mathbf{r}}{r}, \quad v^2 = \mathbf{v} \cdot \mathbf{v}, \quad \dot{r} = \mathbf{n} \cdot \mathbf{v}.
\end{equation}

For later convenience, I split the acceleration into four pieces by first separating non-spinning and spin-dependent terms and then dividing each of those into conservative and radiation-reaction contributions:
\begin{equation}
    \mathbf{a} = \mathbf{a}_{\rm ns}^{\rm cons} + \mathbf{a}_{\rm ns}^{\rm RR} + \mathbf{a}_{\rm spin}^{\rm cons} + \mathbf{a}_{\rm spin}^{\rm RR}.
\end{equation}
Each contribution is then expanded in PN order as
\begin{align}
    \mathbf{a}_{\rm ns}^{\rm cons} &= \mathbf{a}_{0{\rm PN}} + \mathbf{a}_{1{\rm PN}} + \mathbf{a}_{2{\rm PN}}, \\
    \mathbf{a}_{\rm ns}^{\rm RR} &= \mathbf{a}_{2.5{\rm PN}}^{\rm RR} + \mathbf{a}_{3.5{\rm PN}}^{\rm RR} + \mathbf{a}_{4{\rm PN}}^{\rm RR}, \\
    \mathbf{a}_{\rm spin}^{\rm cons} &= \mathbf{a}_{\rm SO}^{1.5{\rm PN}} + \mathbf{a}_{\rm SO}^{2.5{\rm PN}} + \mathbf{a}_{\rm SS}^{2{\rm PN}} + \mathbf{a}_{\rm SS}^{3{\rm PN}}, \\
    \mathbf{a}_{\rm spin}^{\rm RR} &= \mathbf{a}_{\rm SO}^{2.5{\rm PN,RR}} + \mathbf{a}_{\rm SO}^{3.5{\rm PN,RR}}.
\end{align}
In this notation, $\mathbf{a}_{\rm spin}^{\rm cons}$ and $\mathbf{a}_{\rm spin}^{\rm RR}$ collect the spin-orbit (SO) and spin-spin (SS) contributions. The spin vectors precess with angular frequency $\boldsymbol{\Omega}_A$, which I write as
\begin{equation}
    \boldsymbol{\Omega}_A = \boldsymbol{\Omega}_{A, \rm SO}^{\rm NLO} + \boldsymbol{\Omega}_{A, \rm SS}^{\rm LO} + \boldsymbol{\Omega}_{A, \rm SS}^{\rm NLO} + \boldsymbol{\Omega}_{A, \rm QM},
\end{equation}
This decomposition contains geodetic and LT precession through $\boldsymbol{\Omega}_{A, \rm SO}$, spin-spin couplings through $\boldsymbol{\Omega}_{A, \rm SS}$, and quadrupole-monopole self-spin coupling through $\boldsymbol{\Omega}_{A, \rm QM}$. Here LO denotes the leading-order contribution and NLO the next-to-leading correction.

For quasi-circular orbits, the initial tangential speed $v_\phi$ is obtained by imposing radial force balance using only the conservative sector:
\begin{equation}
    a_r^{\rm cons} + \frac{v_\phi^2}{r} = 0.
\end{equation}
The resulting PN system is then integrated forward in time so that the orbit and spin vectors evolve together.

\subsection{Superposed Kerr-Schild Metric for the Binary}
\label{app:sks}

I adopt the superposed Kerr-Schild construction \citep{2021ApJ...913...16L}, in which the covariant metric is written as a Minkowski background plus Kerr-Schild perturbations sourced by two boosted Kerr black holes,
\begin{equation}
g_{\mu\nu}
=
\eta_{\mu\nu}
+
\sum_{A=1}^{2} H_A\,l^{(A)}_\mu l^{(A)}_\nu,
\end{equation}
where $H_A$ is the Kerr-Schild scalar associated with black hole $A$ and $l^{(A)}_\mu$ is the corresponding ingoing null covector. Orbital motion is included by Lorentz-boosting the Kerr-Schild data for each hole from its instantaneous comoving frame into the simulation frame,
\begin{equation}
l^{(A)}_\mu
=
(\Lambda_A)_\mu{}^\alpha\,l^{(A,{\rm rest})}_\alpha,
\end{equation}
where $\Lambda_A$ is the Lorentz transformation built from the instantaneous PN velocity of that hole.

\subsection{GRRT Formulation and Electron Thermodynamics}
\label{app:grrt}

The GRRT post-processing is performed with \texttt{BHOSS} \citep{2012A&A...545A..13Y, 2020IAUS..342....9Y} in the fast-light approximation. Along each null geodesic, the general relativistic transfer equation is solved,
\begin{equation}
\frac{d\mathbf{S}}{d\lambda}=\mathbf{j}-\mathbf{K}\mathbf{S},
\end{equation}
where $\mathbf{S}=(I,Q,U,V)^{\top}$ is the Stokes vector, $\mathbf{j}$ is the emissivity, and $\mathbf{K}$ contains the absorption and Faraday coefficients. The electron temperature is specified through the $R$-$\beta$ prescription \citep{Moscibrodzka2016},
\begin{equation}
\frac{T_i}{T_e}
=
R_{\rm high}\frac{\beta^2}{1+\beta^2}
+
R_{\rm low}\frac{1}{1+\beta^2},
\end{equation}
where $\beta$ is the plasma beta. In the fiducial models I adopt $(R_{\rm high},R_{\rm low})=(10,1)$, guided by comparisons with two-temperature GRMHD calculations \citep{2023MNRAS.522.2307J, 2024A&A...688A..82J}.

\section{Additional Appendix Material}

\subsection{Additional Flow Slices and Shock Structure}\label{sec: shock_3D}

\begin{figure*}
\centering 	
\includegraphics[height=\linewidth]{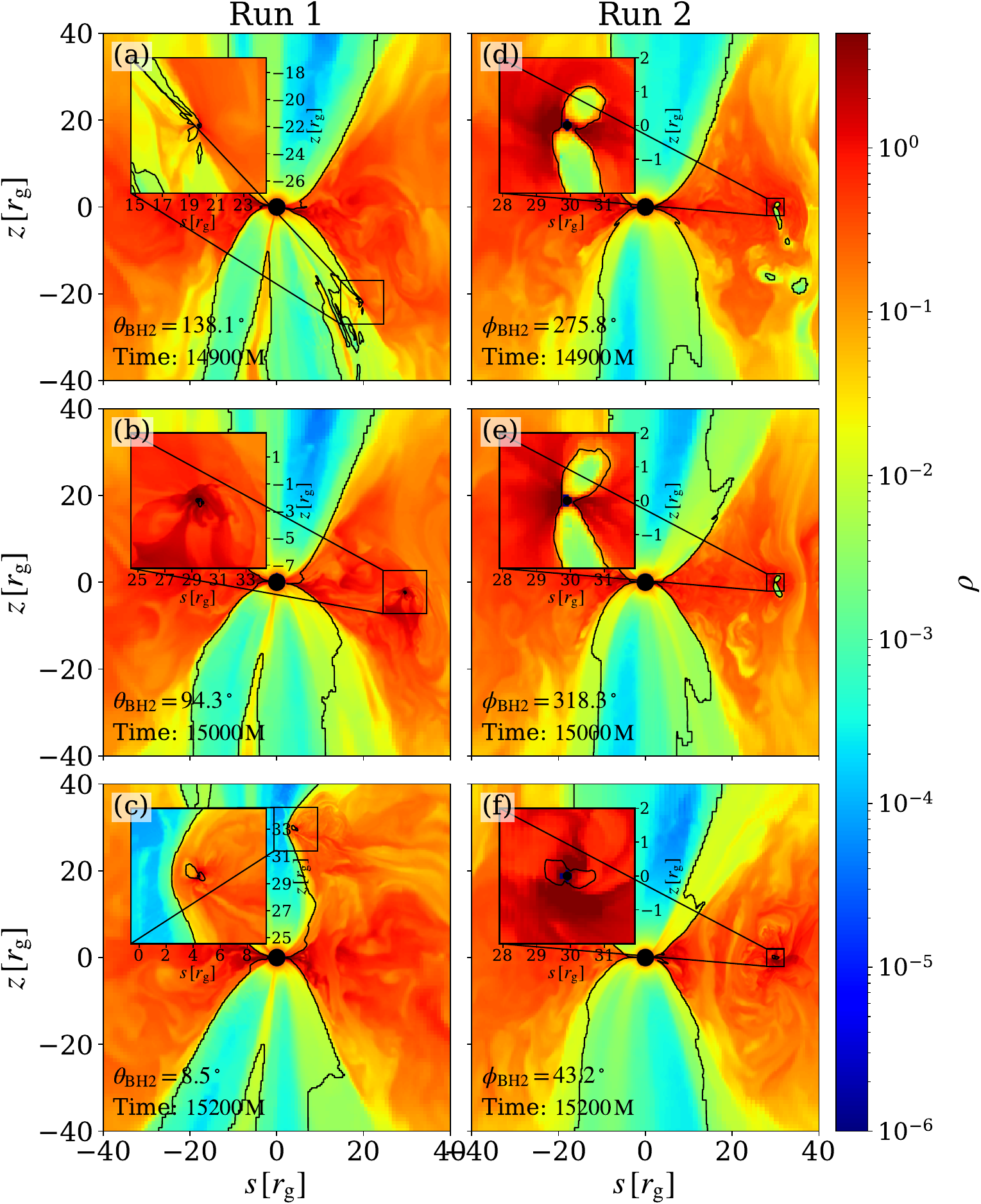}
\caption{Time sequence of gas-density maps ($\rho$) for Run~1 in the left column and Run~2 in the right column at $t=14,900\,M$, $15,000\,M$, and $15,200\,M$. Each panel is a vertical $(s,z)$ slice centered on the secondary. The coordinate $s$ follows the instantaneous BH-BH separation vector, so the primary remains fixed at $s=0$, while $z$ is the vertical height in units of $r_g$. Density is shown on a logarithmic color scale, the solid black contour indicates $\sigma=1$, and the insets zoom in on the immediate environment of the secondary. Every frame is annotated by the simulation time and by the instantaneous orbital phase, $\theta_{\text{BH2}}$ for Run~1 and $\phi_{\text{BH2}}$ for Run~2.}
\label{fig: 2D_slice}
\end{figure*}

\begin{figure*}
\centering 	
\includegraphics[height=.41\linewidth]{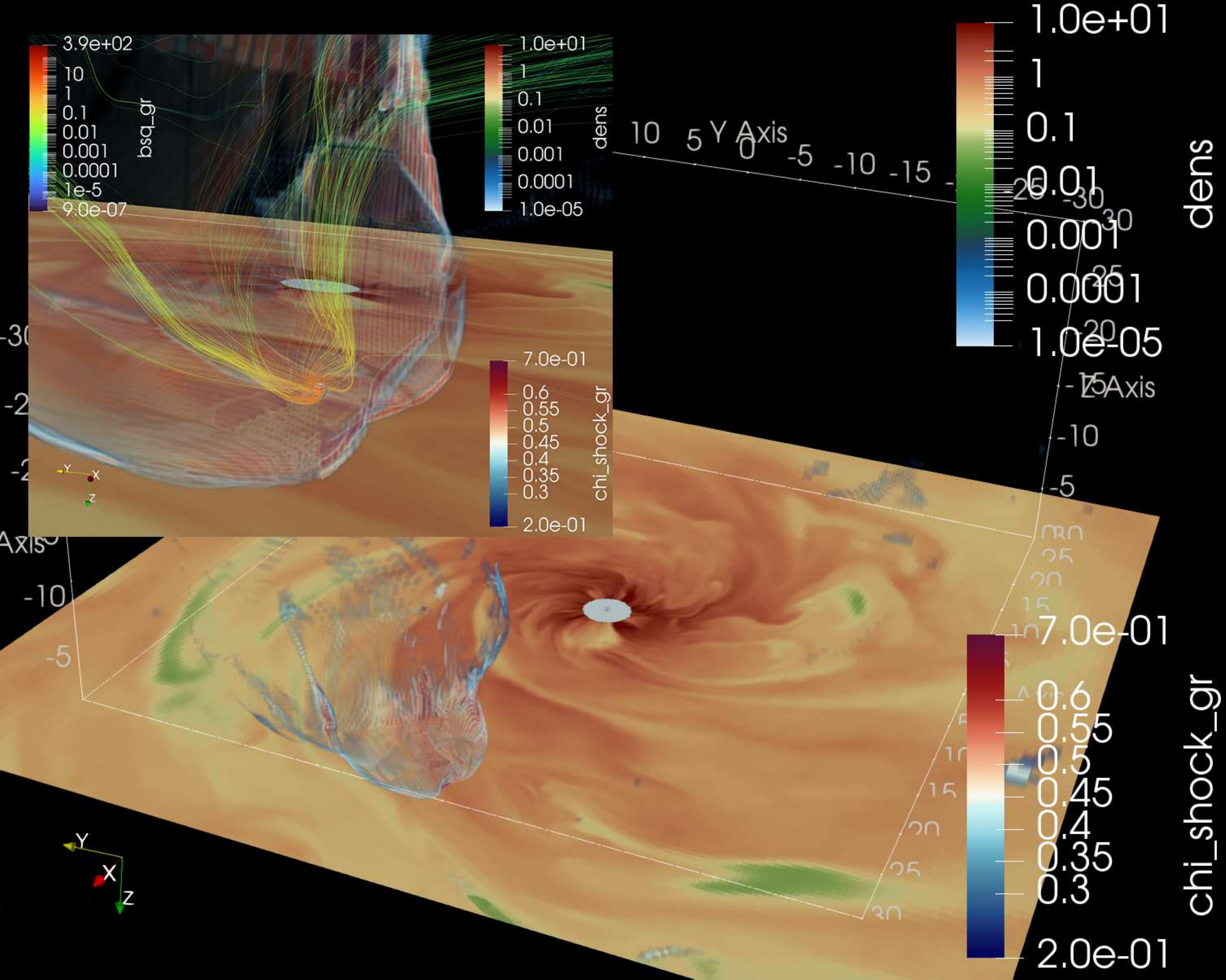}
\includegraphics[height=.41\linewidth]{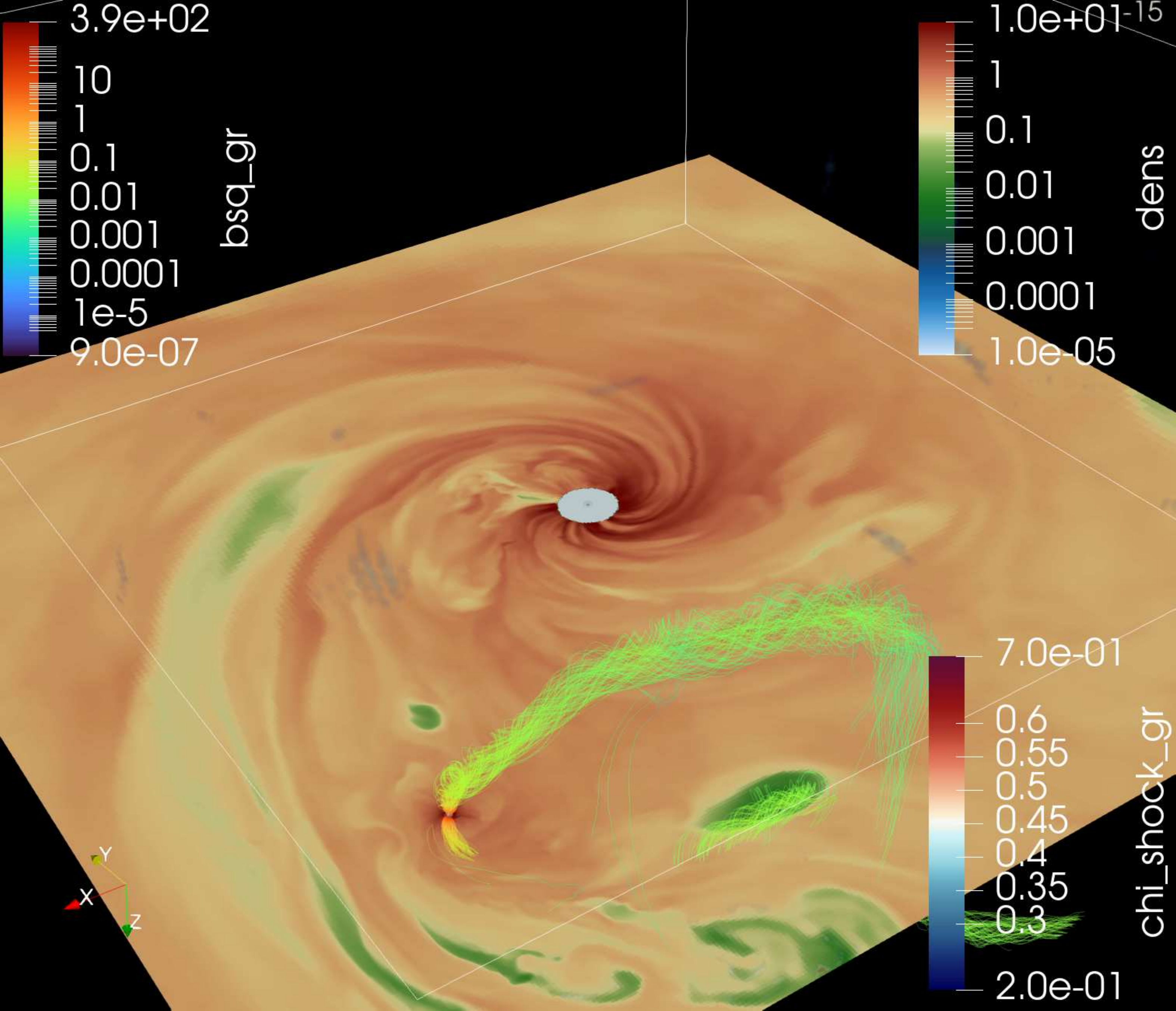}
\caption{Three-dimensional rendering of the shock diagnostic and surrounding plasma near the secondary black hole. Each panel combines a density slice with a volume rendering of the shock indicator $\chi_{\rm shock}$. The left panel corresponds to a representative Run~1 snapshot, whereas the right panel shows a representative Run~2 snapshot. Insets and color bars indicate the plotted ranges of density and $\chi_{\rm shock}$, together with the magnetic-field overlay where applicable.}
\label{fig: shock_bfield}
\end{figure*}

Figure~\ref{fig: 2D_slice} shows a time sequence of vertical density slices for Run~1 and Run~2. For each frame, the slice lies in the instantaneous plane spanned by the BH-BH separation vector and the global $z$-axis. Because that plane co-rotates with the binary, the horizontal coordinate $s$ continues to follow the line joining the two black holes. The primary is therefore fixed at $s=0$, whereas the secondary appears at positive $s$ and is tagged by its instantaneous phase, $\theta_{\rm BH2}$ in Run~1 and $\phi_{\rm BH2}$ in Run~2. The black contour marks $\sigma=1$, which I use as an operational boundary for the primary jet and for any mini-jet driven by the secondary. In Run~1, the sequence captures a transient disk-impact episode: the secondary passes through dense gas, emerges from the torus, strongly compresses the surrounding plasma, and carries with it a compact overdense clump that is later stripped as it encounters the primary jet. Over the same time interval, Run~2 keeps the secondary embedded inside the torus, so the companion experiences a steadier fuel supply. As a result, the slices show mini-jet activity that is more persistent, though still intermittent, and that weakens again by $t=15,200\,M$.  

Figure~\ref{fig: shock_bfield} offers a complementary three-dimensional view of the same interaction, emphasizing both compression and magnetic structure around the companion. I quantify the compression with the dimensionless shock diagnostic
\begin{equation}
\chi_{\rm shock} \equiv \frac{\max\left(-\nabla_i v^i,0\right)}{\ell_{\rm proper}/c_{\rm fast}},
\end{equation}
where $\nabla_i v^i$ is the covariant velocity divergence, $\ell_{\rm proper}$ is the local proper cell size, and $c_{\rm fast}$ is the local fast magnetosonic speed. Operationally, $\chi_{\rm shock}$ compares the local compression time with the time needed for a fast-mode wave to cross the cell. Regions with $\chi_{\rm shock}\ll1$ are only mildly compressed, whereas values $\chi_{\rm shock}\gtrsim1$ identify genuinely shock-like behavior.

In the Run~1 snapshot (Fig.~\ref{fig: shock_bfield}a), the largest values of $\chi_{\rm shock}$ are confined to a compact zone centered on the secondary and aligned with the dominant disturbance. The zoomed panel and the magnetic-field overlay show strong bending and draping of the field lines, as expected when the secondary undergoes a violent vertical encounter with torus gas under flux freezing. In the Run~2 snapshot (Fig.~\ref{fig: shock_bfield}b), by contrast, the high-$\chi_{\rm shock}$ volume is much less extended, showing that the flow around the companion is far less compressive. There the secondary co-moves much more closely with the surrounding disk material, which reduces the relative speed and suppresses strong shocks. Even so, the continued fueling of the rapidly spinning secondary ($a=0.9375$) still permits intermittent mini-jet launching, and that outflow is visibly twisted by the ambient torus and the large-scale magnetic field.

\subsection{Evolution of Orbit, Spin, and Torus Orientation}
\label{sec:orientation}

\begin{figure}
    \centering
    \includegraphics[width=0.45\linewidth]{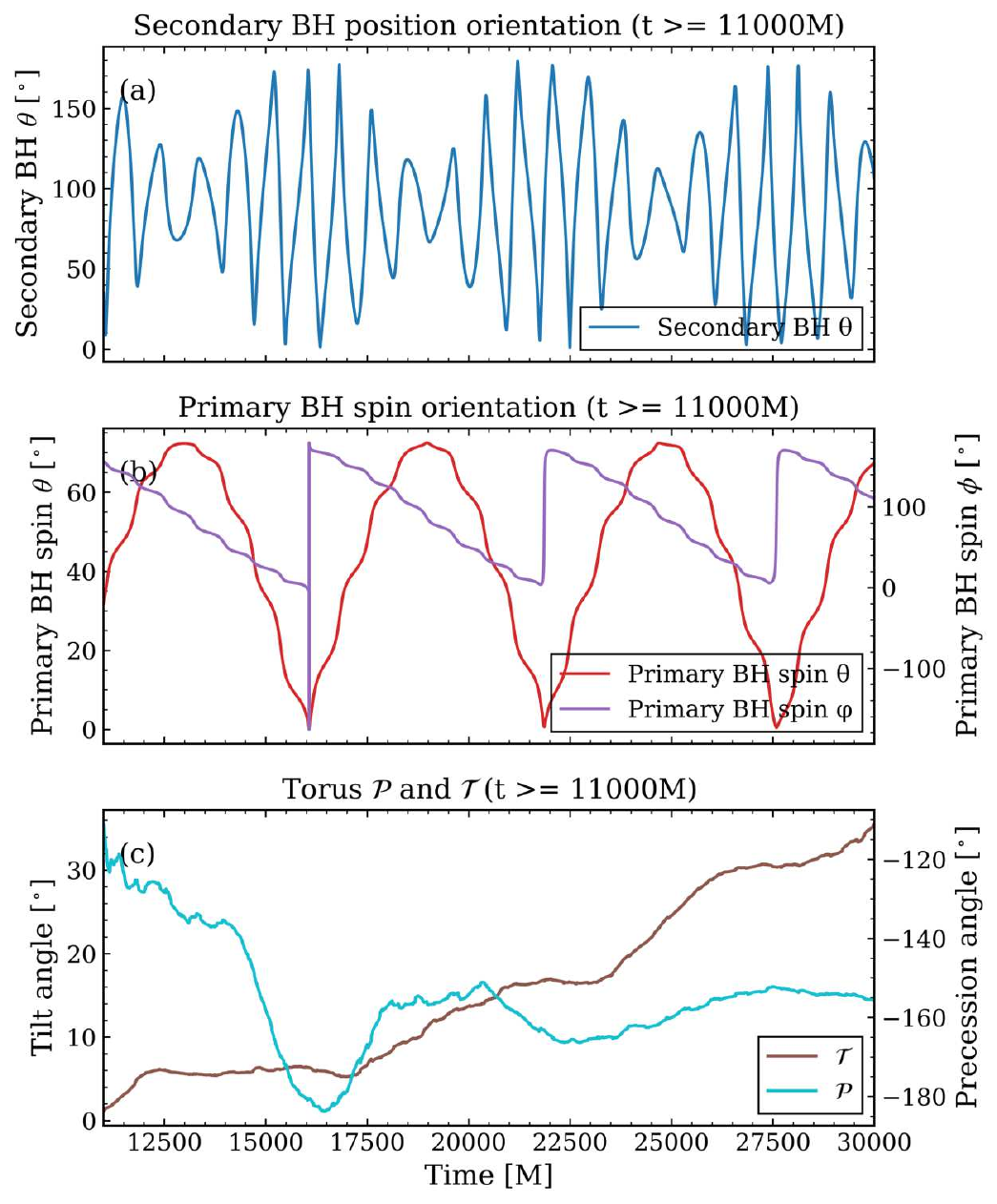}
    \includegraphics[width=0.53\linewidth]{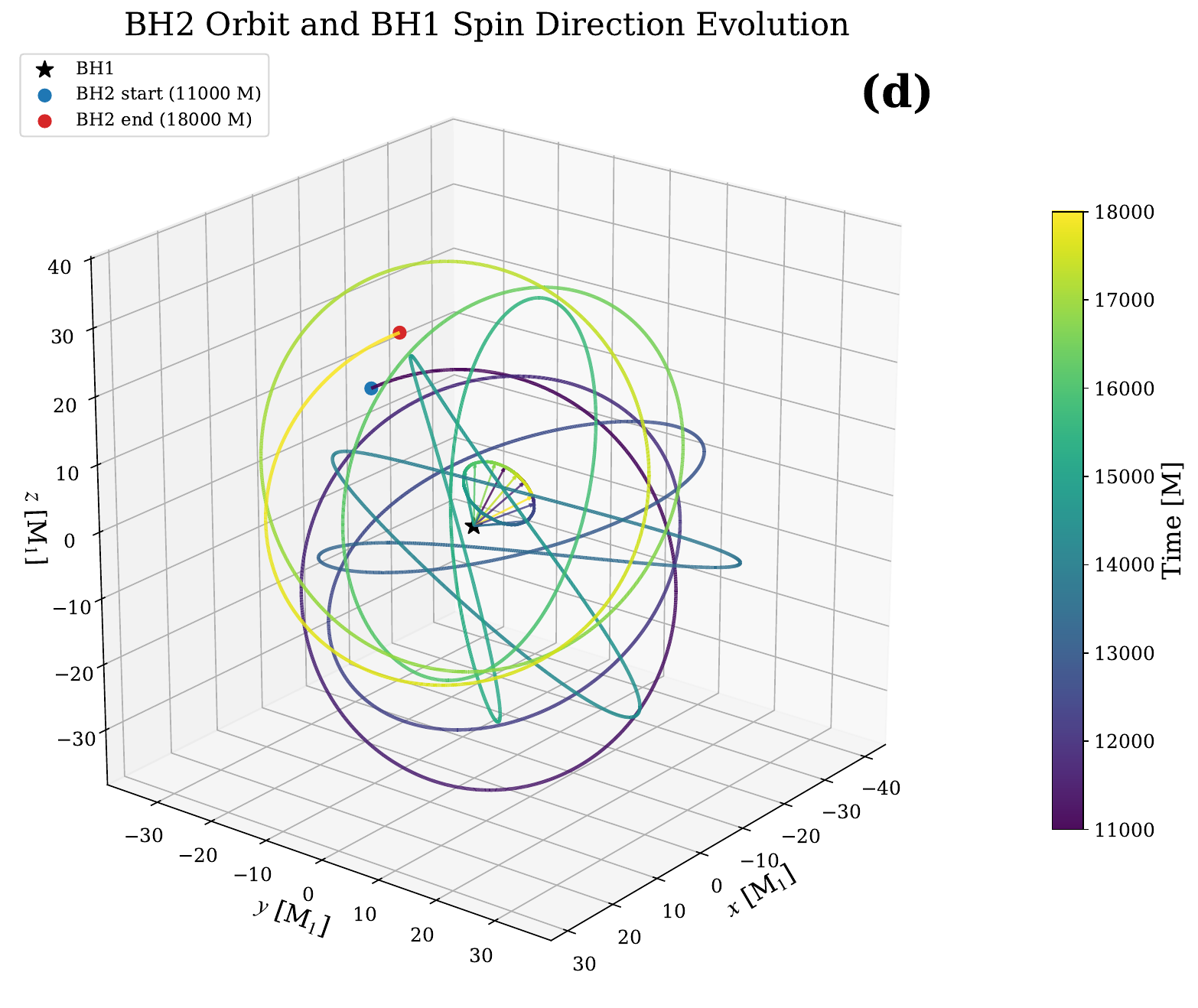}
    \caption{Orientation evolution in Run~3 for $t\ge 11{,}000\,M$. Panel (a) shows the polar angle $\theta$ of the secondary position vector. Panel (b) gives the primary-spin orientation through $\theta_{\rm spin}$ (red; left axis) and $\phi_{\rm spin}$ (purple; right axis). Panel (c) shows the torus angular-momentum direction through the tilt $\mathcal{T}$ (brown; left axis) and precession angle $\mathcal{P}$ (cyan; right axis), measured following \citep{2005ApJ...623..347F, 2025ApJ...995..112J}. Panel (d) combines the orbital and spin evolution in three dimensions, with the colored trajectory tracing the secondary motion and the short vector attached to the primary marking the instantaneous primary-spin direction.}
    \label{fig:orientation}
\end{figure}

To understand where the precessing jet morphology of Run~3 comes from, I follow the evolution of the black-hole positions and spin directions together with the orientation of the torus. Figure~\ref{fig:orientation} compiles the relevant orbital, spin, and torus angles.

Panel (a) follows the polar angle $\theta$ of the secondary position vector. The system starts with the companion orbit tilted by $90^\circ$ relative to the torus midplane. After $t\geq11,000\,M$, $\theta$ undergoes large oscillations, which indicate strong nodal precession of the orbital plane driven by LT torques from the spinning primary. The orbit therefore alternates between configurations reminiscent of vertical torus crossings and phases that are closer to coplanar motion.

Panel (b) shows the primary-spin evolution through $\theta_{\rm spin}$ and $\phi_{\rm spin}$. Although the primary outweighs both the disk and the companion, the orbiting secondary still exerts a cumulative torque that produces a secular drift of the spin axis. The primary spin thus precesses, and the two angular coordinates vary in a coupled fashion.

Panel (c) characterizes the torus orientation using the tilt angle $\mathcal{T}$ and the precession angle $\mathcal{P}$, measured with the same procedure as in \citep{2025ApJ...995..112J}. The tilt increases secularly from only a few degrees at $t\simeq11,000\,M$ to about $\mathcal{T}\sim30^\circ$ by $t\simeq30,000\,M$, so the global misalignment relative to the $+z$ axis steadily grows. At the same time, $\mathcal{P}$ evolves rapidly early on and then slows, indicating that the azimuthal reorientation becomes less abrupt after the initial adjustment stage. Comparing panels (b) and (c) suggests that the torus follows the same overall sense of motion as the primary spin, consistent with a response to the evolving torque geometry. The torus reacts more slowly than the spin, however, so it lags behind the most rapid components of the primary-spin precession.

Panel (d) combines the orbital and spin evolution in three dimensions. Between $t\simeq11,000$ and $18,000\,M$, the secondary does not remain confined to a single plane; instead, it repeatedly shifts between more vertical and more coplanar trajectories, in agreement with panel (a). Over the same interval, the primary spin sweeps around an inclined axis, while panel (c) records the concurrent growth of the torus tilt. Taken together, these trends indicate that the torus angular momentum is gradually torqued toward the evolving black-hole configuration, but with a delayed response that makes its precession lag behind the primary spin.
}


\chapter{Conclusions and Future Prospects}

\section{Summary of the Thesis}

This dissertation has navigated the complex intersection of general relativistic magnetohydrodynamics simulations, covariant radiative transfer (GRRT), and multi-wavelength observations to decode the electromagnetic signatures of black holes in extreme environments. Over the course of this research, the physical scope of our numerical models has systematically evolved from the fundamental algorithm development for isolated Kerr black holes to the intricate multi-wavelength comparison of Sgr A$^*$, and ultimately to the highly dynamic shock mechanisms inherent in supermassive binary black hole (SMBBH) systems. By bridging the gap between first-principles plasma dynamics and interferometric observables, this work provides a comprehensive framework for interpreting the horizon-scale physics of low-luminosity active galactic nuclei (LLAGNs).

\paragraph{Part I: Flares and Precession in Kerr Spacetime} In the first part of this thesis, I critically examined the role of magnetic topology in governing the accretion state and time-domain variability of Sgr A$^*$. By initializing GRMHD simulations with multiple, alternating-polarity poloidal loops (with wavelengths ranging from $\lambda_r = 6~r_g$ to $80~r_g$), I demonstrated that continuous magnetic dissipation prevents the rapid accumulation of horizon-penetrating flux, thereby sustaining a Standard and Normal Evolution (SANE) state. This configuration naturally triggers violent magnetic reconnection at the current sheets between loops. I tracked the formation of macroscopic flux ropes within these regions, which are characterized by high electron temperatures ($\Theta_e \gtrsim 10$) and strong local magnetization ($\sigma \gtrsim 0.5$). Utilizing a hybrid thermal and non-thermal kappa electron distribution function calibrated by Particle-in-Cell physics, our GRRT post-processing confirmed that these flux ropes are the physical progenitors of the near-infrared (NIR) flares observed by the GRAVITY instrument. Furthermore, I provided a self-consistent physical mechanism for the observed multi-frequency time lags. As the highly energized plasma expands and shifts outward, the evolving optical depth $\tau$ dictates that the $43$~GHz and $86$~GHz radio flares lag behind the $138$~THz NIR peak by up to $50$ minutes, perfectly matching observational constraints. I also uncovered a novel dynamical regime in misaligned systems: magnetically driven retrograde precession. I established that when a tilted disk achieves a Magnetically Arrested Disk (MAD) state, the immense magnetic torque generated by the horizon-threading poloidal field completely overrides the relativistic Lense-Thirring torque, forcing both the accretion disk and the jet to precess in the opposite direction of the black hole's spin.

\paragraph{Part II: Accretion and Shadows in Complex Spacetimes} In the second part of this thesis, I extended our rigorous numerical pipeline beyond stationary Kerr spacetimes. I first utilized the horizon-scale resolution of the Event Horizon Telescope (EHT) to constrain the nature of spacetime itself. By simulating accretion flows within a regular Loop Quantum Black Hole (LQBH) metric, I discovered that the loop quantum gravity (LQG) effect systematically enlarges the photon ring and alters the circular polarization pattern. Through direct comparison with EHT data, I placed stringent upper limits on the polymeric function, restricting $P \lesssim 0.2$ for Sgr A$^*$ and $P \lesssim 0.07$ for M~87$^*$. I further demonstrated that the LQG modifications effectively increase the horizon angular frequency, thereby amplifying the toroidal magnetic field and enhancing the Blandford-Znajek jet power.

In the final part of this thesis, I develop and apply a time-dependent superposed Kerr–Schild metric to carry out global 3D GRMHD coupled with GRRT simulations of a large mass ratio SMBBH ($q=0.1$) interacting with a primary BH with a magnetically arrested disk. Across three representative orbital configurations vertical impact, coplanar embedded, and an eccentric high-spin precessing case, the thermal synchrotron observables show that dynamical periodicity does not generically translate into periodic flares: recurrent shocks and bursts in $\dot{M}_{\rm BH2}$ are typically subdominant to the stochastic variability of the primary MAD at submillimeter frequencies, with clear signatures emerging only during brief line-of-sight alignments or rare episodes of exceptionally strong shock heating. The simulations instead point to robust multi-wavelength diagnostics. I find a frequency-dependent emission hierarchy in which the primary dominates the submillimeter while the hotter secondary can dominate the near-infrared, and I identify rapid, high-contrast flares produced by binary self-lensing in coplanar, edge-on geometries. In the high-spin configuration, spin–orbit coupling drives strong Lense–Thirring precession, imprinting twisted, wobbling jet morphologies reminiscent of those inferred in sources such as OJ~287. Together, these results motivate a shift toward time-domain, polarimetric, and multi-wavelength strategies for identifying embedded SMBBH candidates when direct imaging remains under-resolved.

\section{Statement of Originality and Core Contributions}

This dissertation delivers original contributions across plasma microphysics, magnetically dominated accretion dynamics, strong gravity phenomenology, and computational astrophysics. Building upon the foundational physics outlined in the summary, this work establishes the following core advancements.

First, I made a novel unified model for Sgr~A$^*$ flares and their multifrequency time delays. By explicitly linking macroscopic magnetic polarity inversion events to localized nonthermal particle acceleration, this framework provides a complete causal explanation for the correlated multiwavelength time delay for the flares without relying on artificial hotspot prescriptions. 

Second, I discovered the entirely new dynamical regime of retrograde precession in misaligned Magnetically Arrested Disks. This finding upends the standard assumption that Lense-Thirring frame dragging alone dictates the orientation and precession of relativistic jets. It provides a brand new theoretical basis for interpreting precessing jets in microquasars and active galactic nuclei.

Third, I established a rigorous simulation grounded methodology for testing alternative metric geometries via horizon scale observations. By modeling accretion within Loop Quantum Black Hole spacetimes, I identified concrete observational signatures in the photon ring morphology and jet power that allow observers to place active constraints on beyond Kerr metric deviations.

Fourth, I developed an observational prediction for identifying unresolved supermassive binary black holes by connecting global 3D GRMHD and GRRT predictions to multi-frequency diagnostics. In a MAD flow with $q=0.1$, I showed that strong binary-driven shocks can be largely hidden by stochastic accretion variability, while edge-on self-lensing alignments produce rapid, high-contrast flares that cleanly trace conjunctions. I also found a band-dependent emission hierarchy (primary dominates sub-mm, secondary can dominate near-IR), and demonstrated that spin–orbit/Lense–Thirring torques can drive precessing, twisted jets—together enabling targeted time-domain, imaging, and polarimetric searches for SMBBHs.

Finally, to make these astrophysical discoveries possible, I contributed major advancements to the underlying computational infrastructure. I developed and incorporated a time dependent superposed binary black hole metric directly into the $\tt{AthenaK}$ code. I also built a sophisticated interface between $\tt{AthenaK}$ and the radiative transfer code $\tt{BHOSS}$ to smoothly process complex adaptive mesh grids. Furthermore, I greatly optimized the $\tt{AthenaK}$ code by implementing exact metric vectorization. These computational breakthroughs significantly reduced the overhead of complex metric transformations and enabled the extremely high resolution and long duration simulations required to capture the full multiphysics of these extreme spacetimes.

\section{Future Outlook}

The methodologies and physical insights developed in this thesis lay the groundwork for an expansive research trajectory. The transition from isolated, stationary spacetimes to dynamic, multiscale binary environments marks only the beginning of a broader effort to construct a universal theoretical framework for the evolution and observation of low luminosity active galactic nuclei. 

To further elevate our physical model of Sgr~A$^*$ flares and directly confront the exquisite astrometric precision of GRAVITY observations, future work will move beyond idealized torus initial conditions. We plan to incorporate realistic mass feeding mechanisms by linking our horizon scale simulations to galaxy scale cosmological runs or by directly modeling the stellar winds from orbiting Wolf-Rayet stars in the Galactic Center, e.g.  \cite{2024ApJ...973..141G, 2023ApJ...946...26G, 2020ApJ...896L...6R, 2020MNRAS.492.3272R}. This multiscale approach will allow us to simulate the self consistent accumulation of magnetic flux and the natural generation of macroscopic flaring loops. Furthermore, this large scale feeding provides a rigorous pathway to study tilted disks. Rather than artificially initializing a misaligned torus, we will model the self consistent mass injection from asymmetric inflows. This will enable us to track the long term stability of misaligned accretion and observationally validate the magnetically driven retrograde precession mechanism under continuous external feeding.

During my upcoming postdoctoral fellowship at the University of Hong Kong, my immediate focus will be on leveraging the \texttt{AthenaK} architecture to execute global simulations of supermassive binary black holes across a vastly expanded parameter space. While current models rely on the radiatively inefficient approximation applicable to Sgr~A$^*$ and M~87$^*$, many candidate binaries like OJ~287 operate at much higher accretion rates. Future efforts will integrate radiative cooling and radiation pressure directly into the dynamic evolution of the GRMHD solver to capture geometrically thinner, radiation dominated circumbinary disks. In parallel, we will explore a wide variety of binary disk configurations to investigate orbital resonances strictly within General Relativity. Understanding how strong gravity resonances drive disk eccentricity, modulate mass accretion across the cavity, and trigger distinct electromagnetic bursts will be crucial for predicting unique multimessenger signatures.

Ultimately, the research inaugurated in this dissertation is fundamentally aligned with the vanguard of international astronomical mega projects. By continuing to refine our understanding of polarized radiative transport, complex dynamic spacetimes, and nonthermal electron thermodynamics, our forward modeling framework will directly inform target selection and data interpretation for the next generation Event Horizon Telescope\cite{Tiede2022grp}, the Imaging X-ray Polarimetry Explorer (IXPE) \cite{2022JATIS...8b6002W}, and the LISA mission \cite{2019BAAS...51g..77T}. As observational capabilities reach unprecedented levels of sensitivity and resolution, the numerical laboratories developed herein will be essential for decoding the most extreme gravitational environments in the universe.

\printbibliography[heading=bibintoc]


\begin{acknowledgements}
Throughout my academic journey since my undergraduate years at Sichuan University, I have benefited from the guidance and generosity of many people. Although I cannot name everyone here, I am sincerely grateful to all who offered their time and advice, including those who replied warmly to my emails even though we had never met. This thesis reflects not only my own efforts but also the support I have been fortunate to receive from many people.

First and foremost, I would like to express my deepest gratitude to my supervisor, Professor Yosuke Mizuno. In March 2021, shortly after he joined the Tsung-Dao Lee Institute and held his first workshop there, I reached out to him as an undergraduate with little background in the field. He replied with remarkable warmth and kindly supported my participation in the workshop. That first exchange became the starting point that eventually led me to join his group as his first student. After joining TDLI, I especially appreciate the academic freedom and trust he gave me, as well as his generous support that allowed me to attend conferences and grow into an independent researcher.

I am deeply grateful to Professor Dong Lai. His mentorship was instrumental in my transition from a code developer to a physicist. He provided direct and crucial support for my research, offering the physical insight needed to address several difficult problems central to this dissertation. I also thank him for broadening my scientific perspective beyond numerical implementation, and for his continued encouragement and support regarding my professional development.

I am also grateful for the encouragement and opportunities provided by several professors in the community. I would like to express my sincere gratitude to Professor Feng Yuan for his pivotal support during my internship in 2020, which provided my first systematic exposure to numerical simulations, and to Professor Xuewen Liu for his early guidance during my undergraduate studies at Sichuan University. I thank Professor Shude Mao at Westlake University, as well as Professors Bing Zhang and Jane Lixin Dai at the University of Hong Kong, for inviting me to present my work and for their thoughtful feedback during my visits. I also thank Professor Lile Wang at KIAA for invited talk and for several valuable discussions. I am grateful to Professor Xinyu Li for his interest in my work and for our ongoing collaboration. These exchanges helped sharpen my scientific perspective and have been an important source of encouragement as I prepare for the next stage of my career at HKIAA.

I thank Dr. Indu Kalpa Dihingia for teaching me the numerical details of our simulation codes and for spending so much time revising some of my early poorly written papers, which helped me learn how to write scientific papers and develop into an independent researcher. I also thank Dr. Ben Prather for developing and sharing the open-source GRMHD code \texttt{KHARMA} and for his warm and patient replies to my many technical questions, which helped me resolve issues efficiently and get to learn GPU-based GRMHD code and the Kokkos framework.

I would like to express my sincere gratitude to my collaborators for their essential contributions to the projects included in this thesis. I thank Professor Ziri Younsi, Professor Antonios Nathanail, and Professor Christian Fromm for their expertise and for the many productive discussions we shared. Their insights into general relativistic radiative transfer and jet physics significantly enriched my research. I also thank Professor Tao Zhu for the stimulating collaboration on the physics of loop quantum black holes, which allowed me to explore exciting new directions in strong-field gravity.

I also thank the members of our group for their support and companionship. I had many productive discussions and enjoyable collaborations with Mr. Xufan Hu, Cheng Liu, and Jingze Xia, as well as many informative conversations and happy times with Dr. Hongzhe Zhou and Yangyang Cai. I am also grateful to my friends Dr. Xi Lin and Dr. Yihuan Di for their discussions and friendship. I am grateful to Mr. Peng Huang, Haoda Li, Yihuan Li, Zhuokun Wang, Daqian Li, Hengjia Hu, Su Conghao, and Xiyuan Liu for their constant encouragement and for the emotional support and help they have given me.

I owe my most profound thanks to my parents Shengang Jiang and Manjin tang. I am deeply grateful for the absolute freedom and unconditional trust they have given me throughout my life. 

Finally, I would like to express my heart felt gratitude to my foreign grandma, Marjorie Atkinson. I am forever indebted to her for the unique perspective she shared and for the immense kindness she showed me throughout the years. Most importantly, I am profoundly grateful for her absolute trust and the financial support for my interest on astronomy she provided when I was in middle school.
I also owe my deepest thanks to Yishan Wang. We met when I was applying to graduate school, and she has accompanied me through almost the entire journey of my PhD. Her love, patience, encouragement, and companionship carried me through many moments of uncertainty, pressure, and self-doubt. I know I have not always been the person I should have been, but I will always be grateful for the kindness, strength, and warmth she gave me along the way.
\end{acknowledgements}

\end{document}